\documentclass[twocolumn]{aastex631}

\usepackage{microtype}
\usepackage{amsmath}
\usepackage{xcolor}

\newcommand\nuk[2]{$\rm ^{\rm #2} #1$}
\newcommand{\msun}{$\rm M_{\odot}$}

\shorttitle{Zero and extremely low metallicity rotating massive stars}
\shortauthors{Roberti et al.}

\graphicspath{{./}{arxiv_figures/}}

\begin{document}
\doublespace

\title{Zero and extremely low metallicity rotating massive stars: evolution, explosion, and nucleosynthesis up to the heaviest nuclei}
\author[0000-0003-0390-8770]{Lorenzo Roberti}
\affiliation{Konkoly Observatory, Research Centre for Astronomy and Earth Sciences, E\"otv\"os Lor\'and Research Network (ELKH), Konkoly Thege Mikl\'{o}s \'{u}t 15-17, H-1121 Budapest, Hungary}
\affiliation{CSFK, MTA Centre of Excellence, Budapest, Konkoly Thege Miklós út 15-17, H-1121, Hungary}
\affiliation{Istituto Nazionale di Astrofisica—Osservatorio Astronomico di Roma, Via Frascati 33, I-00040, Monteporzio Catone, Italy}

\author[0000-0002-3164-9131]{Marco Limongi}
\affiliation{Istituto Nazionale di Astrofisica—Osservatorio Astronomico di Roma, Via Frascati 33, I-00040, Monteporzio Catone, Italy}
\affiliation{Kavli Institute for the Physics and Mathematics of the Universe, Todai Institutes for Advanced Study, University of Tokyo, Kashiwa, 277-8583 (Kavli IPMU, WPI), Japan}
\affiliation{INFN. Sezione di Perugia, via A. Pascoli s/n, I-06125 Perugia, Italy}

\author[0000-0002-3164-9131]{Alessandro Chieffi}
\affiliation{Istituto Nazionale di Astrofisica—Istituto di Astrofisica e Planetologia Spaziali, Via Fosso del Cavaliere 100, I-00133, Roma, Italy}
\affiliation{Monash Centre for Astrophysics (MoCA), School of Mathematical Sciences, Monash University, Victoria 3800, Australia}
\affiliation{INFN. Sezione di Perugia, via A. Pascoli s/n, I-06125 Perugia, Italy}

% ========== ABSTRACT
\begin{abstract}

We present the evolution and the explosion of two massive stars, 15 and 25 \msun, spanning a wide range of initial rotation velocities (from 0 to 800 km/s) and three initial metallicities: Z=0 ([Fe/H]=$-\infty$), $3.236\times10^{-7}$ ([Fe/H]=--5), and $3.236\times10^{-6}$ ([Fe/H]=--4). A very large nuclear network of 524 nuclear species extending up to Bi has been adopted. Our main findings may be summarized as follows: a) rotating models above Z=0 are able to produce nuclei up to the neutron closure shell at N=50, and in a few cases up to N=82; b) rotation drastically inhibits the penetration of the He convective shell in the H rich mantle, phenomenon often found in zero metallicity non rotating massive stars; c) vice versa rotation favors the penetration of the O convective shell in the C rich layers with the consequence of altering significantly the yields of the products of the C, Ne, and O burning; d) none of the models that reach the critical velocity while in H burning, loses more the 1 \msun\ in this phase; e) conversely, almost all models able to reach their Hayashi track exceed the Eddington luminosity and lose dynamically almost all their H rich mantle.

These models suggest that rotating massive stars may have contributed significantly to the synthesis of the heavy nuclei in the first phase of enrichment of the interstellar medium, i.e., at early times.
   
\end{abstract}

% ================================= INTRODUCTION #2 =================================
\section{Introduction}

% observations
The most iron poor low mass stars (M $\sim$ 0.8 M$_{\odot}$) are probably {\it second generation stars}, i.e., stars formed out of gas clouds enriched in metals by the ejecta of the first supernova explosions during the very early epochs of the evolution of the Universe. In the last years large-scale surveys, often followed by high-resolution spectroscopy, allowed detailed measurements of the surface chemical composition of Fe-poor stars (\textit{Stellar Abundances for Galactic Archaeology} (SAGA) Database \cite{saga1}, \cite{saga2}, \cite{saga3}, \cite{saga4}; \textit{Joint Institute for Nuclear Astrophysics} Database (JINAbase) \cite{jina}; \cite{yong:12}, \cite{roederer:14}, \cite{hansen:15}, \cite{starkenburg:18}, \cite{frebel:19}). The scenario that emerges out of this large amount of data shows (among the other things) that (1) the Fe distribution function (MDF) of Fe-poor stars peaks around $\rm [Fe/H]\sim-2.7$; (2) the number of stars decreases sharply for $\rm [Fe/H]<-3$; (3) there is a small number of objects with $\rm [Fe/H]<-4$; (4) below [Fe/H]=--2 there is a progressive percentage increase of stars that show an "anomalous" high C abundance with respect to Fe that, in many cases, is also associated to large over-abundances of N and O; (5) the vast majority of the stars having $\rm [Fe/H]<-4$ has $\rm [C/Fe]\gg0$; (6) a number of very metal poor stars show an enhancement of $s$-process nuclei (identified by a $\rm [Ba/Eu]>1$), of $r$-process nuclei (i.e. $\rm [Ba/Eu]<0$) or both ($\rm 0<[Ba/Eu]<0.5$). In particular, Ba and Eu are widely used as representative proxies of the amount of matter synthesized by the $s$-process and the $r$-process nucleosynthesis, respectively, because their abundance can be measured relatively easily in the optical spectra of metal poor stars. In the JINAbase sample, 10$\%$ of EMP stars shows the signature of neutron capture elements, the percentage increasing to 19$\%$ if one considers the whole catalogue. It has been also possible to identify stars having $\rm 0<[La/Eu]<0.6$, that could represent the signature of an \textit{intermediate} process ($i$-process) between the $s$- and the $r$-process \citep[see, e.g.,][and references therein]{cowan:77,hampel:16}. 

% production site
While the main site where the bulk of the $r$-process nucleosynthesis occurs is still uncertain (although the $r$-process nucleosynthesis signature has been observed in the case of the kilonova associated to the detection of GW170817, see \citealt{pian:17}), there is a general consensus that the $s$-process nucleosynthesis mainly takes place in intermediate mass stars $\rm (1\leq M_{ZAMS}(M_{\odot})\leq 7)$. The main reason is that the efficiency of the $s$-process nucleosynthesis (for each fixed initial Fe abundance) depends mainly on the neutron exposure (which is simply the total number of neutrons available) that is too low in massive stars to appreciably produce the bulk of the $s$-process nuclei. Massive stars may, in fact, contribute only to the synthesis of nuclei up to the first neutron closure shell. The canonical scenario proposed to explain the observed abundance of Ba in several EMP stars foresees a binary system formed by stars born out of gas enriched only by massive stars \citep{gallino:98,suda:04,cristallo:09,straniero:14,choplin:21}. This binary system should be composed of a low mass and an intermediate mass stars. The AGB phase experienced by the intermediate mass star would be responsible for the $s$-process nucleosynthesis that would then be poured on the surface of the low mass companion through a wind or a common envelope phase. Let us remind that it is now widely accepted that the neutron flux responsible of the $s$-process nucleosynthesis in AGB stars has a primary origin as a consequence of the penetration of some protons below the base of the convective envelope at the end of the third dredge-up and the subsequent activation of the \nuk{C}{12}(p,$\gamma$) first and of the \nuk{C}{13}($\alpha$,n) later \citep[see][]{busso:99}. Given the low abundance of the seed nuclei, the neutron to seed ratio is high enough in low metallicity stars to push matter up to quite high atomic mass numbers \citep{goriely:01,travaglio:01a,cristallo:09,cristallo:11}.

% rotating stars
In the last 20 years another scenario, not necessarily alternative to that of the AGB, has been proposed. In fact, a number of papers have shown that rotating massive stars can considerably boost the primary nucleosynthesis of several nuclei, namely \nuk{N}{14}, \nuk{F}{19}, and also elements beyond the Fe peak.
\cite{meynet:06}, \cite{hirschi:07}, and \cite{ekstroem:08} were the first to show that the slow mixing of matter caused by rotation instabilities between the (central) He burning and the H burning shell (that we call entanglement between these two burning regions) in low metallicity stars leads to an important production of \nuk{N}{14} and they also suggested that such an entanglement could be responsible also for the synthesis of heavy nuclei. 
\cite{pignatari:08} explored the possibility that a large amount of trans-Fe nuclei may be synthesized as a consequence of the entanglement between the He and the H burning zones, by means of a post-processing that simulated the entanglement by injecting fresh \nuk{N}{14} in an He burning environment up to a level of roughly 1\% in mass fraction. This work showed that it is possible to produce a large amount of nuclei heavier than Zn in rotating stars but also that the very large neutron to seed ratio leads to the full destruction of the Fe seeds. Few years later, \cite{frischknecht:12} and \cite{frischknecht:16} presented an homogeneous set of stellar models in which for the first time it was studied the $s$-process nucleosynthesis by including a network with 737 nuclear species in the Geneva stellar evolution code. These computations extended up to the end of the central O burning and therefore missed the possible interplay between the O convective shell and the C rich layers, the passage of the supernova shock wave and the determination of the mass of the remnant, mass that determines which fraction of the mass of the star is ejected and which remains locked in the remnant. Note that, though it is certainly true that most of the $s$-process nucleosynthesis occurs in central He burning, the physical evolution of the CO core is crucial because the ashes of the central He burning are located within the CO core, i.e. behind the He convective shell, that therefore must be well traced.

% summary of the paper
In this paper we present a preliminary set of evolutionary sequences that follow the hydrostatic evolution of the stars from the pre-main sequence up to the collapse of the core and then the formation and passage of the shock wave of the supernova explosion through the structure. The main goal is that of exploring the neutron capture nucleosynthesis as a function of both the initial Fe abundance and the initial equatorial rotational velocity. In particular, we aim to determine the minimum metallicity and the corresponding initial rotation velocity that allows a large production of the typical $s$-process elements observed in EMP stars. This work is organized as follows: Section \ref{sec:network} is dedicated to the description of the setup of the nuclear network to calculate the neutron capture nucleosynthesis in rotating massive stars; Sections \ref{sec:stellar_models} and \ref{sec:expl} discuss the hydrostatic evolutionary properties of zero and very low metallicity rotating massive stars and the explosion, respectively; the $s$-process nucleosynthesis yields as a function of the initial velocity and metallicity are presented in Sections \ref{sec:yields} and \ref{sec:xsuo}; Section \ref{sec:comparison_observation} shows a comparison between our theoretical predictions and a sample of representative observations of $s$-process elements; Section \ref{sec:comp} shows a comparison with similar computations available in the literature; eventually Section \ref{sec:dico} summarizes our results.

% ================================= NETWORK & CODE =================================
\section{FRANEC code and nuclear network} \label{sec:network}

The hydrostatic evolution of all the models discussed in this work has been computed by means of the version of the \verb|FRANEC| (Frascati Raphson Newton Evolutionary Code) presented in detail in \cite{CL13} and \cite{LC18} (hereafter CL13 and LC18). A summary of the main physical and numerical assumptions adopted in the code is presented in the Appendix \ref{app:franec}. The major effort in the present set up of the code was the choice of the nuclear network together to the database of nuclear cross sections. In this section we discuss at some extent the main choices we made. Let us note that the update of the nuclear cross sections was frozen in October 2019.   

\subsection{Strong interactions}
In general, we always preferred the experimental nuclear cross sections with respect to the theoretical ones. The nuclear cross section of each process was taken from KADoNiS v0.3 \citep{dillmann:06} for the neutron captures and from STARLIB \citep{starlib} in all the other cases. If a nuclear cross section was not present in any of the two databases, it was searched for in the NACRE \citep{angulo:99} or NACRE II \citep{nacreii} compilations first and eventually in the Jina Reaclib \citep{reaclib} database. The only exceptions to this procedure were the nuclear cross sections of the reactions \nuk{O}{17}($\alpha,\gamma$)\nuk{Ne}{21} and \nuk{O}{17}($\alpha$,n)\nuk{Ne}{20}, taken from \cite{Best11} and \cite{Best13}. Particular attention was paid to the computation of the nuclear cross sections of the reverse processes. As the temperature increases above roughly 2 Gk, each process $a(i,j)b$ tends to be counterbalanced by the reverse process $b(j,i)a$ so that the net rate of each pair of processes as well as its contribution to the total nuclear energy generation rate depends on the very small difference between the forward and the reverse process. For this reason we decided not to tabulate the nuclear cross sections of the reverse processes but to compute them at runtime for the exact value of the forward process. The advantage of this technique is that in this way the nuclear cross section of the reverse process is exactly linked to the value of the forward process, providing a more stable and precise computation of the nuclear energy generation rate. We considered as forward process of each pair the reaction having a positive Q-value.

\subsection{Weak interactions}
The databases adopted for these processes include \cite{fuller:82}, \cite{oda:94}, \cite{pruet:03}, \cite{langanke:00}, \cite{suzuki:16}, and \cite{li:16}. In this case the leading databases were those of \cite{langanke:00} and \cite{fuller:82} because they are the ones that cover the widest ranges both in temperature and density. If a process is not present neither in the first nor in the second database, the most recent value among \cite{suzuki:16}, \cite{oda:94}, and \cite{pruet:03} is taken. If the reaction is not present in any of the above mentioned databases its decay rate is assumed to be the terrestrial decay. The only exception to this procedure was the decay rate of \nuk{Fe}{59} into \nuk{Co}{59} because in this case we decided to adopt the decay rate provided by \cite{li:16}. We furthermore checked that all the decay rates progressively approach the terrestrial value at low temperature and low density. 

\begin{deluxetable}{lccc|lccc}[t!]
\tabletypesize{\scriptsize}
\tablecaption{The nuclear network. $\rm A_{min}$ and $\rm A_{max}$ are the minimum and maximum atomic weight of each element. Elements between Dy (Z=66) to Au (Z=79)  are assumed to be at local equilibrium under neutron captures.\label{tab:nuclearnetwork}}
\tablehead{
\colhead{Element} & \colhead{Z} & \colhead{$\rm A_{min}$} & \colhead{$\rm A_{max}$} & \colhead{Element} & \colhead{Z} & \colhead{$\rm A_{min}$} & \colhead{$\rm A_{max}$}
}

\startdata 
  n  & 0  & 1   & 1  & Br & 35 & 79  & 85 \\
  H  & 1  & 1   & 3  & Kr & 36 & 78  & 89 \\
  He & 2  & 3   & 4  & Rb & 37 & 84  & 89 \\
  Li & 3  & 6   & 7  & Sr & 38 & 84  & 93 \\
  Be & 4  & 7   & 10 & Y  & 39 & 89  & 95 \\
  B  & 5  & 10  & 11 & Zr & 40 & 90  & 97 \\
  C  & 6  & 12  & 14 & Nb & 41 & 92  & 98 \\
  N  & 7  & 13  & 16 & Mo & 42 & 92  & 103\\
  O  & 8  & 14  & 19 & Tc & 43 & 97  & 103\\
  F  & 9  & 17  & 20 & Ru & 44 & 96  & 107\\
  Ne & 10 & 19  & 23 & Rh & 45 & 102 & 108\\
  Na & 11 & 21  & 24 & Pd & 46 & 102 & 114\\
  Mg & 12 & 23  & 27 & Ag & 47 & 107 & 114\\
  Al & 13 & 25  & 28 & Cd & 48 & 108 & 119\\
  Si & 14 & 27  & 32 & In & 49 & 112 & 119\\
  P  & 15 & 29  & 34 & Sn & 50 & 112 & 128\\
  S  & 16 & 31  & 37 & Sb & 51 & 121 & 131\\
  Cl & 17 & 33  & 38 & Te & 52 & 122 & 133\\
  Ar & 18 & 35  & 41 & I  & 53 & 126 & 135\\
  K  & 19 & 37  & 44 & Xe & 54 & 126 & 137\\
  Ca & 20 & 39  & 49 & Cs & 55 & 133 & 139\\
  Sc & 21 & 41  & 49 & Ba & 56 & 134 & 141\\
  Ti & 22 & 44  & 51 & La & 57 & 138 & 143\\
  V  & 23 & 45  & 52 & Ce & 58 & 138 & 145\\
  Cr & 24 & 47  & 55 & Pr & 59 & 141 & 146\\
  Mn & 25 & 49  & 57 & Nd & 60 & 142 & 149\\
  Fe & 26 & 51  & 61 & Pm & 61 & 147 & 149\\
  Co & 27 & 53  & 62 & Sm & 62 & 147 & 155\\
  Ni & 28 & 55  & 65 & Eu & 63 & 151 & 155\\
  Cu & 29 & 57  & 66 & Gd & 64 & 152 & 159\\
  Zn & 30 & 60  & 71 & Tb & 65 & 159 & 159\\
  Ga & 31 & 62  & 72 & Hg & 80 & 202 & 205\\
  Ge & 32 & 64  & 77 & Te & 81 & 203 & 206\\
  As & 33 & 74  & 77 & Pb & 82 & 204 & 209\\
  Se & 34 & 74  & 83 & Bi & 83 & 208 & 209\\
\enddata   

\end{deluxetable}

\subsection{Nuclear network}

The nuclear network is an extension of the one adopted in LC18. The main upgrades are, 1) a wider extension towards larger atomic masses in order to properly manage higher neutron densities and 2) a larger number of elements (in particular we added explicitly all elements between Tc and I). 

The extension (in atomic mass) of the nuclear network is mainly dictated by the neutron density because it is this parameter (together to the temperature and density) that mainly controls how far matter can move out of the stability valley, because of the neutron captures, before decaying back towards its closest stable daughter. The parameter that controls the balance between beta decay and neutron capture may be written as the ratio $B$ between decay rate and neutron capture rate:

\begin{equation}
B= \frac{\lambda_{\rm i}(T,\rho)}{n_{\rm n} \left \langle \sigma v\right \rangle_{\rm in}(T)}
\label{eq:condition1}
\end{equation}

where $n_n$ is the neutron density, $\lambda_i$ the probability of decay per time unit of the nucleus $i$ and $\left \langle \sigma v\right \rangle_{\rm in}$ the nuclear cross section of the neutron capture process on the nuclear specie $i$. This ratio mainly depends on the neutron density $n_{\rm n}$ and mildly on the temperature $T$ and the density $\rho$. In massive stars, the $s$-process nucleosynthesis predominantly occurs in core and shell He burning and in shell C burning (see, e.g., LC18), and the typical physical conditions in these phases are $T\sim2.5\times10^8\ K$ and $\rho\sim10^3\ g/cm^3$ and $T\sim10^9\ K$ and $\rho\sim10^5\ g/cm^3$, respectively. The maximum neutron densities that are obtained in these conditions (in standard non rotating models) are $n_{\rm n}\sim10^7\ n/cm^3$, in core and shell He burning, and $n_{\rm n}\sim10^{11}\ n/cm^3$ in shell C burning. In order to take into account an even larger neutron production, in this work we considered a limiting neutron density up to $n_{\rm n}\sim10^{8}\ n/cm^3$ for He burning and $n_{\rm n}\sim10^{14}\ n/cm^3$ for C burning.

We therefore determined the maximum atomic mass of each element by requiring that $B\geq$100, i.e. that the probability of decay were at least 100 times larger than the neutron capture:

\begin{equation}
\lambda_{\rm i}(T,\rho)\geq{n_{\rm n} \left \langle \sigma v\right \rangle_{\rm in}(T)} \times 100
\label{eq:condition2}
\end{equation}

We applied this procedure to all the elements with $31\leq Z \leq 65$ and $80\leq Z \leq 83$. The nuclear network that comes out from this procedure is reported in Table \ref{sec:network}: it includes 524 isotopes and more than 3000 reactions fully coupled to the physical evolution of the star. It is important to note that with respect to LC18, we have now only one cluster at local equilibrium corresponding to the elements with $65<Z<80$.

\begin{deluxetable}{cL}
\tablecaption{Initial chemical composition in mass fraction of zero metallicity models. See text.\label{tab:BBN}}
\tablehead{
\colhead{Element} & \colhead{Mass fraction X}
}

\startdata 
$^{ }$H  & 0.7550                 \\
$^{2}$H  & $3.8507\times10^{-5}$  \\
$^{3}$He & $2.4916\times10^{-5}$  \\
$^{4}$He & 0.2449                 \\
$^{7}$Li & $8.4564\times10^{-10}$ 
\enddata   

\end{deluxetable}

\begin{deluxetable}{ccccc}
\tablewidth{0pt}
\tablecaption{The metallicity explored in this work. Z represents the absolute metallicity, i.e., the sum of all the abundances in mass fraction of the elements heavier than He. The abundance of $\rm ^4 He$ is supposed to remain constant through the different metallicity.\label{tab:metallicity}}
\tablehead{
\colhead{Set name} & \colhead{[Fe/H]} & \colhead{Z} & \colhead{$\rm^4He$} & \colhead{number of models}
}

\startdata 
Z & $-\infty$ & 0                   & 0.2449 & 14 \\
%G & --6       & $3.236\times10^{-8}$& 0.2449 &  1 \\
F & --5       & $3.236\times10^{-7}$& 0.2449 & 11 \\
E & --4       & $3.236\times10^{-6}$& 0.2449 &  7 \\
\enddata      

\end{deluxetable}

\subsection{Initial chemical composition}

The initial chemical composition, i.e., the composition of the gas cloud from which the star formed out, adopted in the case of zero metallicity models is not based on the theoretical computation of the Big Bang Nucleosynthesis (BBN) but, on the contrary, on the abundances observed in extremely metal poor environment. In particular, we follow the work of \cite{fields:2020}, in which they select the most reliable environments in which the primordial abundances may still be visible. In particular, the abundance of $\rm ^4 He$ is derived from observations of extragalactic HII emission lines; the abundance of deuterium is estimated through the study of quasar absorption systems; the reference abundance of $\rm ^7 Li$ is taken from the analysis of stellar spectra of low metallicity stars. For $\rm ^3 He$, there are not enough reliable observations, therefore for this isotope we adopt the value provided by \cite{coc:13}, which is derived from the local observations of our Galaxy. The abundances of all the nuclear species having $\rm Z\geq 4$ are put to zero. The initial abundances in mass fraction adopted in this work for the zero metallicity models are shown in \tablename~\ref{tab:BBN}.

For the higher metallicity models we adopt a scaled solar chemical composition with the exception of C, O, Mg, Si, S, Ar, Ca, and Ti, for which we adopt an enhancement with respect to the solar value derived from the observations of low metallicity stars, i.e., [C/Fe]=0.18, [O/Fe]=0.47, [Mg/Fe]=0.27, [Si/Fe]=0.37, [S/Fe]=0.35, [Ar/Fe]=0.35, [Ca/Fe]=0.33, [Ti/Fe]=0.23 \citep[see, e.g.,][]{cayrel:04,spite:05}.
The adopted solar chemical composition is the one provided by \cite{asplund:09}, that corresponds to a total metallicity equal to $\rm Z=1.345\times10^{-2}$. Therefore, as a result of the enhancements mentioned above, the metal fraction corresponding to [Fe/H]=$-N$ is $\rm Z=3.236\times 10^{-(N+2)}$. We also assume that the initial abundance of $\rm ^4 He$ does not significantly vary from the adopted BBN value in the range of the explored metallicity and therefore it is kept constant. \tablename~\ref{tab:metallicity} summarizes the metallicity range explored in this work.

% ================================= MODELS  =================================

\section{Stellar models} \label{sec:stellar_models}

We explored the evolution of two massive stars (15 and 25 \msun) spanning a wide range of initial equatorial surface velocities, from 0 to 800 km/s, that correspond to a range in $\Omega/\Omega_{crit}$ that extends up to $\simeq1$. Three extremely low metallicity were considered: [Fe/H]=$-\infty$, --5, --4.

A summary of the main properties of the computed models is reported in \tablename~\ref{tab:setzgfe}. We use the following convention to refer to the models: the first part of the name is the 3-digit-mass, the second one is a letter which marks the metallicity of the set, the three last digits represent the initial rotation velocity in km/s. The zero metallicity set is labelled as "Z", while the letters "F" and "E" correspond respectively to [Fe/H] = --5 and --4: 025z450, for example, indicates a zero metallicity 25 \msun\ star having an initial equatorial rotation velocity equal to 450 km/s. Each quantity is evaluated at the end of every burning phases. The label "MS" marks the beginning of the main sequence phase. The various columns refer to: (1) the burning phase, (2) the duration of the phase in years, (3) the maximum size of the convective core during the burning phase in solar masses, (4) the logarithm of the effective temperature in Kelvin, (5) the logarithm of the luminosity in solar units, (6) the total mass in solar masses, (7) the size of the He core in solar masses, (8) the size of the CO core in solar masses, (9) the surface equatorial rotation velocity in km/s, (10) the surface angular velocity in $\rm s^{-1}$, (11) the ratio of surface angular velocity to the critical one, (12) the total angular momentum J in unit of $10^{53}~$g cm/s, (13-14-15) the surface mass fractions of H, \nuk{He}{4} and \nuk{N}{14}, respectively. We discuss the evolution of the non rotating models (up to the onset of the core collapse) first, and then the role played by rotation.

%%%%%%%%%%%%%%%%%%%%%%%%%%%%%%%%%%%%%%%%%%%%%%
\subsection{The non rotating models}

\subsubsection{The H burning}
The lack of nuclear species with $\rm Z\geq6$ seriously affects the evolution of a zero metallicity star both because the very low opacity forces these stars to spend their life as compact blue objects and also because the lack of the energy provided by an initial abundance of CNO nuclei forces these stars to contract and raise their central temperature much more than their more metal rich counterparts. Figure \ref{fig:hr} shows the HR diagram of all models presented in this paper. The plots on the left and those on the right refer to the 15 \msun\ and the 25 \msun, respectively. The two zero metallicity models are represented as black lines in the two upper panels. 

The PP chain is able to produce enough energy to replace the one lost from the surface of the star just at the beginning of the central H burning phase. Right after, the continuous increase of the mean molecular weight due to the conversion of protons in $\alpha$ particles leads necessarily to a progressive increase of the temperature. The low dependence of the energy generation rate of the PP chain on the temperature forces these stars to contract and heat more effectively than their more metal rich counterparts, reaching very early a temperature high enough to partially activate the $3\alpha$ nuclear reactions and therefore the synthesis of some \nuk{C}{12}. The full CNO cycle activates immediately and progressively replaces the PP chain as the main energy producer: in both masses the CNO cycle produces 50\% of the nuclear energy when the central H burning is still 0.5 by mass fraction. At $\rm H_c\sim 0.2$ the CNO cycle produces more than 80\% of the energy required to sustain the 15 \msun\ and more than 90\% in the 25 \msun. As the central temperature increases, also the amount of CNO nuclei produced by the 3$\alpha$ increases, so that towards the end of the central H burning phase ($\rm H_c\simeq$0.05) it reaches a mass fraction of $\rm 4\times10^{-10}$ (15 \msun) and $\rm 6\times10^{-10}$ (25 \msun). During the very latest phases of the central H burning, the coupling $3\alpha$ + CNO cycle raises the abundance of \nuk{N}{14} from a few times $10^{-10}$ to roughly $10^{-6}$ in both masses. Once the H is exhausted in the convective core, the H burning shifts in a shell where the $3\alpha$ and the CNO cycle still work together, so that the level of \nuk{N}{14} within the whole H exhausted core settles to a value of the order of $10^{-6}$ by mass fraction (again in both masses). Note that during the central H burning a convective shell forms well outside the convective core in the 25 \msun\ (the Kippenhahn diagrams of the zero metallicity stars are shown in the two upper panels of Figure \ref{fig:kipnorot} while those of the other two metallicity are shown in the second and third row). 

Turning from Z=0 (or, equivalently, [Fe/H]=$-\infty$, Set Z)  to [Fe/H]=--5 (Set F), the initial global abundance of the CNO nuclei ($\rm X_{\rm CNO}=3.24\times10^{-7}$) is already high enough to fully sustain the star without the need of additional production of C in H burning. However, also at this metallicity some C is produced towards the very end of the H burning so that once again the mass fraction of the \nuk{N}{14} within the whole H exhausted core reaches a value of the order of $10^ {-6}$ by mass fraction in both masses.
The initial abundance of the CNO nuclei in Set E ([Fe/H]=--4) is higher than $10^{-6}$ so that the $3\alpha$ processes at this metallicity do not even activate towards the end of the central H burning.

In general, the size of the convective core (that practically corresponds to the initial size of the He core) scales inversely with the initial metallicity because the lower the CNO abundance the larger the size of the convective core (the higher central temperature implies higher energy fluxes). But when the metallicity drops to zero, the PP chain overcomes the CNO cycle, and the lower dependence of the energy flux on the temperature inverts the trend leading to convective cores (and hence He core masses) smaller than those proper of more metal rich stars. This explains why the He core mass increases from 3.6 \msun\ (Set Z) to 3.9 \msun\ (both Set F and E) in the 15 \msun\ and from 7.2 \msun\ (Set Z) to 8.1 \msun\ (both Set F and E) in the 25 \msun.

The last thing worth noting is that both stars practically evolve at constant mass in central H burning because mass loss (that scales directly with Z) is negligible at these extreme metallicities.

\subsubsection{The He burning}\label{sec:nrot:he}

The natural consequence of the partial activation of the $3\alpha$ already in MS, reinforced by the extremely low opacity of the H rich mantle caused by the lack or extremely low abundance of metals, is that all models (Set Z, F, and E) ignite and burn He on the blue side of the HR diagram (black lines in the three rows of Figure \ref{fig:hr}). It is worth noting that the H convective shell that forms in H burning in the zero metallicity 25 \msun\ remains active also through the whole central He burning (upper right panel in Figure \ref{fig:kipnorot}). A modest H convective shell that persists beyond the central He exhaustion (second and third row in the right panel of Figure \ref{fig:kipnorot}) develops also in the 25 \msun\ of Set F and E.

At variance with the H burning phase, the physical evolution of the He core in He burning (and beyond) does not depend any more directly on the metallicity because both the nuclear energy generation and the opacity are controlled only by the He abundance and the products of its nucleosynthesis (all primary elements). The initial metallicity influences the evolution of the He core only indirectly through the increase of the mass size of the He core caused by the advancing in mass of the H burning shell and/or its decrease if the H rich mantle is removed as a consequence of mass loss or binary interactions. In both masses the increase of the He core mass in He burning scales monotonically with the metallicity. In the 15 \msun\ case the increase amounts to $\sim$ 0.6 \msun\ in Set Z and to $\sim$ 0.9 \msun\ and $\sim$ 1.2 \msun\ in Set F and E, respectively. In the 25 \msun, the increase amounts to $\sim$ 0.7 \msun\ in Set Z and to $\sim$ 1.1 \msun\ and $\sim$ 1.6 \msun\ in Set F and E. 

The central He burning is powered essentially by the two well known processes, 3$\alpha$ and \nuk{C}{12}($\alpha$,$\gamma$)\nuk{O}{16}, and is characterized by a convective core that progressively advances in mass. The competition between these two processes contributes, together to the behavior of the convective core, to determine the abundance of \nuk{C}{12} left by the He burning. Such a value inversely scales with the final size of the He core and therefore inversely with the metallicity. 
In the present set of models the values obtained in Set Z are $\rm X_C=0.36$ (15 \msun) and $\rm X_C=0.30$ (25 \msun) and reduce down to 0.32 (15 \msun) and 0.26 (25 \msun) at [Fe/H]=--4.

The \nuk{N}{14} left by the H burning is fully converted in \nuk{Ne}{22}, part of which is in turn converted in \nuk{Mg}{25} and \nuk{Mg}{26} towards the end of the He burning, through the two competing exit channels (n and $\gamma$) of the $\alpha$ capture. Neutrons produced by the n channel may activate (in principle) the $s$-process nucleosynthesis, but the extremely low abundances of \nuk{N}{14} and \nuk{C}{13}, coupled to the lack of seed nuclei, completely inhibits the possible formation of nuclei above Zn. In both models of Set Z, the heaviest element whose abundance reaches at least $10^{-4}$ by mass fraction at the end of the central He burning is \nuk{Mg}{24}. Beyond this nucleus the abundances of all others nuclei drops quickly below $10^{-10}$ by mass fraction (see next section).

Once He is exhausted in the center, the He burning shifts in a shell where convection rapidly develops. In stars of lower metallicity ([Fe/H]$<$--4 ), the outer border of the He convective shell may penetrate the base of the H rich mantle soon after the central C exhaustion bringing protons inside, at very high temperatures, where they ignite violently. The abrupt release of nuclear energy forces the outer border of the He convective shell to further and quickly move outward in mass, eventually engulfing most of the H rich envelope. In the present case such a peculiar behavior of the He convective shell occurs in the 25 \msun\ of both Set Z and F. Figure \ref{fig:kipnorot} shows clearly the penetration of the He convective shell in the H rich mantle and the formation of a huge convective shell that covers most of the mantle. It goes without saying that the penetration of the He convective shell into the H rich mantle produces a burst of production of \nuk{N}{14} (and related H burning products) because it brings fresh protons in a C rich region where the temperature is of the order of 200 MK or even more.  

\subsubsection{The advanced burning}
\begin{figure*}[p]
\epsscale{1.0}
\plottwo{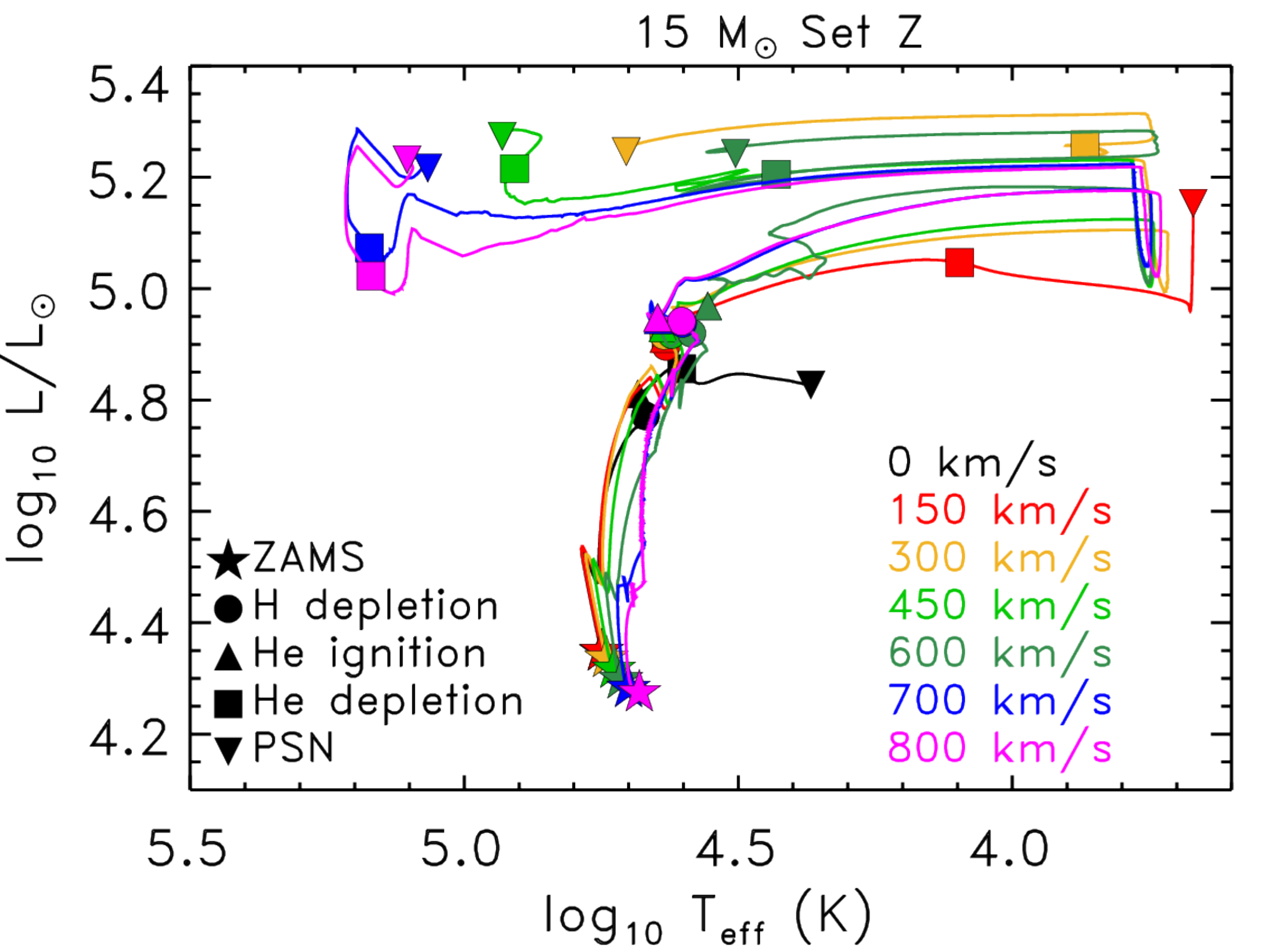}{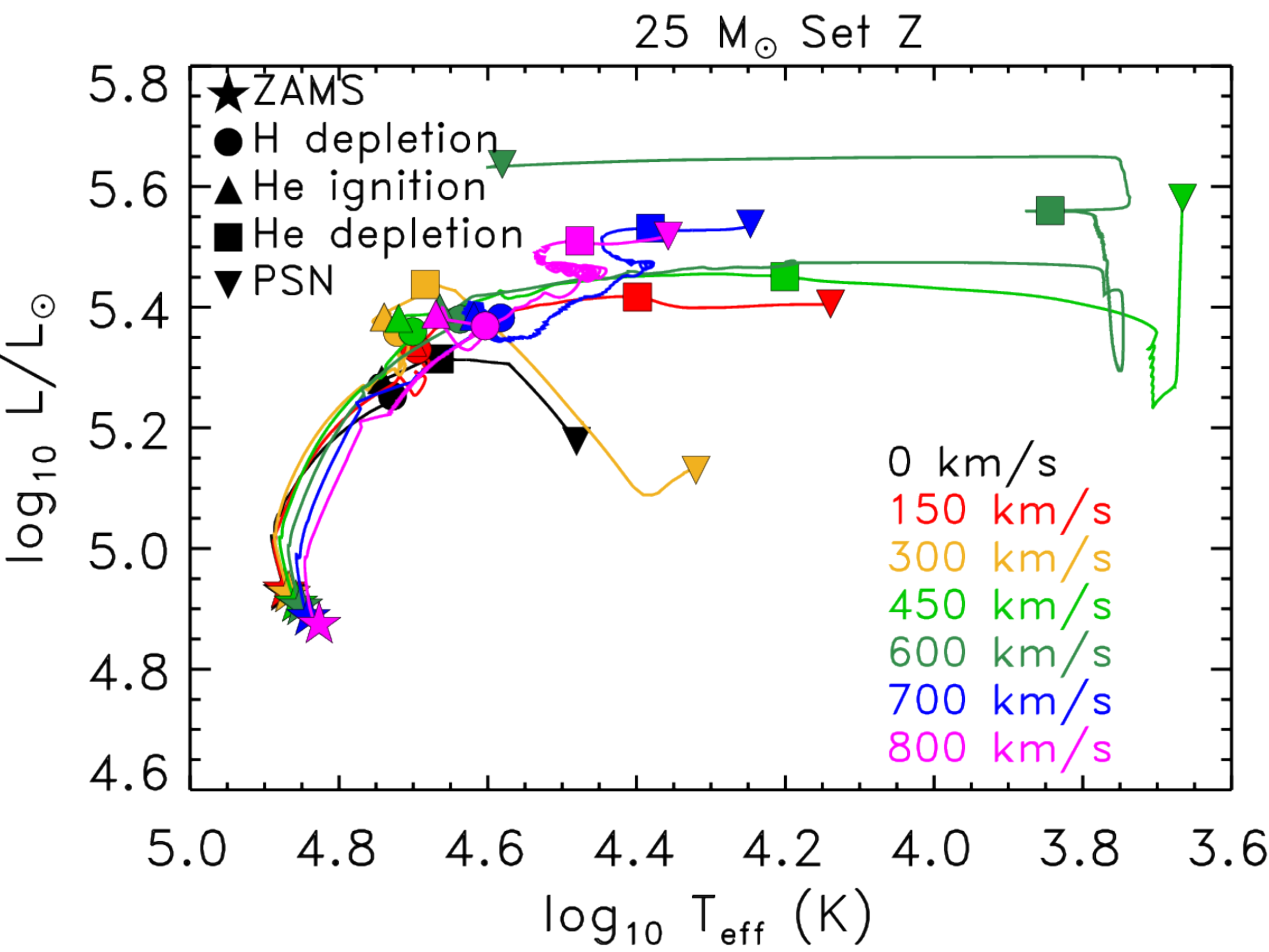}
\plottwo{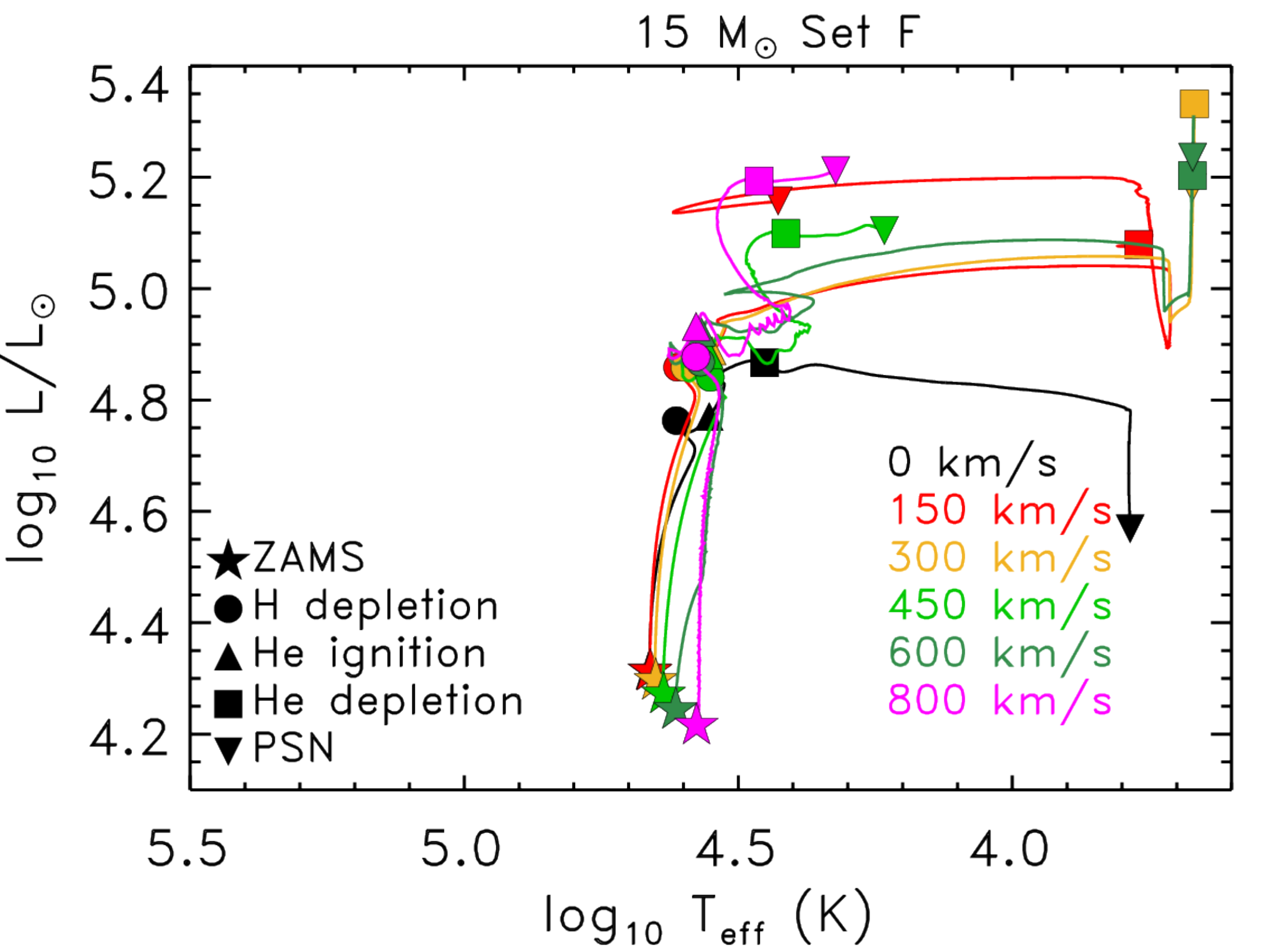}{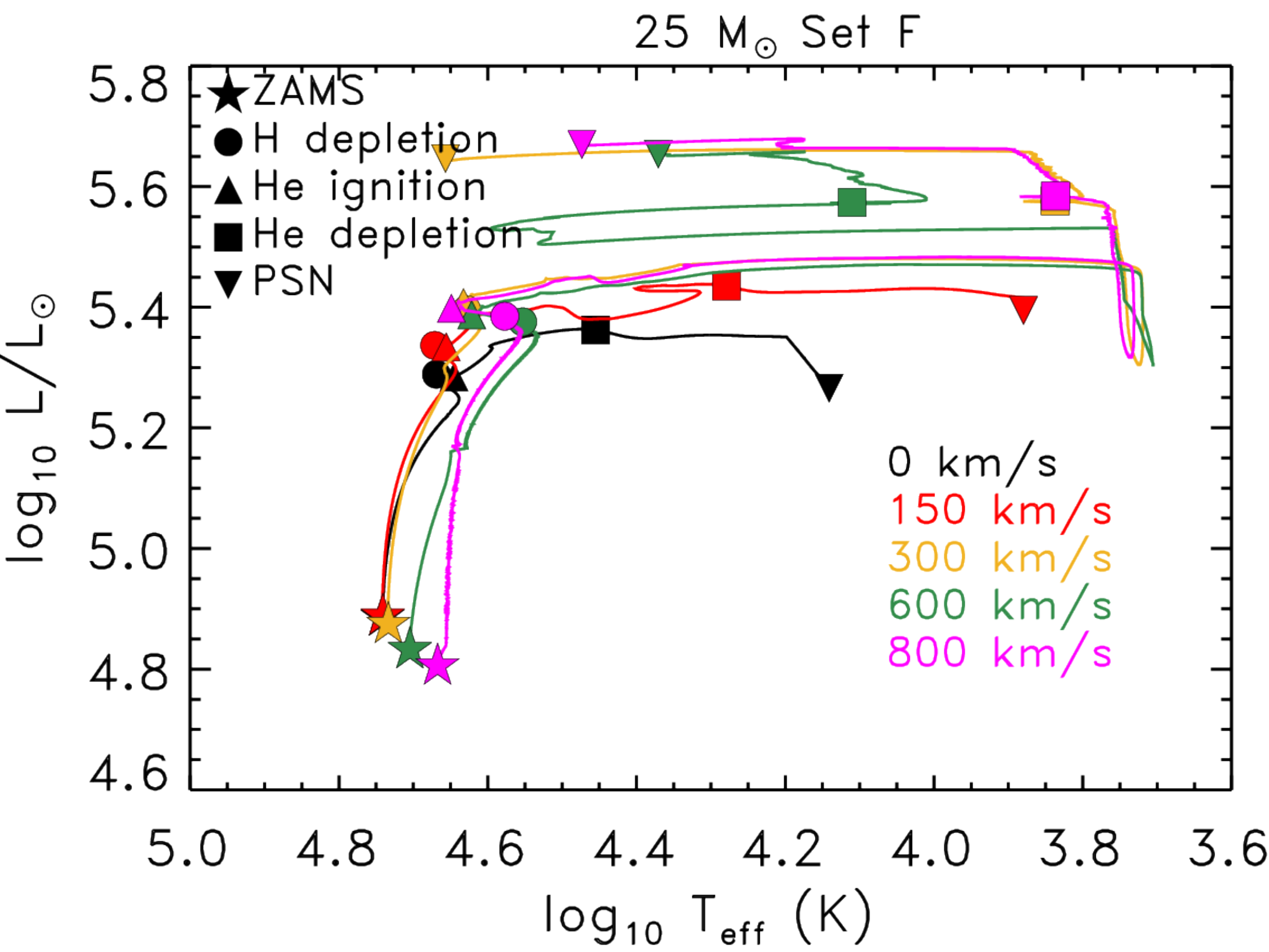}
\plottwo{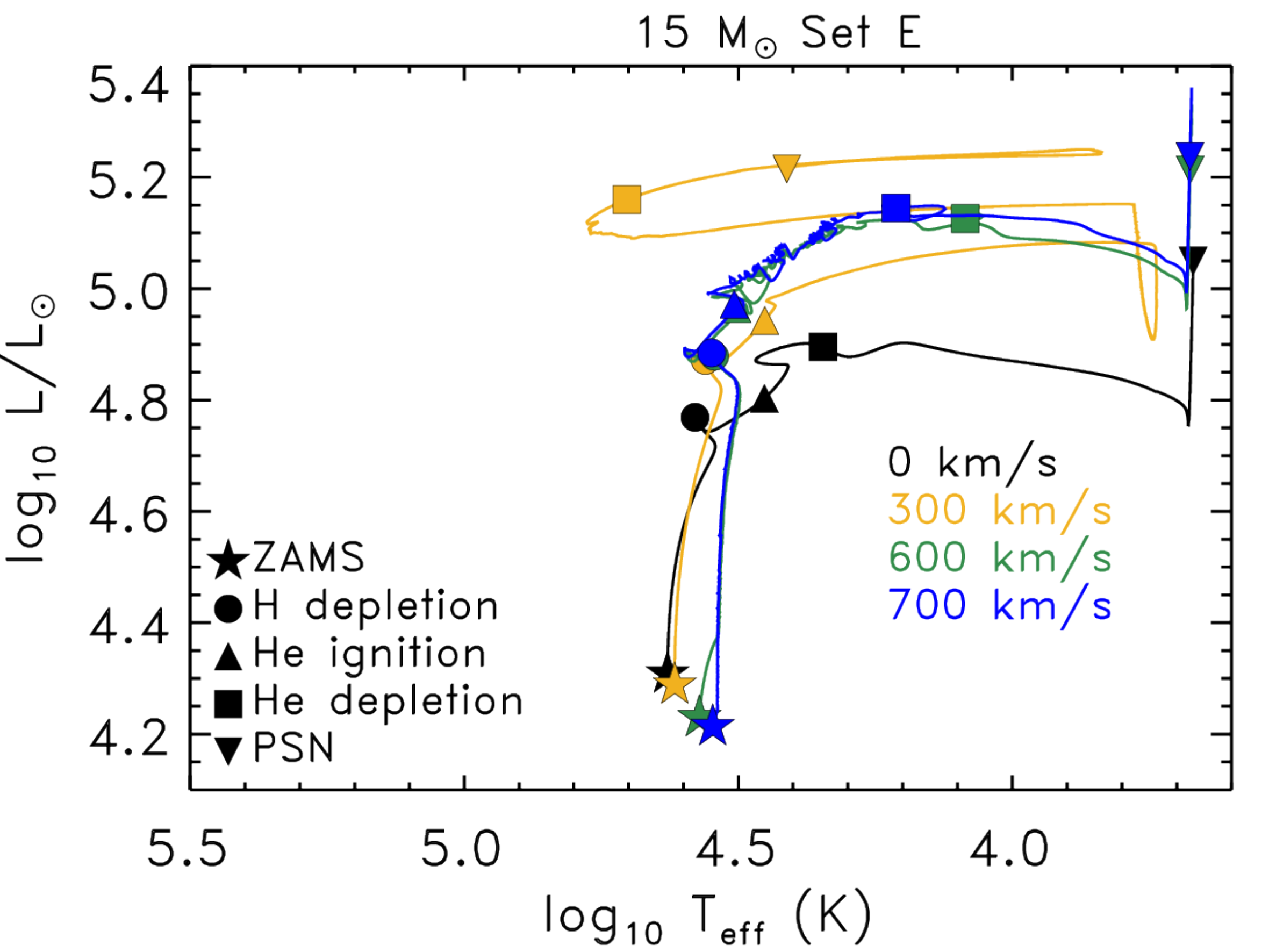}{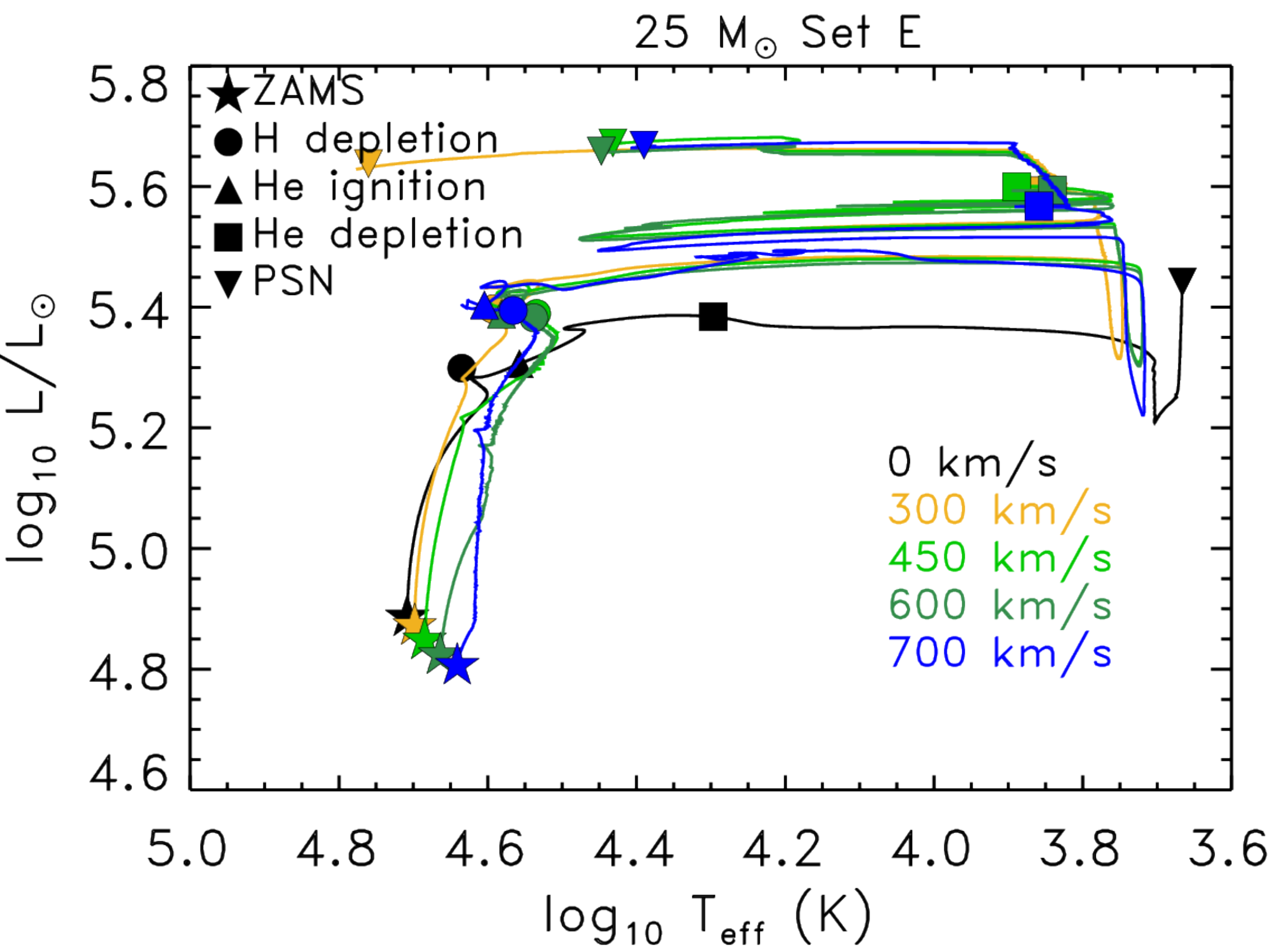}
\caption{Evolution on the Hertzsprung-Russell (HR) diagram of 15 (left panels) and 25 \msun (right panels) rotating massive stars. The different colors indicate the various initial equatorial velocity. The filled symbols mark different evolutionary stages.
\label{fig:hr}}
\end{figure*}

As in the case of core He burning, also the physical evolution of the CO core of a massive star does not depend on the initial metallicity. Its further evolution depends on both the CO core mass size and the \nuk{C}{12} abundance left by the previous burning stage \citep{chieffi:20,chieffi:21}. The CO core mass plays the same role the initial mass has in H burning, while the abundance of \nuk{C}{12} determines the amount of fuel available to the C burning and therefore controls the presence/size of the C convective core as well as the number and extensions of the C convective shell. As far as the behavior of the H rich mantle is concerned, the present set of models satisfy the general rule that the lower the initial metallicity the more difficult is for the star expand the H rich envelope up to the Hayashi track. In fact both zero metallicity stars remain at very high temperature all along the advanced burning phases, the models of Set F move somewhat towards the red without really reaching the Red Giant Branch while the initial metallicity of models of Set E is high enough to allow both models to reach the RGB soon after the central He exhaustion (see Figure \ref{fig:hr}, third row).

\begin{figure*}[p]
\epsscale{1.0}
\plottwo{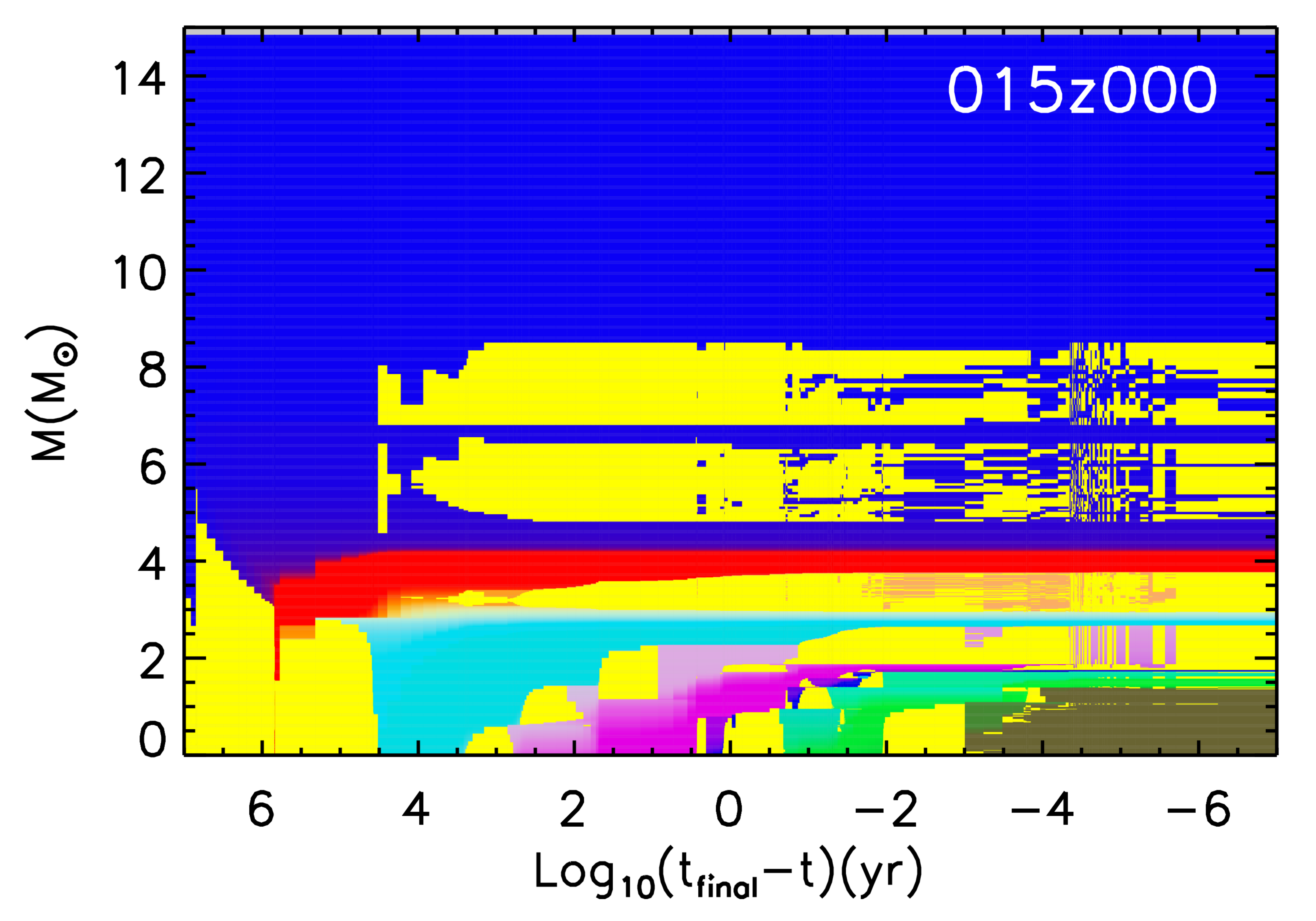}{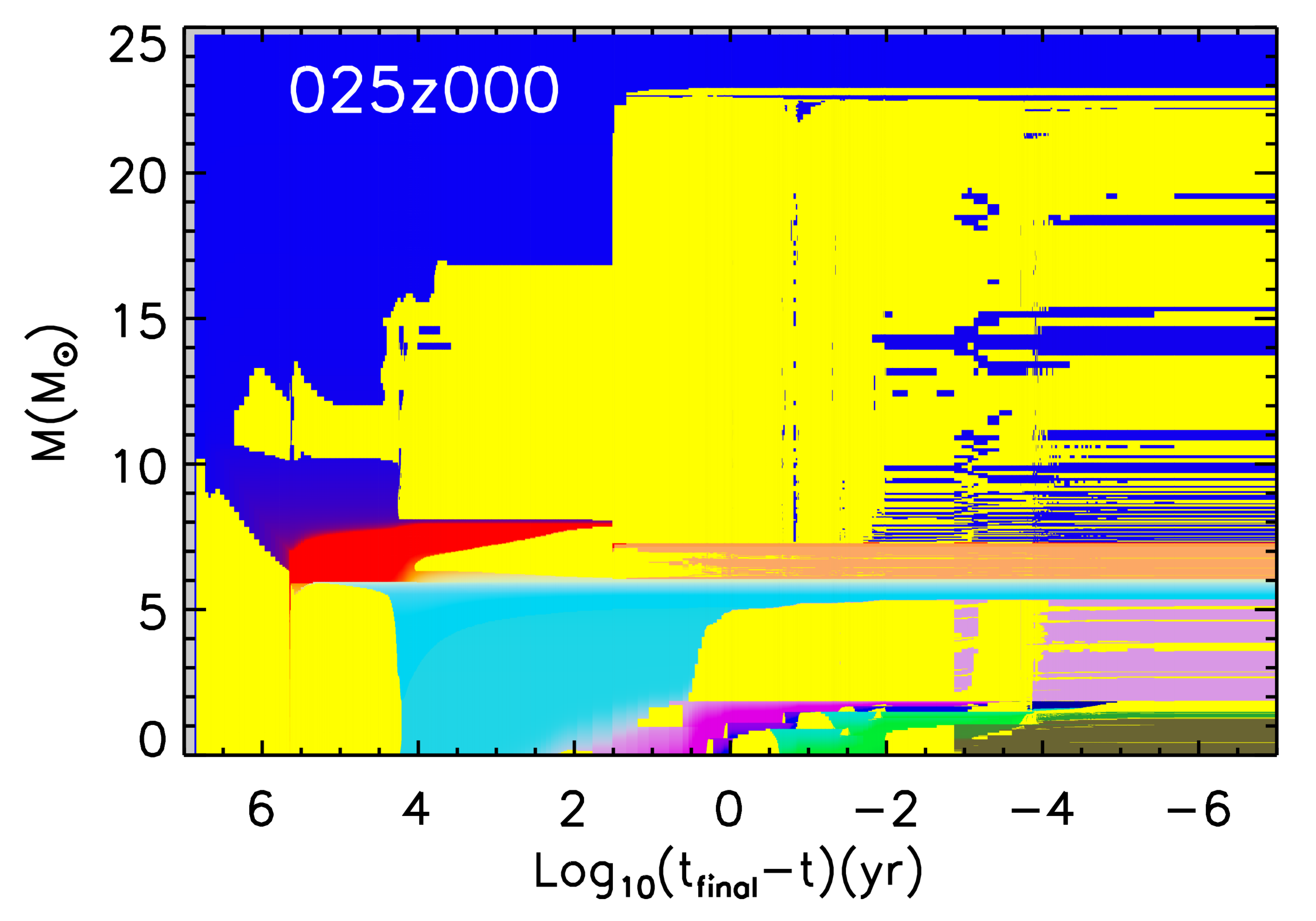}
\plottwo{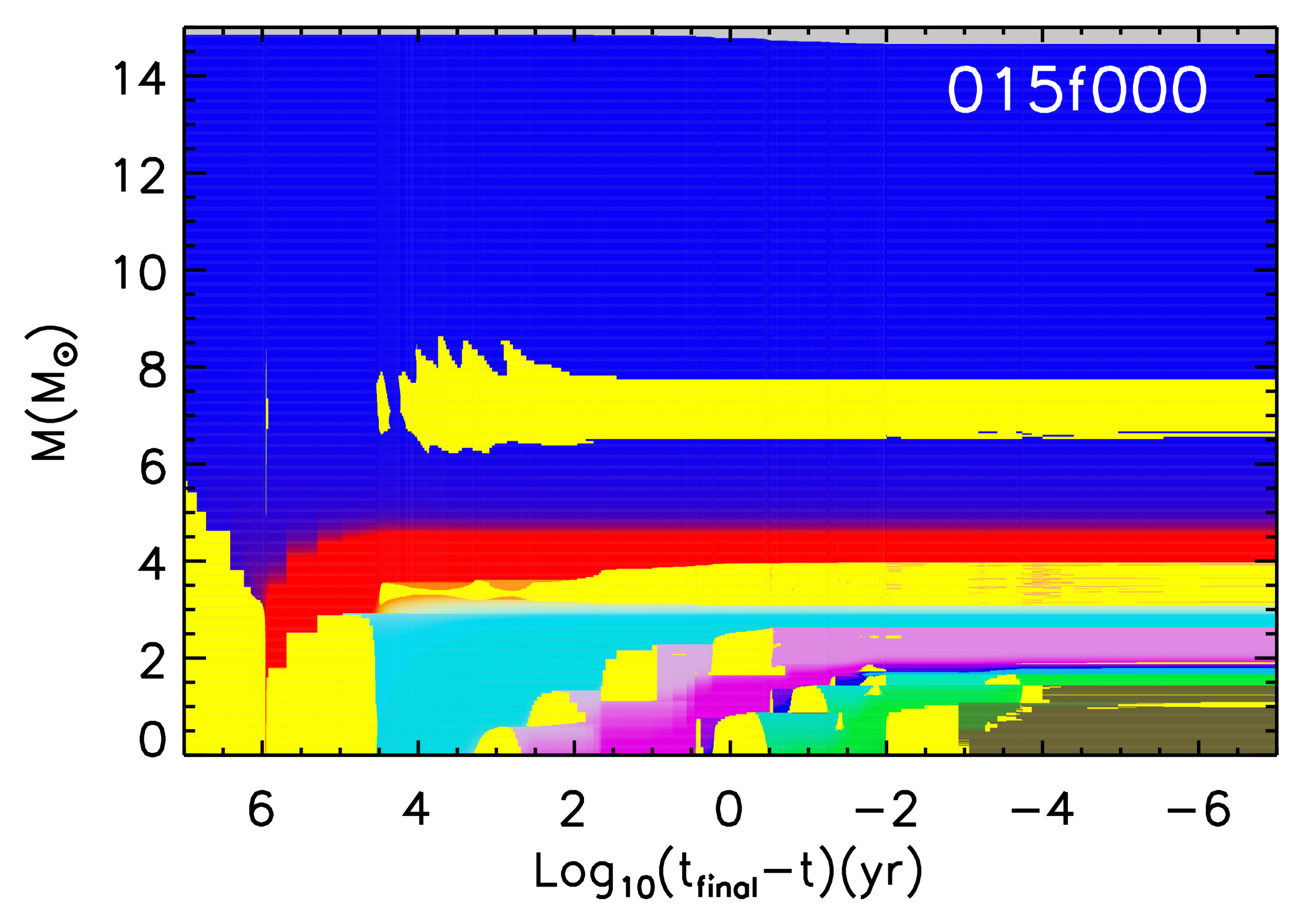}{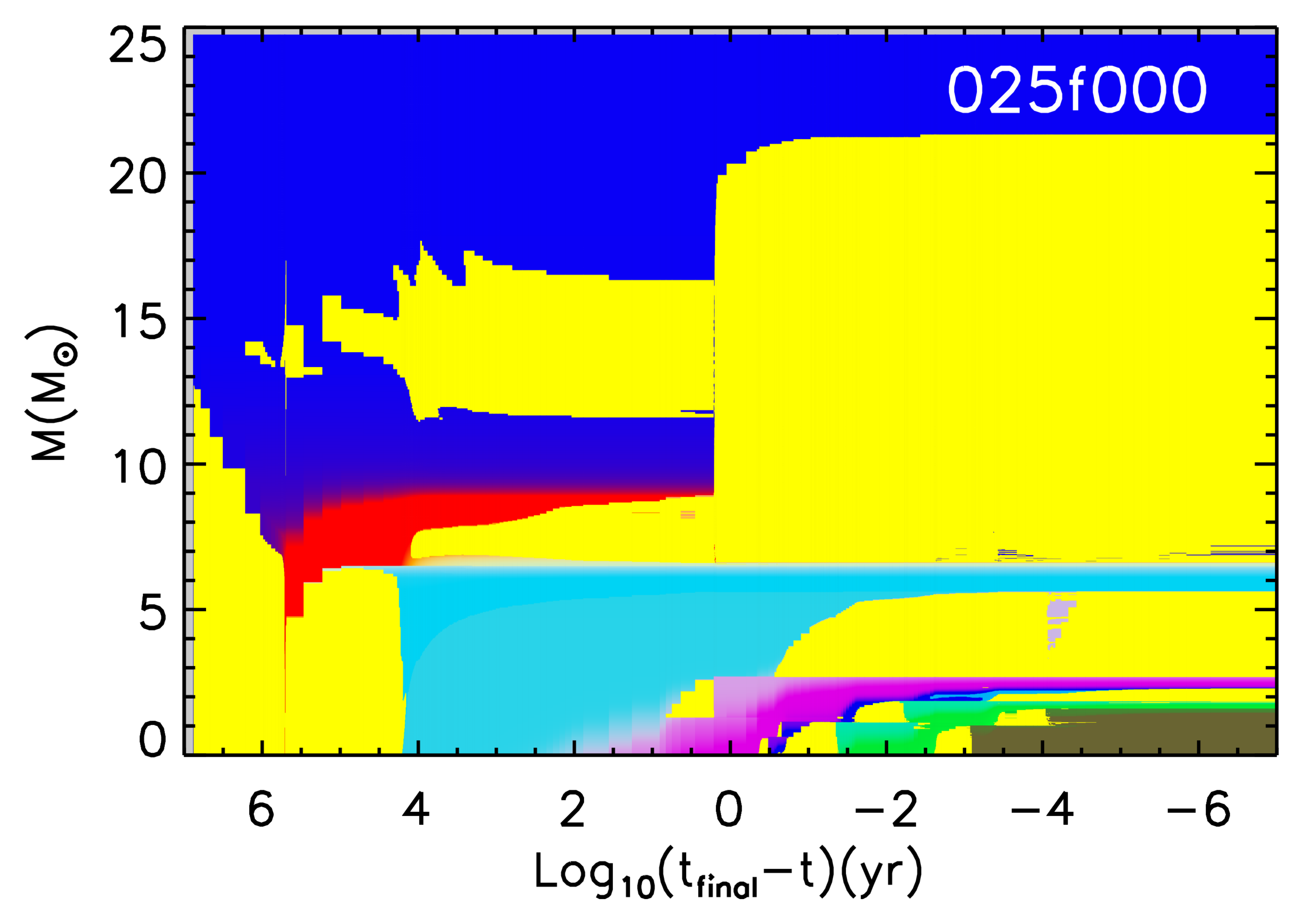}
\plottwo{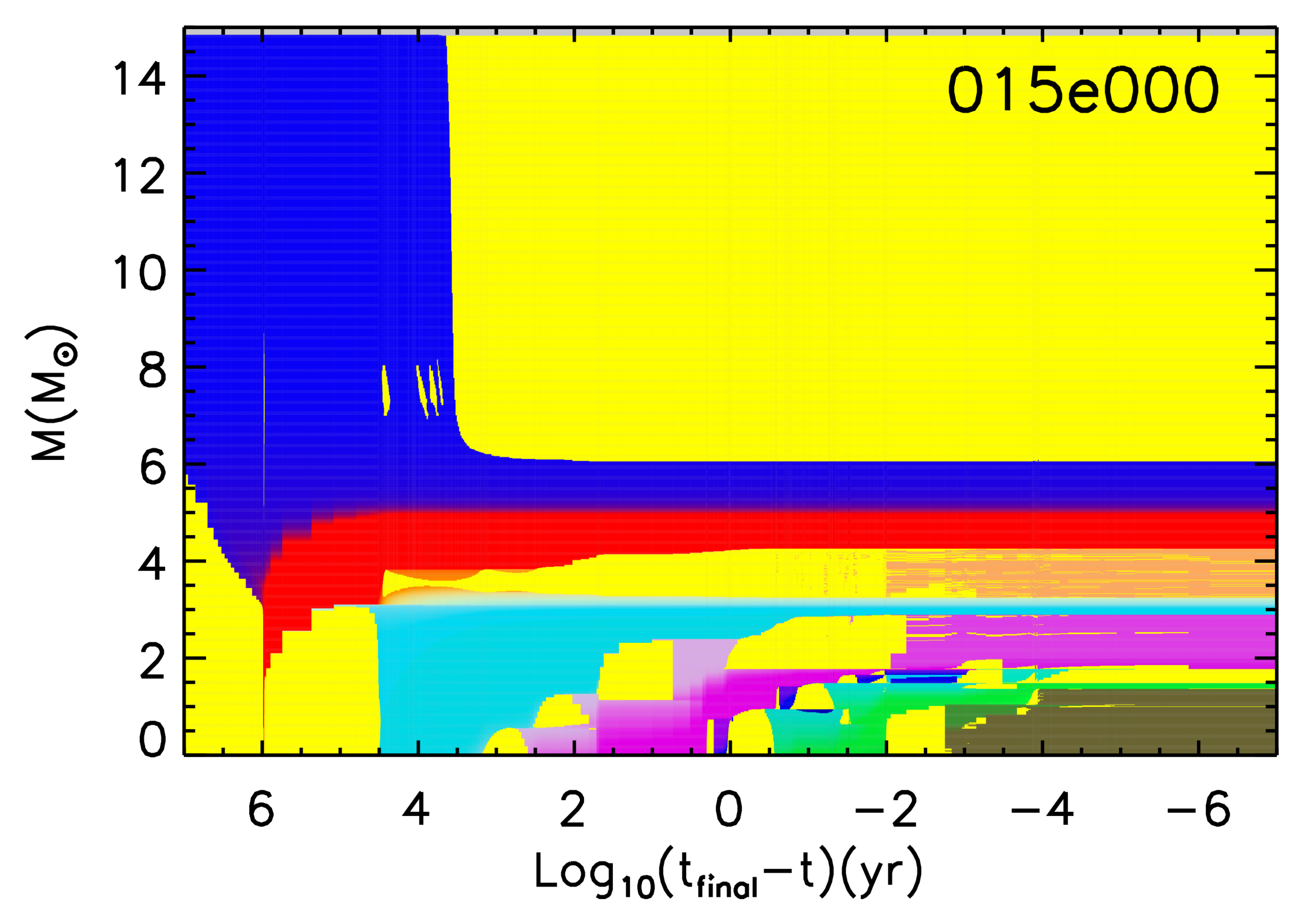}{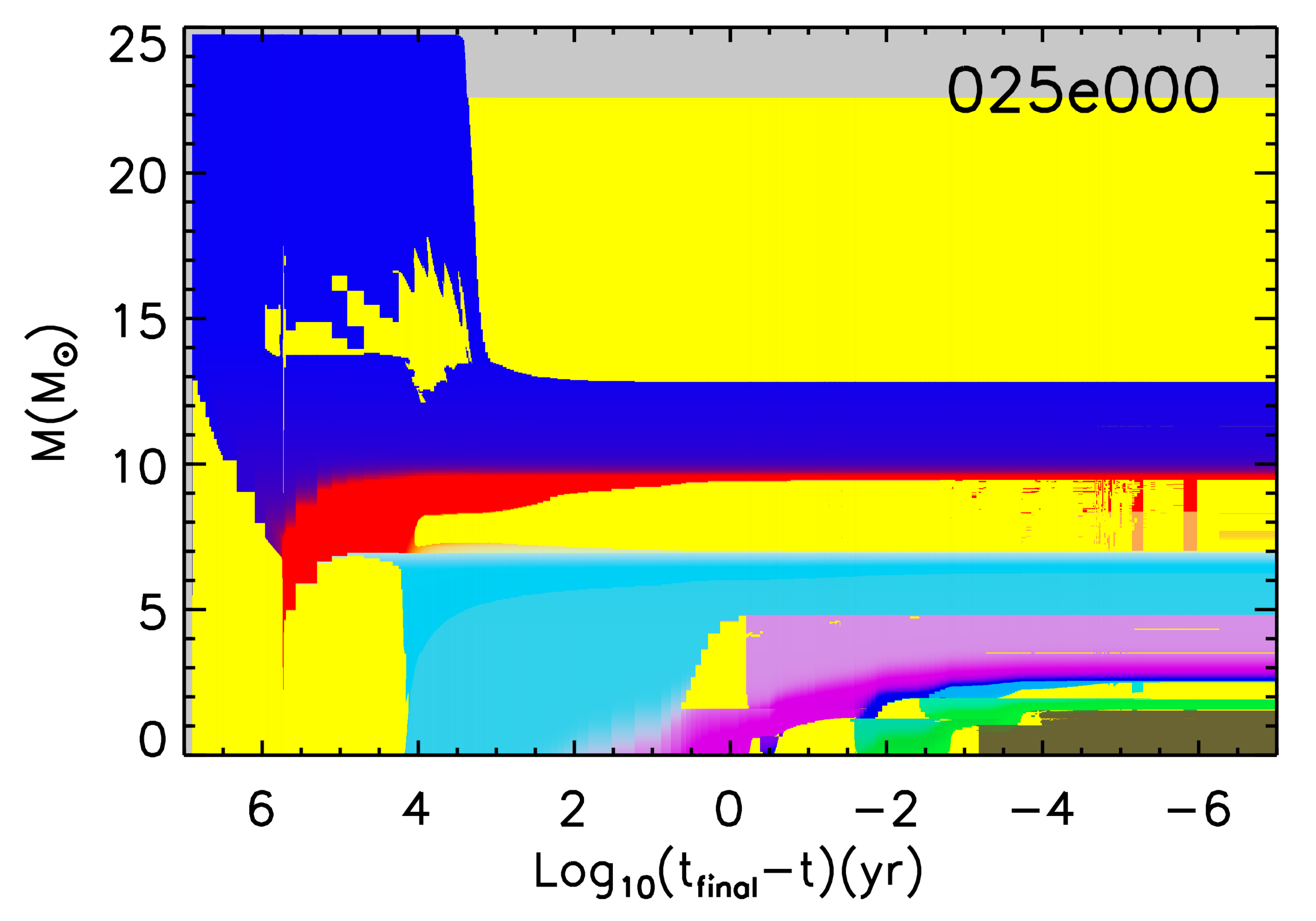}
\caption{Kippenhahn diagrams of the non rotating models. Top to bottom: Z, F, E. The x-axis represents the time prior the collapse of the Fe core, marked as $t_{final}$ in the plot. The blue area marks the H rich layers, the red one the pure He zone, the cyan the C/O region, the magenta the Ne/O region, the violet the O rich layers, the green area the Si rich zones while the brown zone the Fe core. The yellow area marks the convective zones. 
\label{fig:kipnorot}}
\end{figure*}

All three 15 \msun\ share a similar CO core mass ($\sim$ 3.1 \msun) and abundance of \nuk{C}{12} ($\sim$ 0.34). Hence the advanced evolutionary phases are quite similar: all of them form a C convective core of similar size and three C convective shell), left column in Figure \ref{fig:kipnorot}. In the case of the 25 \msun, the two most metal poor models share similar values of the two latter parameters ($M_{\rm CO}\sim$6 \msun~ and \nuk{C}{12}$\sim0.28$) and hence show a similar evolution during the advanced burning stages: radiative central C burning and two similar C convective shells. The 25 \msun\ of Set E, on the contrary, forms a slightly larger CO core mass ($\sim$7 \msun) and forms just one extended C convective shell. The way in which the C burning develops, in particular the number and extensions of the various C convective shells, influences the more advanced burning and plays a major role in sculpting the final density profile, or, equivalently, the mass-radius relation, at the core bounce: in other words it largely controls the final run of the binding energy of the mantle of the star, i.e., all the zones above the Fe core, as a function of the mass coordinate \citep[see, e.g.,][]{chieffi:20,chieffi:21,boccioli:23}. This quantity obviously plays a crucial role in determining the {\it explodability} of a star because it is the main obstacle to the advancing of the shock wave in its way out. A larger or smaller binding energy of the mantle may lead to a failed or successful explosion. Instead of comparing the full M-R relation of different models, \cite{oconnor:11} proposed to use a single parameter to quantify the compactness of a star, the so called $\xi$ parameter, defined as $\rm \xi_i=M_i(M_\odot)/R_i(1000~km)$. According to these authors, the most reliable mass coordinate at which this parameter should be evaluated is $\rm M=2.5$ \msun. Over the years it has become clear that this parameter {\it per se} is not very reliable in predicting the real final fate of a star but it is anyway useful because it is a synthetic way to describe at least the final compactness of a star through a single number \citep[see, e.g.,][and references therein]{ertl:16,burrows:19,chieffi:20}. The main physical properties of the stars at the onset of the core collapse, including the $\xi_{2.5}$ parameter are shown in \tablename~\ref{tab:final}.

\begin{figure*}[p]
\epsscale{1.0}
\plotone{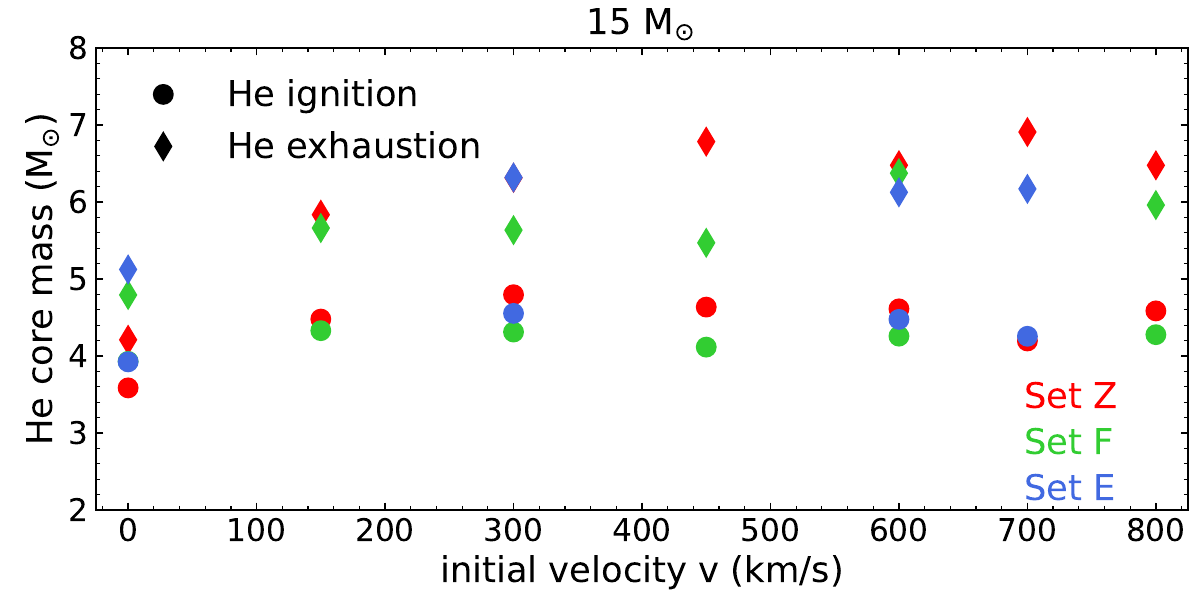}
\plotone{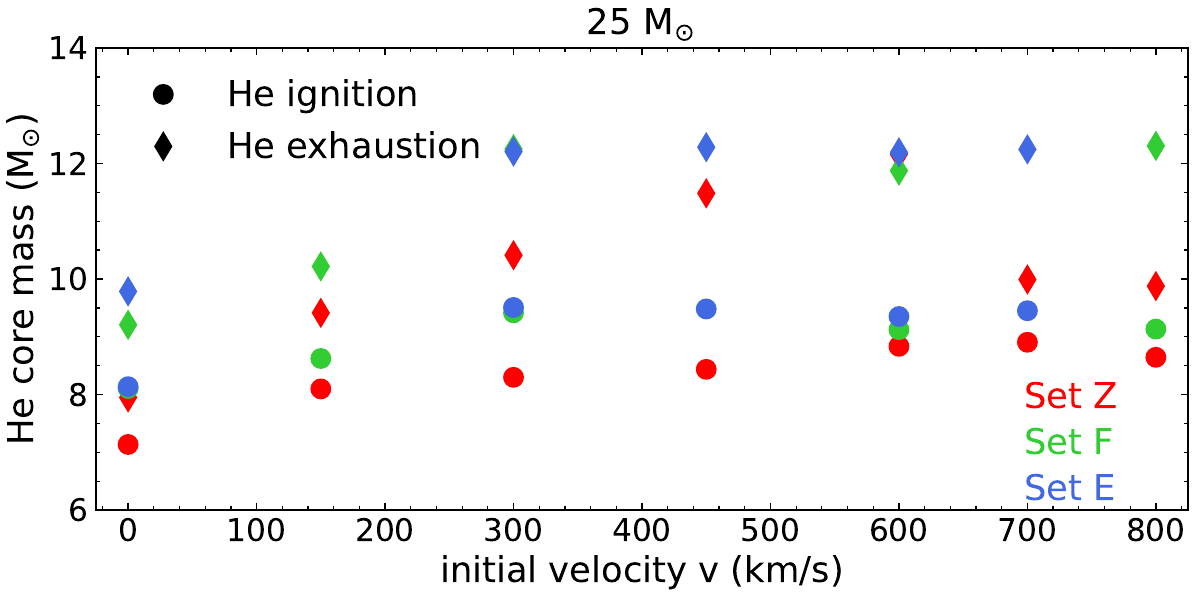}
\caption{He core mass in the 15 (top panel) and 25 \msun\ models (bottom panel) at central He ignition (filled dots) and exhaustion (filled diamonds).
\label{fig:mhe}}
\end{figure*}

\begin{figure*}[p]
\epsscale{1.0}
\plotone{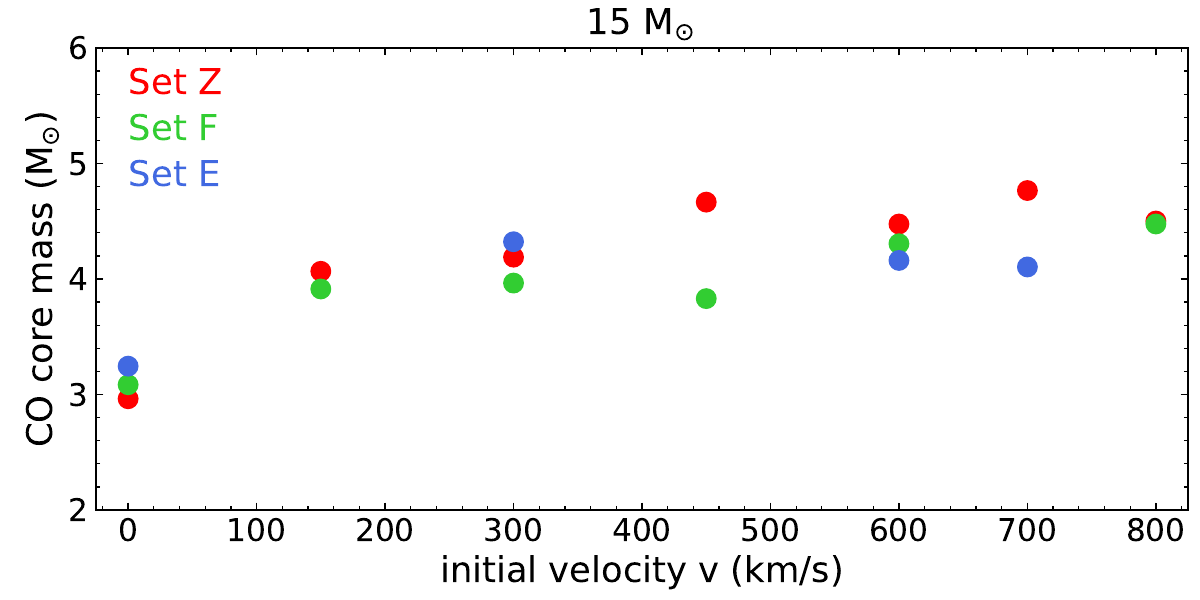}
\plotone{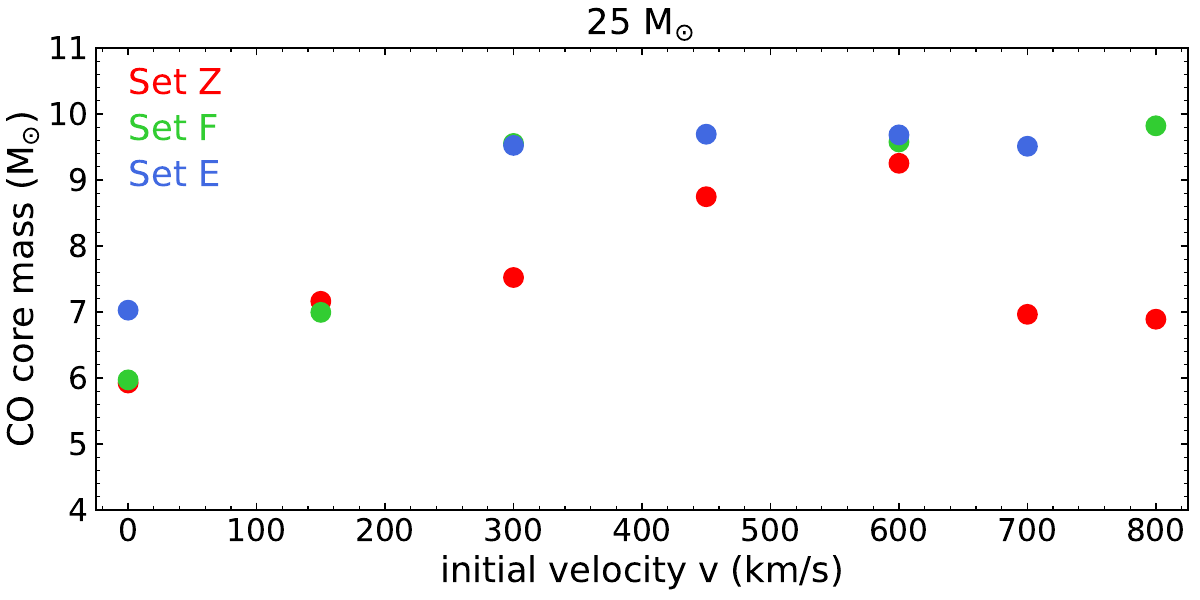}
\caption{CO core mass at central He depletion in the 15 (top panel) and 25 \msun\ models (bottom panel). \label{fig:mco}}
\end{figure*}

\begin{figure*}[p]
\epsscale{1.0}
\plotone{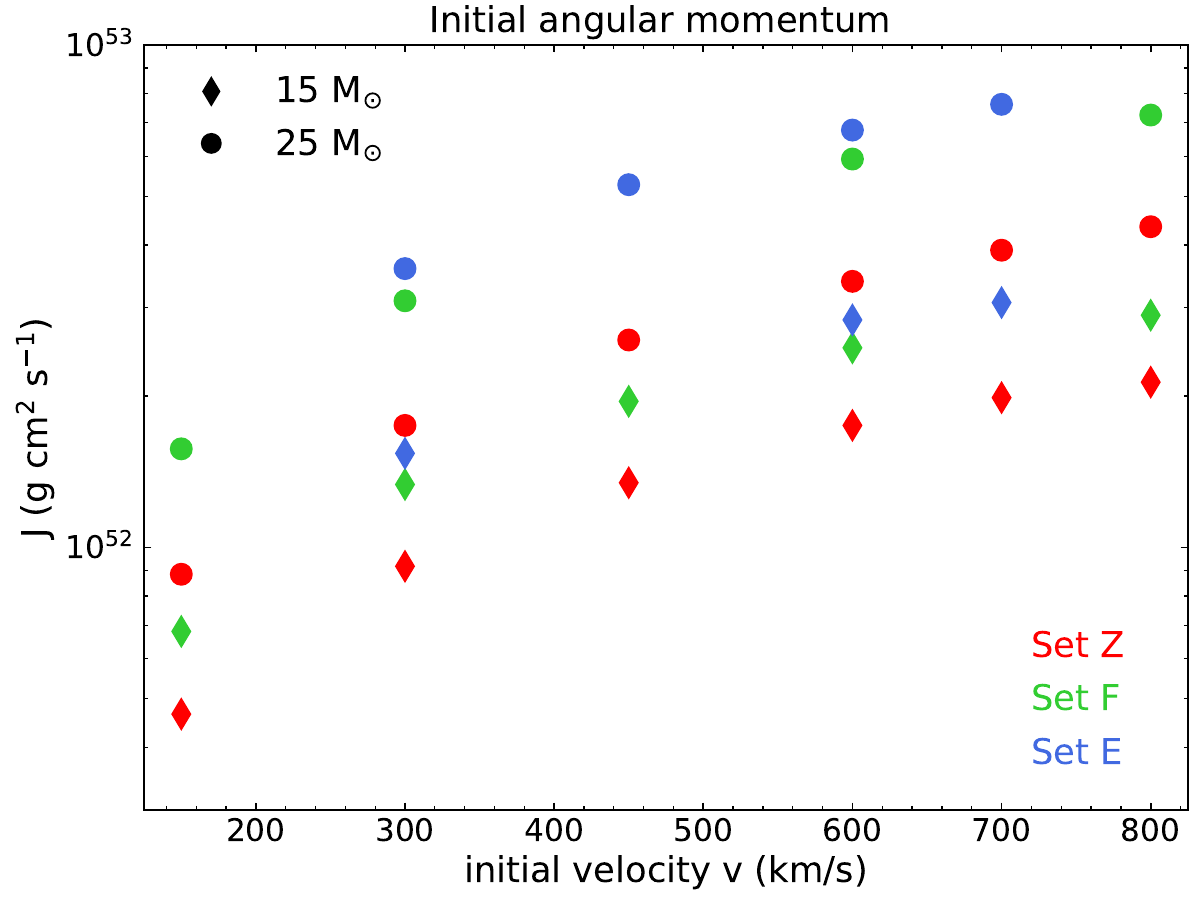}
\caption{The total angular momentum $J$ at the onset of the main sequence phase in all the computed models. Diamonds represent 15 \msun\ models, dots represent 25 \msun\ models. \label{fig:ini_momang}}
\end{figure*}

\begin{figure*}[p]
\epsscale{1.0}
\plottwo{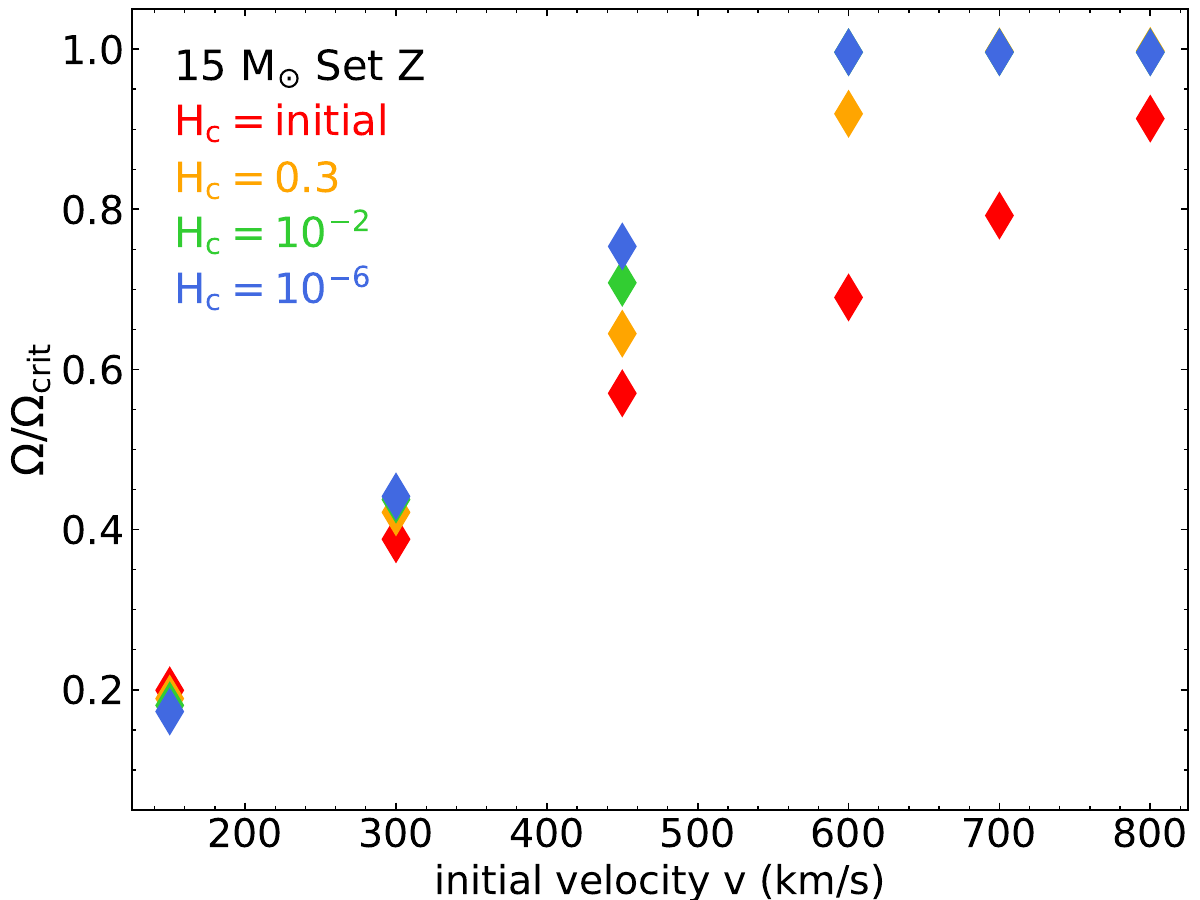}{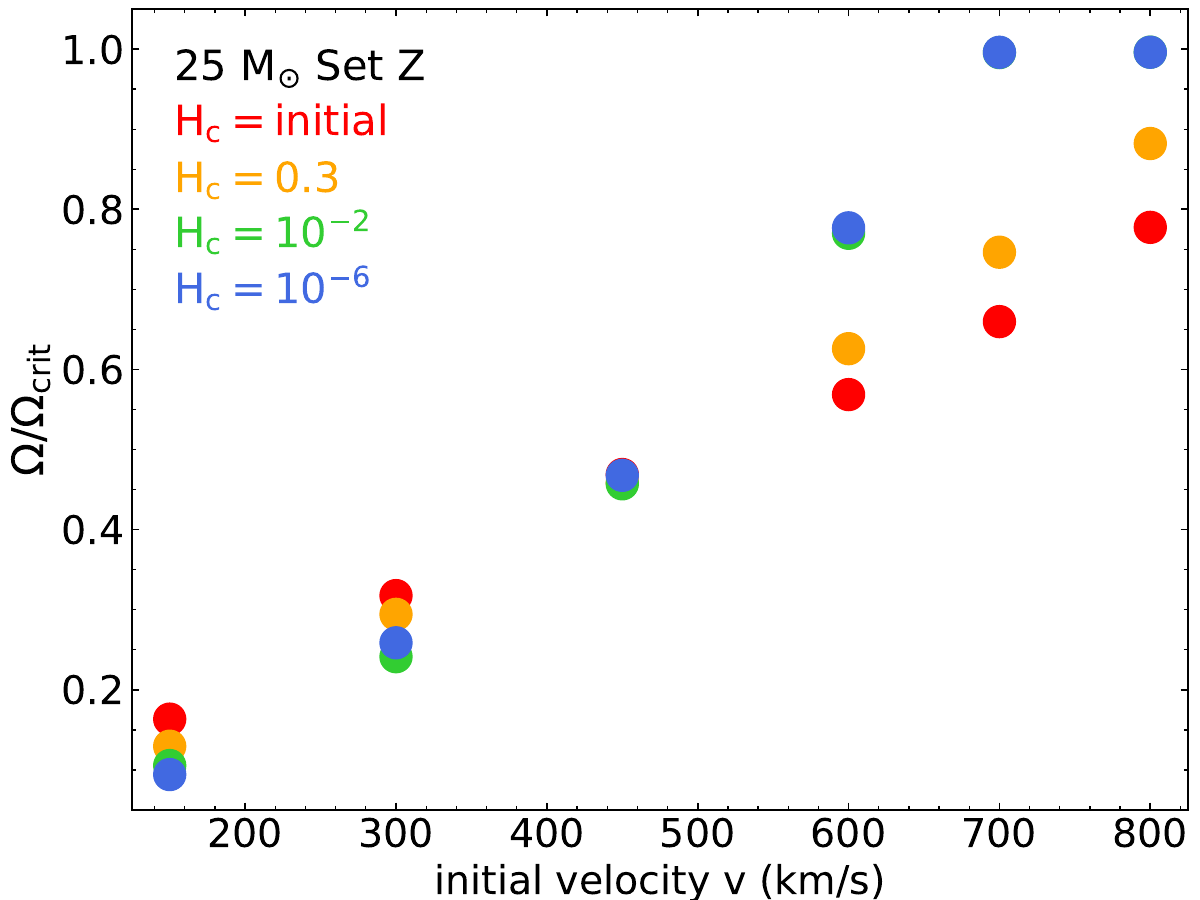}
\plottwo{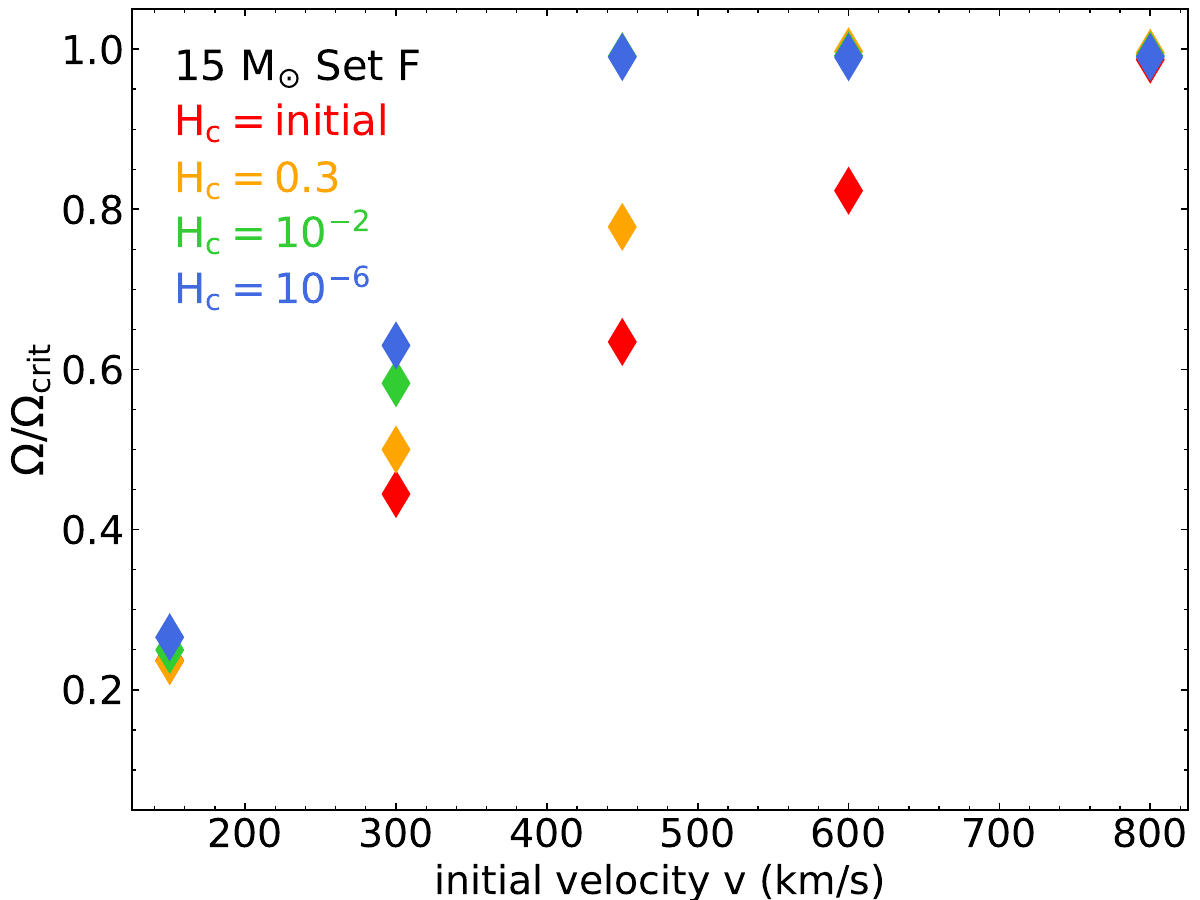}{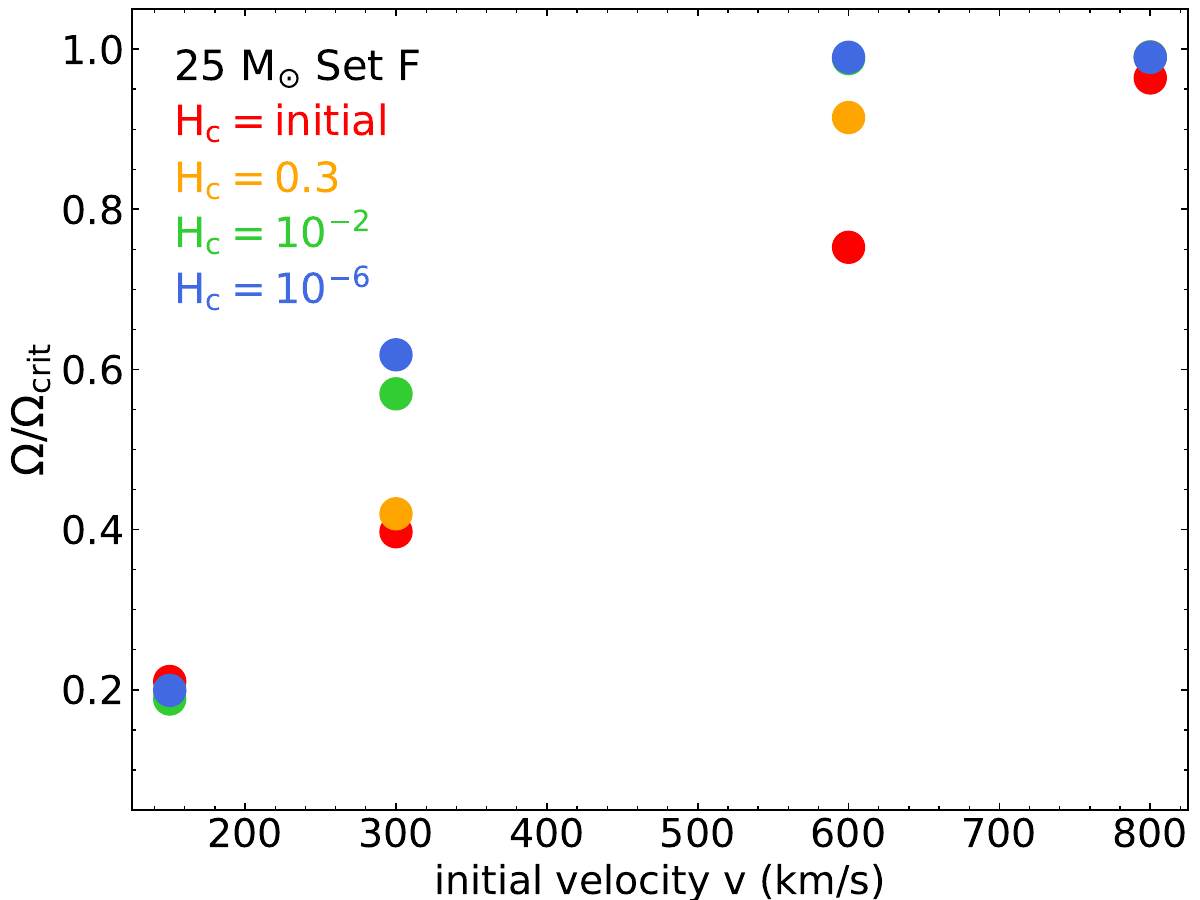}
\plottwo{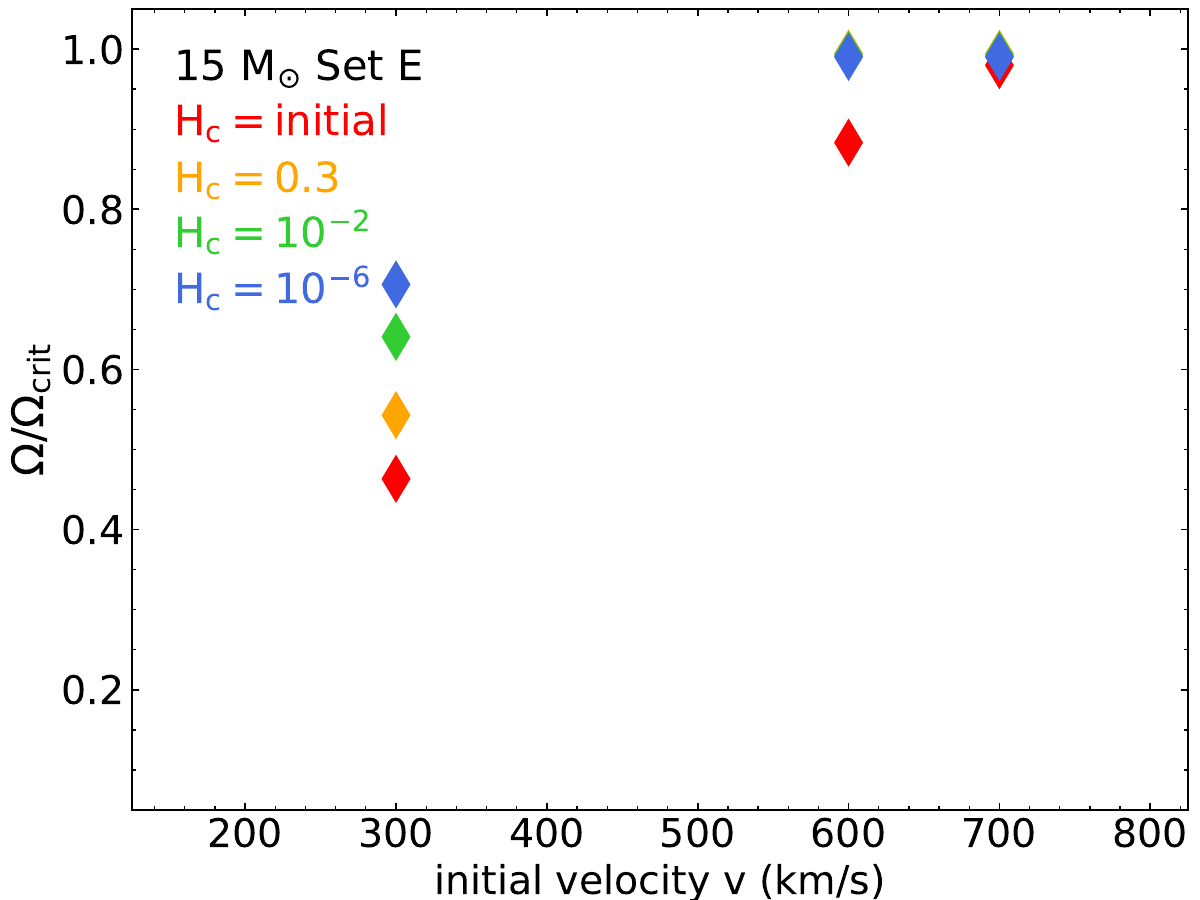}{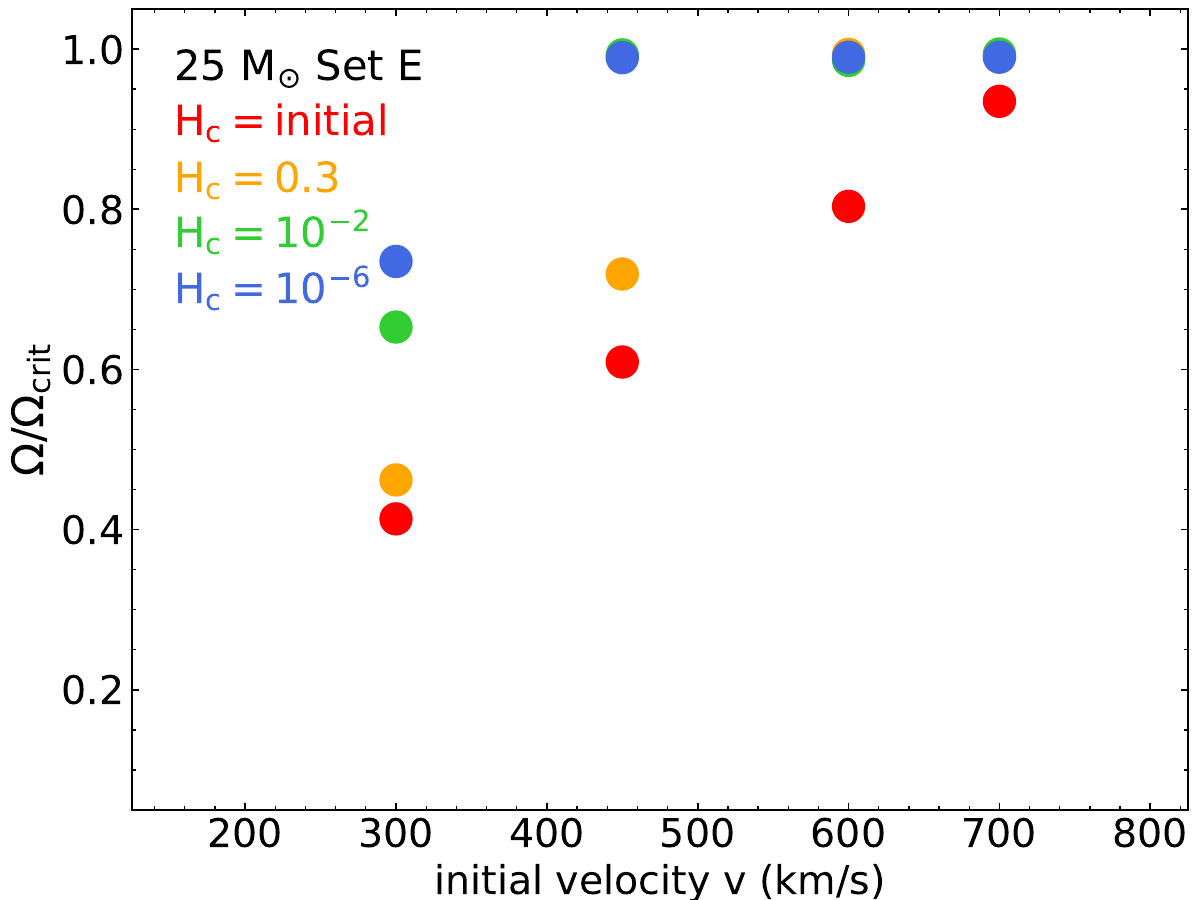}
\caption{Ratio of surface angular velocity to critical velocity in the 15 \msun\ (left panels) and 25 \msun\ stars (right panels) in Set Z (first row), Set F (second row), and Set E (third row) for four characteristic central H abundances in mass fraction: initial (red), 0.3 (yellow), $10^{-2}$ (green) and $10^{-6}$ (blue). \label{fig:osuoc}}
\end{figure*}

\begin{figure*}[p]
\epsscale{1.0}
\plottwo{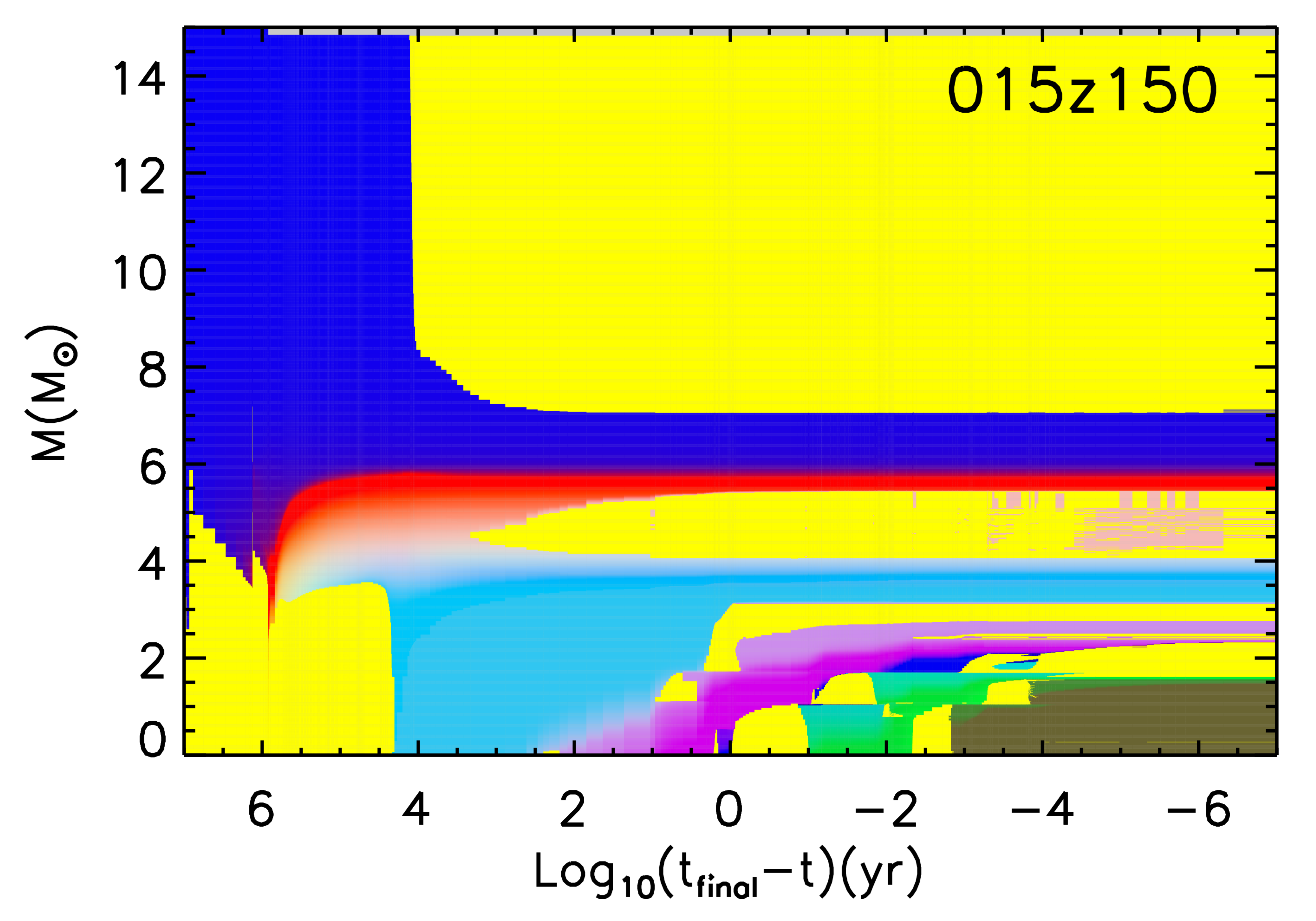}{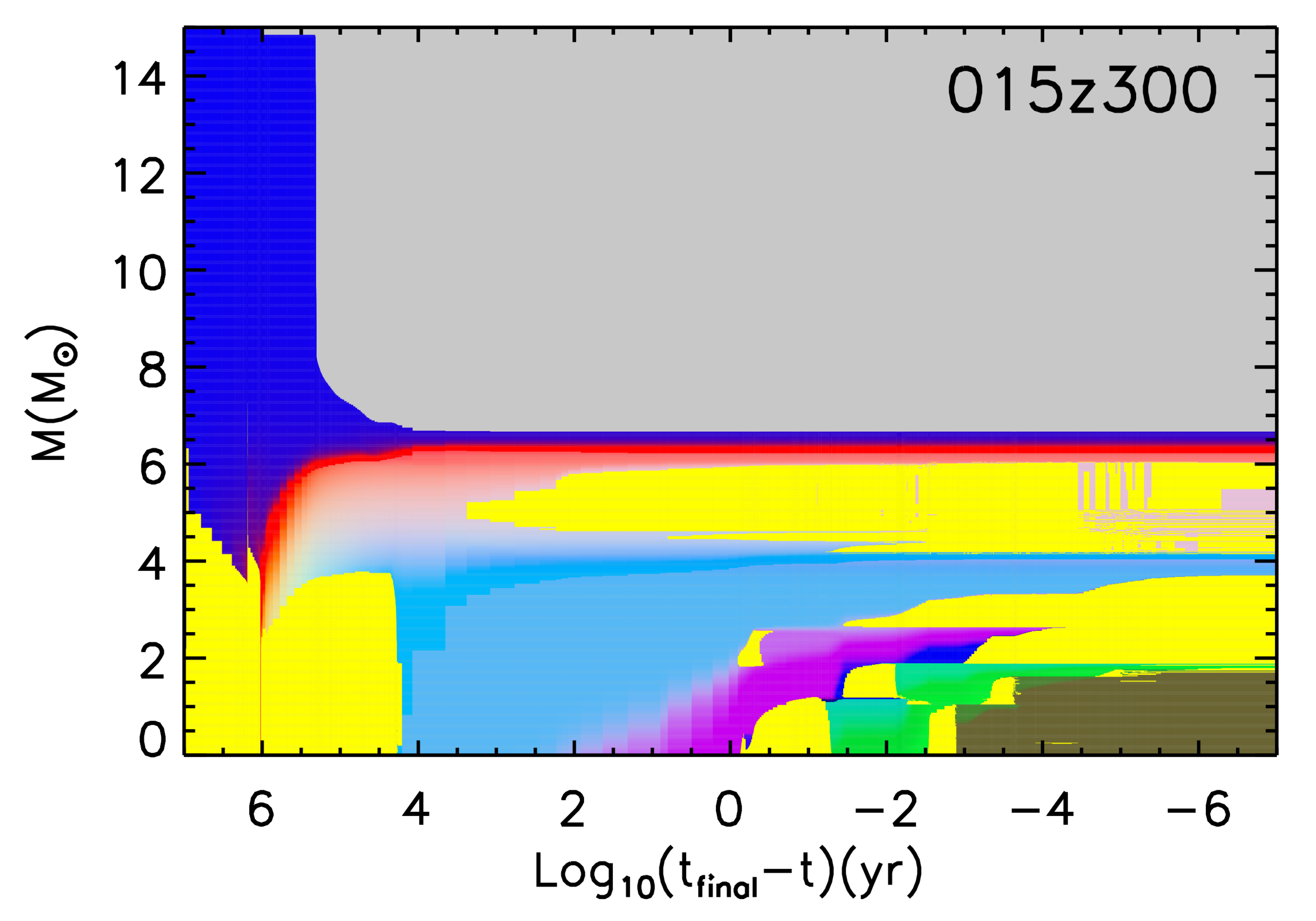}
\plottwo{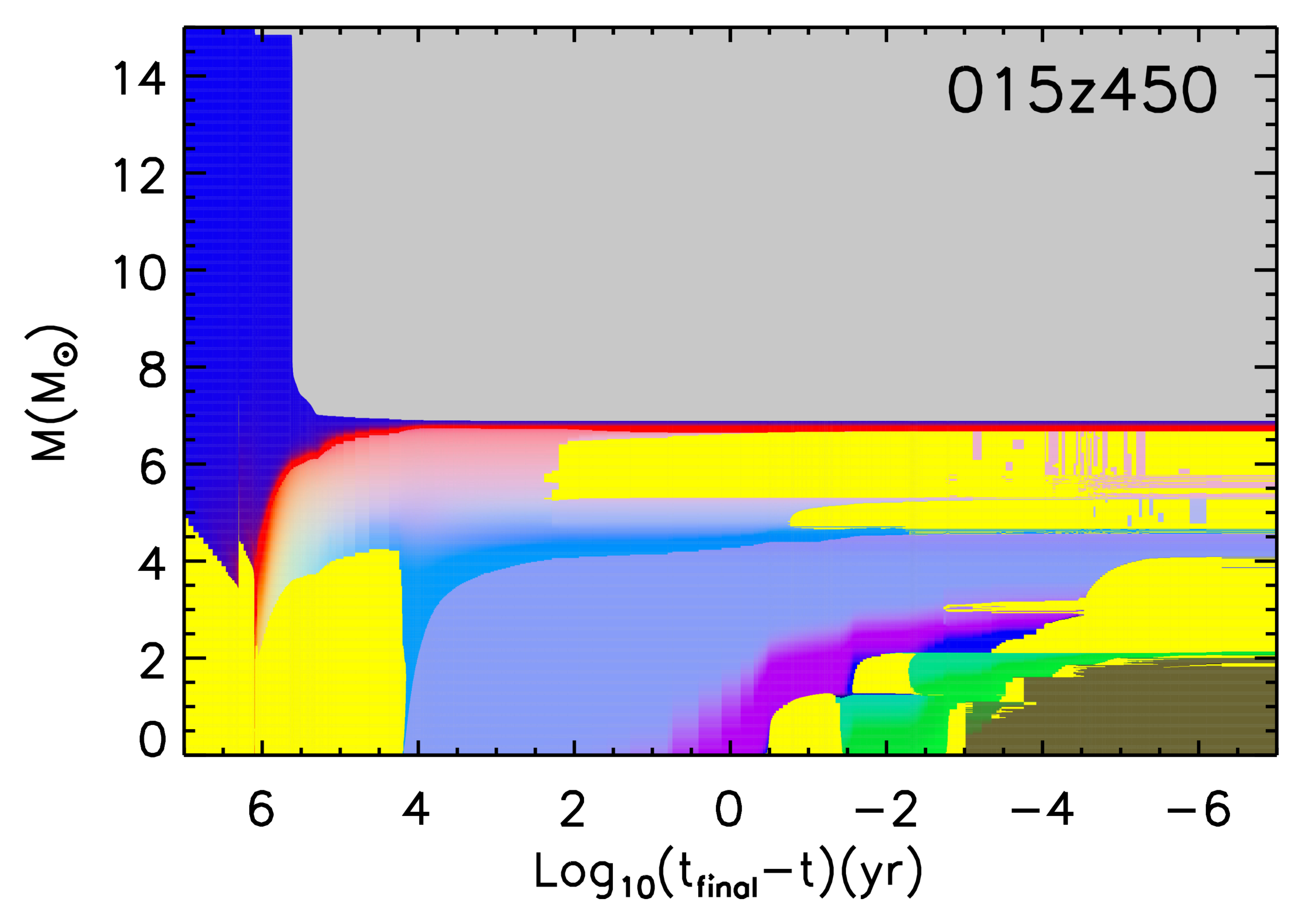}{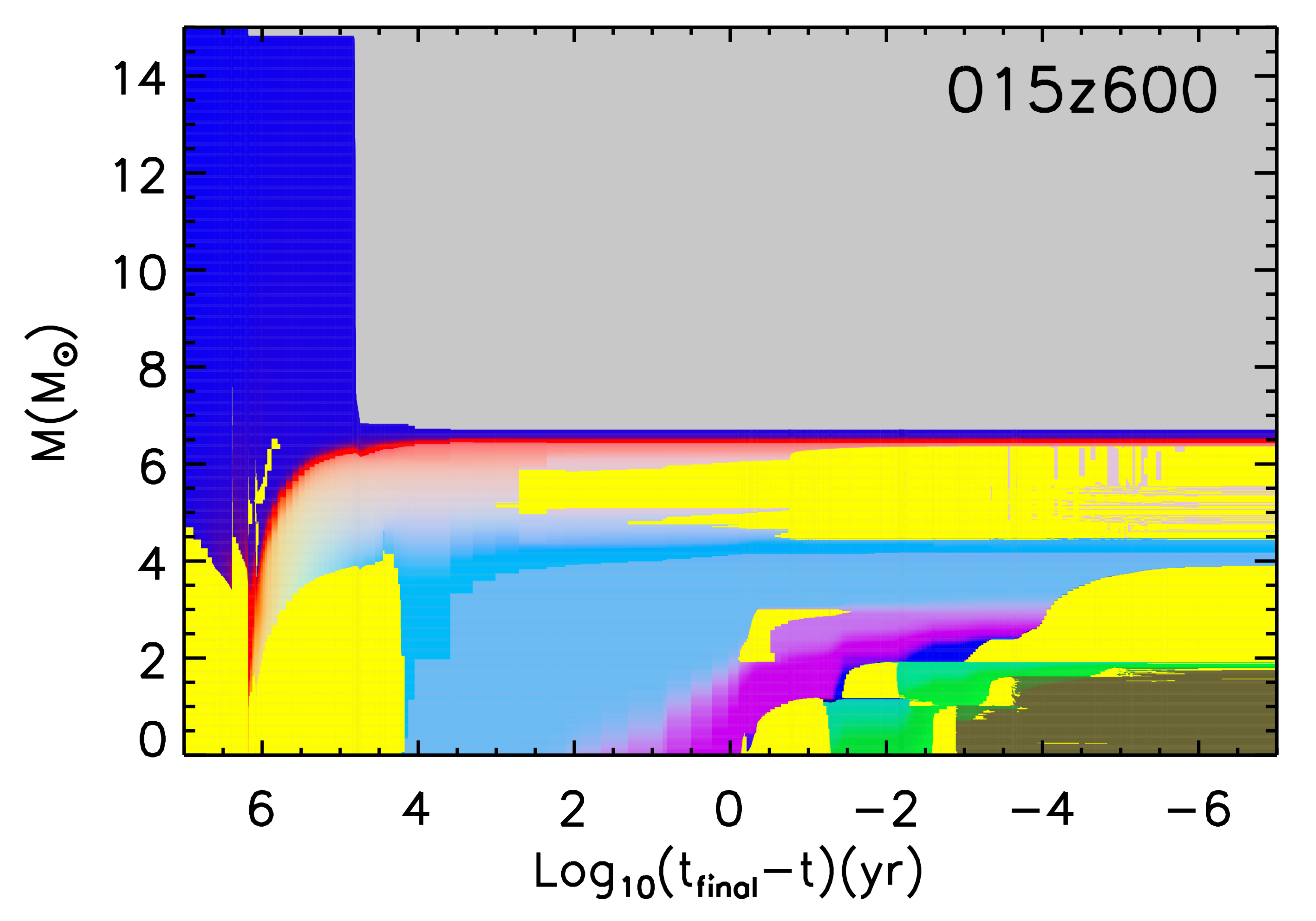}
\plottwo{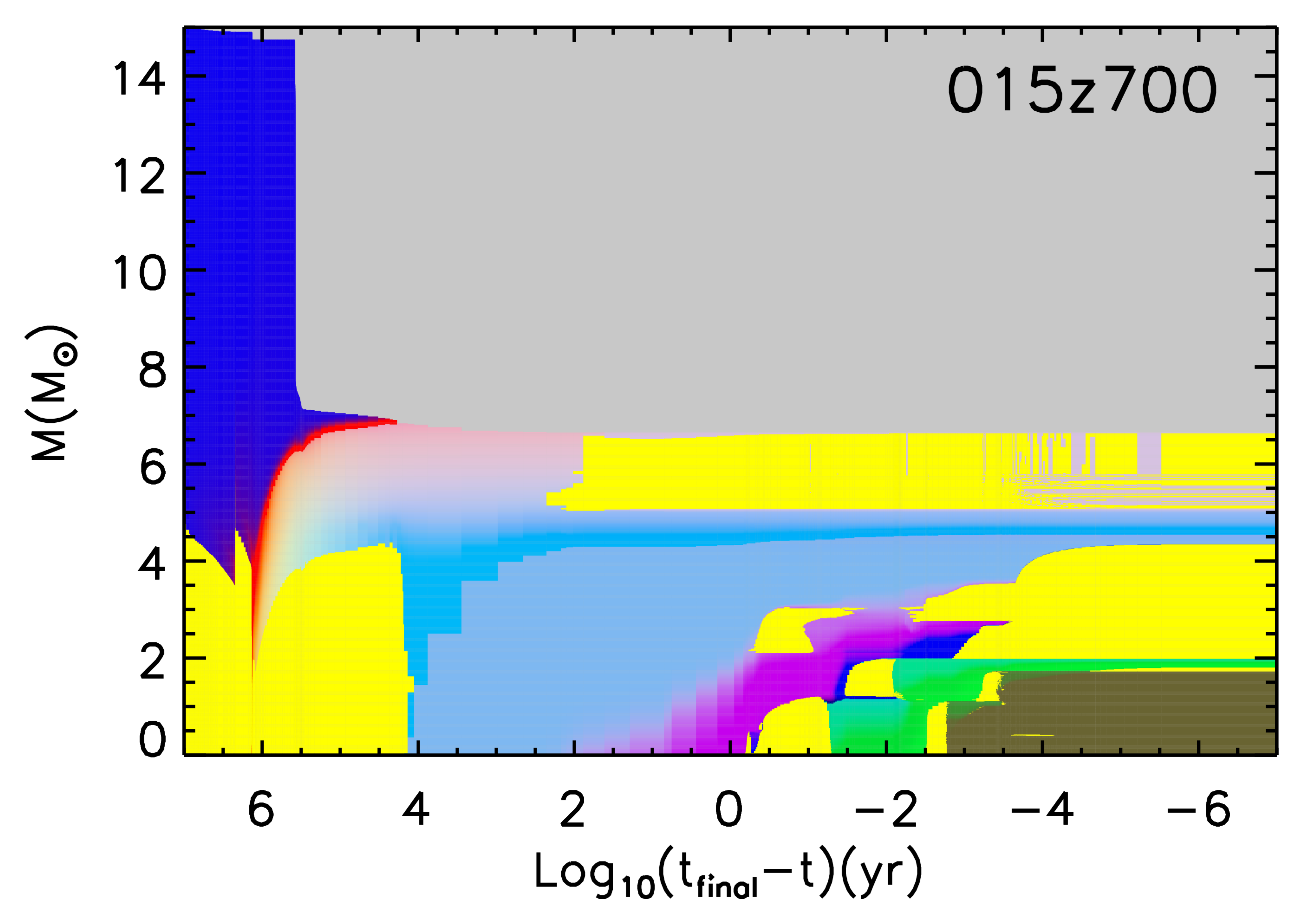}{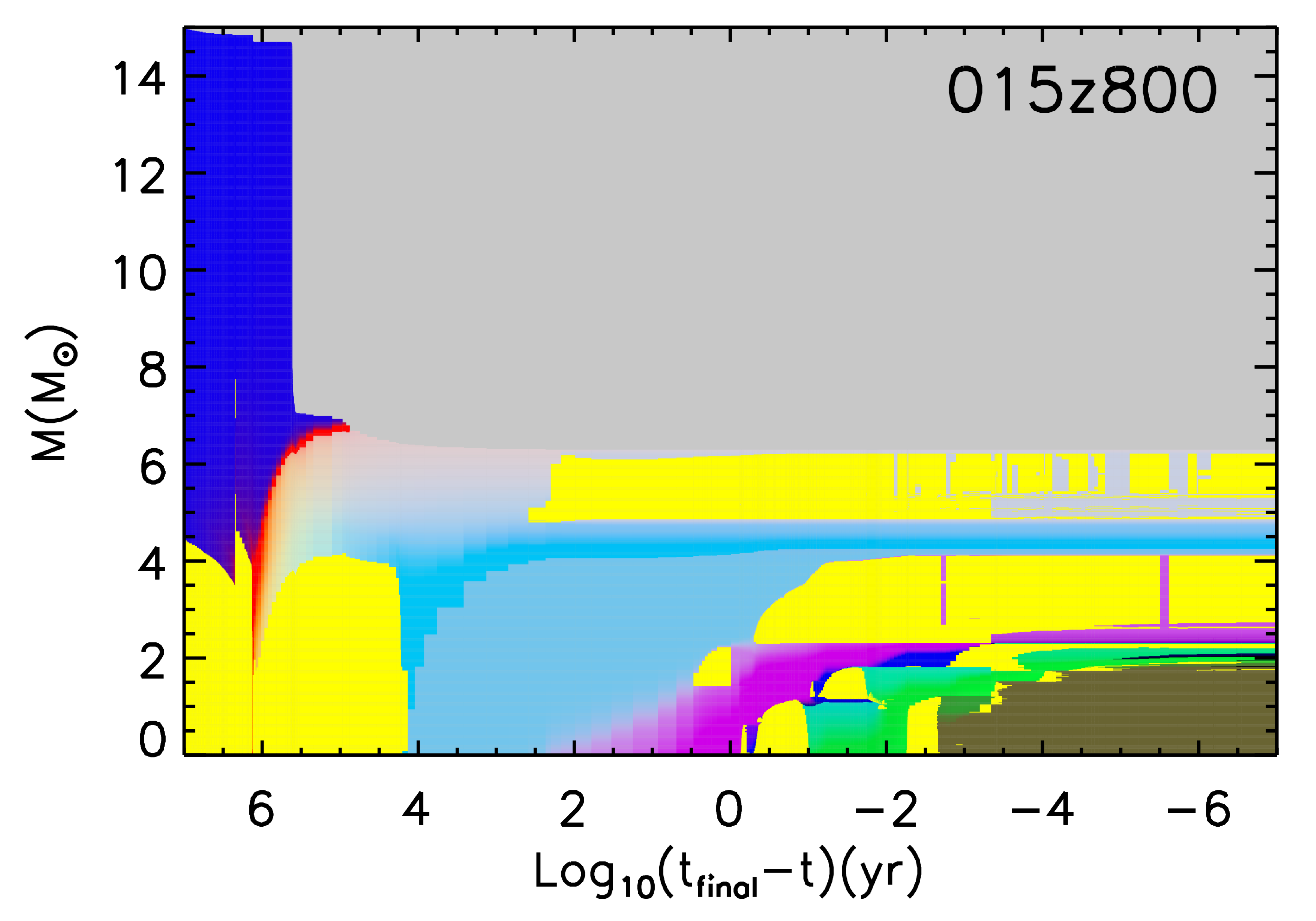}
\caption{Kippenhahn diagrams of the rotating 15 \msun\ zero metallicity models. The color coding is the same as in Figure \ref{fig:kipnorot}.
\label{fig:kiprotz15}}
\end{figure*}

\begin{figure*}[p]
\epsscale{1.0}
\plottwo{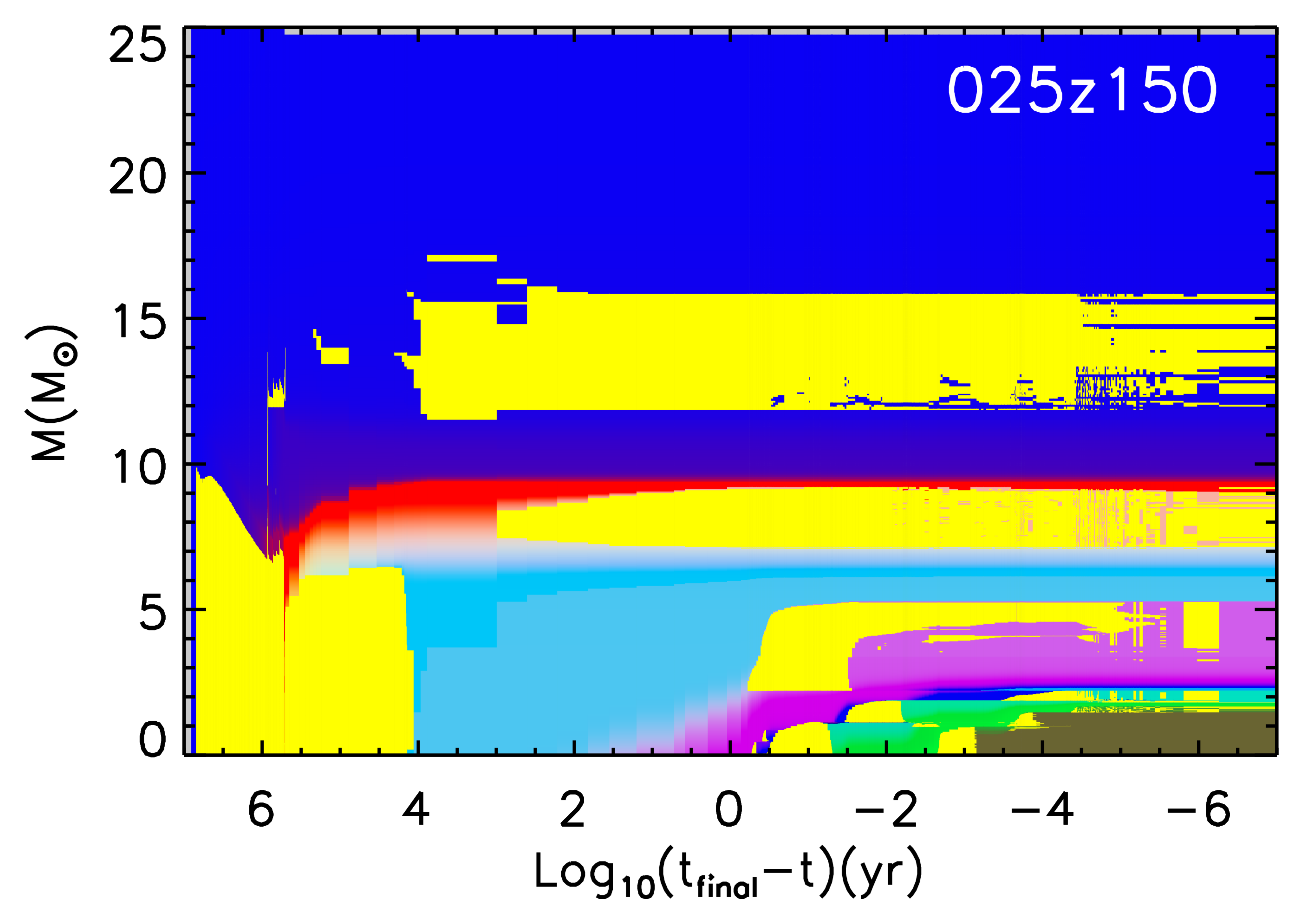}{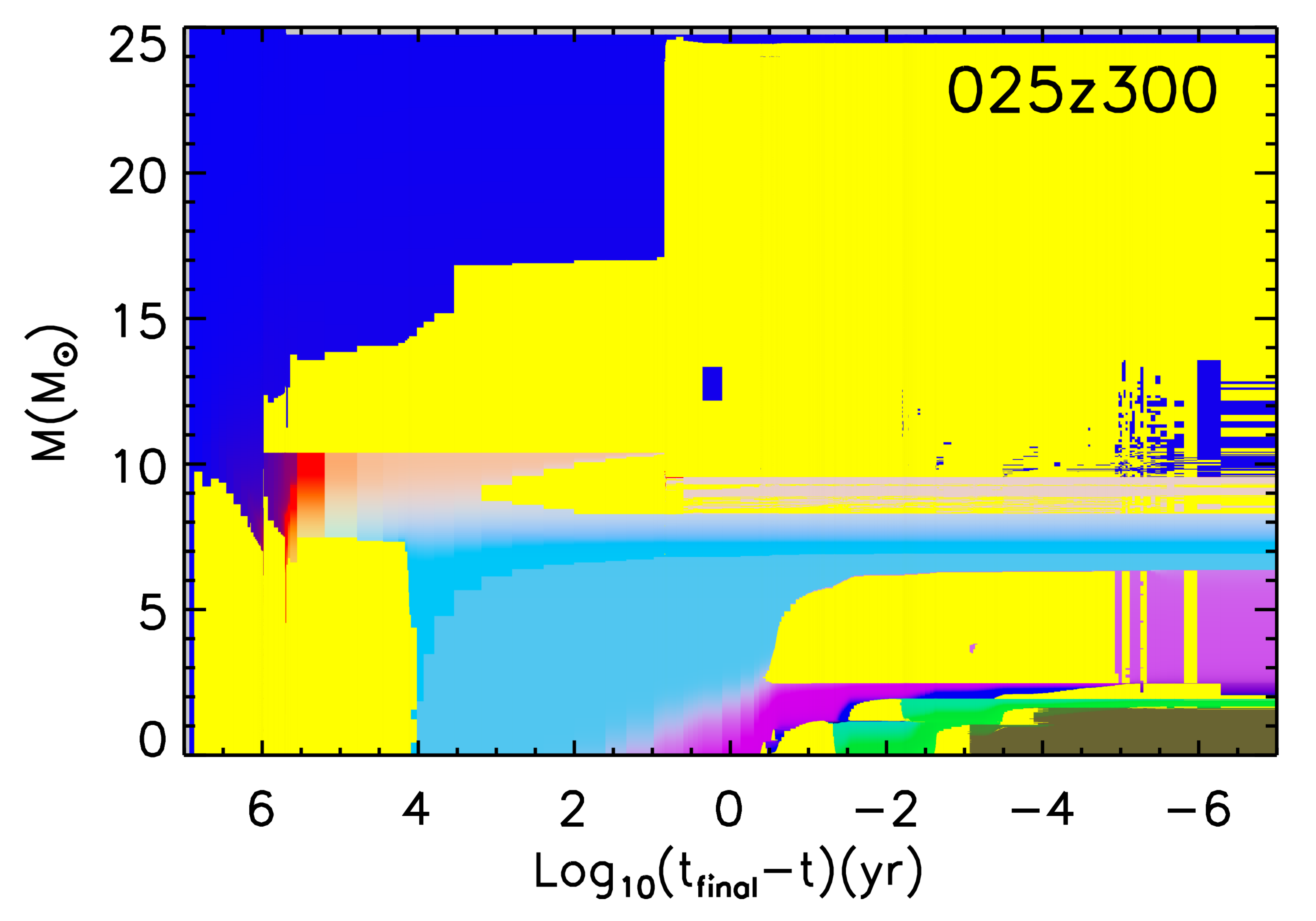}
\plottwo{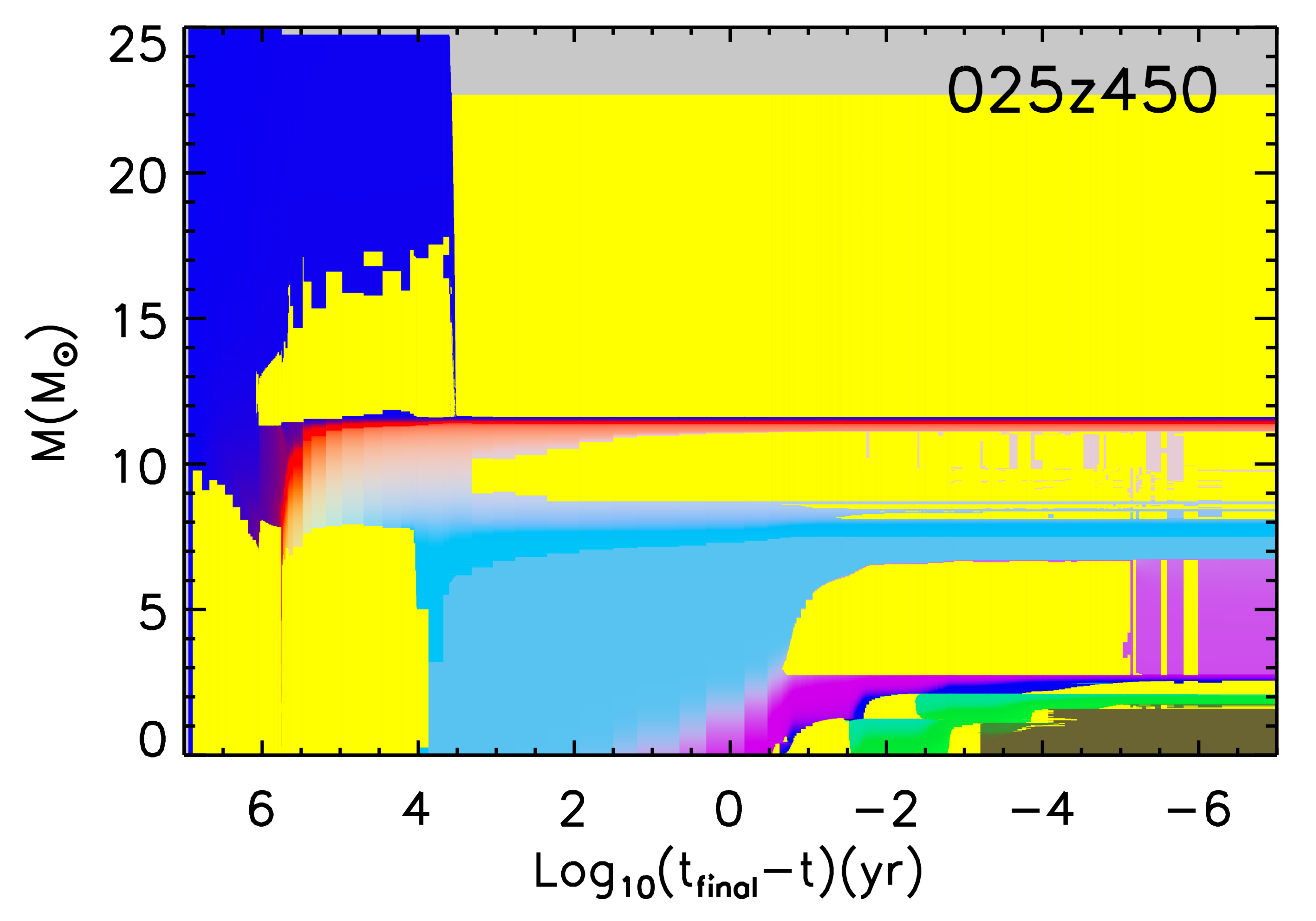}{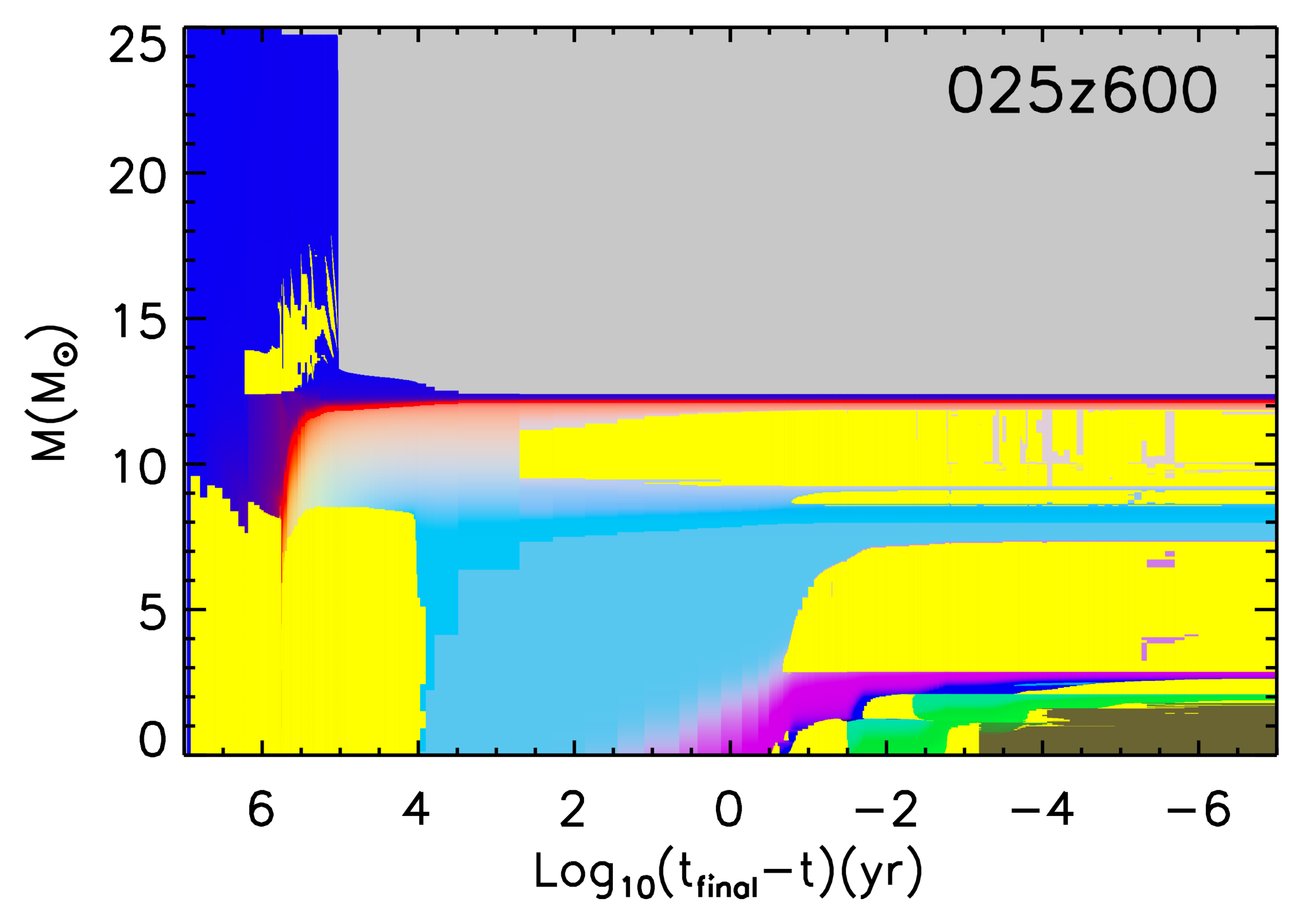}
\plottwo{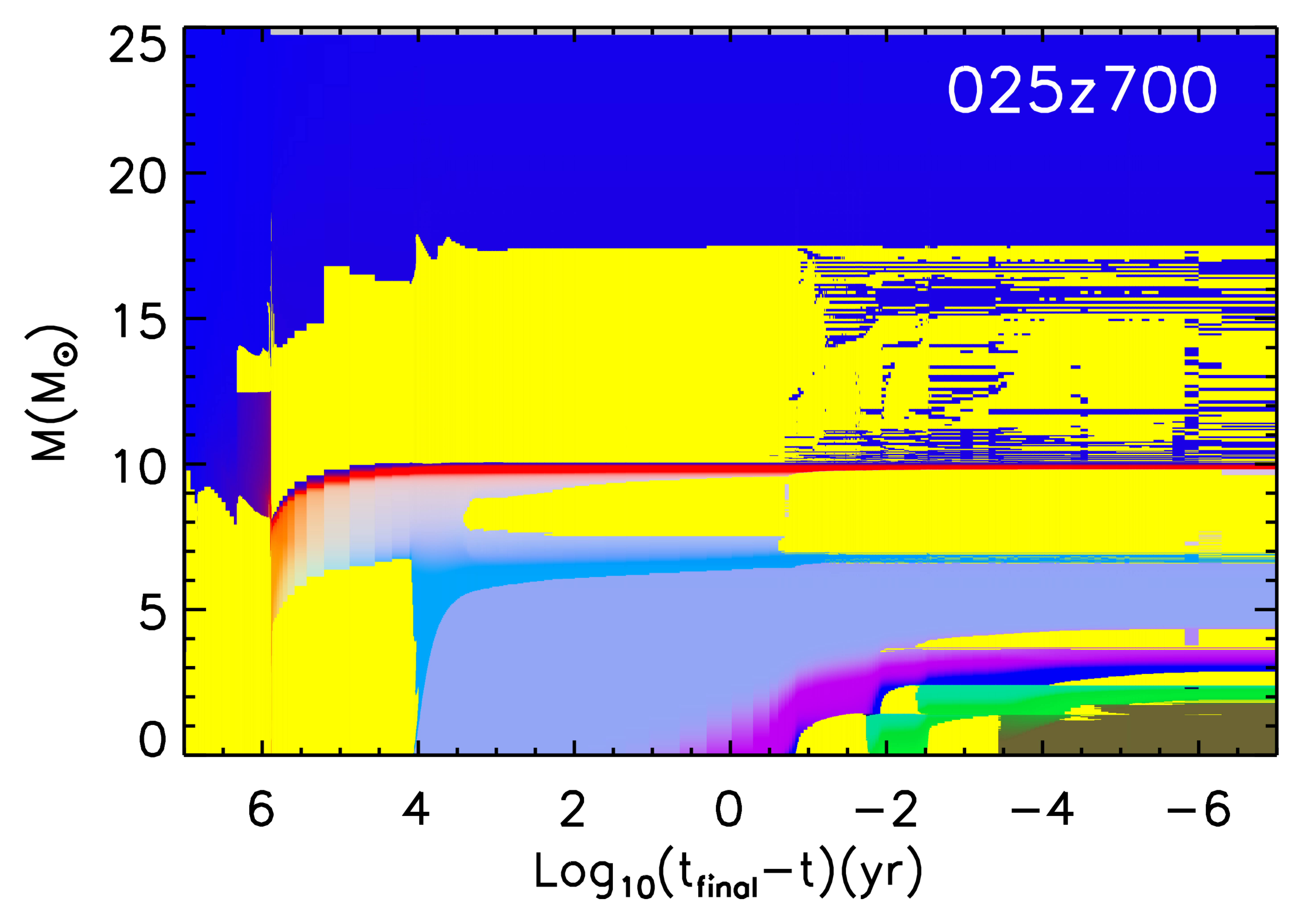}{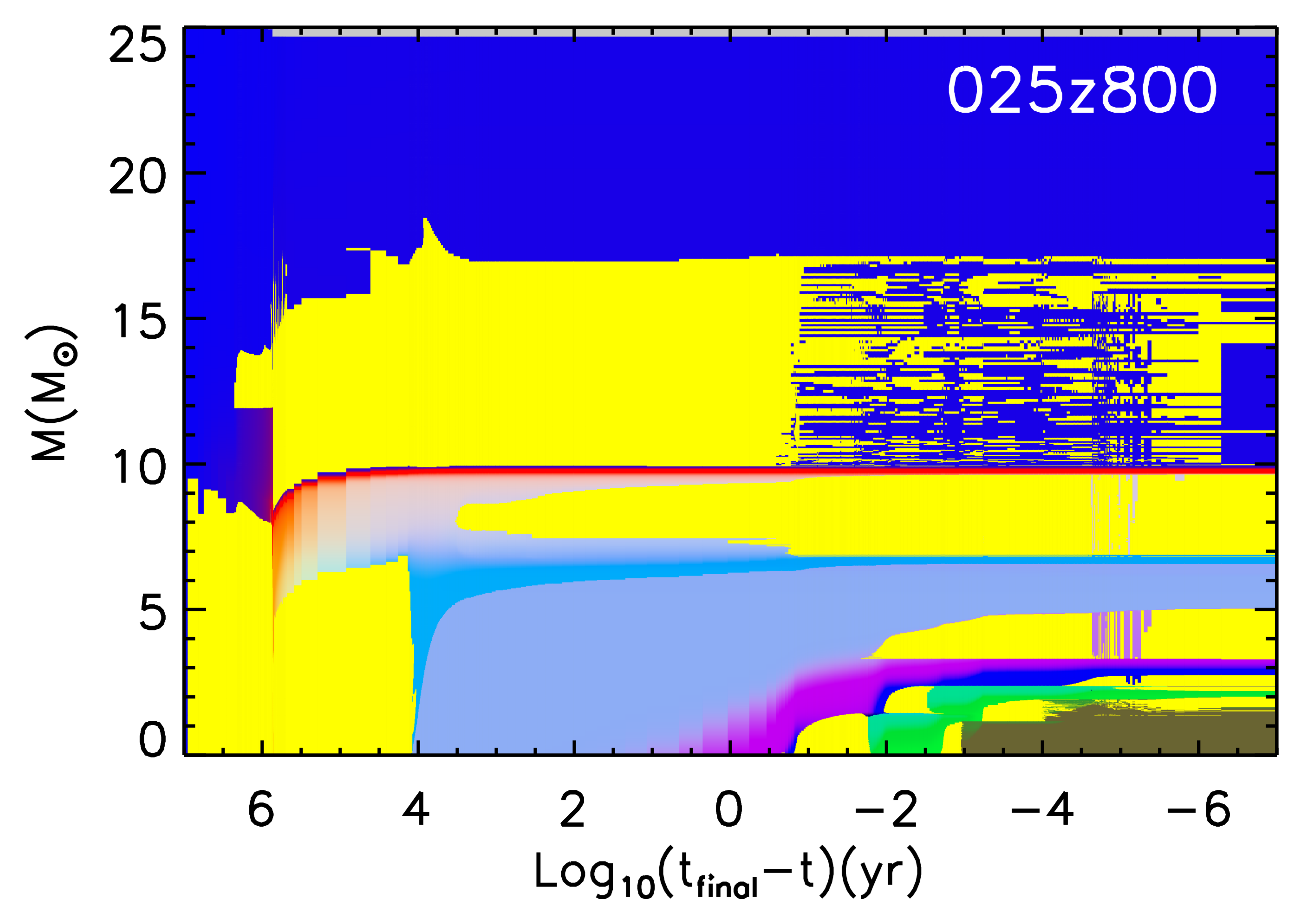}
\caption{Kippenhahn diagrams of the rotating 25 \msun\ zero metallicity models. The color coding is the same as in Figure \ref{fig:kipnorot}.
\label{fig:kiprotz25}}
\end{figure*}

\begin{figure*}[p]
\epsscale{1.0}
\plottwo{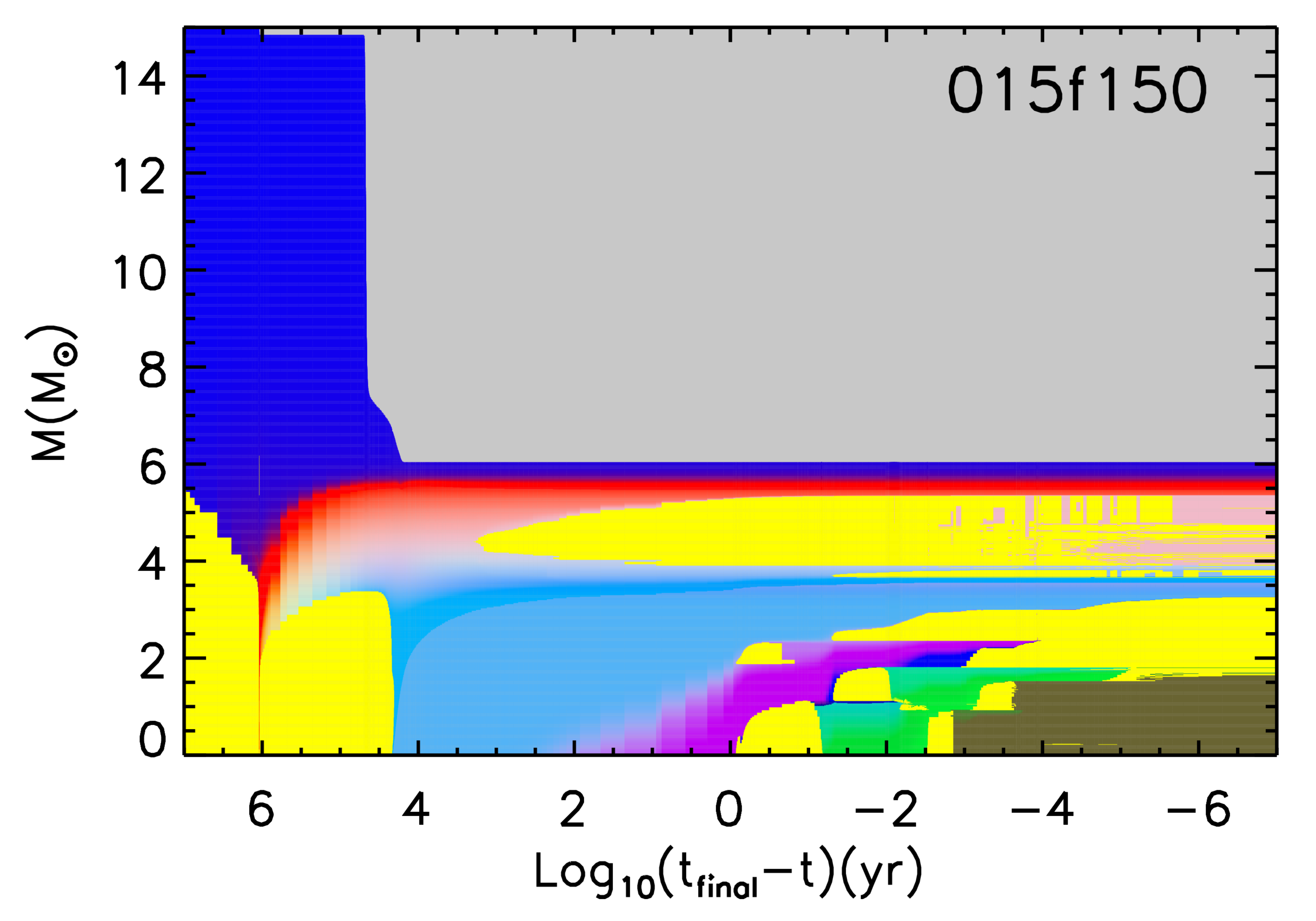}{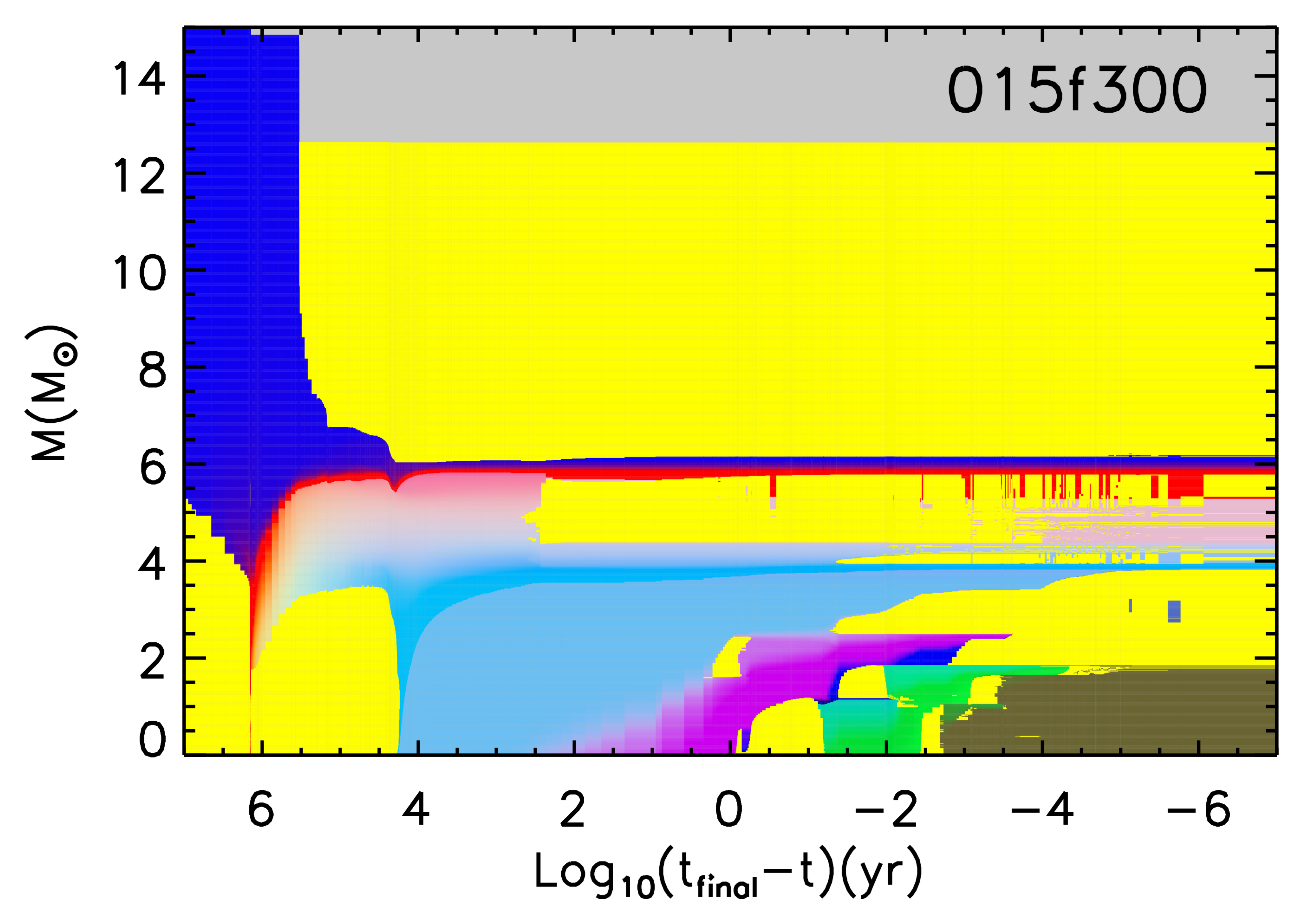}
\plottwo{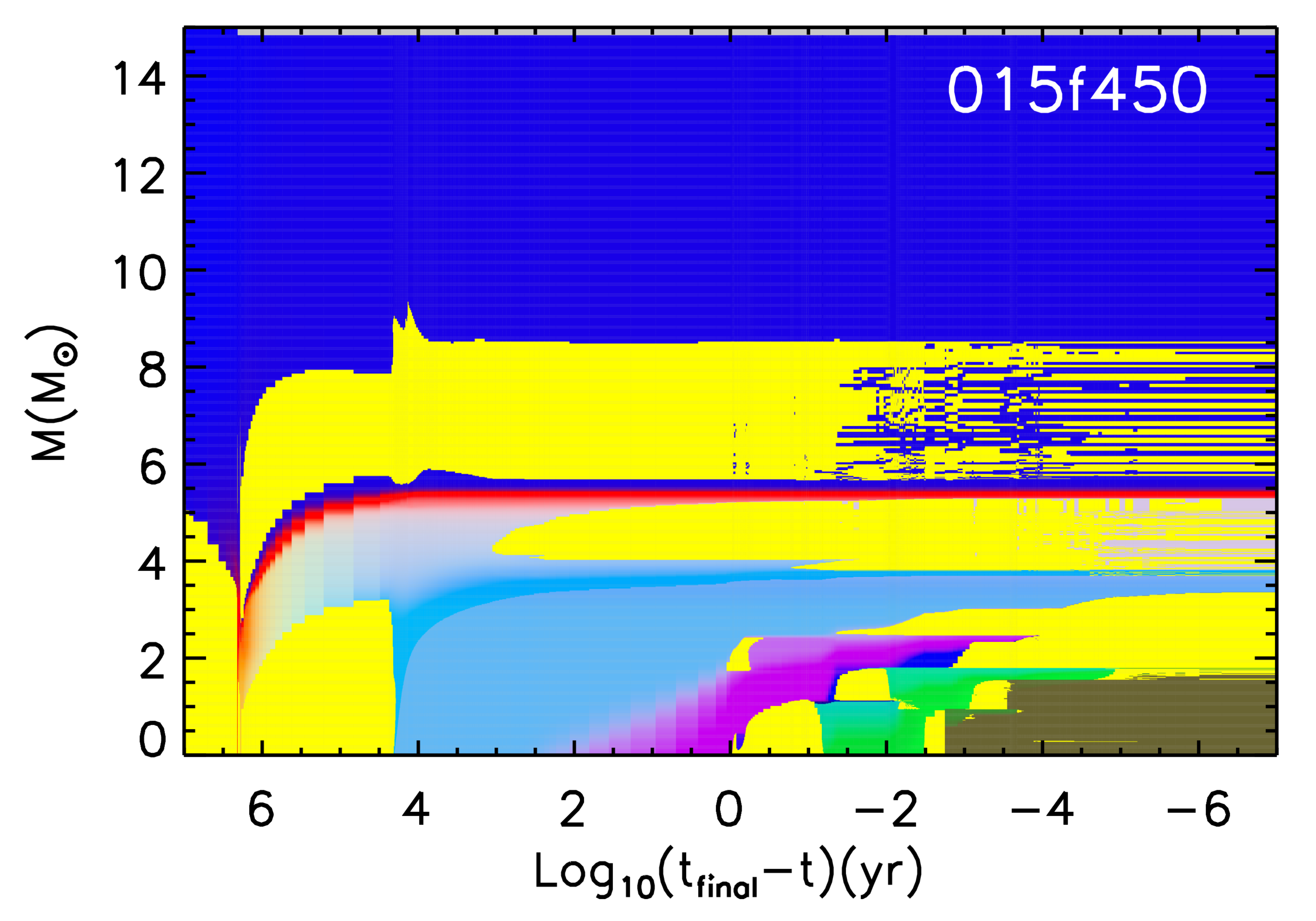}{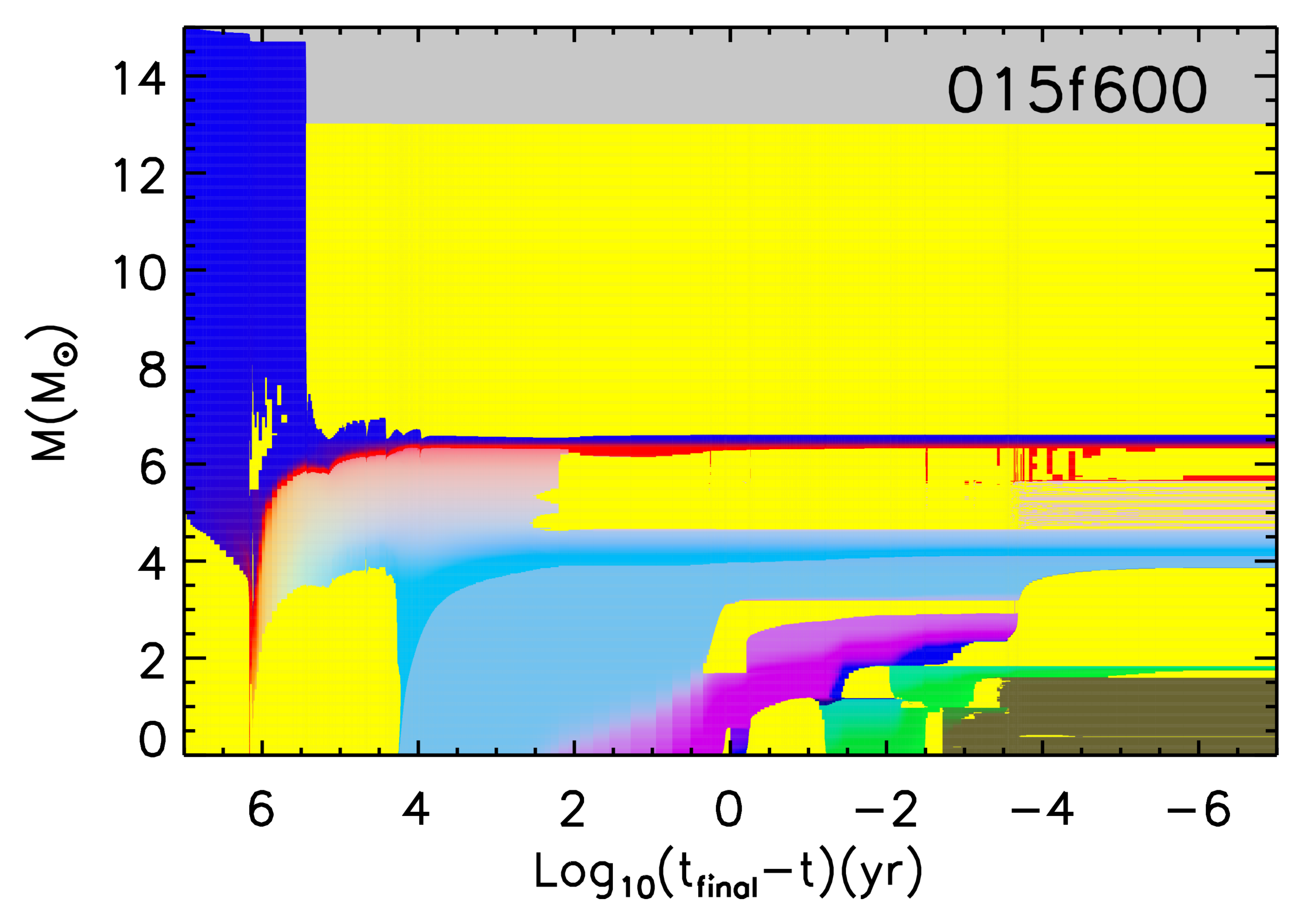}
\epsscale{0.5}
\plotone{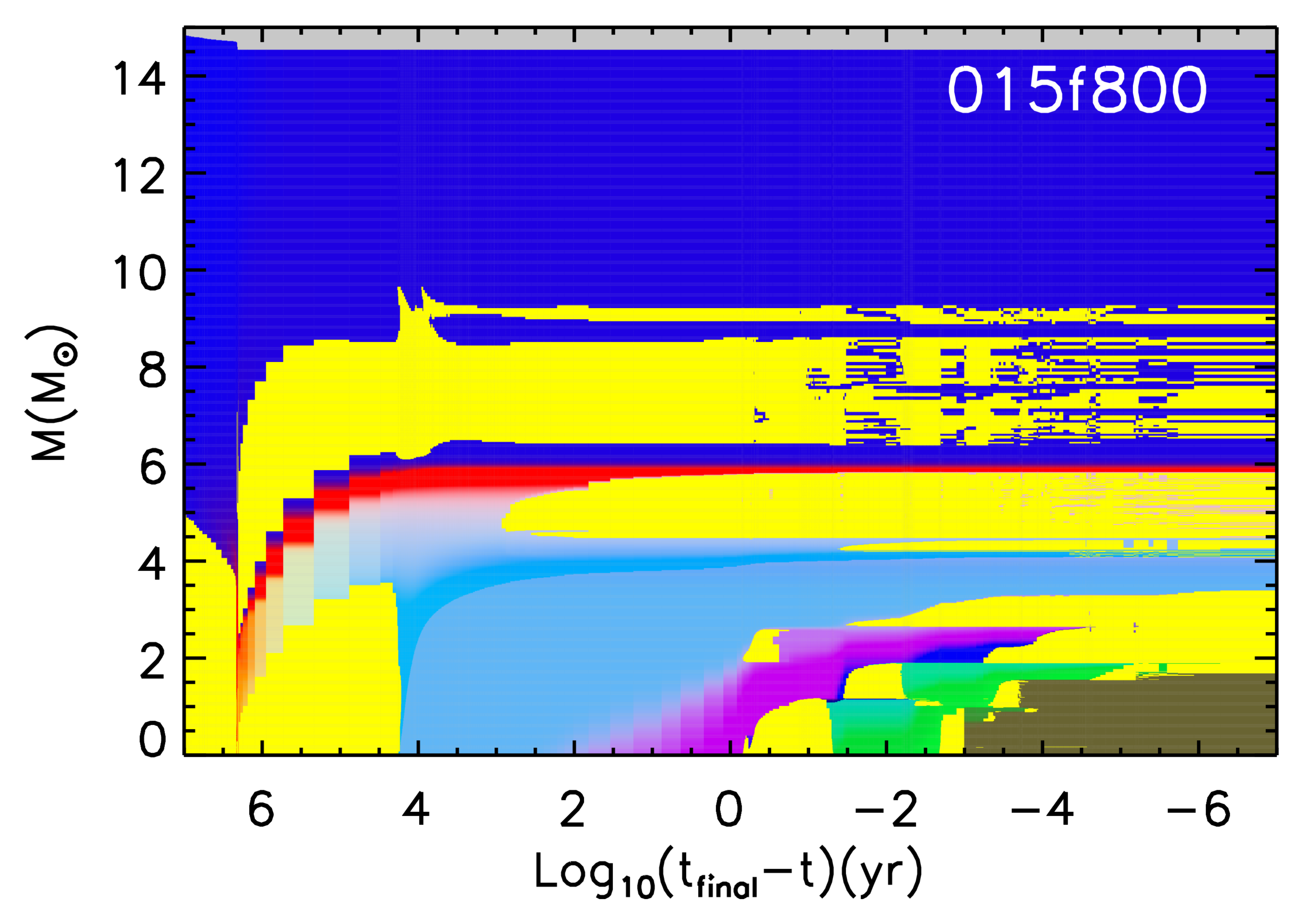}
\caption{Kippenhahn diagrams of the rotating 15 \msun\ models of [Fe/H]=--5. The color coding is the same as in Figure \ref{fig:kipnorot}.
\label{fig:kiprotf15}}
\end{figure*}

\begin{figure*}[p]
\epsscale{1.0}
\plottwo{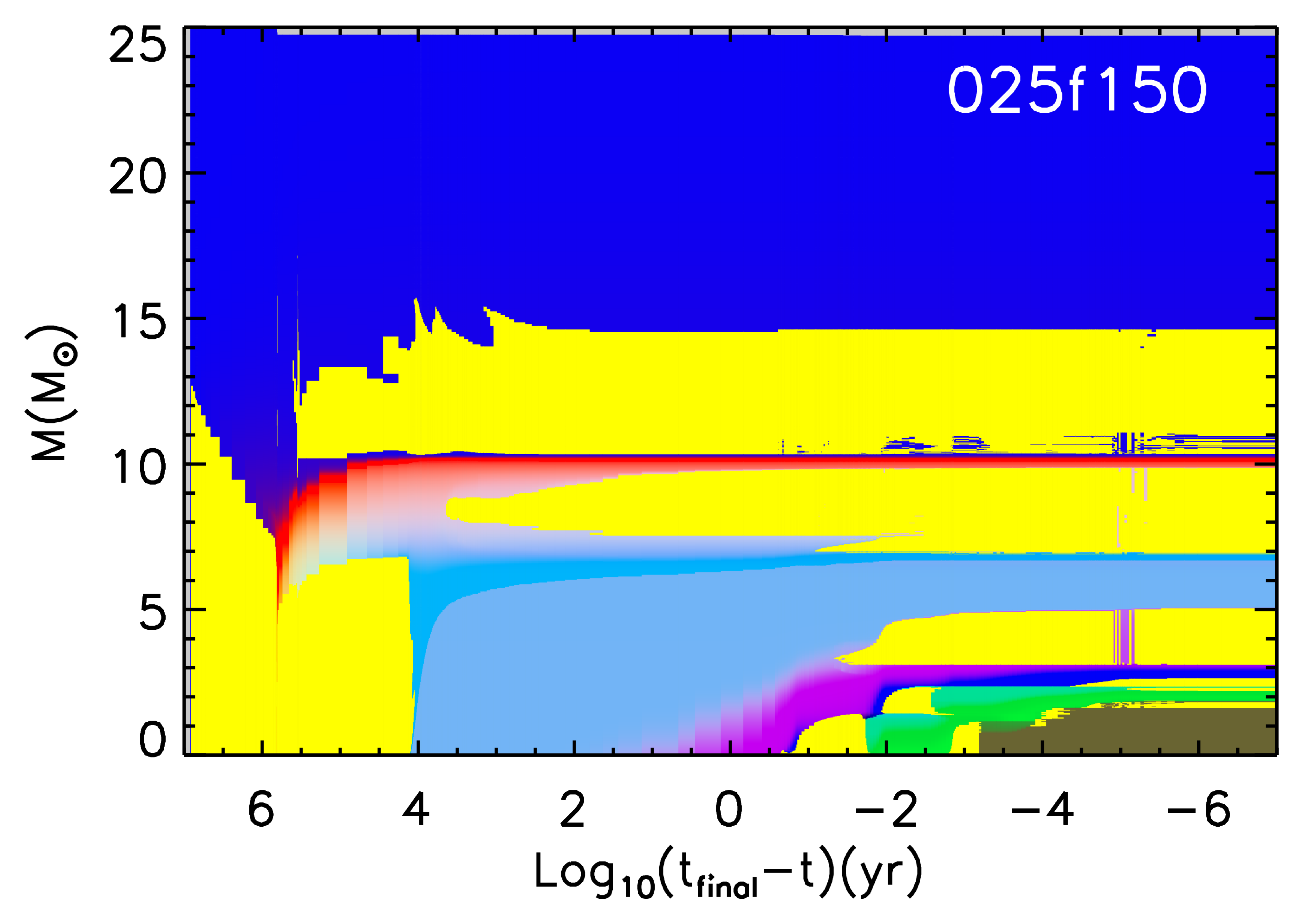}{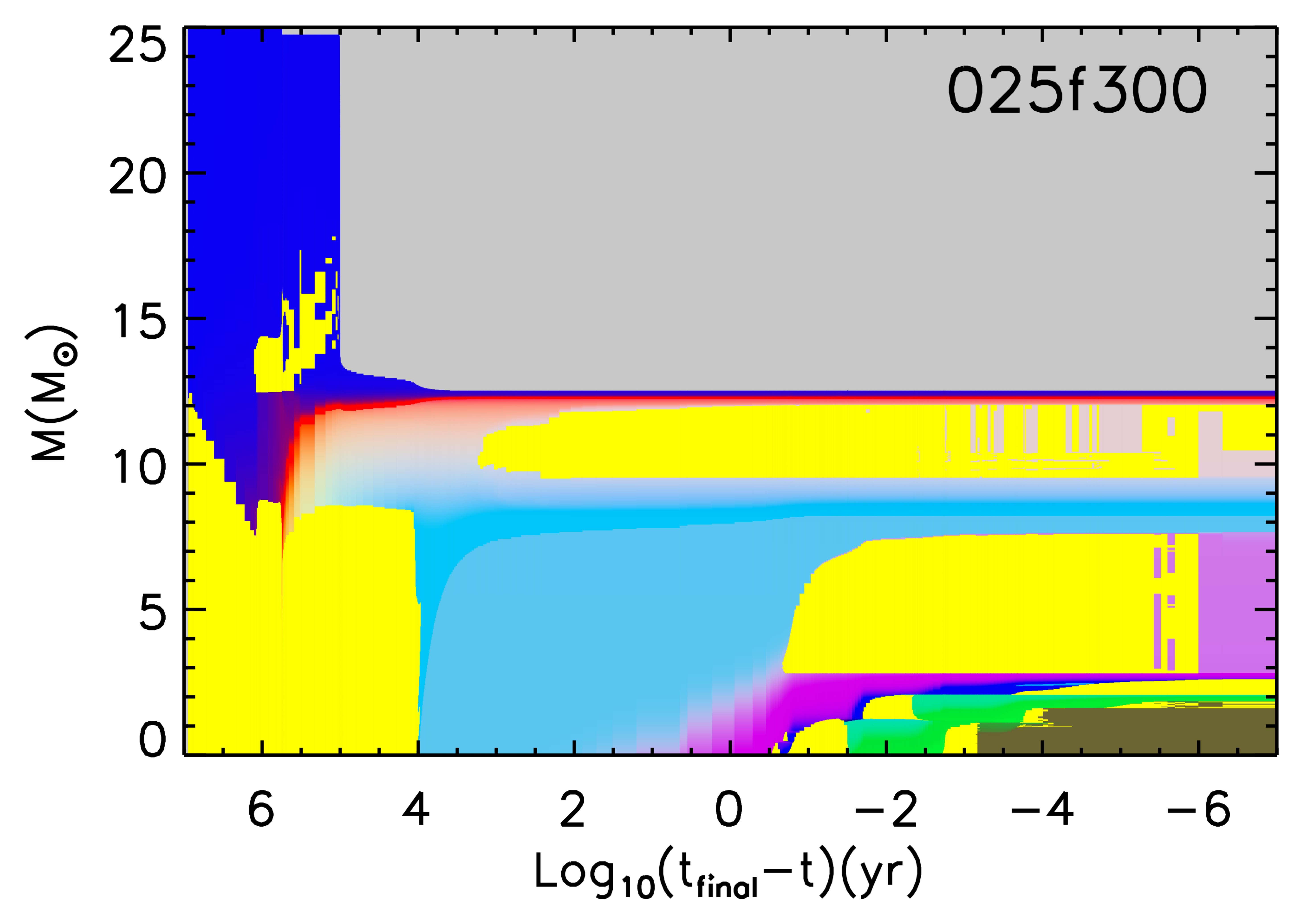}
\plottwo{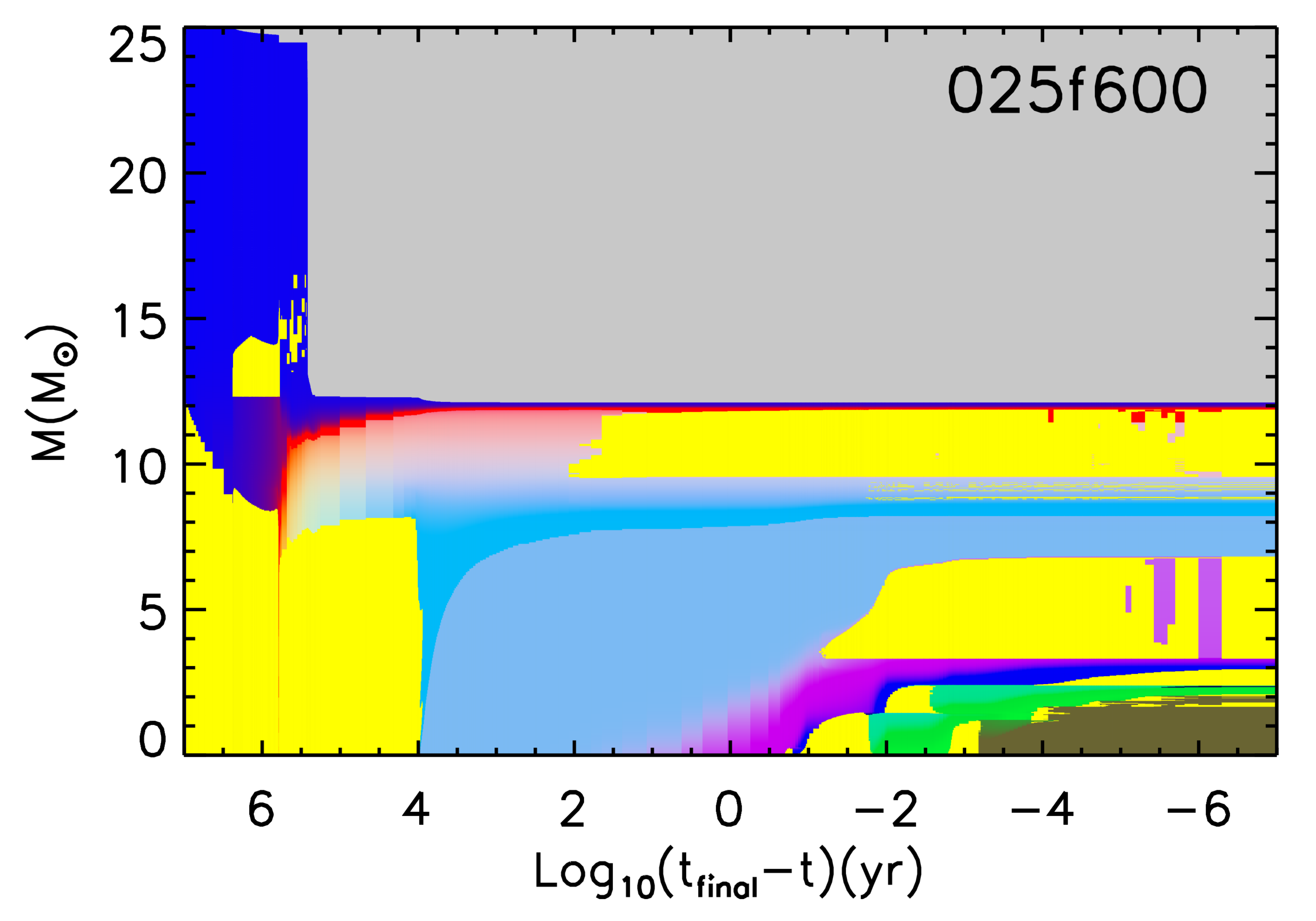}{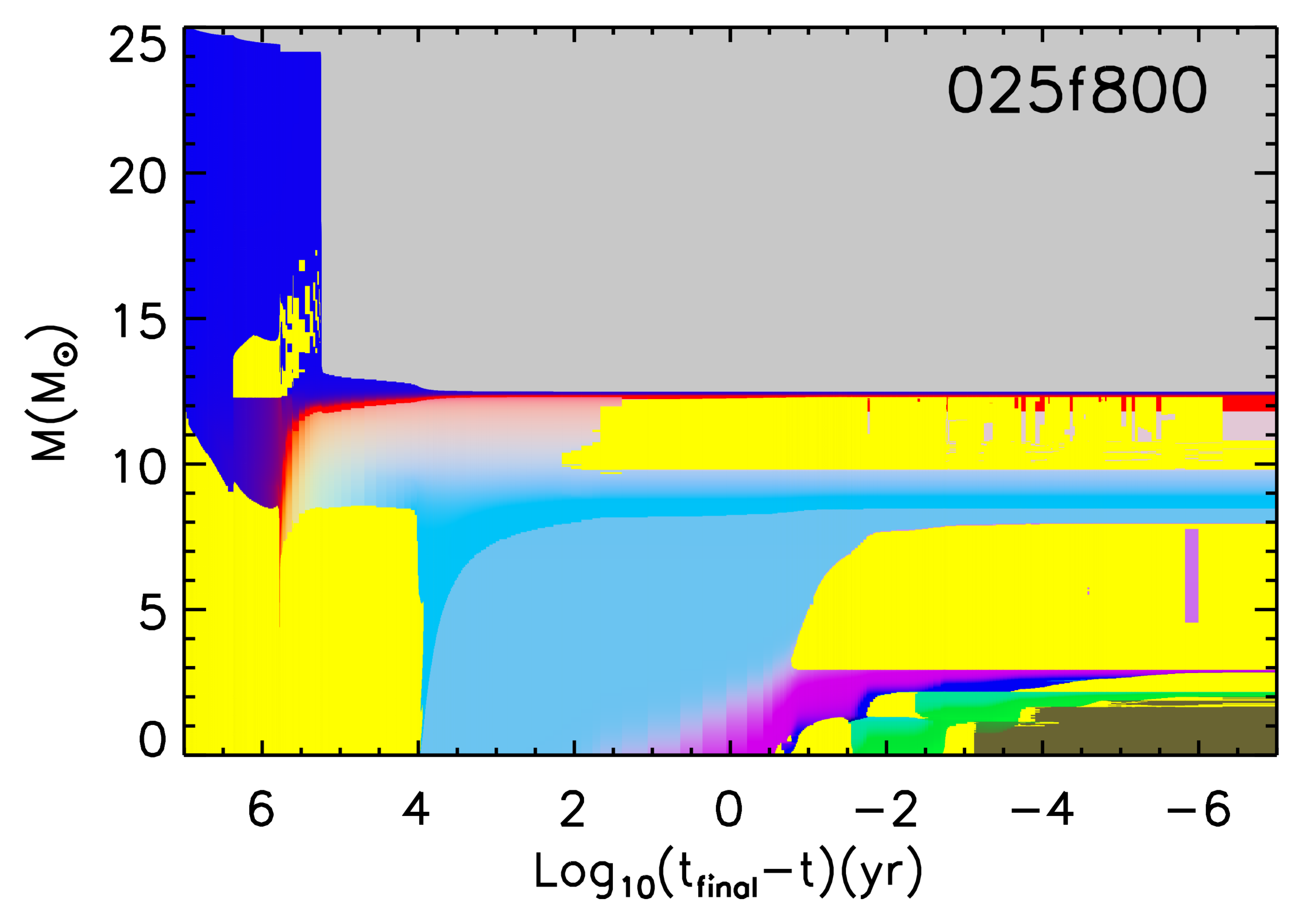}
\caption{Kippenhahn diagrams of the rotating 25 \msun\ models of [Fe/H]=--5. The color coding is the same as in Figure \ref{fig:kipnorot}.
\label{fig:kiprotf25}}
\end{figure*}

\begin{figure*}[p]
\epsscale{1.0}
\plottwo{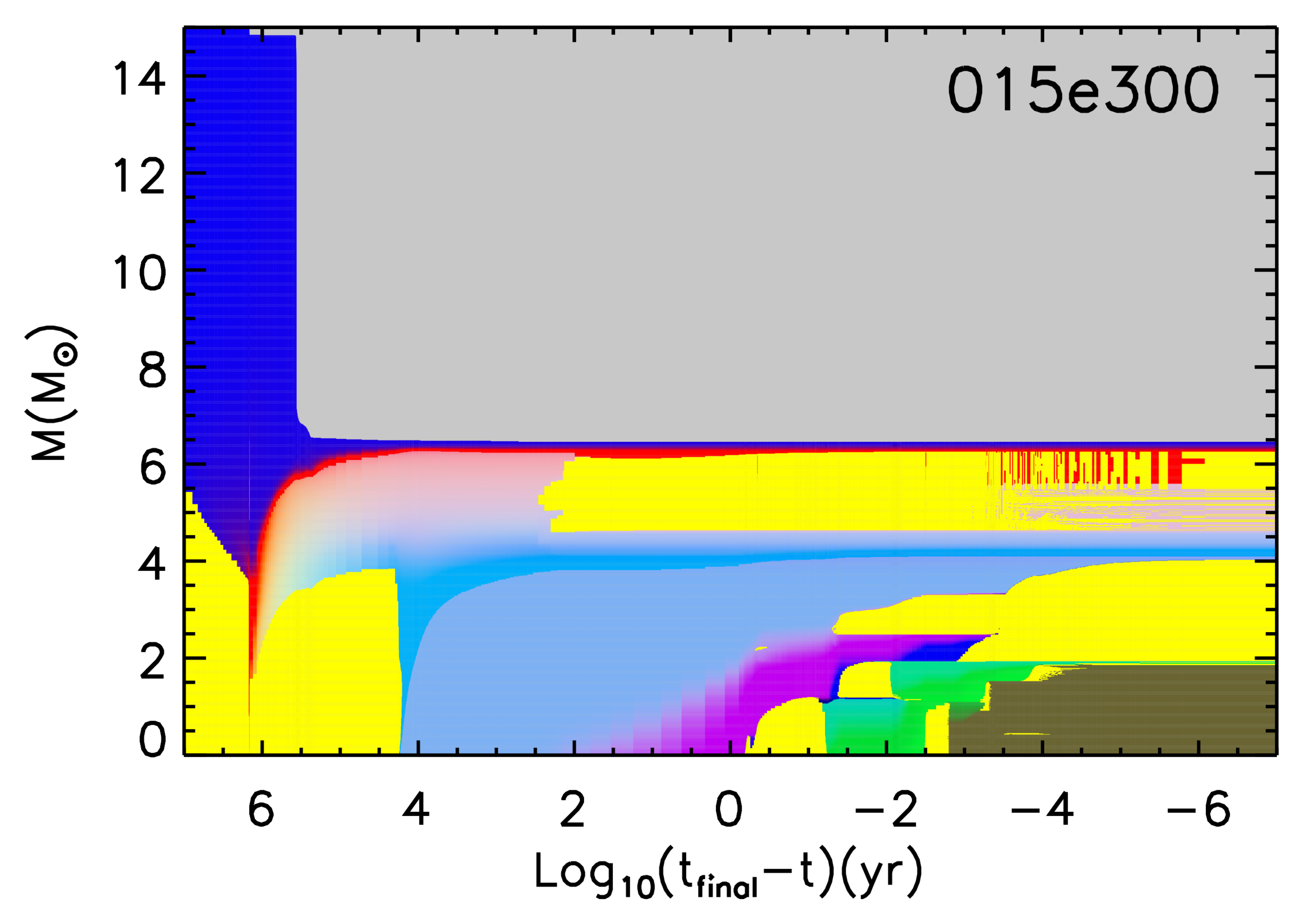}{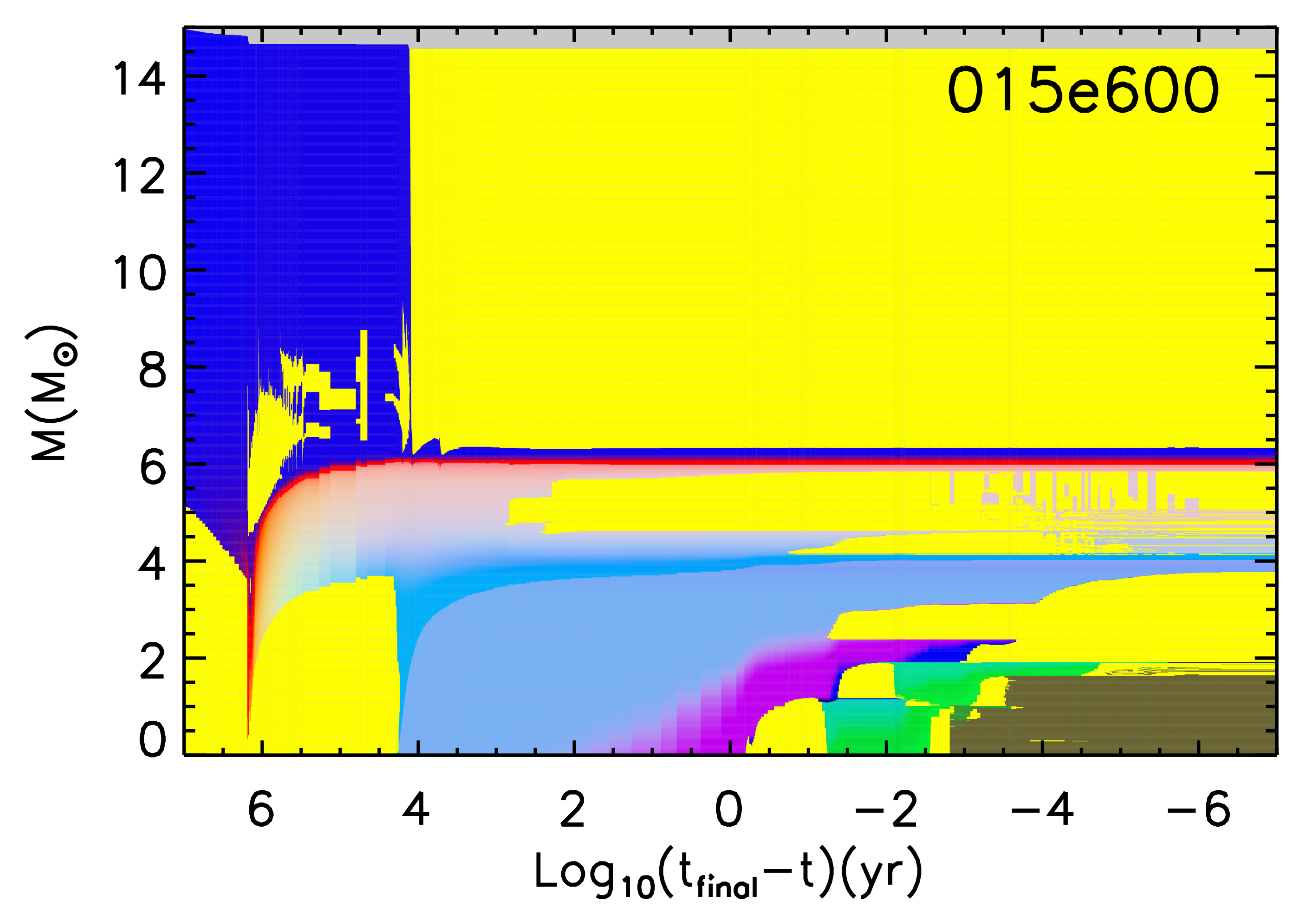}
\epsscale{0.5}
\plotone{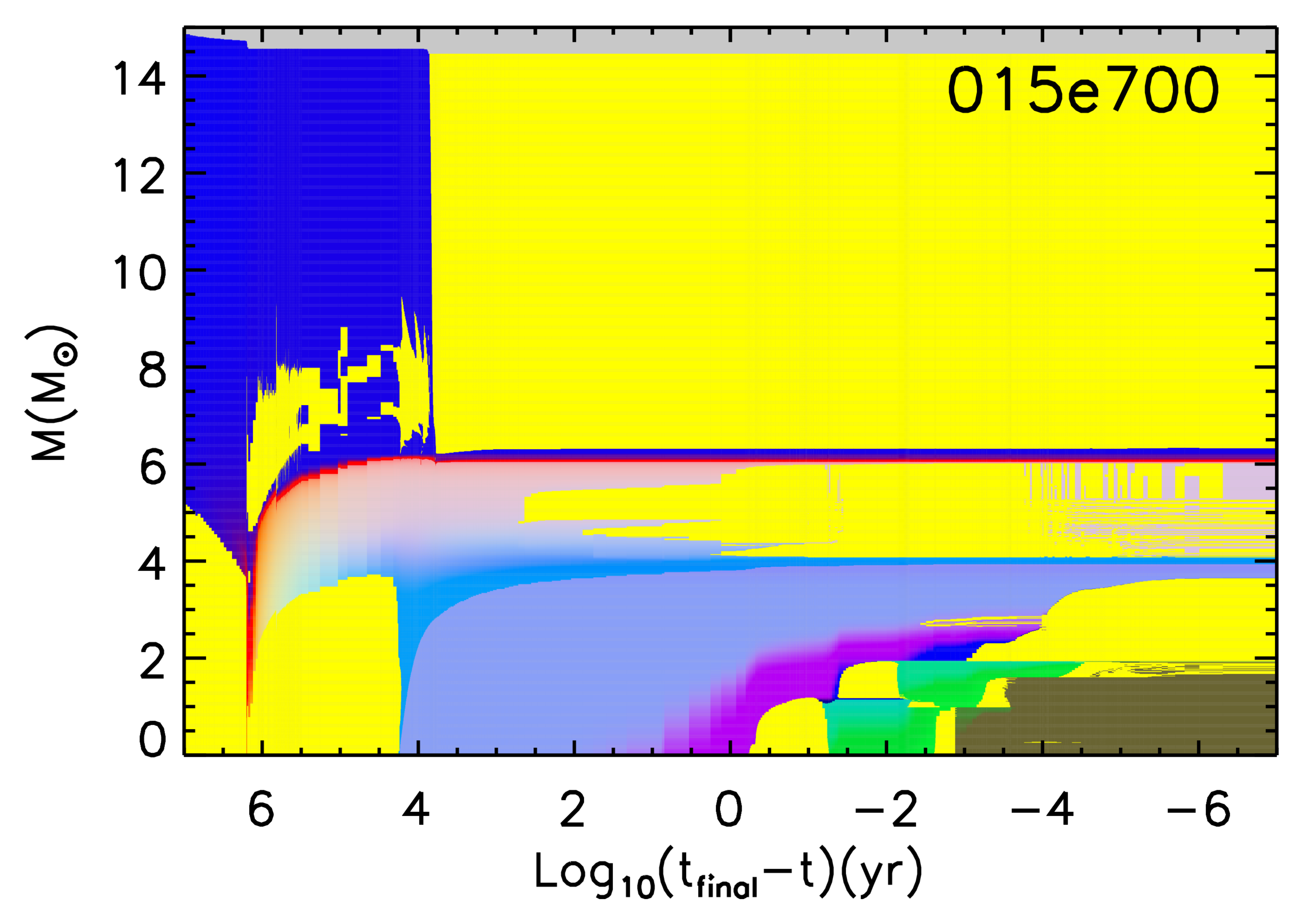}
\caption{Kippenhahn diagrams of the rotating 15 \msun\ models of [Fe/H]=--4. The color coding is the same as in Figure \ref{fig:kipnorot}.
\label{fig:kiprote15}}
\end{figure*}

\begin{figure*}[p]
\epsscale{1.0}
\plottwo{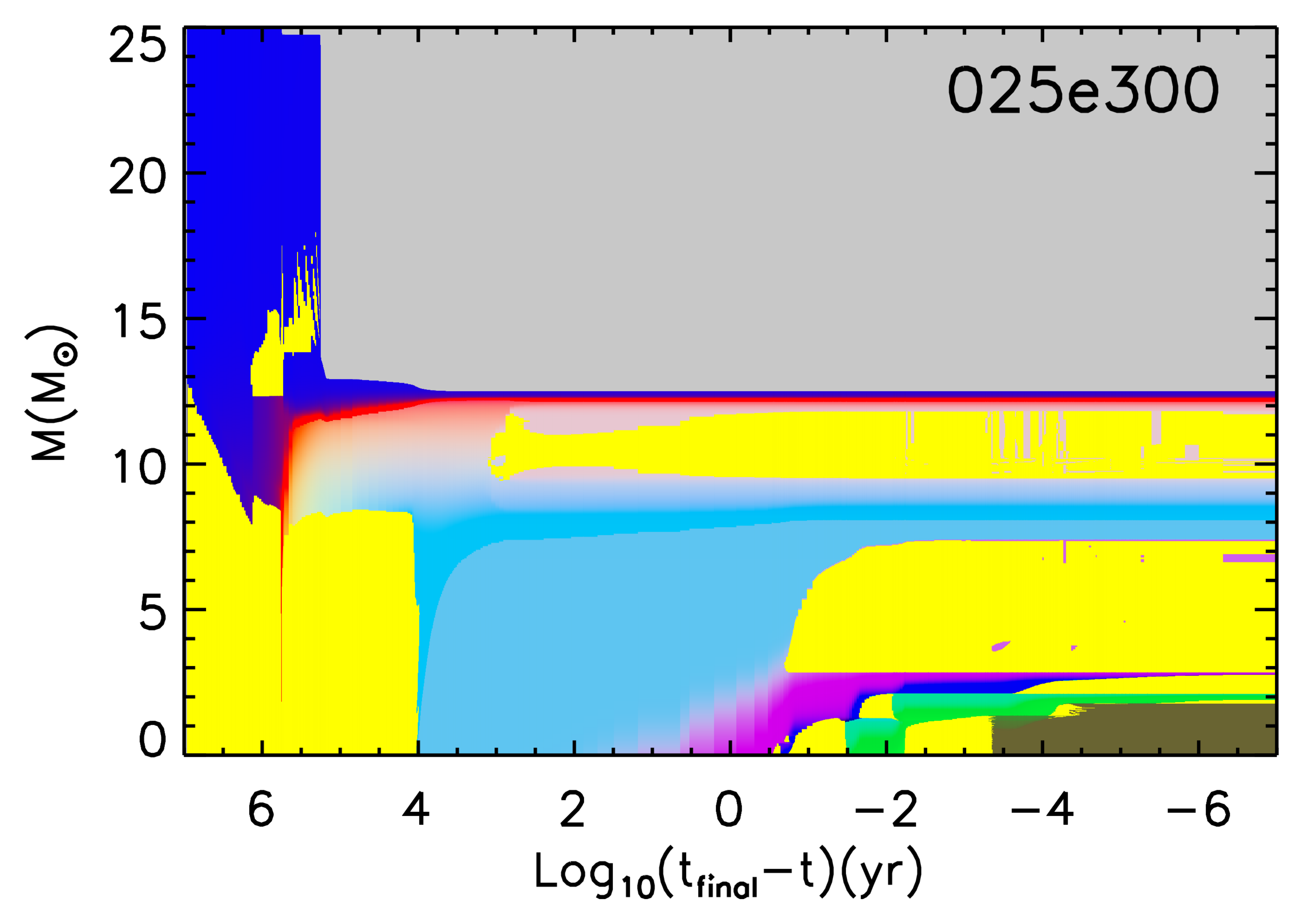}{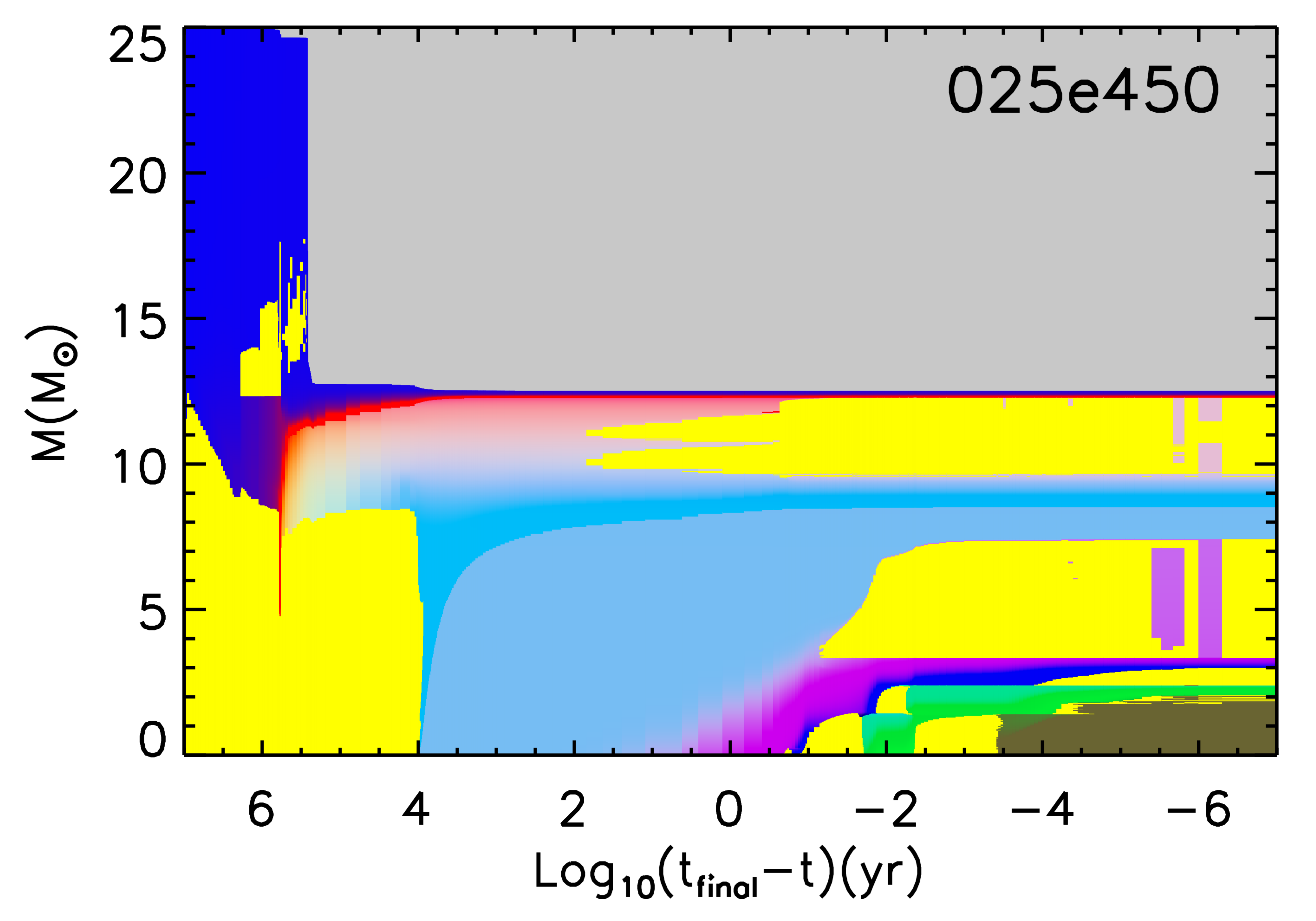}
\plottwo{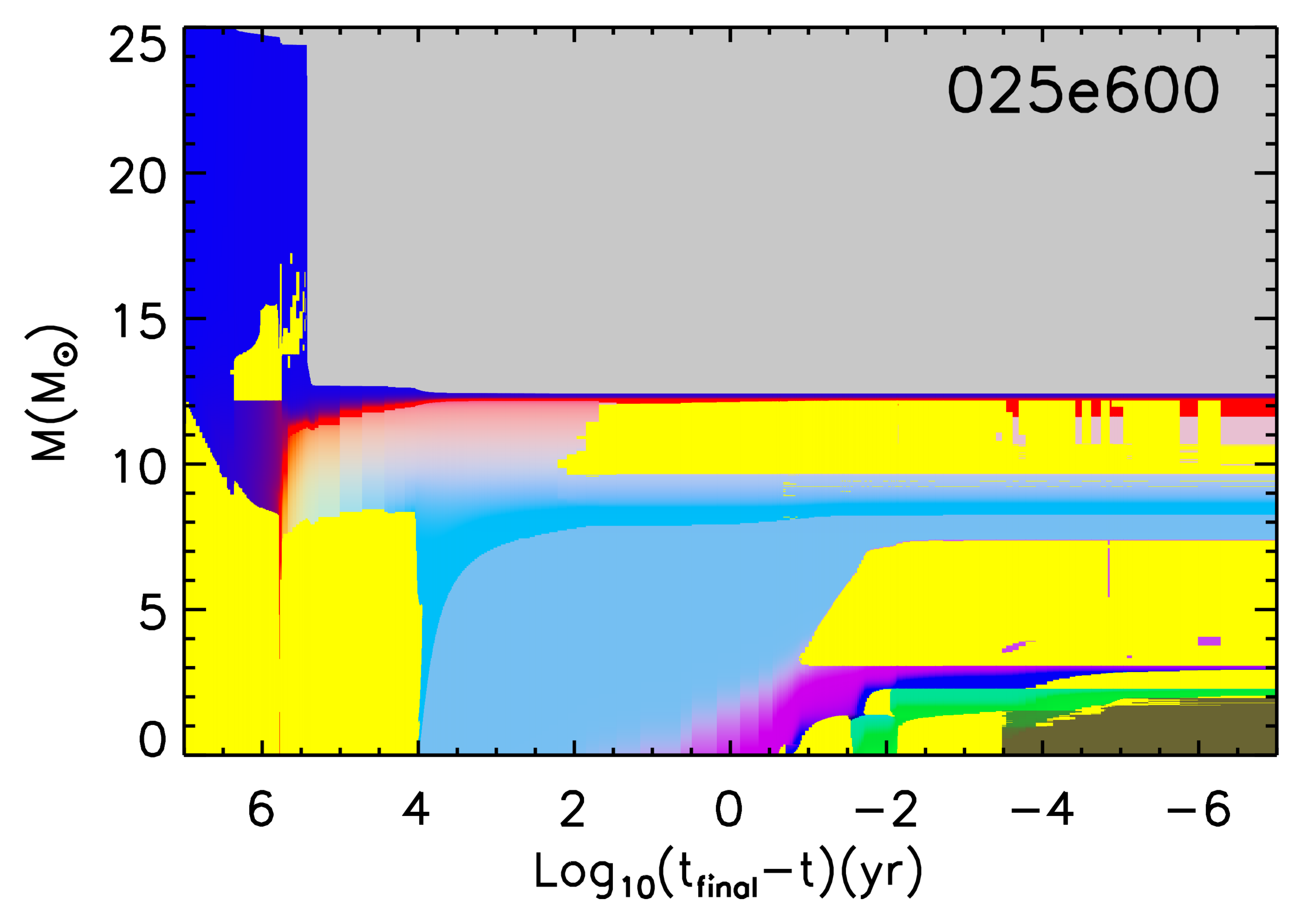}{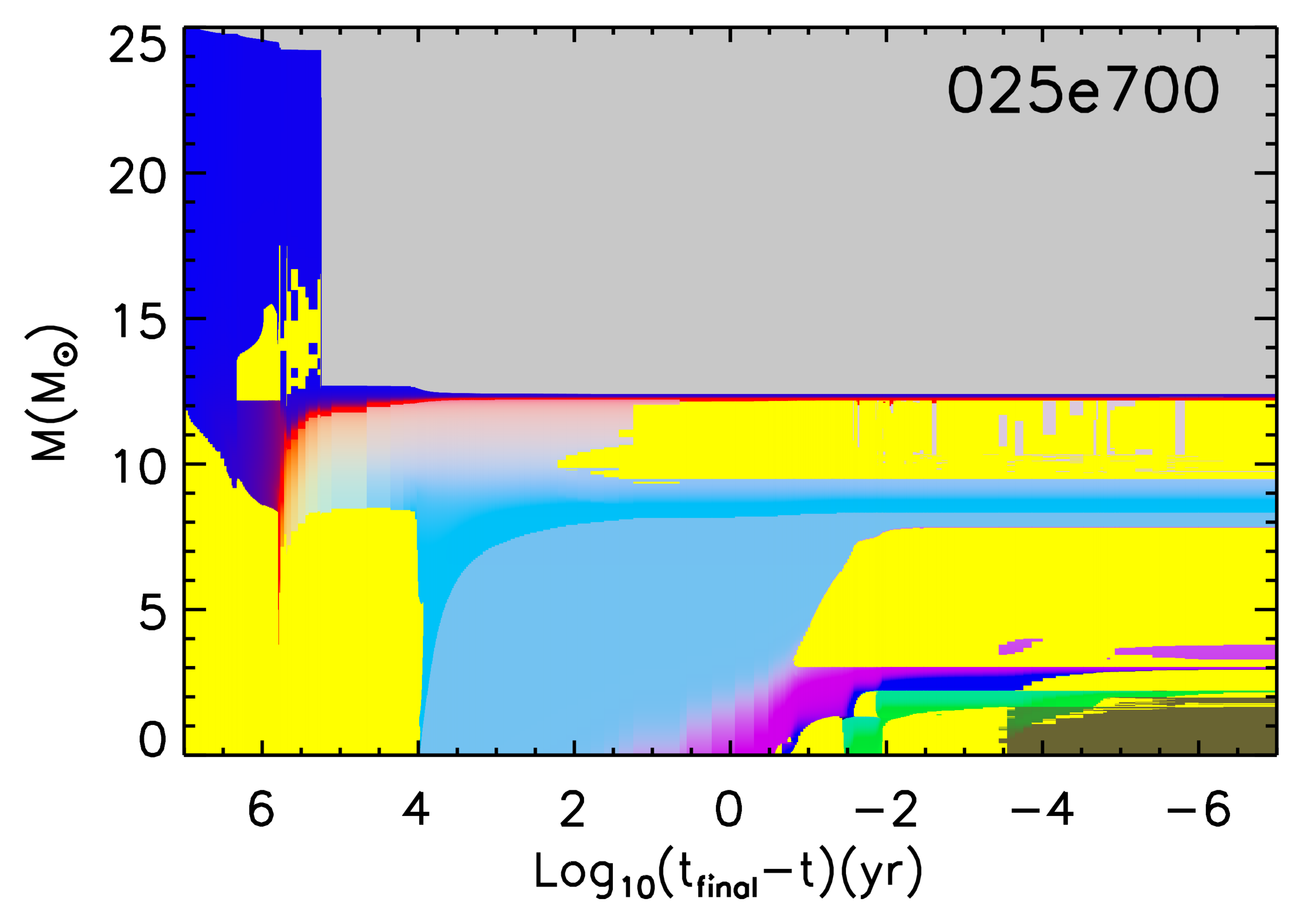}
\caption{Kippenhahn diagrams of the rotating 25 \msun\ models of [Fe/H]=--4. The color coding is the same as in Figure \ref{fig:kipnorot}.
\label{fig:kiprote25}}
\end{figure*}

\subsection{The role of rotation} \label{sec:rotation}

As already extensively discussed in literature \citep[see, e.g.,][CL13, LC18, and references therein]{meynet:06}, the inclusion of some angular momentum changes many evolutionary properties of the stars; among the others, it activates instabilities that drive some mixing of the matter in layers that would otherwise be in radiative equilibrium, forces a star (through the centrifugal force) to inflate its mantle much more (and much faster) than its non rotating counterpart at the end of the central H burning phase and triggers episodes of (dynamical) mass loss.
The red (Set Z), green (Set F) and blue symbols (Set E) in Figure \ref{fig:ini_momang} show the relation between the total angular momentum injected in each model and the initial rotational velocity (at the beginning of the central H burning phase). Note that the amount of angular momentum injected in stars with the same rotational velocity scales directly with the metallicity because the higher the metallicity the more expanded the star in Main Sequence.
The diamonds refer to the 15 \msun\ while the dots to the 25 \msun. The corresponding initial values of $\Omega / \Omega_{crit}~$ are shown in Figure \ref{fig:osuoc}. The same Figure shows that this ratio increases in central H burning and that the fastest rotating models reach the break up velocity before the central H exhaustion. Note that the expansion of the surface that occurs in central H burning would lead, per se, to a decrease of $\Omega / \Omega_{crit}$ in absence of transport of the angular momentum from the interior towards the surface. It is the outward transport of angular momentum that forces $\Omega / \Omega_{crit}$ to increase. In stars that are metal rich enough to lose a substantial amount of mass in MS, $\Omega / \Omega_{crit}$ is kept well below 1 by the efficient angular momentum removal operated by mass loss, while the lack of an efficient mass loss at zero and extremely low metallicity leads to the pile up of angular momentum to the surface and hence to the continuous increase of $\Omega / \Omega_{crit}$. The most external layers in which such a ratio reaches the value of one become unbound and they are lost in the interstellar medium. However, the total amount of mass lost by this phenomenon is quite modest because the models reach the break up velocity only towards the end of the central H burning phase (Figure \ref{fig:osuoc}). In all explored cases, the amount of mass lost by the models due to such a phenomenon never exceeds 1 \msun. It is worth noting that none of our models approaches the condition of quasi homogeneous mixing \citep[see, e.g.,][]{yoon:12,banerjee:19}, though a few of them start their evolution with a rotation velocity very close to the break up one.

The continuous stirring of matter caused by the rotation induced instabilities leads to more extended H convective cores %(simulating in some sense the overshooting) 
and also to the diffusion of the products of the central burning in the H rich mantle. The trend of the He core mass versus the initial rotation velocity is shown in Figure \ref{fig:mhe}. There is a general increase of the He core mass with the initial equatorial rotation velocity $\rm v_{\rm ini}$, even if with some exceptions due to the complex interplay among the various convective zones that develop during the evolution of the stars. The synthesis of fresh C by the $3\alpha$ reaction in rotating stars during H burning is quite similar to that occurring in the non rotating models: zero metallicity stars synthesize an abundance of C of the order of a few times $10^{-10}$ by mass fraction while no models of Set F and E require the activation of the $3\alpha$ during most of the central H burning. Similarly to what happens in the non rotating case, towards the end of the central H burning, models of both Sets Z and F synthesize enough C to increase the N abundance up to a value of the order of $10^{-6}$ by mass fraction in the whole He core, independently on the initial rotation velocity. Also rotating models of Set E behave like their non rotating counterparts, therefore the \nuk{N}{14} present in the He core of these models is simply the one that descends from the initial CNO abundance.

\begin{figure*}[p]
\epsscale{1.0}
\plotone{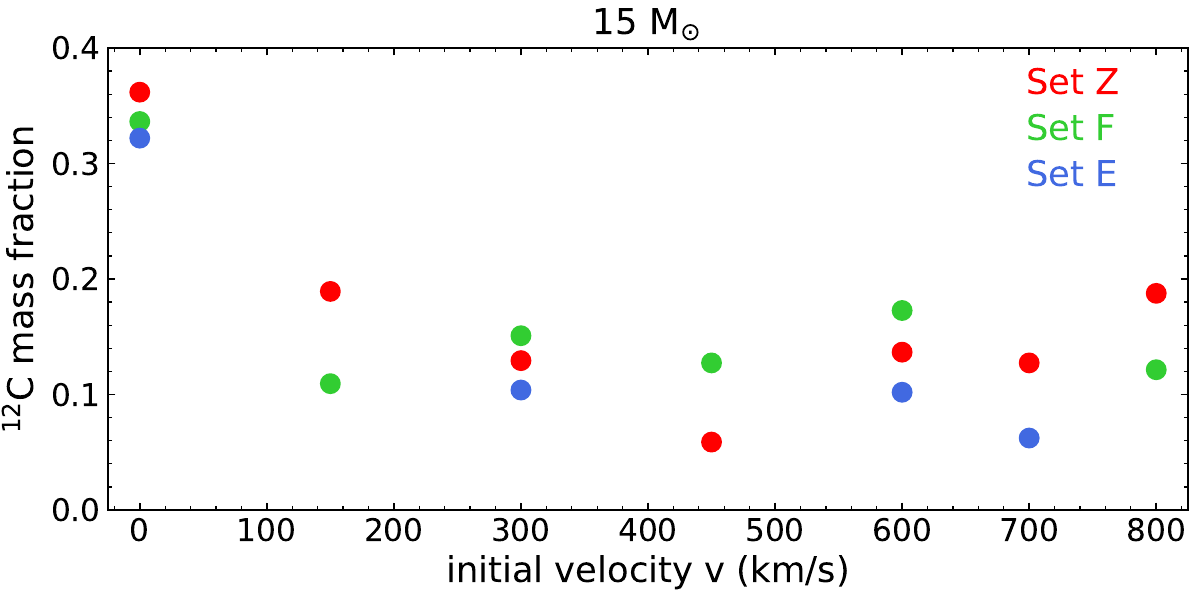}
\plotone{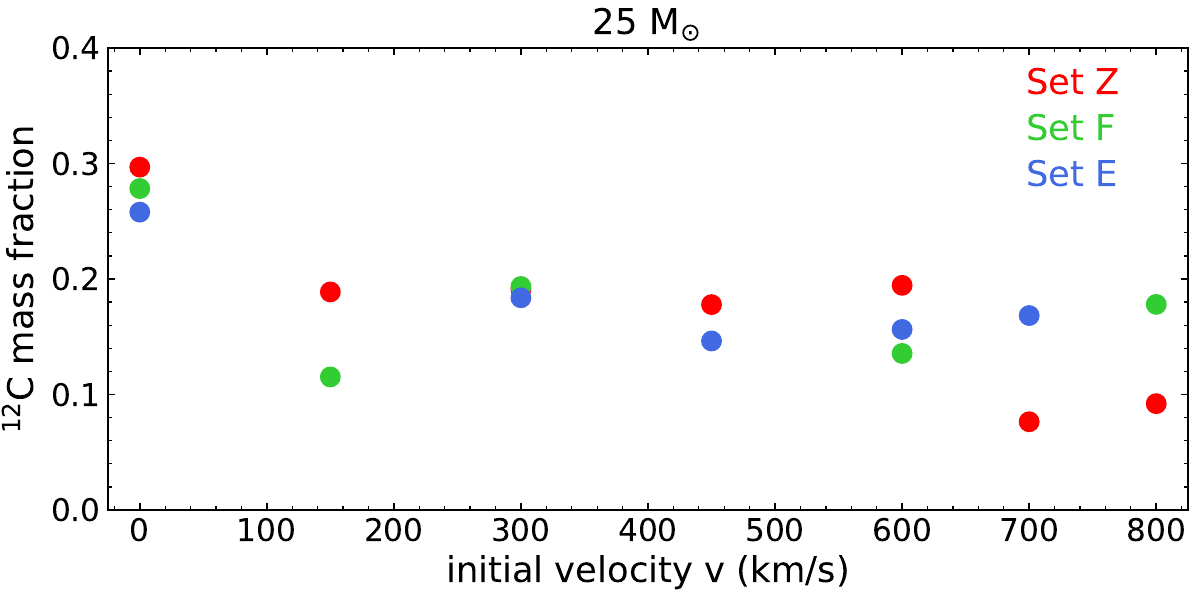}
\caption{\nuk{C}{12} abundance in mass fraction at the central He depletion as a function of the initial equatorial rotation velocity in the 15 (top panel) and 25 \msun cases (bottom panel).\label{fig:c12}}
\end{figure*}

The very high temperature at which the H burning occurs in models of Set Z and F leads to the He ignition and to the formation of a convective core immediately after the central H exhaustion when the models are still on the Blue side of the HR diagram. The rotation induced instabilities, that in H burning led to both an increase of the H convective core and to the diffusion of the products of the nuclear burning into the H rich mantle, in He burning continuously stir matter between the He convective core and the H burning shell. The first consequence of such a stirring is that, similarly to what happens in H burning, the final CO cores are more massive than in the non rotating case (Figure \ref{fig:mco}). The second one is that the continuous ingestion of fresh He in the convective core leads to a lower final abundance of \nuk{C}{12} in the He exhausted core. Figure \ref{fig:c12} shows the final \nuk{C}{12} abundance as a function of the initial rotation velocity for both masses. The third one is that, as already found by \cite{meynet:02a,CL13} a continuous slow exchange of matter between the He convective core and the H burning shell leads to a large enhancement of the products of the CNO (chiefly \nuk{N}{14} and \nuk{C}{13}) in both the He convective core and the whole radiative zone that separates the border of the convective core from the H burning shell.  We call such a strong continuous reprocessing of matter between the He and the H burning, entanglement. The consequences of this entanglement will be discussed in Sect.\ref{sec:nucleosynthesis}. 

If the initial rotation is fast enough, a significant mixing of C might drive the formation of H convective shell, that grows progressively in mass producing a shape that in the Kippenhahn diagram looks like an eagle's beak. Such a peculiar H convective shell forms in the 025z700, 025z000, 015f450, 015f800, 015e600, and 015e700 models. Though there is a clear general trend that shows that such a convective shell forms in the fastest rotating stars, one should not expect a strict correlation because rotation leads simultaneously to an expansion of the H mantle (lower densities and temperature imply lower H burning rates), to the increase of the C abundance in the H rich mantle (which implies higher H burning rates) and to larger He core masses (that would speed up the evolution of the He core, reducing the time available for the secular instabilities to bring freshly synthesized C in the H rich mantle). The Kippenhahn diagrams of all the rotating models are shown in Figures \ref{fig:kiprotz15} to \ref{fig:kiprote25}.

The path of all rotating models in the HR diagram reflects the complex interplay occurring among the various convective zones that form during the evolution of these stars and is shown in the various panels of Figure \ref{fig:hr}. Also the rotating models ignite He as Blue Super Giant (BSG) but, at variance with their non rotating counterparts that spend all their central He burning lifetime as BSG, they are able (in many cases) to almost reach their Hayashy track on a nuclear timescale. The rotating 15 \msun\ of Set Z, for example, reach their Hayashy track towards the end of the central He burning, consuming up to 20\% of the He as a Red Super Giant (RSG). By the way, this means that extremely metal poor stars may populate the Hertzsprung Gap in central He burning. All models able to approach their Hayashi track, however, overcome their Eddington luminosity when their surface temperature drops below $\sim8000$ K, lose most of their H rich mantle on a dynamical timescale and turn back to a BSG configuration on a thermal timescale. Note that the super-Eddington regime is always reached in the atmosphere of the star, within $10^{-4}$ \msun\ from the surface, well before the formation of an extended convective envelope, in layers where the superadiabatic gradient practically coincides with the radiative one. This is a quite new result since the extremely metal poor stars (including the Pop III) have never been expected to experience mass loss events but, on the contrary, they were believed to evolve at constant mass unless their rotational velocity reaches the break up velocity. The model with $\rm v_{ini}=150~km/s~$ is the only one whose luminosity never overcomes the Eddington one and hence remains a RSG up to the final collapse. The rotating 25 \msun\ of Set Z, on the contrary, remain to the blue side of the HR diagram expanding only moderately during the central He burning. The reason is the presence of an H convective shell that slows down or even prevents the expansion of the models towards their Hayashy track. 
Only the models with an initial rotation velocity equal to 450 and 600 km/s reach their Hayashi track because of the smaller mass size of the H convective shell 
and only the one started with $\rm v_{ini}=600$ km/s exceeds its Eddington luminosity, loses most of its envelope becoming again a BSG. The path in the HR diagram of the rotating models of both Set F and E is quite similar to that of the models of Set Z. All of them ignite He as BSG and move on a nuclear timescale towards their Hayashy track. Once again, only the models that do not form an extended H convective shell become RSG, the others remain BSG all along their central He burning phase (see Figure \ref{fig:hr}). All rotating models that are able to reach their Hayashi track before the end of the central He burning overcome their Eddington luminosity when their surface temperature drops to 8000 K, start losing dynamically a large fraction of their H rich mantle becoming again BSG. 

We have already discussed in Sect \ref{sec:nrot:he} that the physical evolution of any model beyond the central He burning is basically independent on the initial mass and metallicity since the two main drivers of the evolution in the advanced burning phases are the CO core mass and the amount of \nuk{C}{12} left by the He burning. 
In addition to this, the advanced burning phases are almost independent also on the initial rotation velocity because a) the CO core is so compact that the centrifugal force is not able to distort significantly the shape of the star with respect to a spherical configuration and b) the rotation induced instabilities do not have time to grow so that nor the angular momentum nor the chemicals are transported any more by the rotation driven instabilities. The only exception is the transport of the angular momentum in the convective regions: since we assume instantaneous mixing of the angular momentum in all the convective regions, we force $\Omega$ to become flat in each convective region even in the advance burning (as discussed in CL13).

There are two additional distinctive feature caused by rotation. The first one is that rotation inhibits the penetration of the He convective shell in the H rich mantle \citep[feature that in non rotating models leads to the synthesis of a large amount of primary \nuk{N}{14};][and reference therein]{heger:10,LC12}. Basically all authors who published over the years models of zero or extremely low metallicity found that in some mass intervals the He convective shell penetrates the H rich mantle whose main effect is to produce a large amount of primary \nuk{N}{14}. In the non rotating models presented in this work such a merging occurs in both the 25 \msun\ of Set Z and F (see Figure \ref{fig:kipnorot}). Vice versa, none of the corresponding rotating models shows such a merging (with one exception). The main reason is that the stirring of the matter driven by the rotation leaves a quite smooth He profile at the central He exhaustion (instead of the essentially vertical profile typical of the non rotating models, see also CL13 and LC18). As a consequence, the He convective shell forms in a region of variable He and therefore its abundance is much lower than in the non rotating case (where it is of the order of 1); such a lower abundance reduces significantly the energy produced by the $3\alpha$ and therefore the capability of the outer border of the He convective shell to penetrate into the H rich layers. 
It must also be considered that the distance, in mass, between the He core and the CO core masses increases with the initial rotation velocity, and therefore the base of the H rich mantle is further away from the outer border of the He convective shell than in the case of non rotating models. The only exception to this general trend is the 25 \msun\ of Set Z, rotating at 300 km/s. The reason is easily understood by looking at the upper right panel in Figure \ref{fig:kiprotz25}. In this case, in fact, the H convective shell is particularly wide and extends down to the region of variable He since the beginning of the central He burning. This means that 1) the He convective shell forms much closer to the H rich mantle and 2) the H profile is so steep that any ingestion (even modest) of H immediately leads to a H burning runaway event because of its high concentration (almost the initial one). 

\begin{figure*}[p]
\epsscale{1.0}
\plotone{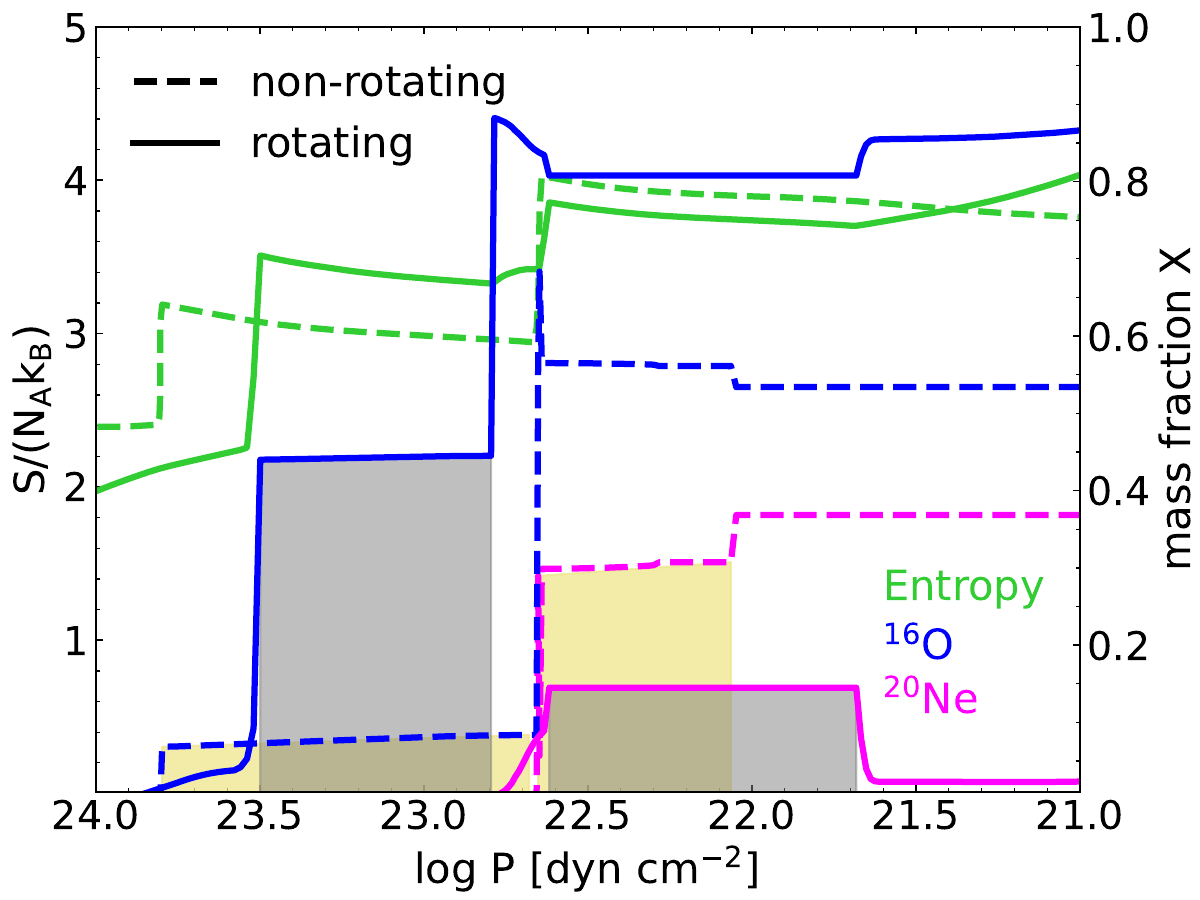}
\caption{Comparison of the entropy profile (per nucleon in units of Boltzmann constant $k_{\rm B}$, green lines) as a function of the internal pressure $P$ between the non-rotating 015e000 model (dashed lines) and the rotating 015e300 one (solid lines). Blue and magenta lines show the abundances in mass fraction of \nuk{O}{16} and \nuk{Ne}{20}, respectively, in the two cases. The yellow and the grey zones show the O and Ne convective regions in the non-rotating and in the rotating models, respectively. \label{fig:entropy}}
\end{figure*}
The second distinctive feature of the rotating models is the penetration of the O convective shell well within the C rich zone soon after the central Si exhaustion in many rotating models. In particular this phenomenon occurs in all the rotating 15 \msun\ models of the three metallicities (except for the models 015z150 and 015z800), while it does not occur at all in the 15 \msun\ non rotating models nor in any of the 25 \msun, rotating or not. The Kippenhahn diagrams of the 15 \msun\ models (Figures \ref{fig:kiprotz15}, \ref{fig:kiprotf15}, and \ref{fig:kiprote15}) show clearly such an occurrence. The reason that leads most of the 15 \msun\ rotating models to experience the penetration of the O convective shell within the Ne and C rich layers is quite difficult to determine robustly because the rotating ones end the central He burning with a much lower C abundance (and hence, in turn, Ne abundance) with respect to their non rotating counterparts and this occurrence affects all the advanced evolutionary phases. However we note that both rotating and non rotating models develop an O convective shell that closely approaches the Ne rich layers but while in the first case the merging occurs, in the second one it does not. A closer look at the models at the time of the merging (i.e., soon after the disappearance of the Si convective shell) shows that the models with a lower Ne abundance (the rotating ones) show a smaller entropy jump at the base of the Ne burning shell with respect to their respective non rotating models. The role of the Ne abundance in determining the entropy jump may be understood by considering that such a jump would vanish as the Ne abundance would turn to zero because in this limiting case any difference, in the nuclear energy generation and in the chemical composition would disappear. Figure \ref{fig:entropy} shows, as a typical example, a comparison between the entropy profiles of the 015e000 (non-rotating, with no merging) and the 015e300 (mildly rotating, with an extended merging). The figure clearly shows the presence of a larger entropy barrier in the non-rotating model with respect to that present in the rotating one at the base of the \nuk{Ne}{20} rich zone (at $\rm log\ P \sim22.7$). To be absolutely clear, we are not saying that the smaller entropy barrier present in the rotating models determines \textit{per se} the penetration of the O convective shell in the Ne rich region but, simply, that a lower entropy barrier certainly favors (i.e., represents a smaller obstacle) the growth of the O convective shell. An additional point worth being stressed here is that also the criterion adopted to determine the outer (actually any) border of the O convective shell plays a pivotal role. The adoption of the Ledoux criterion instead of the Schwarzschild one, e.g., would obviously disfavour the penetration of the outer border of the convective shell across the chemical discontinuity. Note, however, that also in this respect a lower Ne abundance would in any case imply a lower gradient of molecular weight.
The possible occurrence of a merging of the O convective shell in the C rich zone is of great interest because it would reduce the yields of the products of the C burning and increase those of the O burning. The reason is that the merging spreads the products of the C burning down to the base of the O convective shell (where they are easily destroyed by the passage of the shock wave) and, vice versa, it spreads out the products of the O burning where they can survive to the passage of the shock wave \citep[see][and Sect \ref{sec:nucleosynthesis}]{ritter:18a,roberti:23a}. This problem certainly deserves a deeper analysis and we plan to address this problem in the next future. The 25 \msun\ models do not show such a behavior of the O convective because in this case, even in the models with a low \nuk{C}{12} mass fraction (and therefore, in turn, a low \nuk{Ne}{20} mass fraction), the outer border of the O convective shell always remains quite far from the O/Ne discontinuity.

\begin{figure*}
\epsscale{0.85}
\plotone{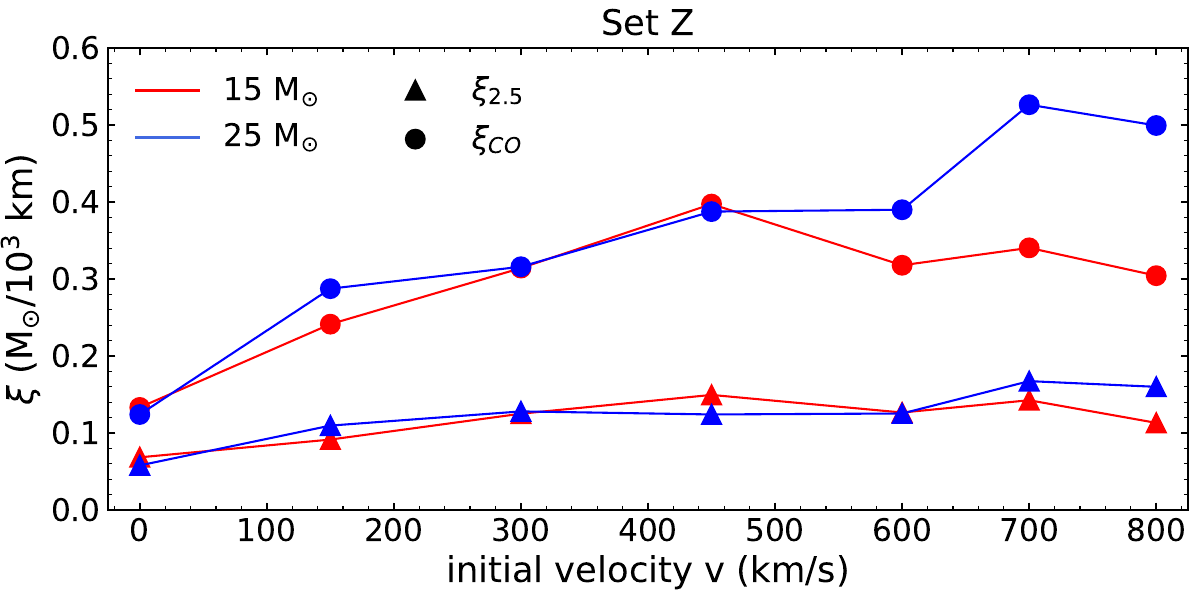}
\plotone{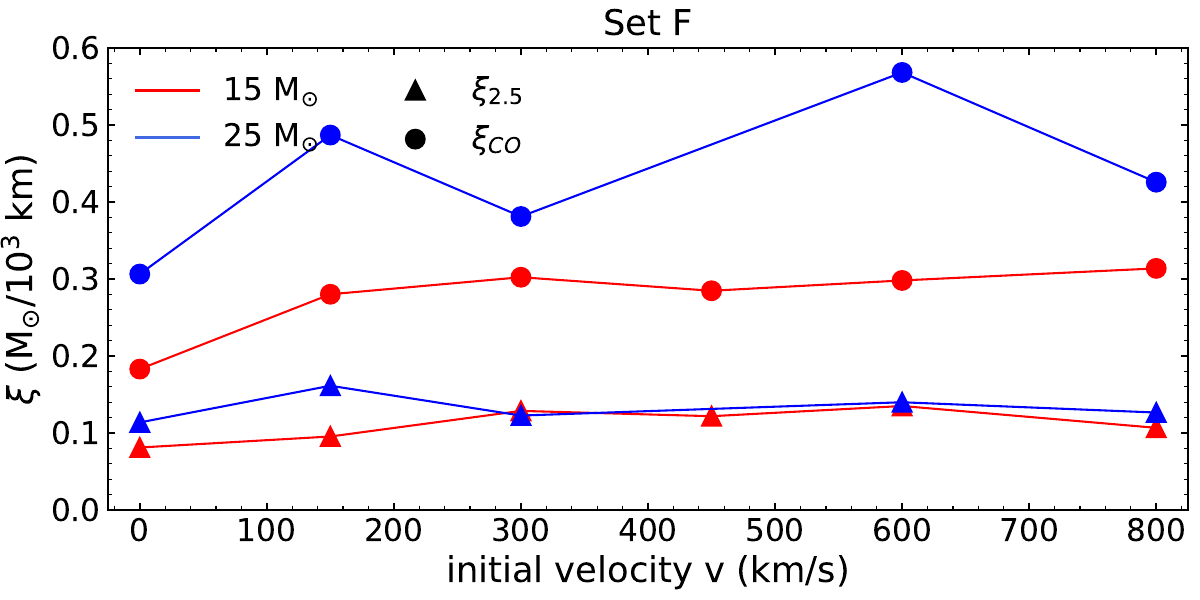}
\plotone{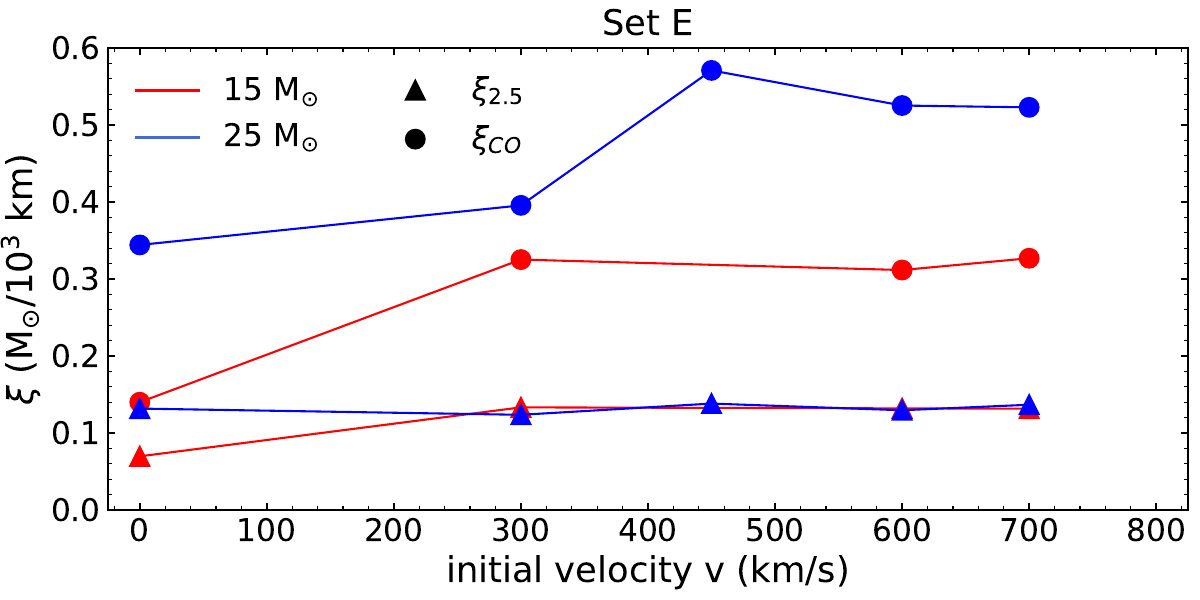}
\caption{The compactness $\xi$ of the star evaluated at 2.5 \msun\ (filled dots) and at the border of the CO core (filled triangles) at the presupernova stage in 15 \msun\ (red points) and 25 \msun\ (blue points)in Set Z (top panel), Set F (central panel), and Set E (bottom panel). \label{fig:csi}}
\end{figure*}

%===== PSN STAGE
The main physical properties of the stars at the onset of the core collapse are shown in \tablename~\ref{tab:final}. The final fate of a massive star, i.e., if a successful explosion will occur or not, depends on the interplay between the amount of energy stored in the shock wave (that would push matter outward) and the binding energy (that opposes to the ejection of the mantle). As already discussed in Sect \ref{sec:nrot:he}, the binding energy at the onset of the collapse is the result of the overlap of the various nuclear burning phases that followed each other in the hydrostatic evolution of the star. Since rotation modifies somewhat the various burning stages, it is interesting to see how rotation modifies the binding energy of a stellar model, that directly reflects, in turn, the final Mass-Radius relation. 
We already introduced the compactness $\rm \xi_{2.5}$. Another interesting point in which is interesting to evaluate the compactness is at the CO core, i.e., $\rm \xi_{CO}$, since it reflects the behavior of the C convective shells in the advanced burning stages. 
\figurename~\ref{fig:csi} shows the trend of the compactness, evaluated in these two points, as a function of the initial rotation velocity and of the metallicity. $\rm \xi_{2.5}$ shows a modest increase with the initial rotation velocity, while $\rm \xi_{CO}$ shows a significantly larger dependence on it. This suggests that, generally speaking, rotating models would \textit{not} be more difficult to explode. However, since the compactness parameter can give at most a preliminary idea of the final fate of a star, we plan to address in the future such a problem with the aid of codes able to follow the collapse and re-bounce of the collapsing core \citep[see, e.g.,][]{boccioli:23}.

\subsection{Nucleosynthesis} \label{sec:nucleosynthesis}

In this section we discuss basically how rotation affects the yields of the various nuclear species at these extreme metallicity. The key differences induced by rotation on the chemical evolution of a star are essentially two: the first one is that rotation leads to larger CO core masses and lower C abundances at the end of the central He burning, and this implies that rotating stars tend to behave as stars of moderately larger mass, and the second one is the entanglement between the central He burning and the shell H burning. The consequence of this phenomenon is the formation of a robust concentration of \nuk{N}{14} (and \nuk{C}{13}) within the He core in all models. The \nuk{N}{14} engulfed in the He convective core is rapidly converted in \nuk{Ne}{22} while \nuk{C}{13} is fully destroyed by the ($\rm \alpha,n$) nuclear reaction becoming therefore a primary neutron source. The \nuk{N}{14} and the \nuk{C}{13} that instead remain locked between the border of the convective core and the base of the H shell are left substantially unaltered because of the relatively low temperature of this region. We call this region, rich in \nuk{N}{14}, \nuk{C}{13}, and in general products of the CNO cycle, \textit{CNO pocket}. Towards the end of the central He burning the \nuk{Ne}{22}($\alpha$,n)\nuk{Mg}{25} nuclear reaction activates in the convective core, providing therefore a quite robust primary neutron source. This primary neutron source depends of course on the efficiency of the entanglement between the He and H burning which, in turn, is strictly connected to the initial rotation velocity. A useful tool able to quantify the efficiency of the chemical mixing between the H and the He burning regions in the rotating models is the $\chi$ parameter, defined in LC18 as: 

%\begin{equation}
%\rm \chi(N,Mg)=\frac{X(^{14}N)}{14}+\frac{X(^{18}F)}{18}+\frac{X(^{18}O)}{18}+\frac{X(^{22}Ne)}{22}\\
%+\frac{X(^{25}Mg)}{25}+\frac{X(^{26}Mg)}{26}.
%\label{eq:chi}
%\end{equation}

\begin{equation}
\begin{split}
\rm \chi(N,Mg)=& \rm \frac{X(^{14}N)}{14}+\frac{X(^{18}F)}{18}+\frac{X(^{18}O)}{18}+\frac{X(^{22}Ne)}{22}\\
	          & \rm +\frac{X(^{25}Mg)}{25}+\frac{X(^{26}Mg)}{26}.
\label{eq:chi}
\end{split}
\end{equation}

\begin{figure*}
\epsscale{1.0}
\plotone{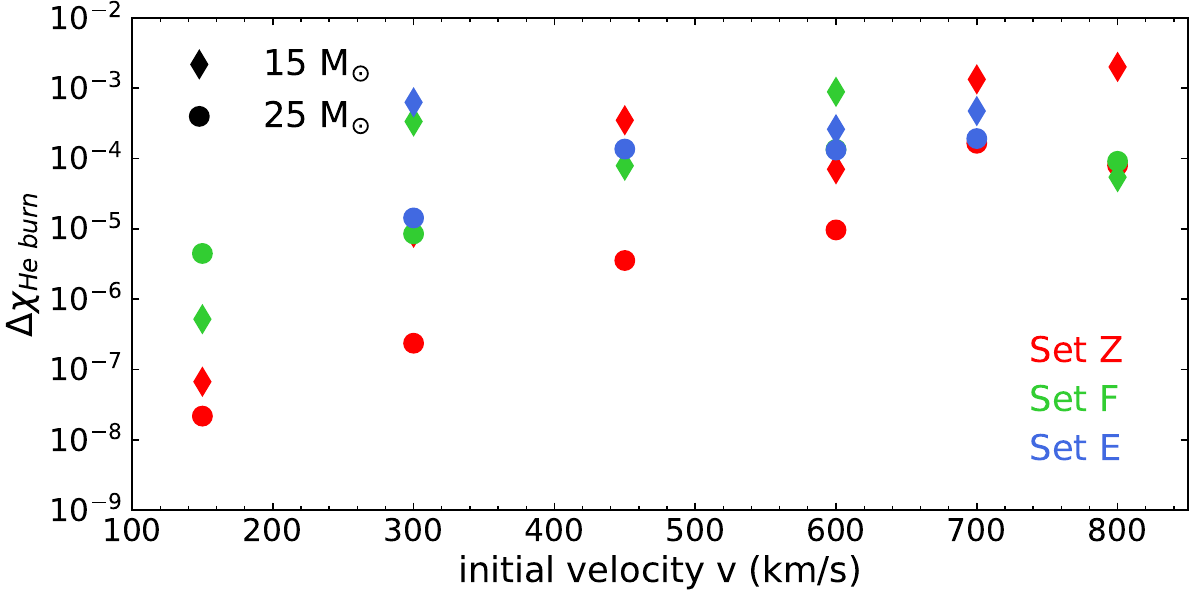}
\caption{ The variation of the parameter $\chi$ (\equationautorefname~\ref{eq:chi}) between central He ignition and exhaustion, as a function of the initial equatorial velocity for the 15 (filled diamonds) and the 25 \msun\ (filled dots) in Set Z (red), Set F (green),  and Set E (blue).\label{fig:chi}}
\end{figure*}

and it is simply the sum by number fraction of all the nuclei between $^{14}$N and \nuk{Mg}{26} involved in the conversion of N into Mg. This parameter remains constant if the He and the H burning regions do not interact because in this case the \nuk{N}{14} burning may populate only nuclei of the chain that goes from the \nuk{N}{14} to the \nuk{Mg}{25}/\nuk{Mg}{26}. Conversely, in rotating stars the amount of \nuk{N}{14} increases continuously because of the progressive conversion of freshly made $\rm ^{12}C$ into \nuk{N}{14}. \figurename~\ref{fig:chi} shows the strong correlation between the variation of the $\chi$ parameter and the initial rotation velocity. Since $\chi$ is basically a proxy of the amount of \nuk{Ne}{22} present in the core, the neutron density in He burning increases as the initial rotation velocity increases, turning from the negligible values of $\rm 10^2-10^4~n/cm^3$ (typical of the non rotating models at these metallicity) up to values of the order of $\sim10^7$ n/cm$^3$, i.e. very close to that of solar metallicity models. 

At zero metallicity this quite large neutron density does not allow in any case the synthesis of nuclei heavier than Zn, not even the $s$--process weak component because of the lack of seed nuclei. The high neutron density, in this case, leads to the synthesis of several neutron rich nuclei like, e.g., \nuk{Mg}{25-26}, \nuk{Na}{23}, and, in progressively faster rotating stars, of \nuk{Al}{27}, \nuk{Si}{28-29-30}, \nuk{P}{31} and \nuk{S}{32-33-34}.

At higher metallicity (Set F and E) the presence of seed nuclei, coupled to the high neutron to seed ratio, allow the synthesis of heavy nuclei up to, in some cases, the strong component (i.e., up to Pb).

In the following, for simplicity, we will mainly discuss Sr, Ba, and Pb to describe the climbing of the matter along the chart of the nuclides as a function of the initial metallicity and rotation velocity.

\begin{figure*}
\epsscale{1.15}
\plottwo{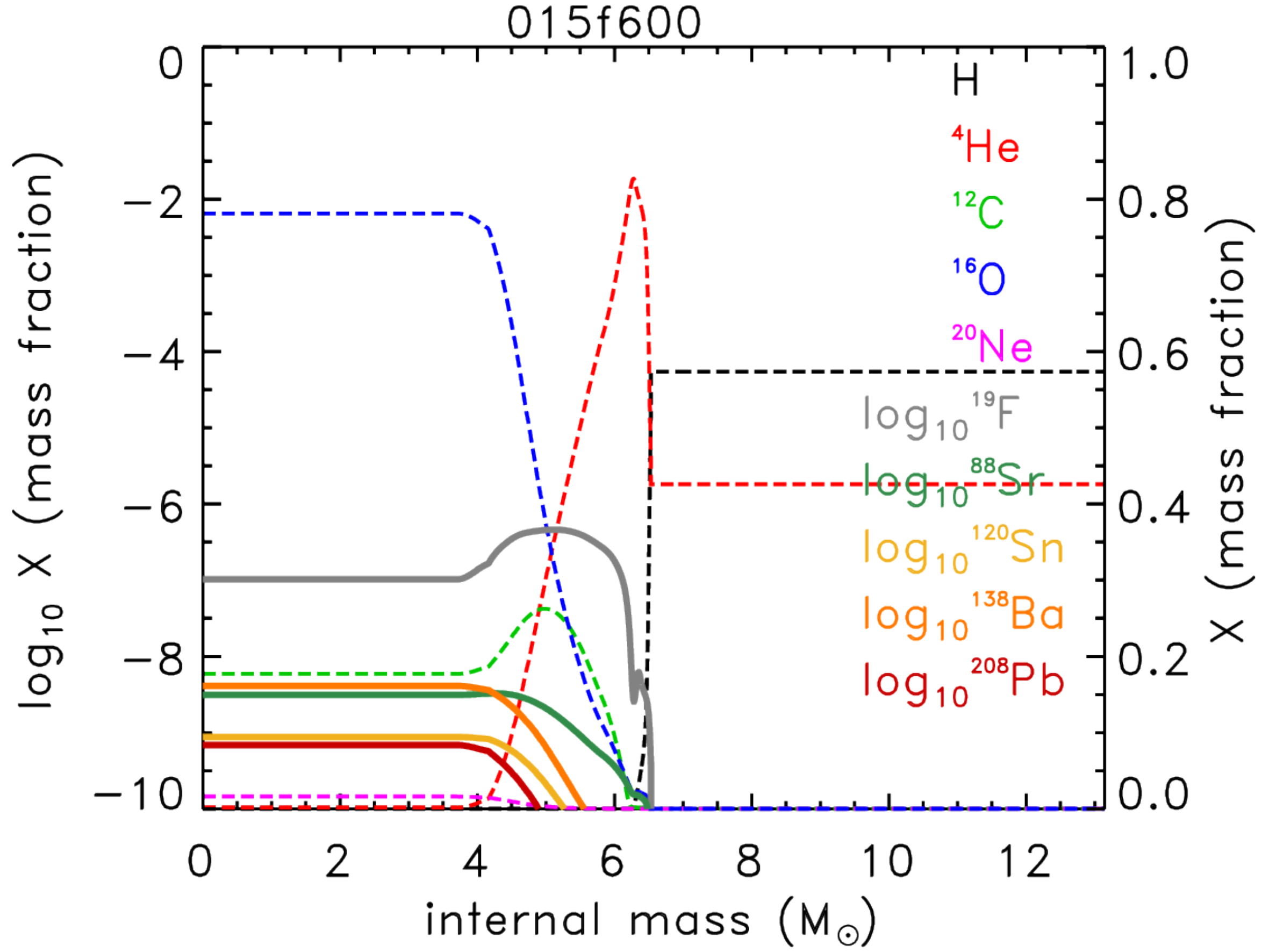}{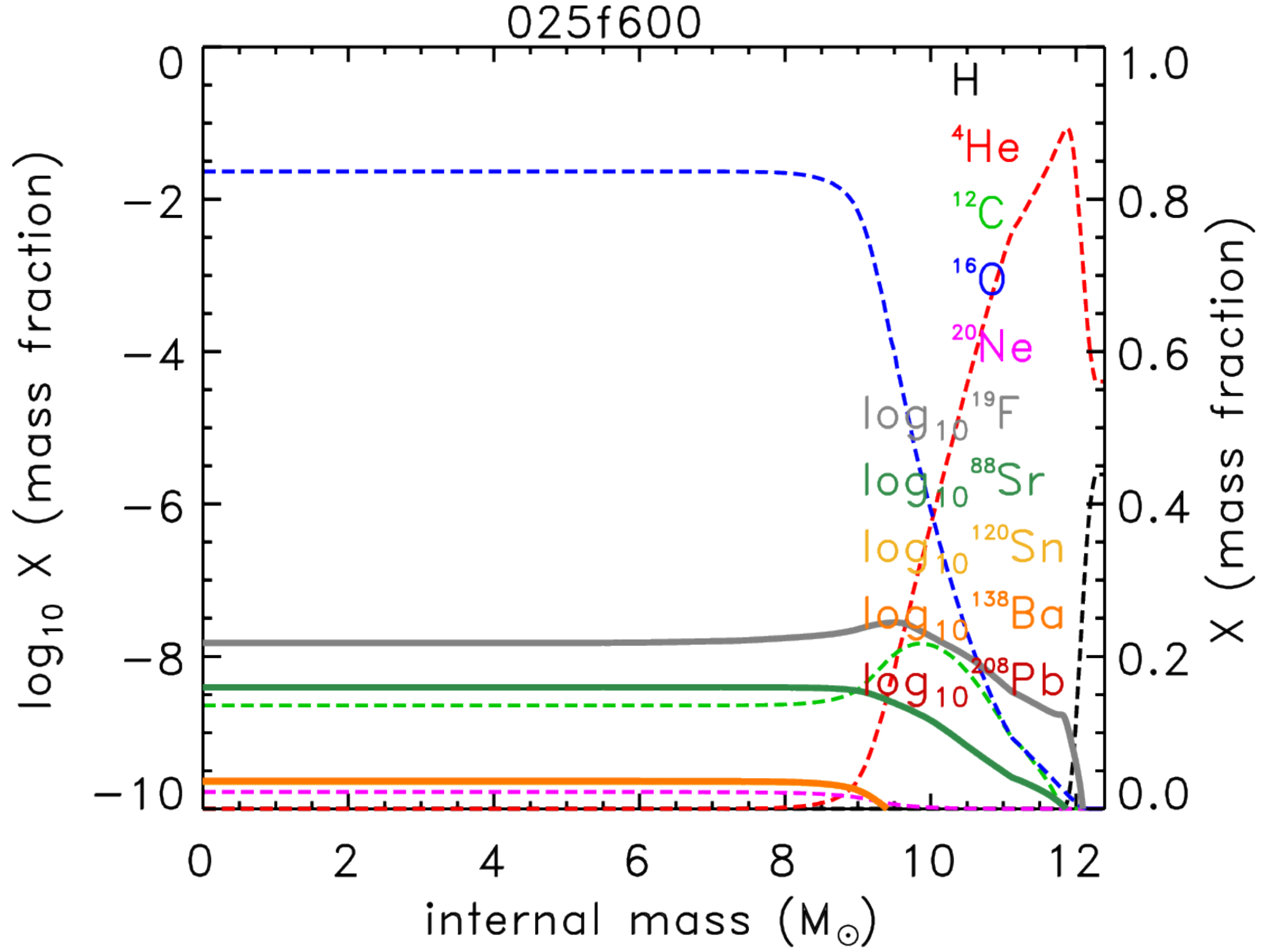}
\plottwo{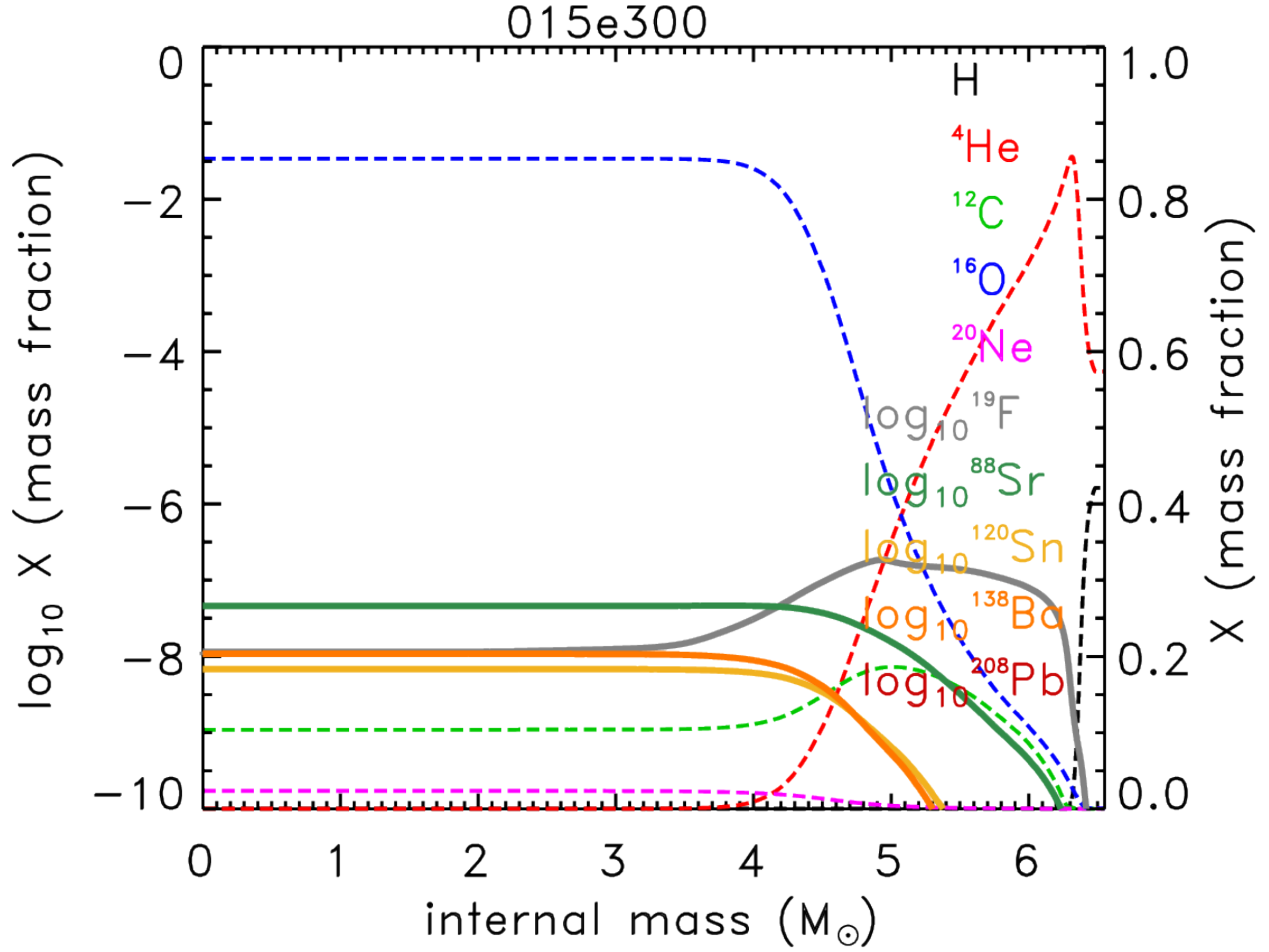}{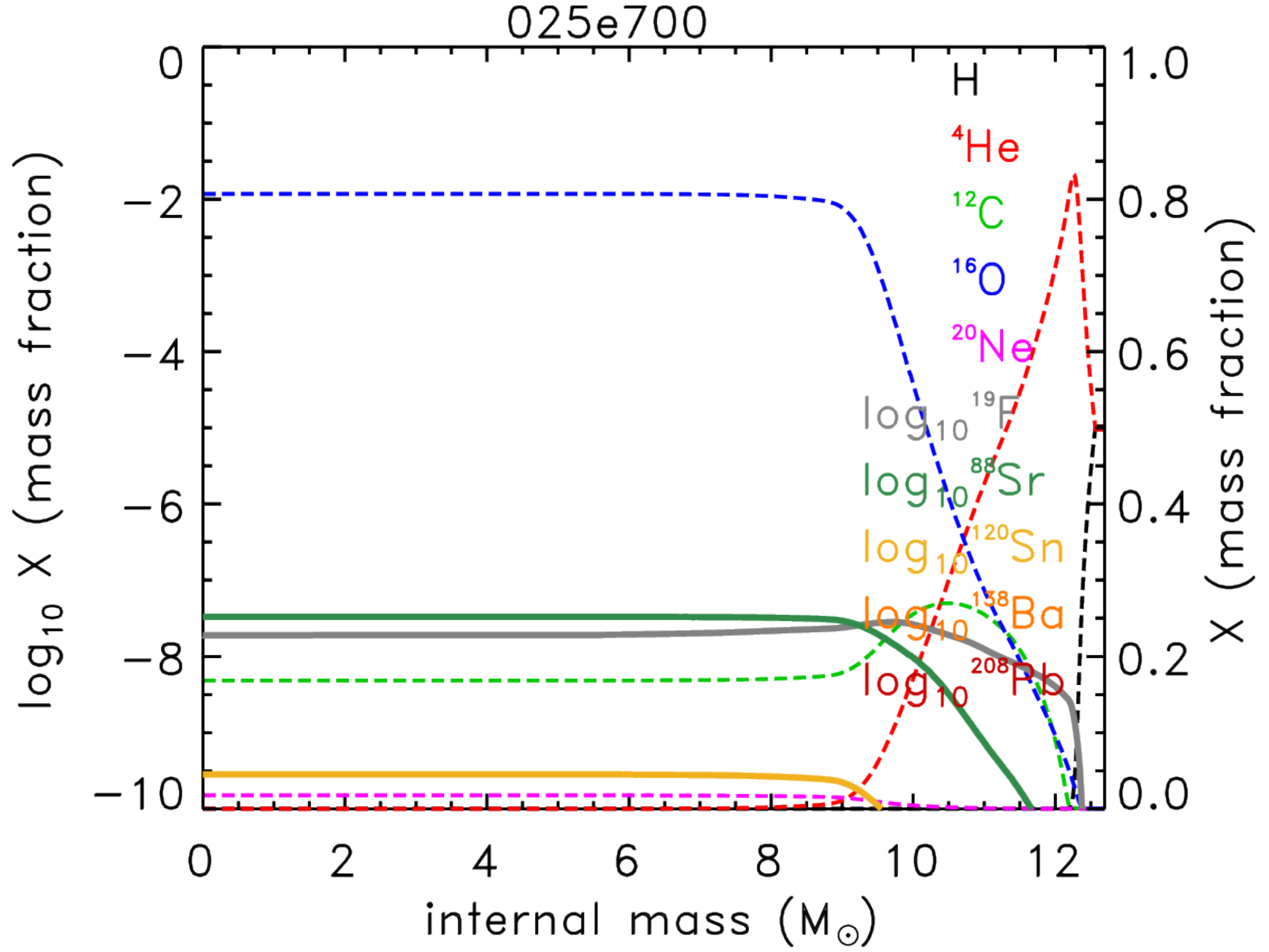}
\caption{Internal structure at core He exhaustion of four representative models 015f600, 025f600, 015e300, and 025e700. The dashed lines (right y-axis) represent the main nuclear species (H, \nuk{He}{4}, \nuk{C}{12}, \nuk{O}{16}, and \nuk{Ne}{20}), while the solid lines (left y-axis) show the internal profile of selected $s$-process isotopes: \nuk{Sr}{88} (weak component), \nuk{Ba}{138} (main component), \nuk{Pb}{208} (heavy component) and one isotope between the neutron magic nuclei N=50 and N=82, \nuk{Sn}{120}. At last, the gray solid line represents the \nuk{F}{19} abundance.\label{fig:shec}}
\end{figure*}

\figurename~\ref{fig:shec} shows a snapshot of the internal structure of four representative models at the end of the central He burning for Set F and E. All four models show a consistent production of the weak component (represented by \nuk{Sr}{88}) and the 15 \msun\ shows an increase of the elements of the main component (represented by \nuk{Ba}{138}). The nucleosynthesis extends up to  \nuk{Pb}{208} in both 15 \msun\ models. Note that also \nuk{F}{19} shows a pronounced peak in both 15 \msun, while the 25 \msun\ does not. It is worth noting that the core He burning phase produces the bulk of the final amount of the $s$-process elements.

\begin{figure*}
\epsscale{1.15}
\plottwo{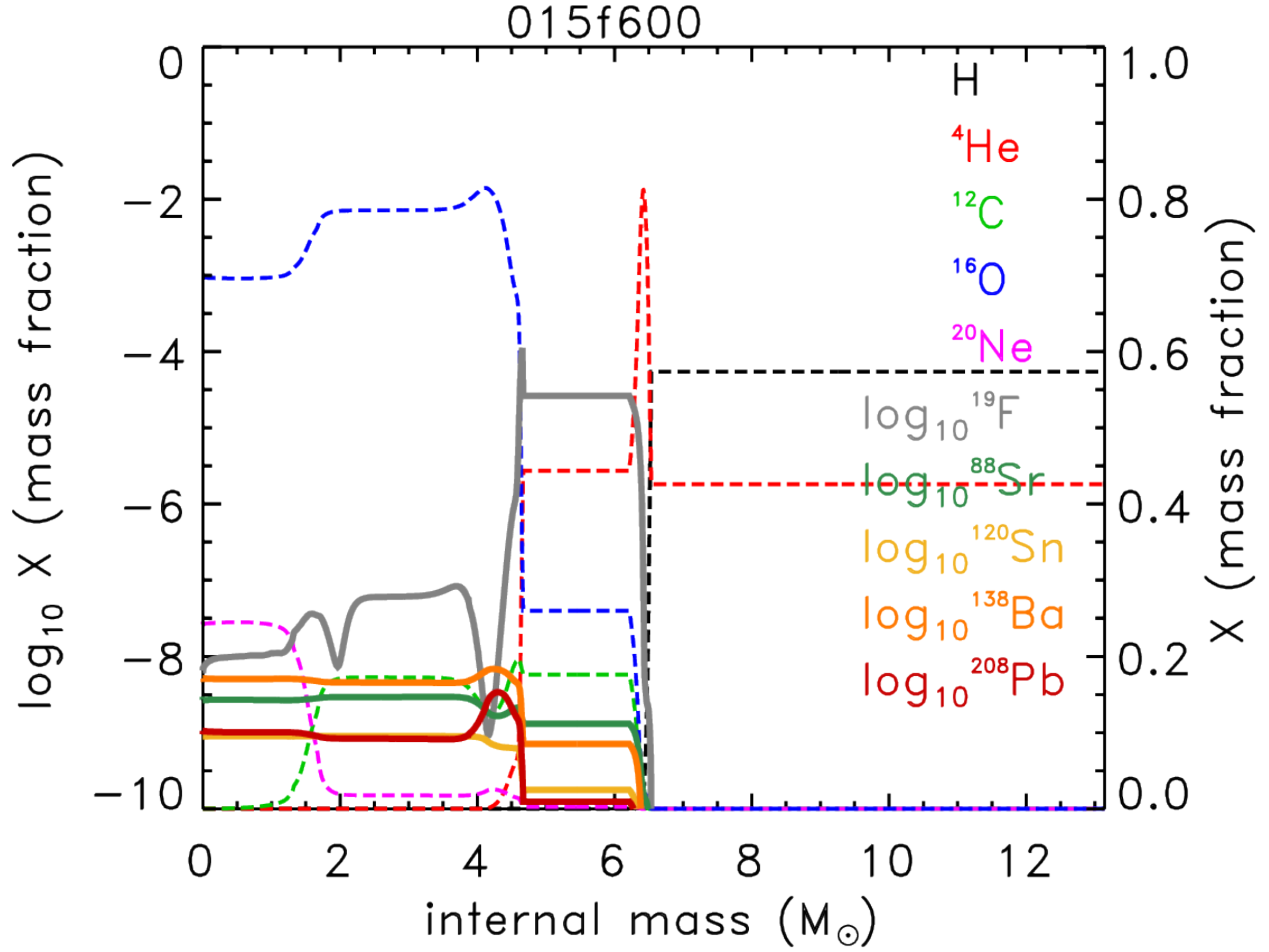}{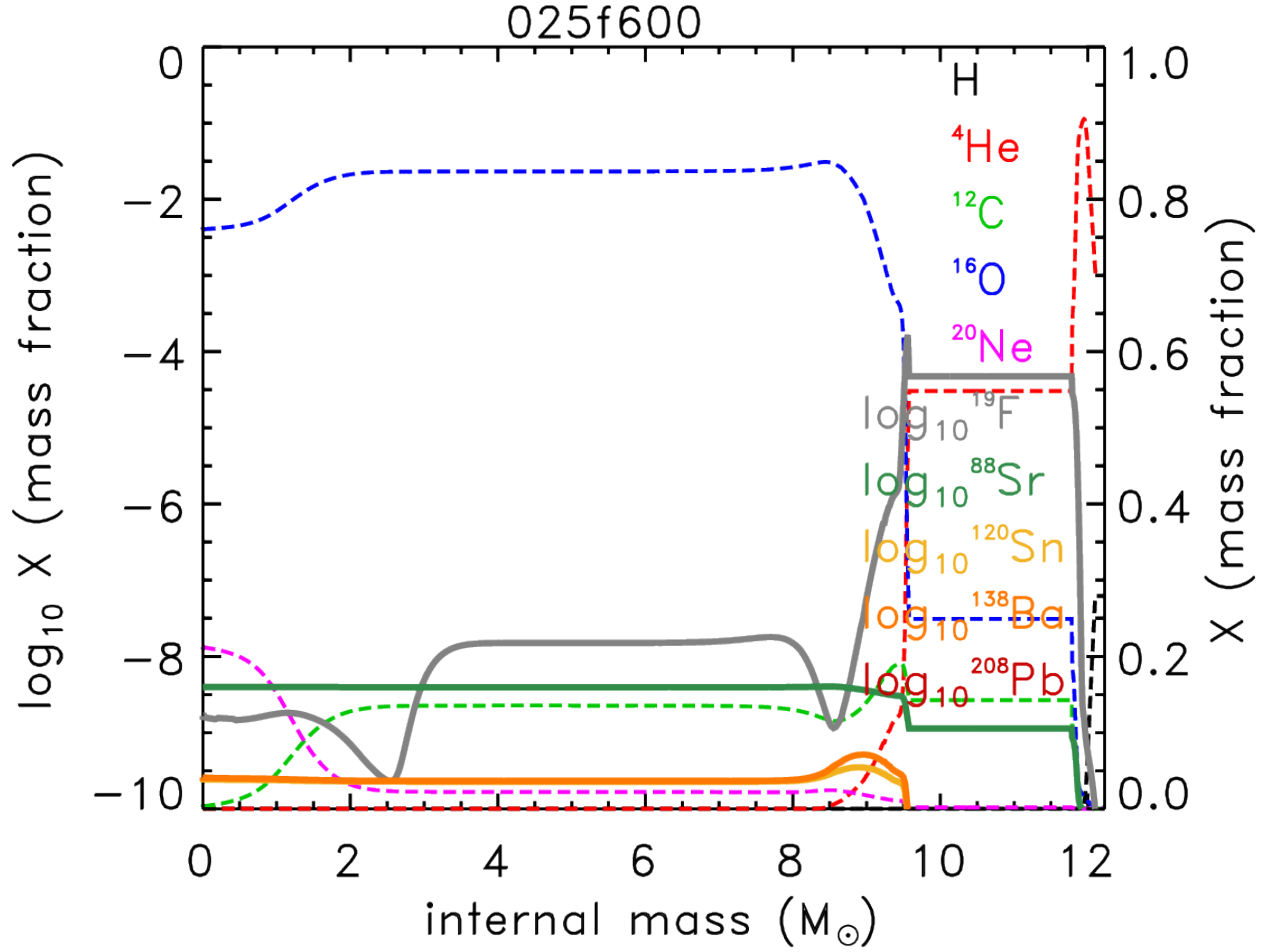}
\plottwo{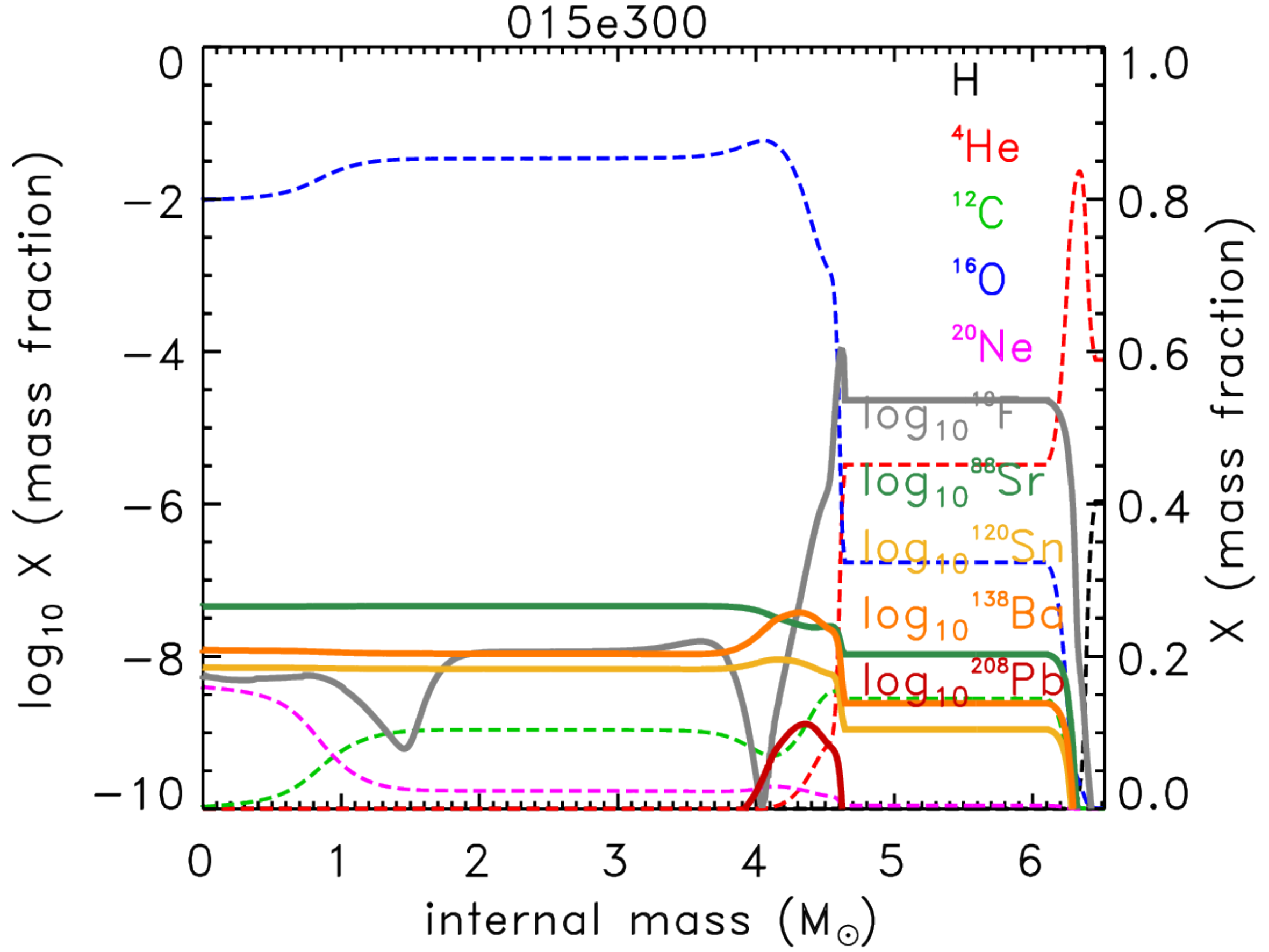}{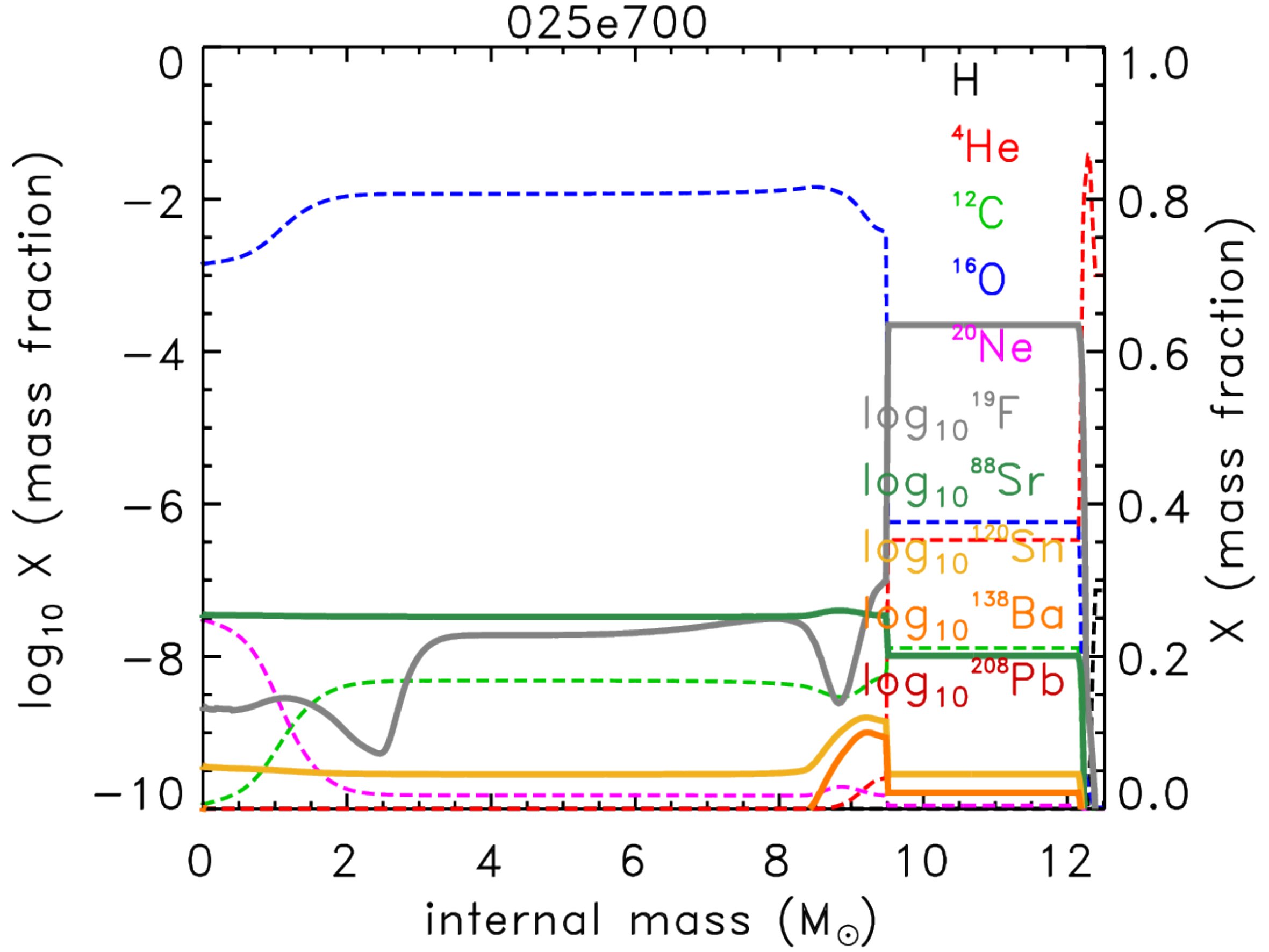}
\caption{Same as \figurename~\ref{fig:shec}, but after the formation of the He convective shell.\label{fig:shes}}
\end{figure*}

Once He is exhausted in the convective core, He burning shifts in shell where an extended convective shell soon develops. Though the neutron density reached at the base of the convective shell reaches values in the range $10^{8-10}$ n/cm$^3$, the overall contribution of the He convective shell to the synthesis of the heavy elements is quite modest. \figurename~\ref{fig:shes} shows a snapshot of the four structures already shown in \figurename~\ref{fig:shec} once the He convective shell is fully developed.

Though the He convective shell does not play an important role in the synthesis of the trans-Fe elements, it is actually the nursery of F. \nuk{F}{19}, in fact, is synthesized by the sequence of reactions firstly identified by \cite{forestini:92}: \nuk{N}{14}($\alpha,\gamma$)\nuk{F}{18}($\beta^+$)\nuk{O}{18}(p,$\alpha$)\nuk{N}{15}($\alpha,\gamma$)\nuk{F}{19}, where the protons necessary to feed the \nuk{O}{18} come from the \nuk{N}{14}(n,p)\nuk{C}{14}, while the neutrons are produced by the \nuk{C}{13}($\alpha$,n). It is not easy to find an environment in which this chain of reactions may work because of the simultaneously need of both neutrons and \nuk{N}{14}. The He burning shell in rotating massive stars perfectly embodies these requirements because of the large buffer of \nuk{N}{14} and \nuk{C}{13} present in the CNO pocket between the CO core and the He core. When the He convective shell develops and extends  within the He core, it engulfs a large amount of \nuk{N}{14} and \nuk{C}{13} that are quickly brought at the He burning temperature where the above quoted reactions may easily occur (see, e.g., C13, LC18). \nuk{F}{19} may reach a concentration as high as $10^{-4}$ in mass fraction in the fastest rotating models.  

\begin{figure*}
\epsscale{1.15}
\plottwo{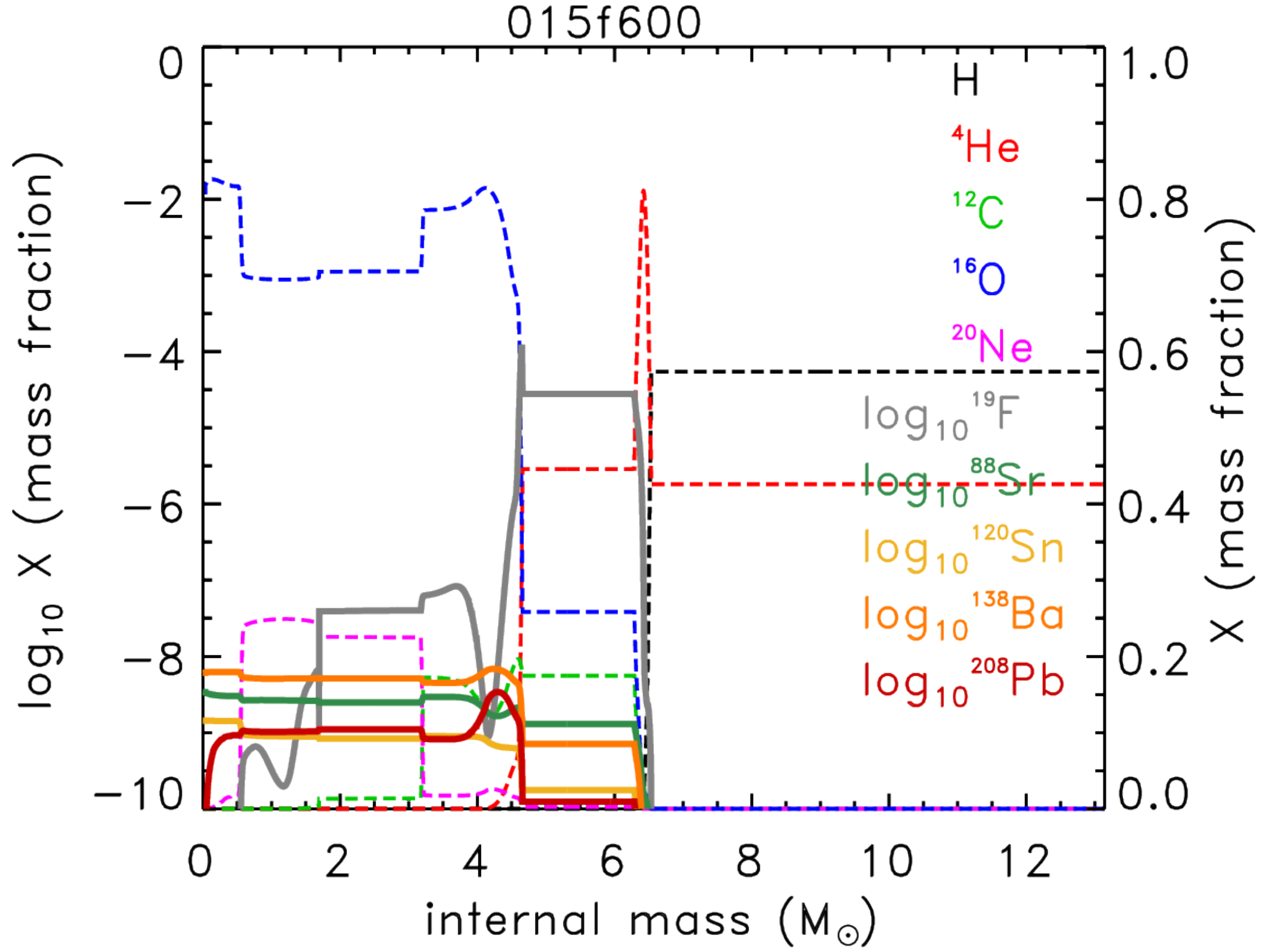}{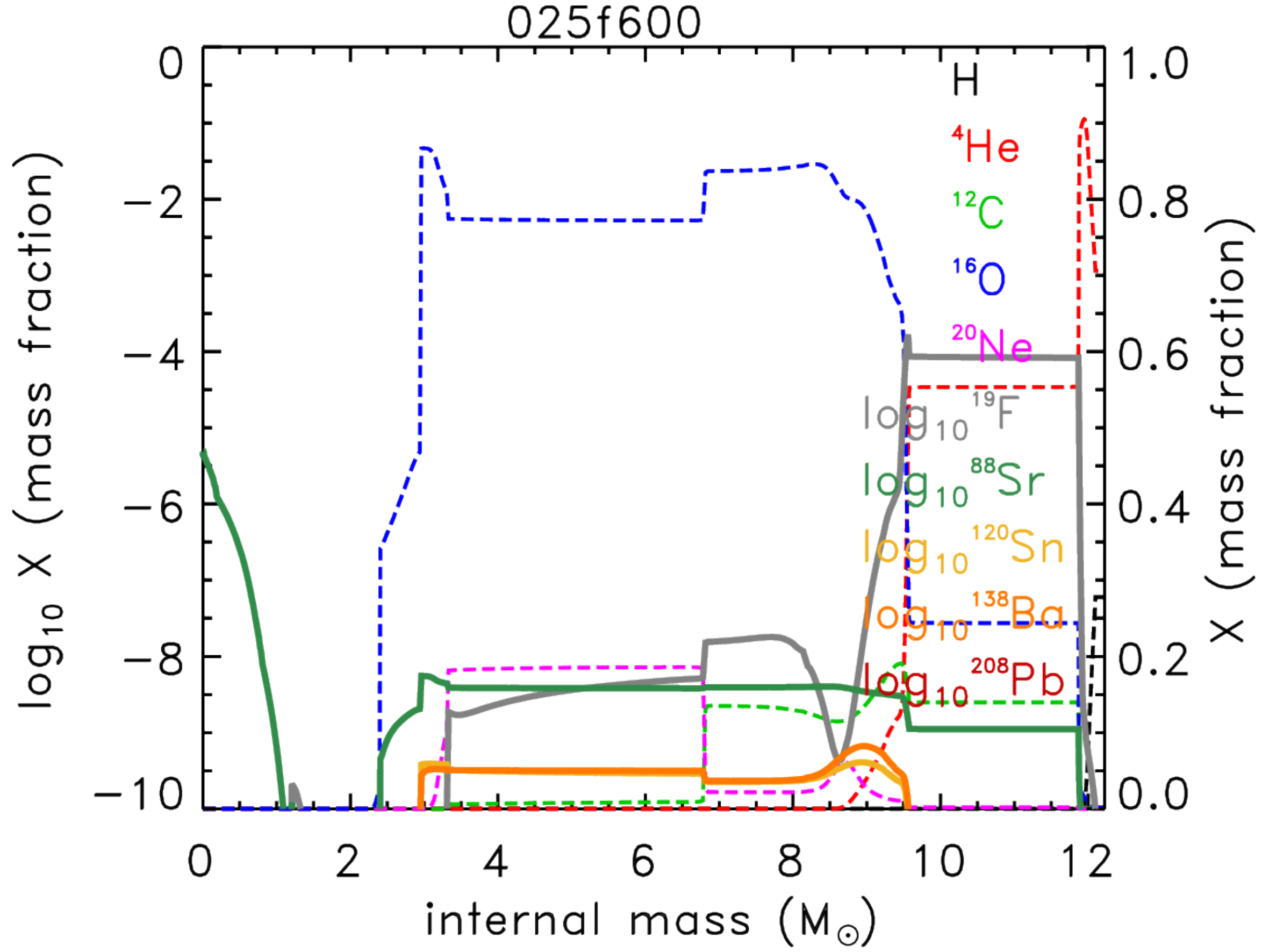}
\plottwo{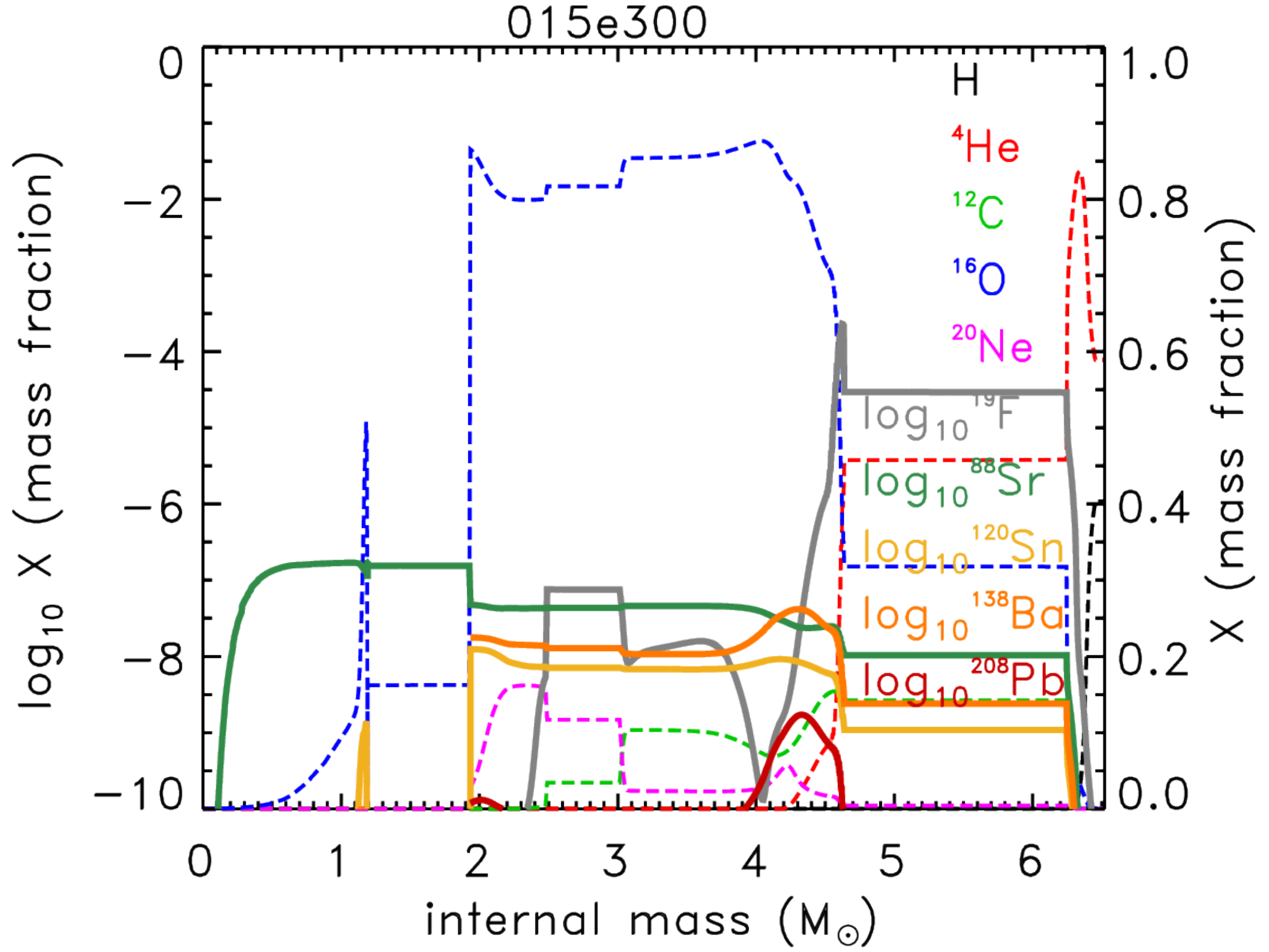}{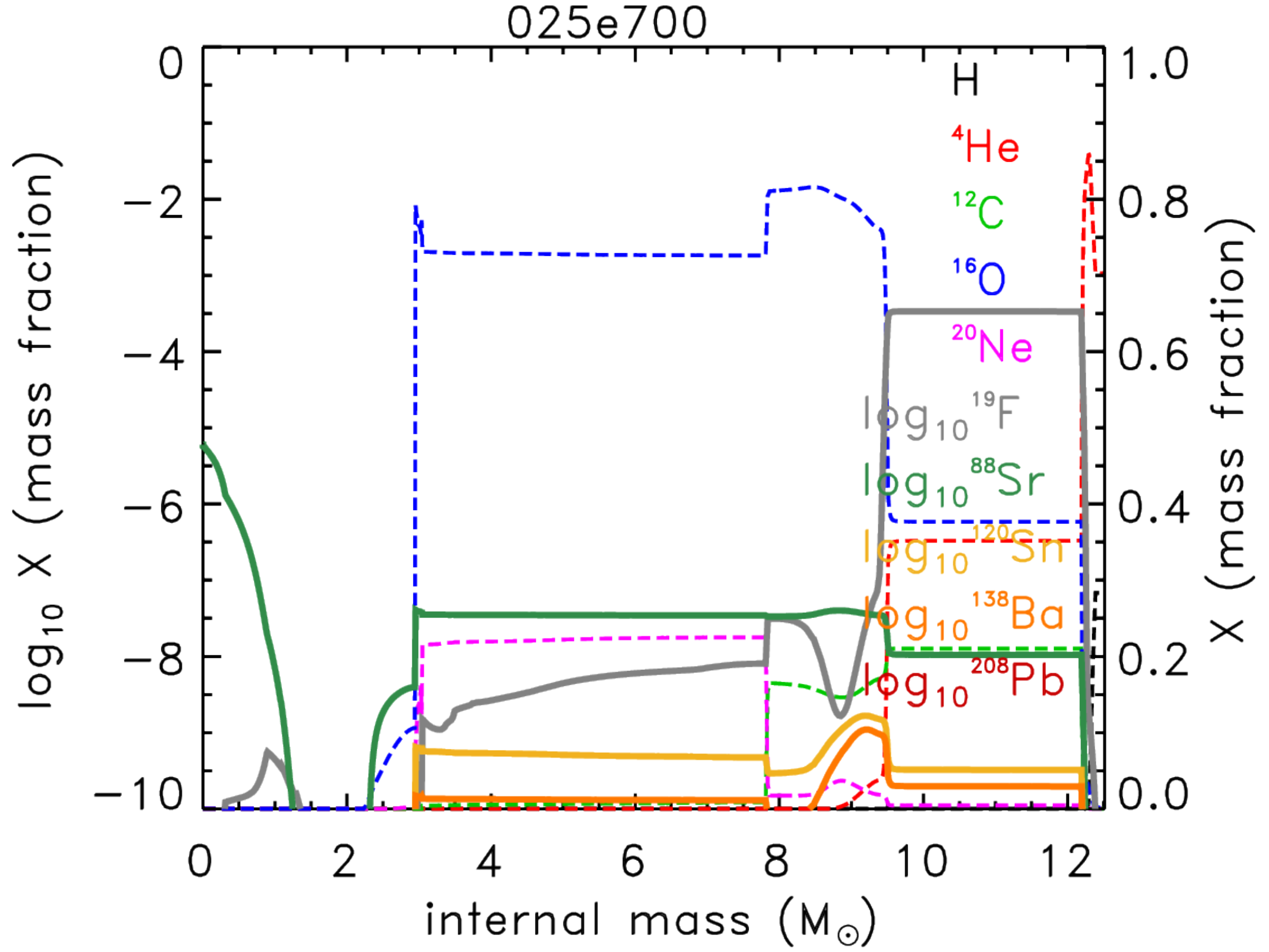}
\caption{Same as Same as \figurename~\ref{fig:shec}, but after the fully development of the C convective shell.\label{fig:scs}}
\end{figure*}

The last neutron production site occurs in C burning, more specifically in the convective C shell burning where the \nuk{Ne}{22} survived the central He burning reacting with the $\alpha$ particles provided by the \nuk{C}{12}(\nuk{C}{12},$\alpha$)$\gamma$ nuclear reaction, pours a substantial amount of neutrons. Though the contribution of the C burning to the synthesis of the elements heavier than Zn is modest, it is worth noting that even if the C convective shell does not increases the yields of the heavy nuclei, it does not even destroy significantly them. Typical temperatures and densities at the base of the C convective shell are $\rm T=1.2 - 1.5~ GK$ and $\rm \rho=1 - 2\times 10^5~ g/cm^3$ and the neutron densities span the range $\rm 10^{11} - 10^{13}~n/cm^3$.

Beyond the C burning, no $s$--process nuclei production is possible because the temperature is high enough that they are progressively and completely photo disintegrated.

\begin{figure*}
\epsscale{1.15}
\plottwo{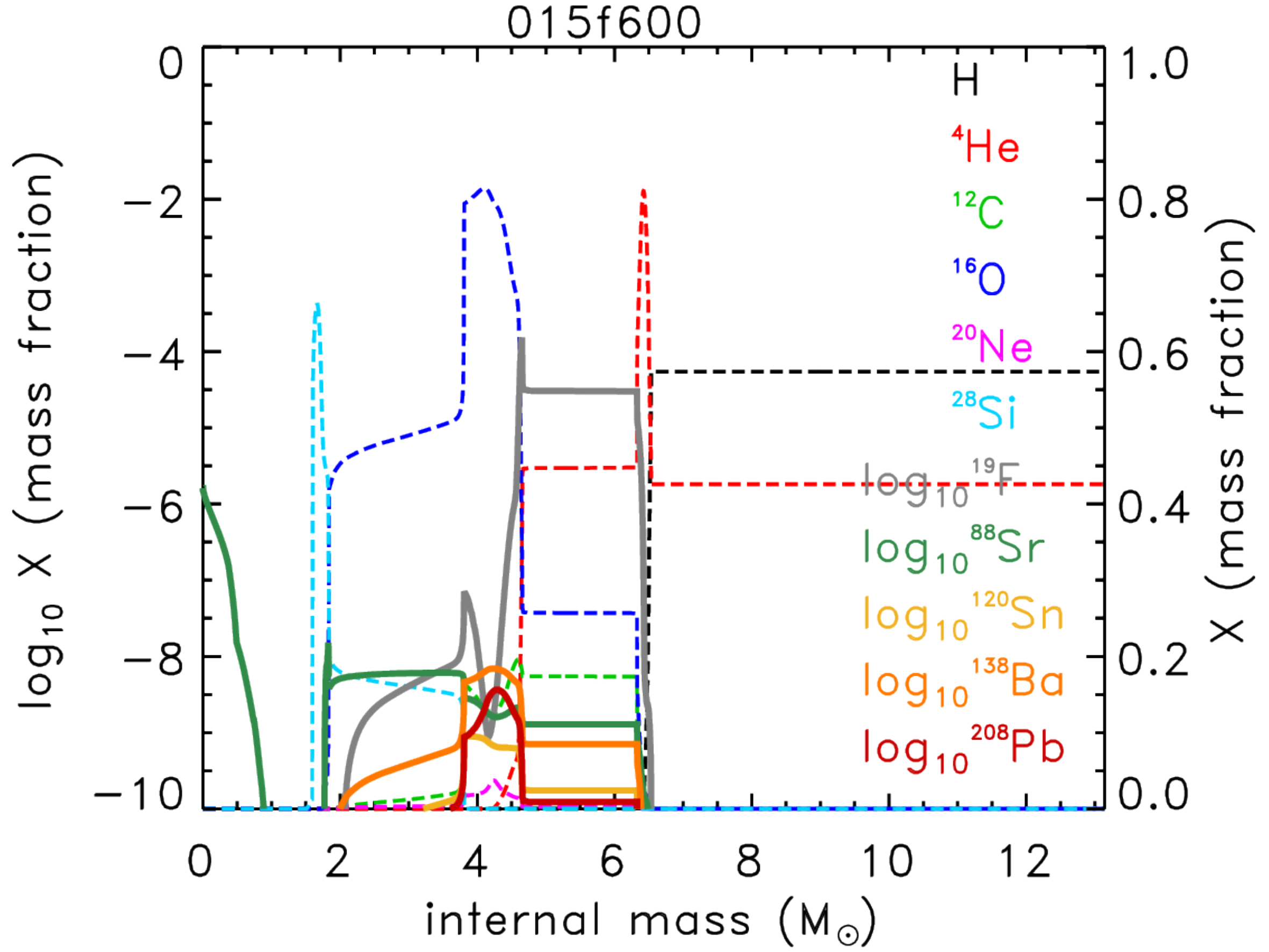}{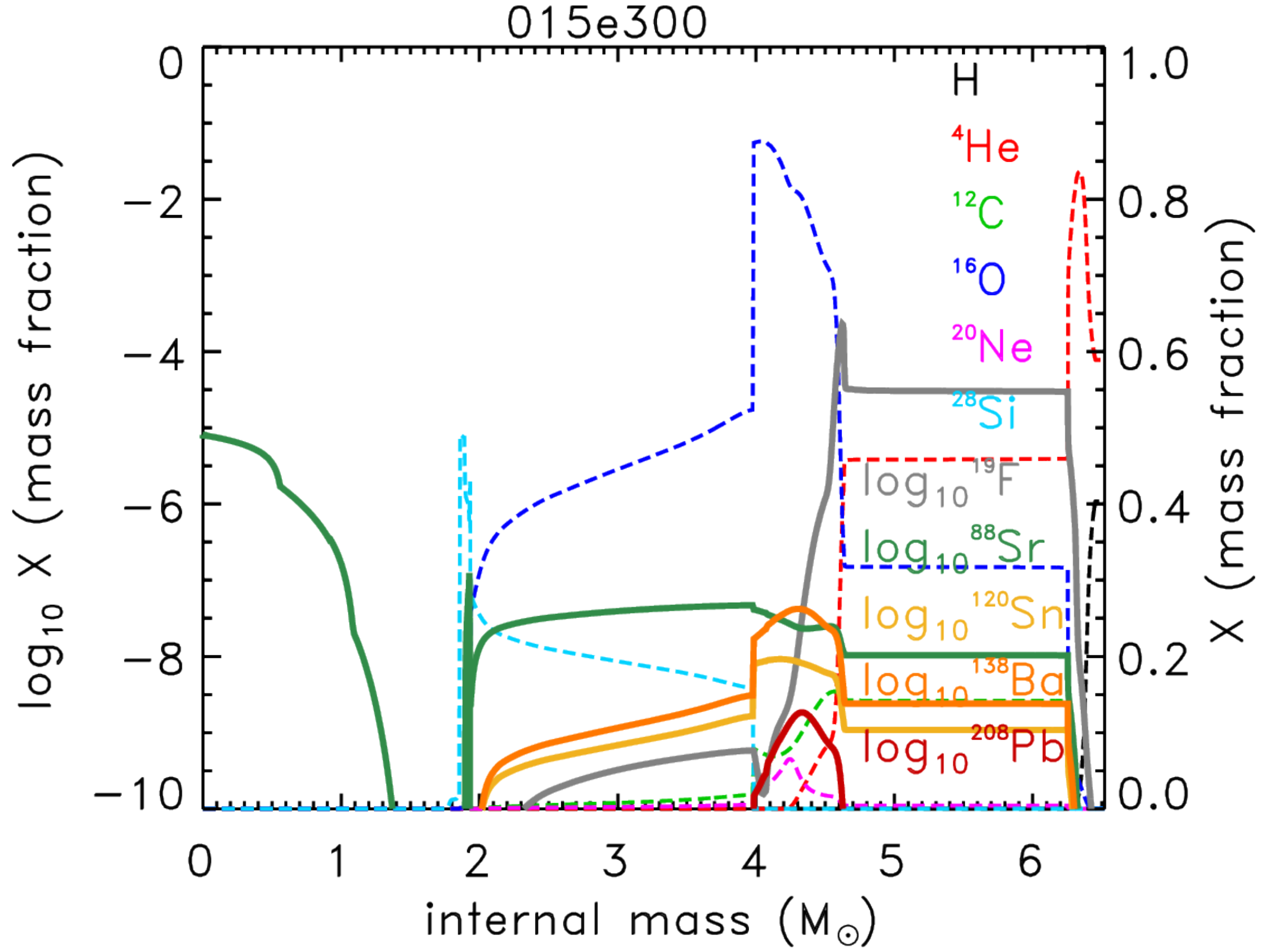}
\caption{Internal structure after the C--O shell merger in two 15 \msun\ rotating stars, i.e., 015f600 (left panel) and 015e300 (right panel). The dashed lines (right y-axis) represent the main nuclear species (H, \nuk{He}{4}, \nuk{C}{12}, \nuk{O}{16}, \nuk{Ne}{20}, and \nuk{Si}{28}), while the solid lines (left y-axis) show the internal profile of selected $s$-process isotopes: \nuk{Sr}{88} (weak component), \nuk{Ba}{138} (main component), \nuk{Pb}{208} (heavy component) and one isotope between the neutron magic nuclei N=50 and N=82, \nuk{Sn}{120}. Finally, the gray solid line represents the \nuk{F}{19} abundance.\label{fig:sco}}
\end{figure*}

There is however another phenomenon the occurs in the 15 \msun\ rotating models. In fact, as discussed in Sect.\ref{sec:rotation}, during the very late stages of their evolution, the O convective shell merges with the C shell and ingests a sizeable amount of fresh \nuk{C}{12}. The result of this phenomenon is the formation of a large O convective shell in which part of the products of the C shell burning are reprocessed. In particular, the most abundant nuclei that emerge from this interaction are the typical products of the O burning (see discussion in Sect.\ref{sec:xsuo}).
This extended convective region preserves these intermediate and Fe peak nuclei from being reprocessed by the explosive nucleosynthesis. In the case of the element above Fe, the effect of such an ingestion is the destruction of the heaviest isotopes and an increase of the abundances of the elements of the weak component in the extended convective shell. \figurename~\ref{fig:sco} shows the result of such a mixing in two representative cases.

\begin{figure*}
\epsscale{1.15}
\plottwo{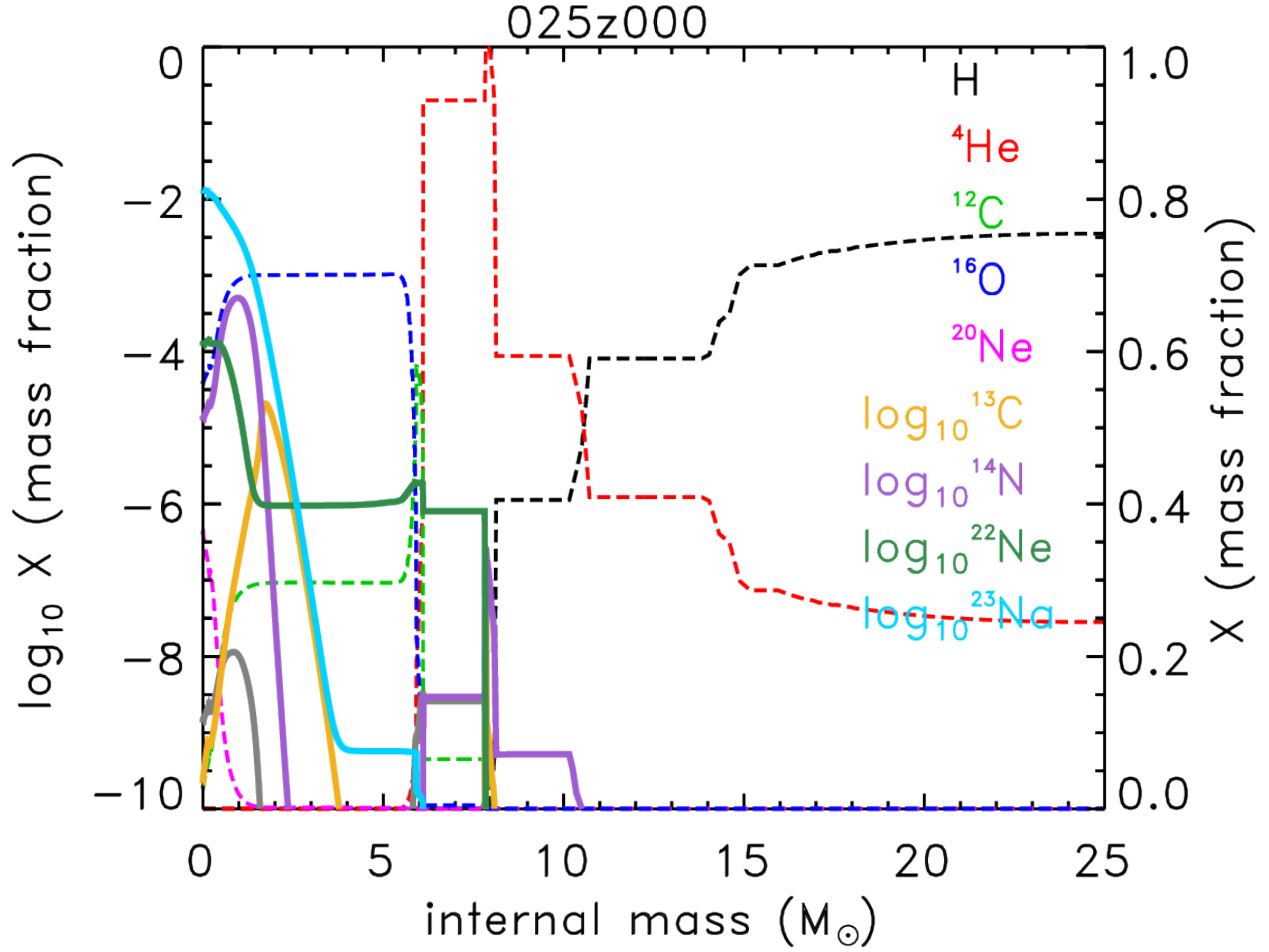}{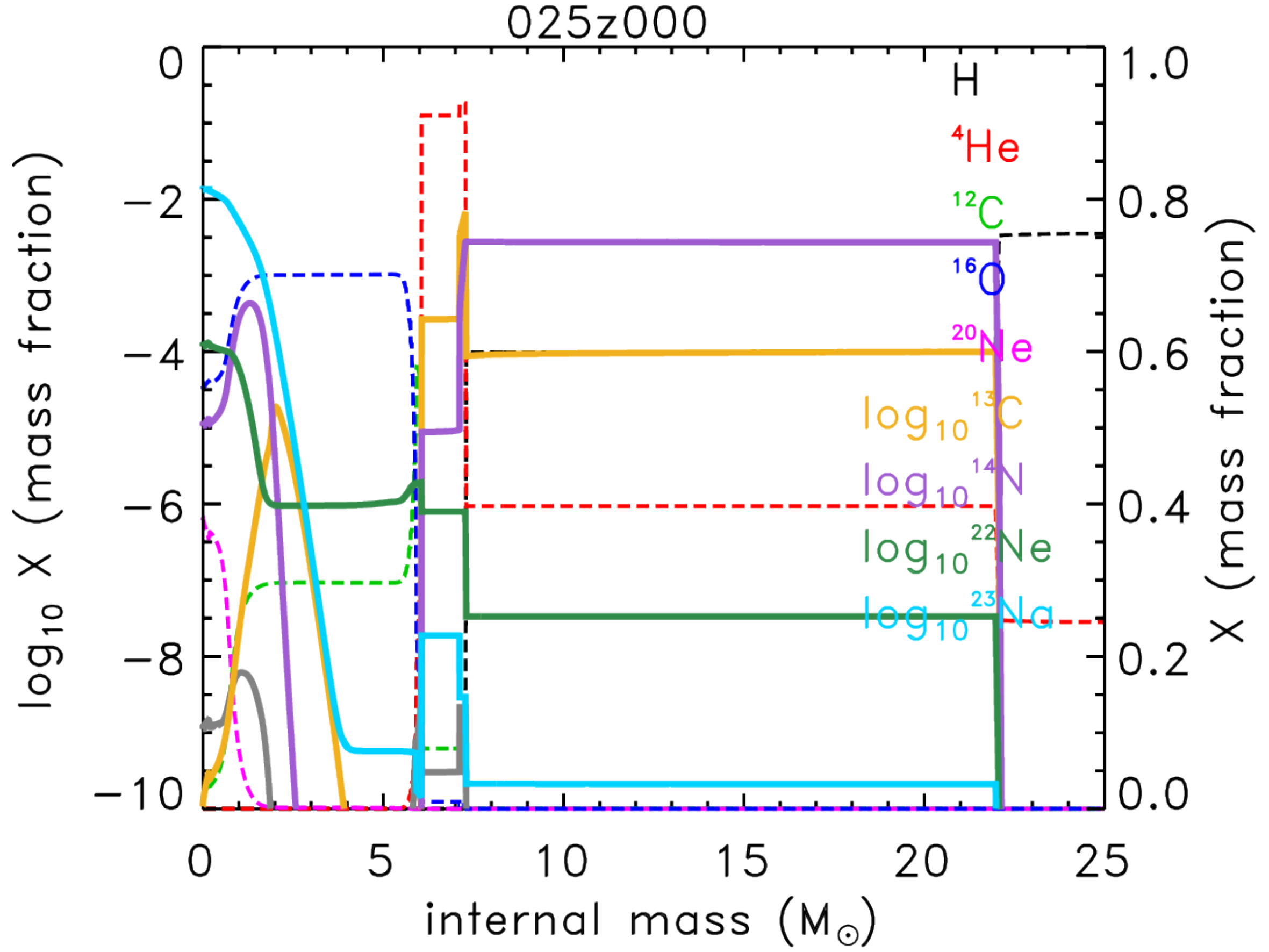}
\plottwo{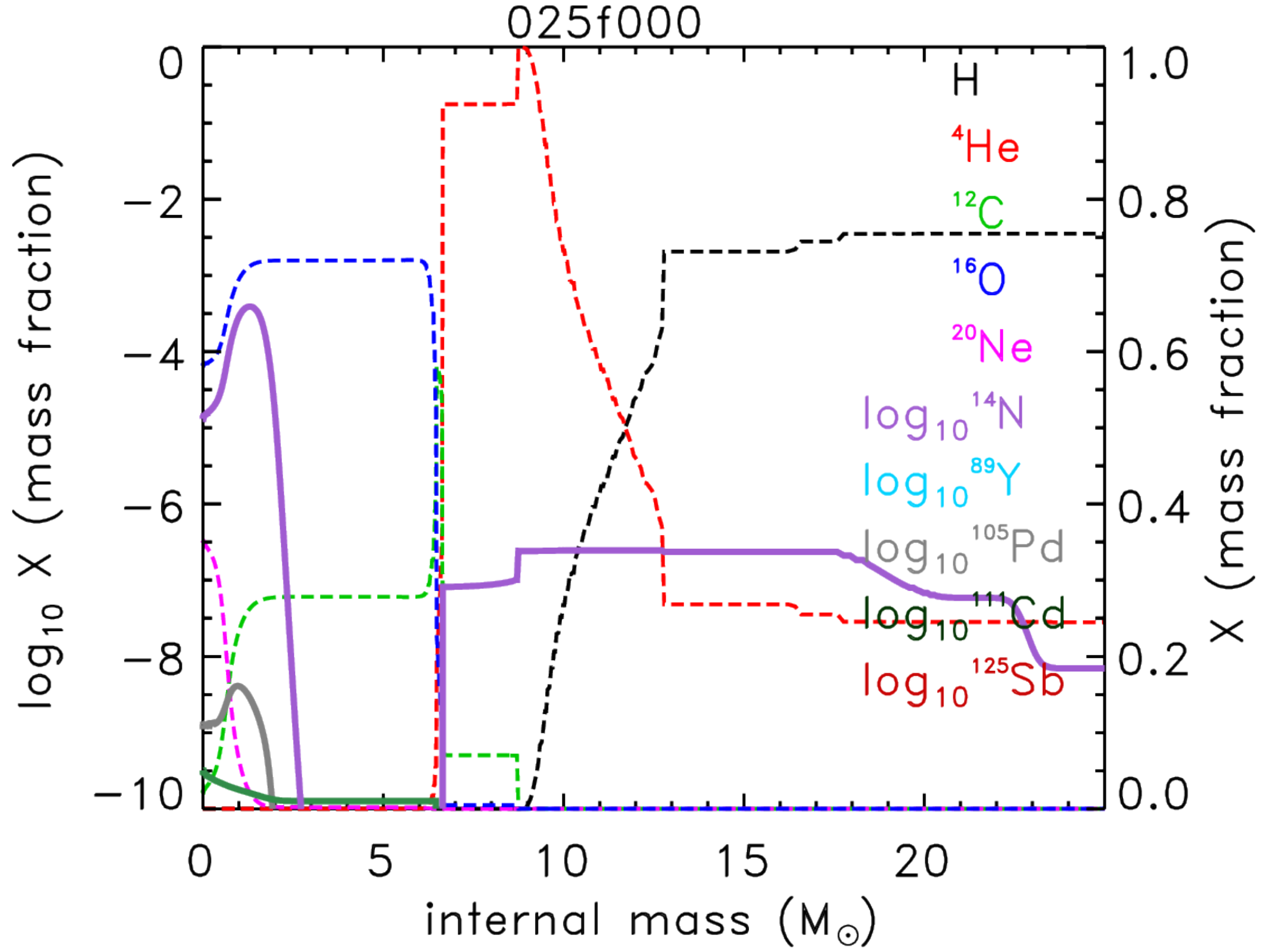}{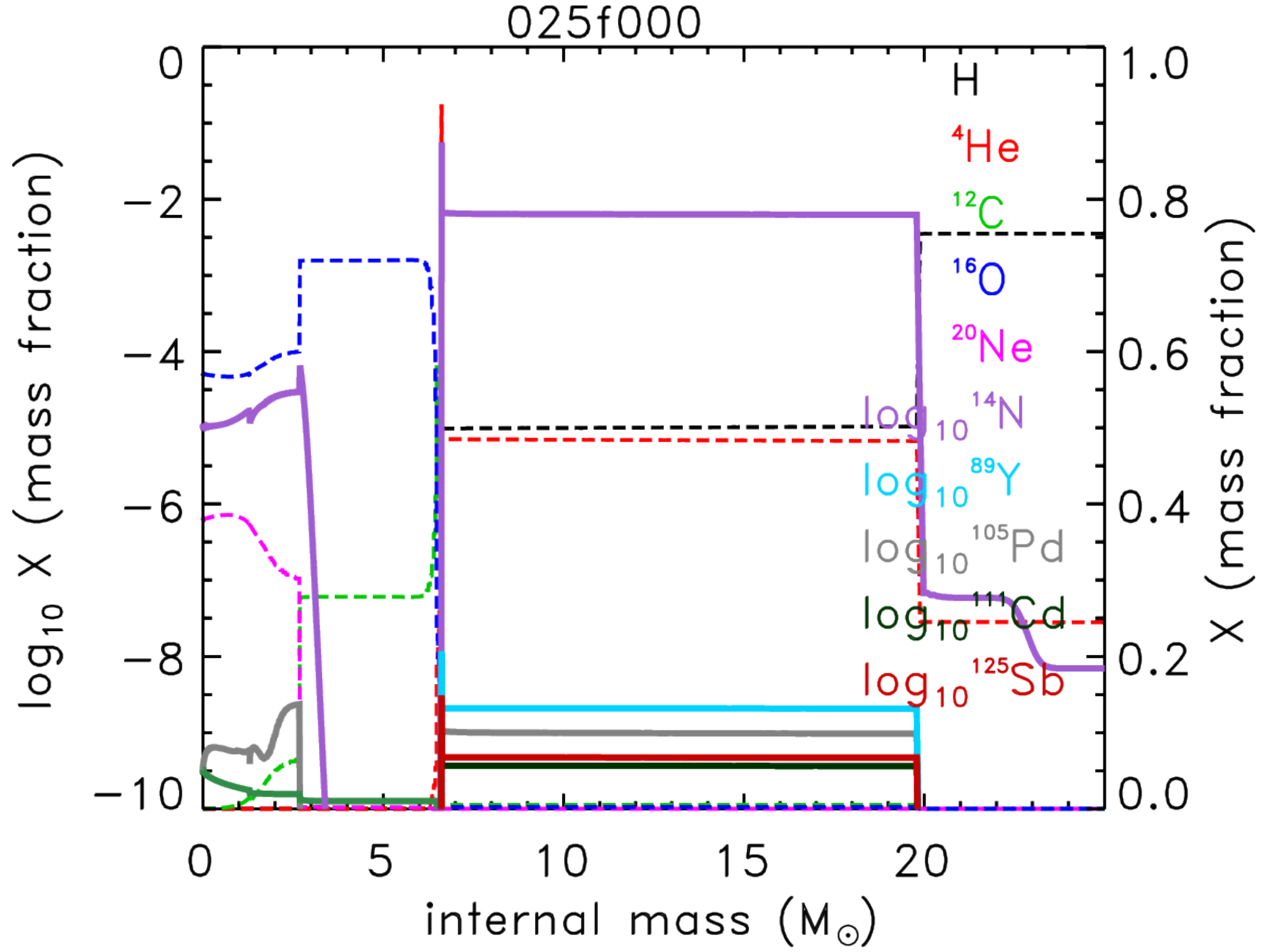}
\caption{Internal structure of the non rotating 25 \msun\ star of Set Z (top panels) and Set F (bottom panels) before (left panels) and after (right panels) the convective H--He shell merging. The black, red, green, magenta and blue solid lines represent the abundance in mass fraction of the principal nuclear species (H, \nuk{He}{4}, \nuk{C}{12}, \nuk{Ne}{20} and \nuk{O}{16}) as a function of the internal mass coordinate and they refer to the secondary y-axis. The other solid lines show the internal distribution of the most produced nuclei produced in the merger between the H and the He convective shells. (in particular, \nuk{N}{14} is the violet solid line) and they refer to the principal y-axis. \label{fig:shhe}}
\end{figure*}

As discussed in the previous sections (Sect.\ref{sec:nrot:he} and \ref{sec:rotation}), in several cases the He convective shell penetrates the H rich region and the extra energy produced by the proton ingestion in the He shell leads to the consequent formation of a large convective region that rapidly covers a large part of the H rich envelope. This phenomenon has been associated to the synthesis of primary \nuk{N}{14} at early times. In both non rotating 25 \msun\ stars, the abundance of \nuk{N}{14} in the extended convective shell reaches a concentration of $\sim3\times10^{-3}$ in mass fraction. But this merging leads also to the production of a strong neutron flux due to $\alpha$ captures on $^{13}\rm C$. Very high neutron densities may be obtained, typical of the so-called \textit{intermediate} neutron capture process (\textit{i-process}), i.e., $\rm n_n \sim 10^{14}~n/cm^3$. In the 025z000 model, similarly to the He burning phase in the rotating case, these neutron are absorbed by lighter nuclei, populating the light and intermediate nuclei up to Ti. The most abundant isotopes in the newly formed H convective shell are \nuk{N}{14}, \nuk{C}{13}, \nuk{Ne}{22}, and \nuk{Na}{23}. In the 025f000 model the most abundant nuclei are \nuk{Y}{89}, \nuk{Pd}{105}, \nuk{Cd}{111}, and \nuk{Sb}{125}, therefore the neutron capture nucleosynthesis favors in this case the synthesis of some nuclei belonging mostly to the weak component. Note that \nuk{Sb}{125} is an unstable nucleus, therefore it will contribute to the abundance of \nuk{Te}{125}. \figurename~\ref{fig:shhe} shows the internal structure before (left panels) and after (right panels) the merging of the H and He convective shell in the non rotating 25 \msun\ stars at [Fe/H]=$-\infty$ (upper panels) and --5 (lower panels).

% ================================= EXPLOSION  =================================
\section{The explosion}\label{sec:expl}

In the previous section we discussed in detail the evolution of models of 15 and 25 $\rm M_\odot$ having different metallicity and initial rotation velocities, starting from the pre-main sequence up to the onset of the iron core collapse and the associated nucleosynthesis. The last step necessary to obtain the chemical composition of the ejecta requires the simulation of the passage of the shock wave within the mantle together to the determination of the mass of the remnant. In this paper the explosion has been computed by means of the \verb|HYPERION| (\textit{HYdrodynamic Ppm Explosion with Radiation diffusION}) code: this is a lagrangian hydrodynamic flux limited diffusion radiation code, designed to calculate the explosive nucleosynthesis, remnant mass, and light curve associated with the explosion of a massive star. This code is presented and extensively described in \cite{LC20}. More specifically, the explosion is triggered by means of a thermal bomb, i.e., by depositing the minimum amount of thermal energy, at a mass coordinate of $0.8~\rm M_\odot$ in the presupernova model (i.e., well inside the Fe core), that leads to the full ejection of the mantle above the Fe core. The remnant mass is fixed by requiring the ejection of 0.07 M$_\odot$ of \nuk{Ni}{56}. No mixing and fall back \citep[see, e.g.,][]{umeda:02,umeda:05} is taken into account. We stop our hydrostatic calculation, and hence define the presupernova model, when the central temperature is $\sim$6 GK. At this stage the fist Si convective shell has already exhausted and the final Fe core is formed.

\begin{figure*}
\epsscale{1.0}
\plotone{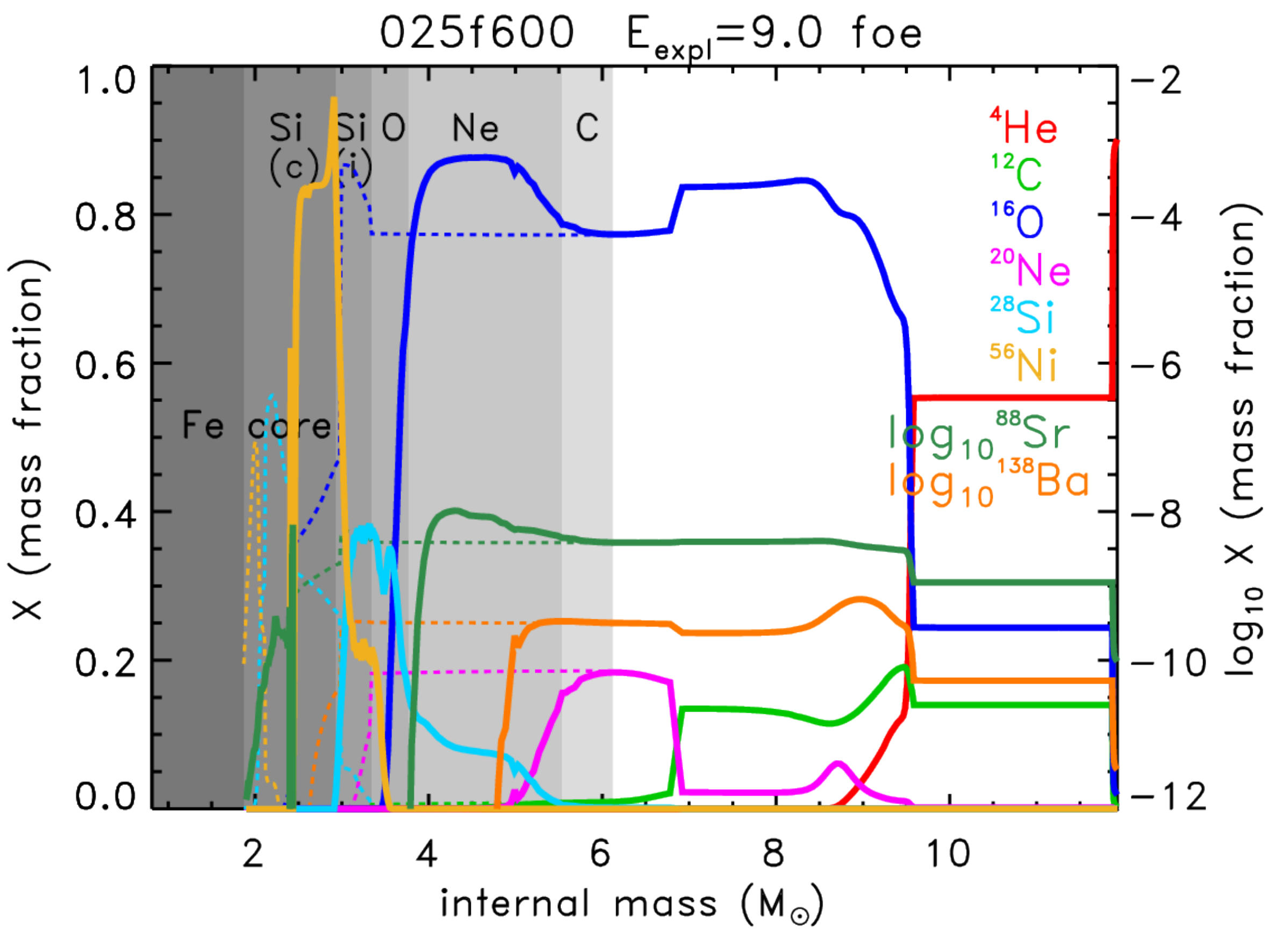}
\plotone{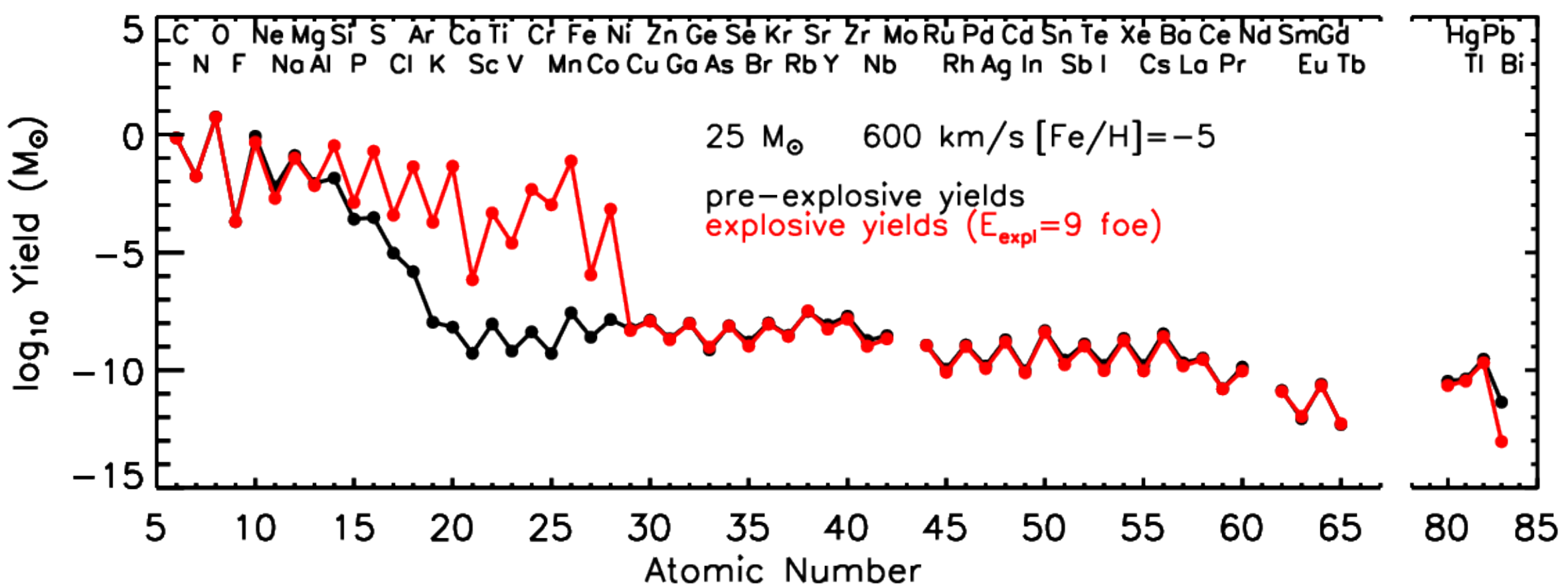}
\caption{Top panel: presupernova (dashed lines) and post-explosive (solid lines) structure of a 25 \msun\ star at [Fe/H]=--5, rotating at 600 km/s. In this case the abundances are presented as a function of the internal mass of the star, from the center up to the base of the H shell. The gray bands in the plot mark each explosive burning stage in the corresponding internal mass coordinate, as discussed in the text. The dashed lines represent the abundances of the isotopes at the pre-supernova stage, while the solid lines represent the isotopic abundances after the explosive nucleosynthesis. Bottom panel: pre-explosive (black dots) and explosive (red dots) yields of the model 025f600.\label{fig:expstrut}}
\end{figure*}

A typical example of the influence of the explosive burning on the final yields is shown in \figurename~\ref{fig:expstrut}. The upper panel shows, in fact, the abundances of the most abundant nuclei, plus the two isotopes \nuk{Sr}{88} and \nuk{Ba}{138} (representative of the $s$-process elements), before (dashed lines) and after (solid lines) the passage of the shock wave. The zones exposed to the various explosive nuclear burning are marked by the gray areas (complete Si burning: T$>5\times10^9$\ K; incomplete Si burning:  T$>4\times10^9$\ K; explosive O burning: T$>3.3\times10^9$\ K; explosive Ne burning: T$>2.1\times10^9$\ K; explosive C burning: T$>1.9\times10^9$\ K). This figure clearly shows that the chemical composition is deeply modified by the explosive nucleosynthesis up to a great fraction of the C convective shell. The two selected isotopes representative of the $s$-process elements are mainly produced by the core He burning so that their final yields are dominated by the hydrostatic component.

In order to have a global information on the effect of the explosion on the $s$-process element abundances we show in the lower panel of \figurename~\ref{fig:expstrut} a comparison between the elemental explosive (red line) and pre-explosive (black line) ejected masses (yields) obtained for the 025f600 model. The pre-explosive yields have been computed by adopting the same mass cut obtained by the simulation of the explosion, i.e., 3.13 \msun\ in this specific case. As expected, the explosion produces iron peak elements as well as intermediate mass elements between Si and Ti. Vice versa the other elements, i.e. those lighter then Al or heavier than Ni, are very mildly or not affected at all by the passage of the shock wave (see, e.g.,CL13, LC18, and references therein).

% ================================= YIELDS  =================================
\section{The yields}\label{sec:yields}

We have seen above that the elements more affected by rotation are N plus all the ones whose synthesis is widely dependent on the presence of a neutron source, i.e., the elements above the Fe group and F. As proxies of all the nuclei above the Fe group we will discuss the yields of Sr, Ba, and Pb since are the ones that mark the three neutron magic numbers with N=50, 82 and 126. A complete table of all the yields obtained in this work is presented in the Appendix \ref{app:yields} and in our online repository O.R.F.E.O. (Online Repository of the Franec Evolutionary Output)\footnote{\url{http://orfeo.iaps.inaf.it}}. 

\subsection{Nitrogen}

\begin{figure*}
\epsscale{1.0}
\plotone{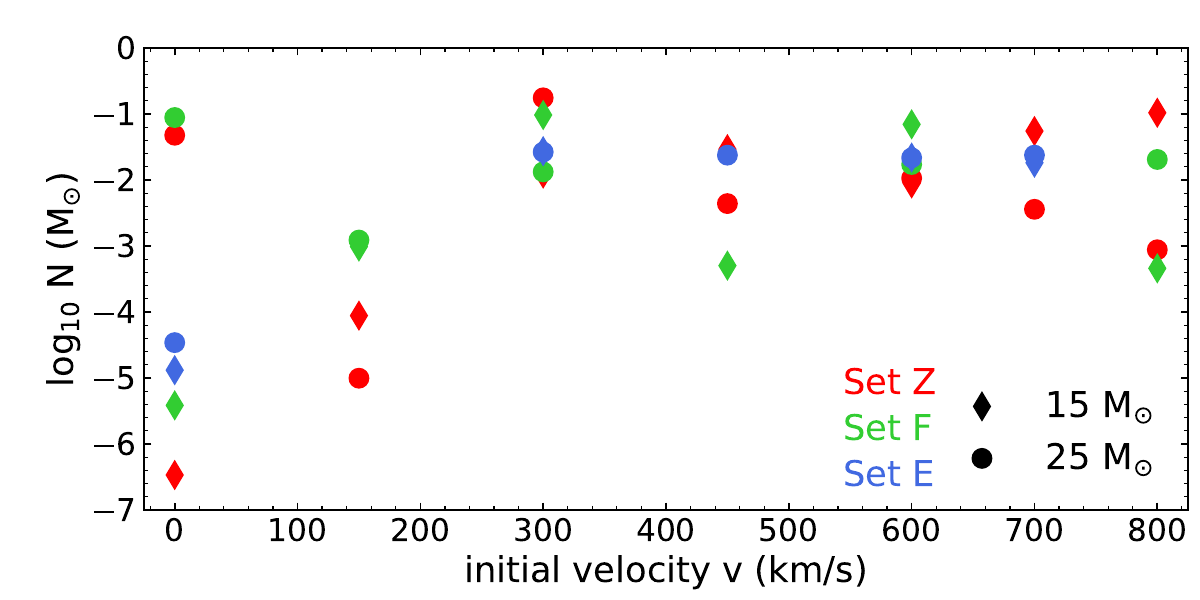}
\caption{Nitrogen yields as a function of the initial rotation velocity.\label{fig:nitro}}
\end{figure*}

Nitrogen is synthesized by the CNO cycle in the hydrostatic H burning stage and it is therefore a typical secondary elements in non rotating stars. The only exception occurs at zero metallicity where the He convective shell engulfs part of the H rich mantle in some models. The abrupt ingestion of protons at temperatures of the He burning in an environment C rich leads to a burst of synthesis of nitrogen (primary). We obtain such an ingestion in the two non rotating models 025z000 and 025f000.
As already discussed in Sect. \ref{sec:rotation} and \ref{sec:nucleosynthesis}, rotation plays a critical role in the production of N. \figurename~\ref{fig:nitro} shows the N yield as a function of the initial rotation velocity for all the computed models. Both masses and all three metallicities show a steep increase of the N yield up to $\rm v_{ini}=$ 300 km/s and a rough plateau for larger initial velocities. It is worth noting that rotation may lead to N yields as high as those provided by the non rotating massive stars where the penetration of the penetration of the He convective shell in the H rich mantle occurs.
Therefore, the rather high and somewhat unexpected N abundance plateau obtained from observations in low metallicity halo stars of the Milky Way \citep{chiappini:05,chiappini:06,grisoni:21} could be explained in terms of the nucleosynthesis of these very low metallicity massive stars (rotating and not). By the way, the temporal evolution of N has been studied over the years by many groups in the context of the chemical evolution of the Milky Way \citep[see, e.g.,][and refereces therein]{grisoni:20,grisoni:21,franco:21}. Of course, a conclusive analysis on this subject would require additional models as well as a quantitative estimate of the contribution of these stars and of the Initial Distribution of their ROtation Velocities (IDROV) to the enrichment of N in the early stages of evolution of the Galaxy and of the Universe in general.

\subsection{Fluorine}

\begin{figure*}
\epsscale{1.0}
\plotone{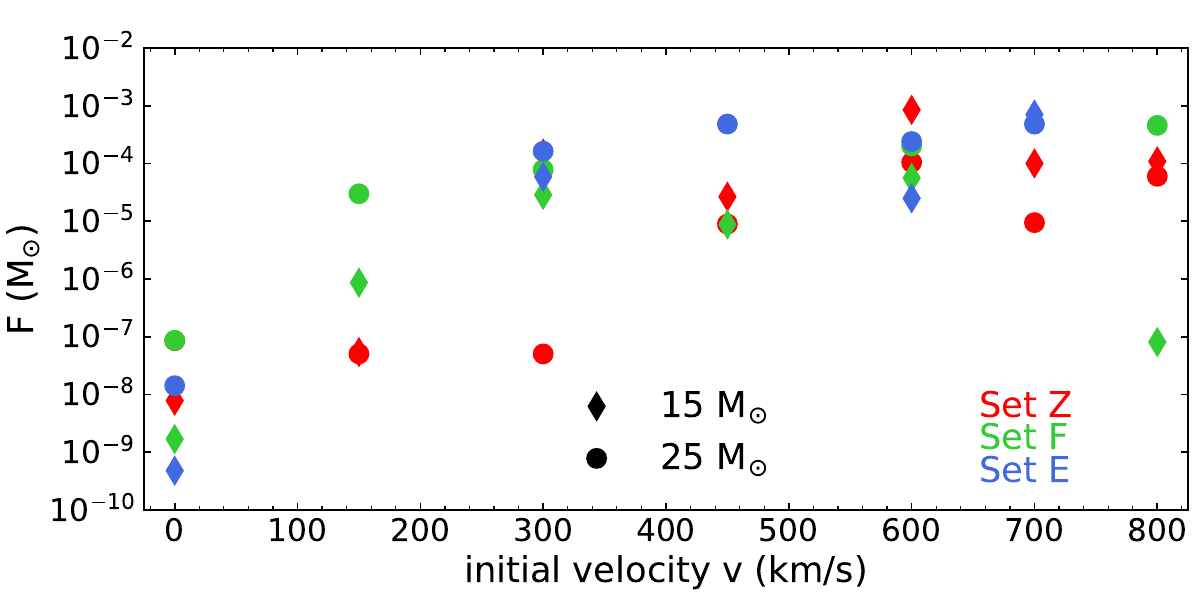}
\caption{Flourine yields as a function of the initial rotation velocity.\label{fig:fluo}}
\end{figure*}

%The main sequence of processes that leads to the synthesis of fluorine is 
%the one identified by \cite{forestini:92} and described in Sect \ref{sec:nucleosynthesis}.
%Rapidly rotating massive stars could significantly contribute or even dominate the fluorine production at least at very low metallicity \citep{grisoni:20,franco:21}. Figure \ref{fig:fluo} shows the dependence of the \nuk{F}{19} yields on the initial rotation velocity for both masses and the three initial rotation velocities. The plot shows a good trend with the velocity, while it is almost independent of the initial metallicity, resembling that of N. In particular it increases with the initial rotation velocity and then bends towards a rough plateau for $v_{ini}$ $>300$ km/s, reaching values as high as $10^{-4}-10^{-3}$ \msun. Note, however, that the non rotating models that in which the H--He shell merging occurs do not contribute significantly to the synthesis of F.

The main sequence of processes that leads to the synthesis of fluorine is the one identified by \cite{forestini:92} and described in Section \ref{sec:nucleosynthesis}. \cite{prantzos:18} have shown that rapidly rotating massive stars can be responsible of the synthesis of fluorine at low metallicity \citep[see also][]{grisoni:20,franco:21}. Here we confirm such a result also at extremely low metallicities. Figure \ref{fig:fluo} shows the dependence of the \nuk{F}{19} yields on the initial rotation velocity for both masses and the three initial rotation velocities. The plot shows a good trend with the velocity, while it is almost independent of the initial metallicity, resembling that of N. In particular it increases with the initial rotation velocity and then bends towards a rough plateau for $v_{ini}$ $>300$ km/s, reaching values as high as $10^{-4}-10^{-3}$ \msun. Note, however, that the non rotating models that in which the H--He shell merging occurs do not contribute significantly to the synthesis of F.

\subsection{The heavy elements}

\begin{figure*}
\epsscale{1.15}
\plotone{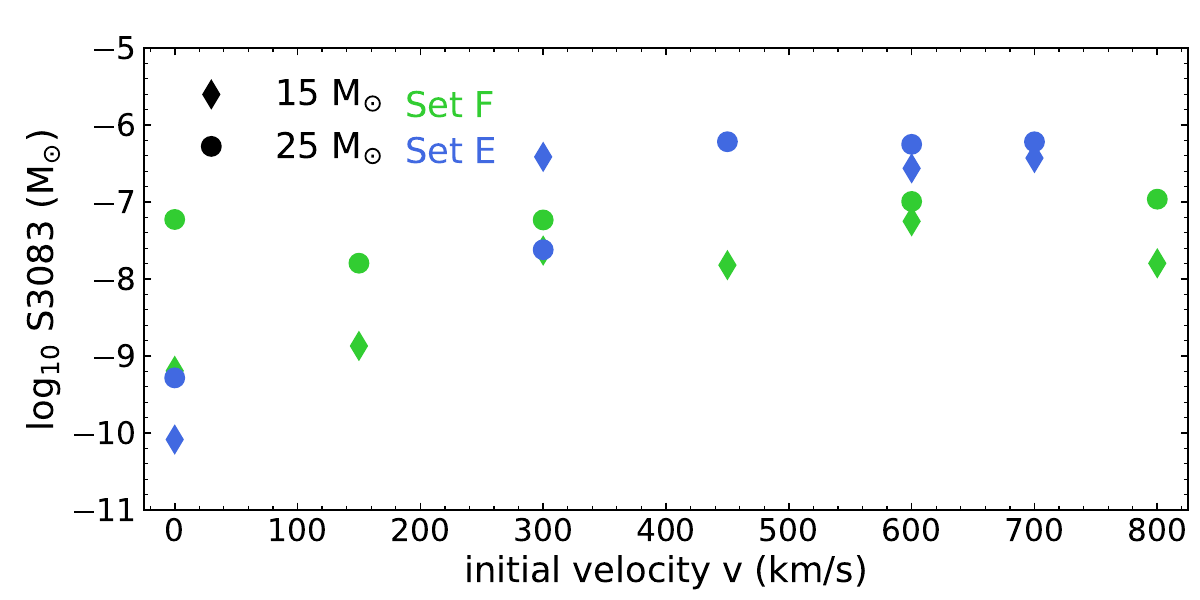}
\caption{Total net yield of elements from Ga up to Bi (S3083, see text) as a function of the initial velocity for all the explored metallicity.\label{fig:sumz}}
\end{figure*}

Let us consider first the sum of the yields of all the elements heavier than Zn. \figurename~\ref{fig:sumz} shows this quantity as a function of the initial rotation velocity for stars of Set F and E. Note that zero metallicity stars do not synthesize any appreciable amount of elements beyond Zn \citep{chieffi:04} and hence do not appear in this plot. The quantity plotted in the figure is the net total yield, defined as the difference between the final and the initial sum of the yields, of all the elements with $\rm Z > 30$ up to $\rm Z = 83$ (hereinafter S3083). An overall look at the figure shows a definite increase of S3083 with the initial rotation velocity. A closer look shows a few other features worth being mentioned. First of all, while three out of the four non rotating models (015f000, 015f000, and 025f000) barely reach a global abundance of the order of $10^{-10}-10^{-9}~$ \msun, the model 025f000 shows a much larger increase of S3083 (of the order of $10^{-7}$ \msun, green dot). Such a high abundance is the consequence of the merging of the He convective shell with the H rich mantle. The abrupt ingestion of protons down to the He burning layers triggers a short burst of neutrons due to the activation of the \nuk{C}{13}($\alpha$,n)\nuk{O}{16} nuclear reaction rate that leads to a non negligible neutron capture nucleosynthesis. The second thing worth noting is that, for each mass, S3083 increases up to a rotation velocity of the order of 300 km/s and then bends considerably for large initial rotation velocities. In the 15 \msun\ of Set F rotation increases S3083 up to a factor of the order of 50 times the initial value while the increase may become ten times larger (i.e. 500 times) in the 15 \msun\ of Set E. In the 25 \msun, S3083 may increase up to a factor 1000 with respect to the non rotating model while it remains roughly flat in the 25 \msun\ of Set E and close to the value obtained in the non rotating case. 

The global picture shown by \figurename~\ref{fig:sumz} does not say anything about the distribution of the elements between Ga and Bi but only how much matter flows above Zn as a function of the mass and initial rotation velocity. In order to understand how matter populates different parts of the chart of the nuclides we will take advantage of the existence of the three magic neutron numbers 50, 82 and 126. Nuclei having these numbers of neutrons have in fact a minimum of the neutron capture nuclear cross section and therefore constitute the main barriers to the flux of the matter towards heavier nuclei. Depending on the neutron to seed ratio and the neutron exposure, matter may stop at the first, second or third neutron closure shell.

\begin{figure*}
\epsscale{1.15}
\plottwo{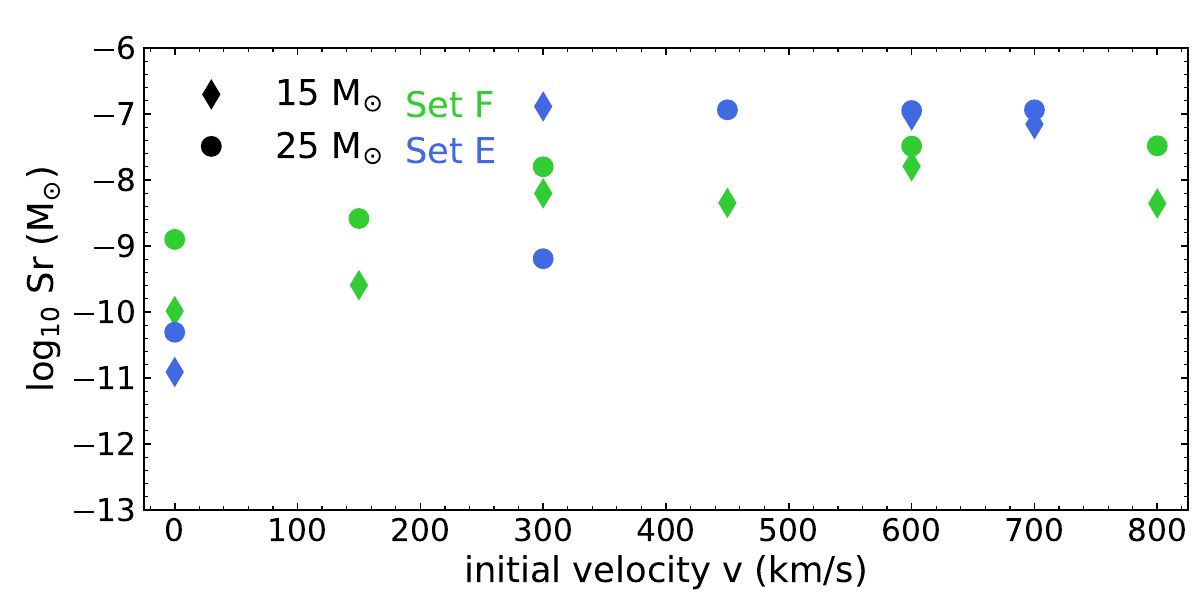}{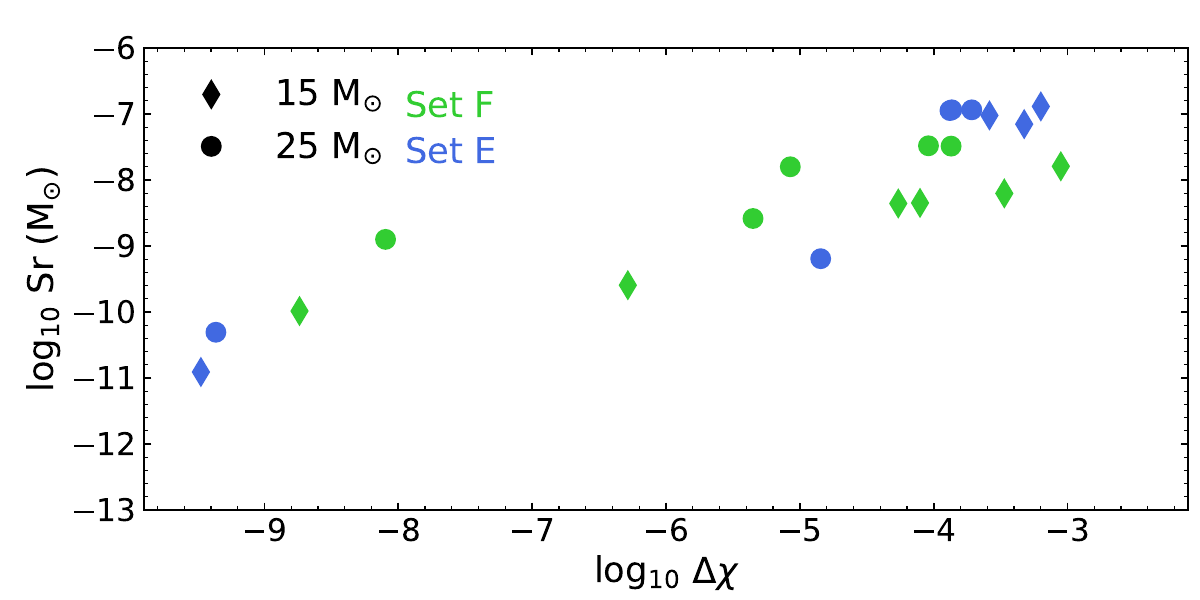}
\plottwo{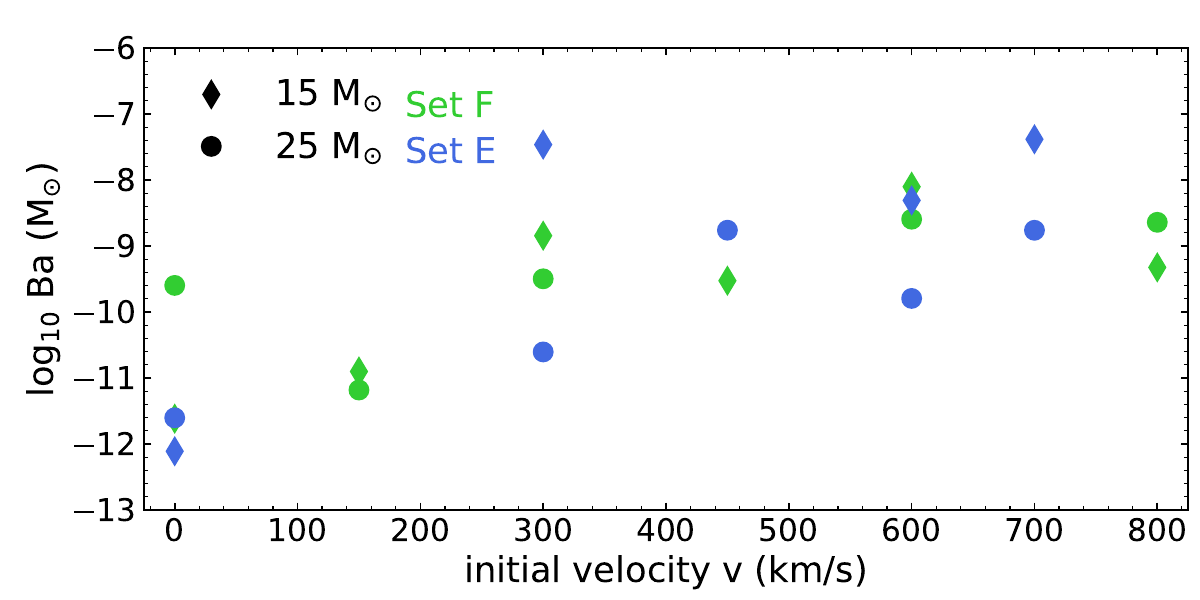}{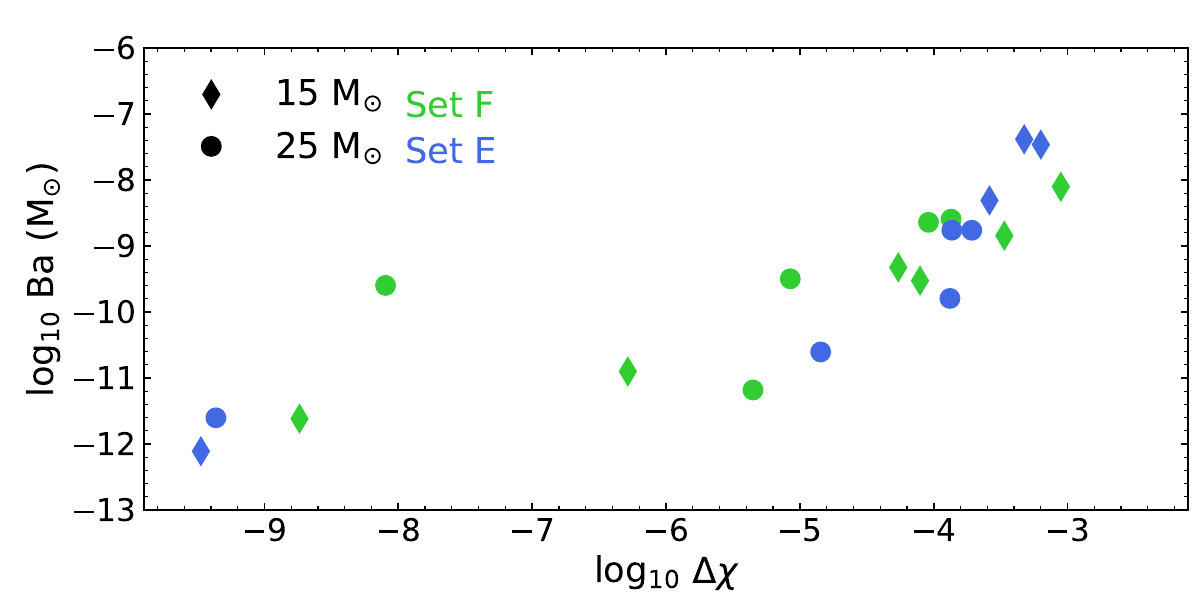}
\plottwo{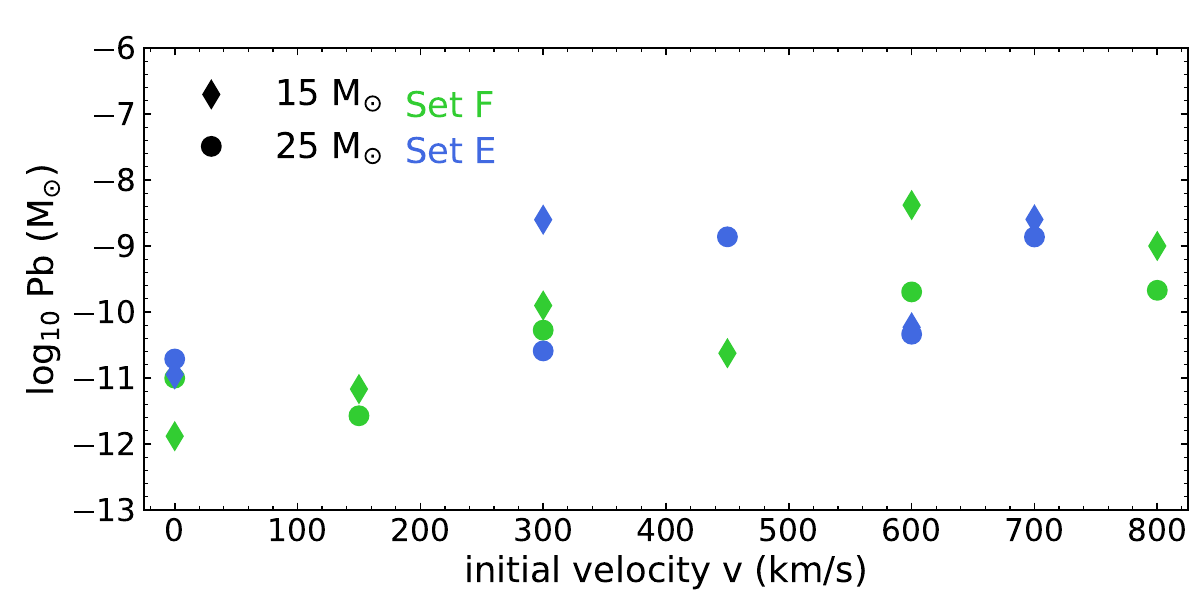}{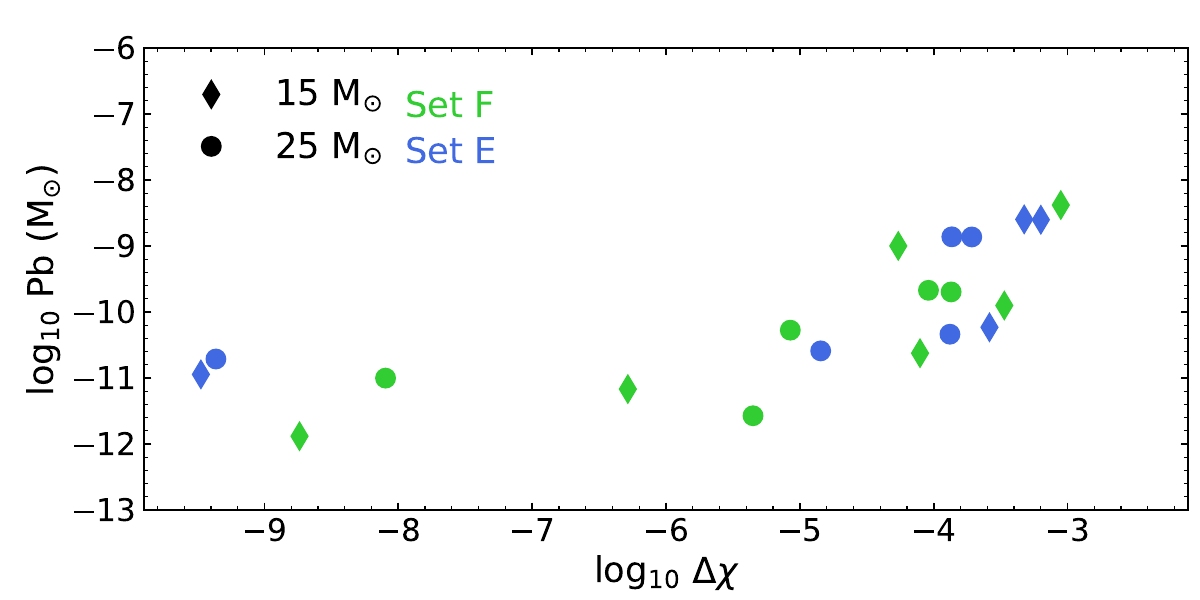}
\caption{Trend of the yields of Sr, Ba, and Pb as a function of the initial rotation velocity (left panels) and of the $\Delta\chi$ parameter (right panels, see text): 15 \msun\ (green diamonds) and 25 \msun (green dots) Set F, 15 \msun\ (blue diamonds) and 25 \msun\ (blue dots) Set E. \label{fig:srbapbvel}}
\end{figure*}

As proxies of the nuclei that have a magic neutron number, we consider in the following three key nuclei: \nuk{Sr}{88}(N=50), \nuk{Ba}{138}(N=82), and \nuk{Pb}{208}(N=126). Figure \ref{fig:srbapbvel} shows in the left column the dependence of the yields of these three nuclei on the initial rotation velocity for both Set F and E, while the trend with respect to the $\Delta\chi$ (which basically quantify the amount of primary \nuk{N}{14} produced by the entanglement between the He and H burning, see Sect \ref{sec:nucleosynthesis}) is shown in the right column. The yields of all three nuclei scale directly with the initial rotation velocity (and $\Delta\chi$), a clear indication of the increase of the neutron flux with the initial rotation velocity. The main increase in the yield of \nuk{Sr}{88} occurs for initial rotation velocities in the range 150-300 km/s, the dependence becomes much weaker above 300 km/s. This behavior may be easily understood by looking at Figure \ref{fig:chi} that shows that the largest increase in the $\Delta\chi$ occurs between 150 and 300 km/s. Set E, in particular, shows a saturation of the Sr yield for both masses. The left panels in Figure \ref{fig:srbapbvel} that refer to Ba and Pb also show an increase of the yields with the initial rotation velocity even if with some scatter while the corresponding right panels show a much tighter scaling. The reason is obviously that the yields directly depend on the $\Delta\chi$ (because it is directly responsible of the neutron flux) and only indirectly on the initial rotation velocity (because it determines the synthesis of primary \nuk{N}{14} in central He burning).

% ================================= [X/O]  =================================
\section{The [X/O]}\label{sec:xsuo}

\begin{figure*}
\epsscale{0.90}
\plotone{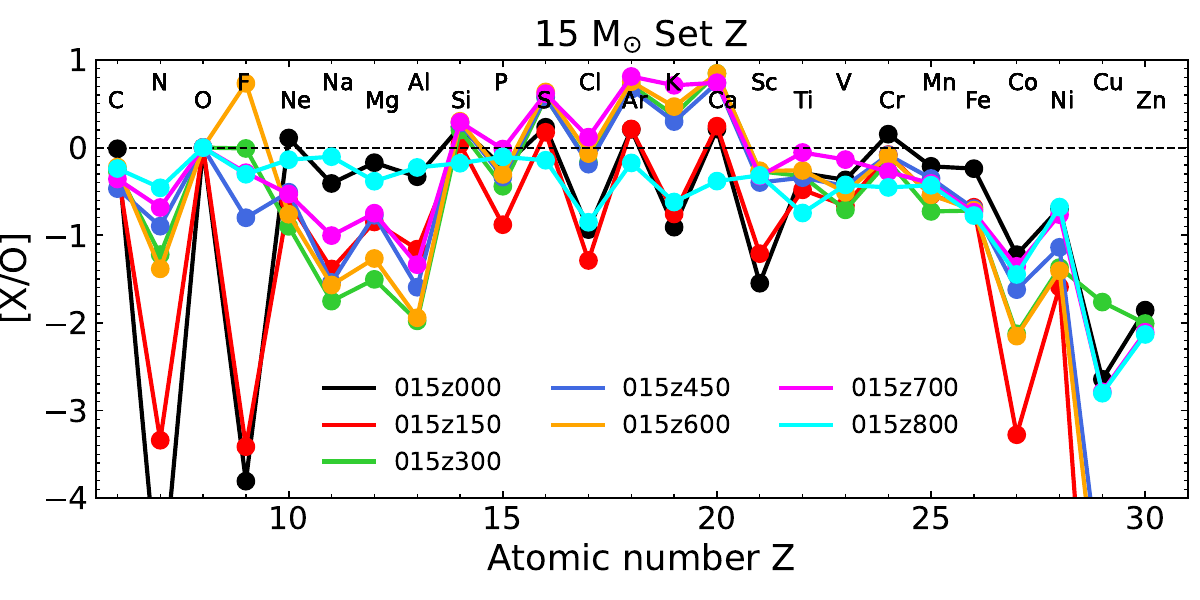}
\plotone{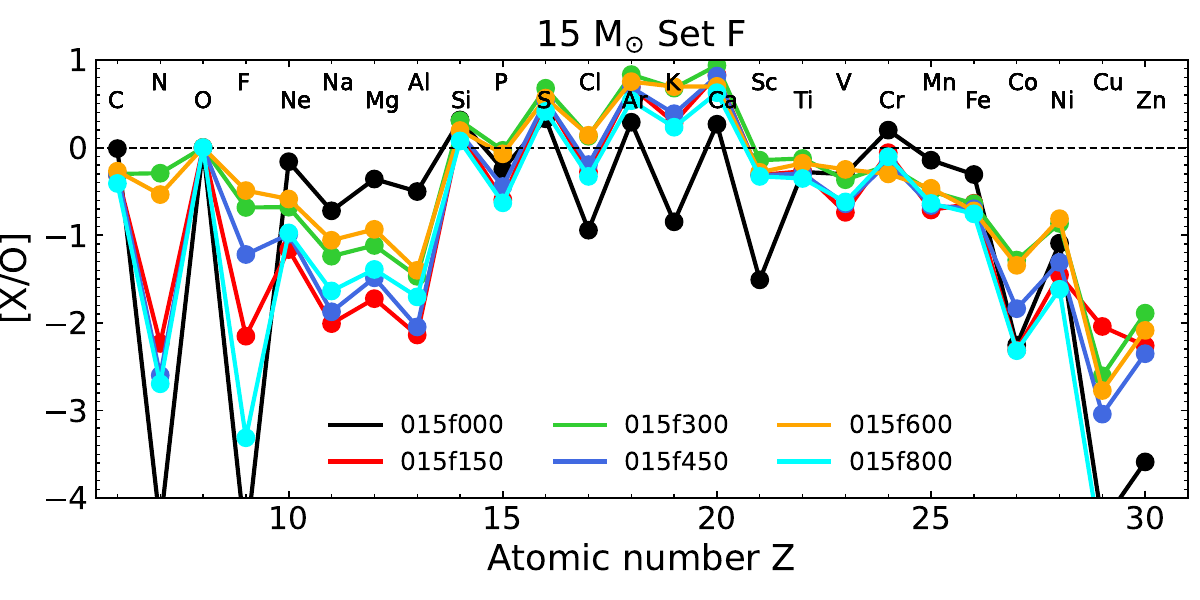}
\plotone{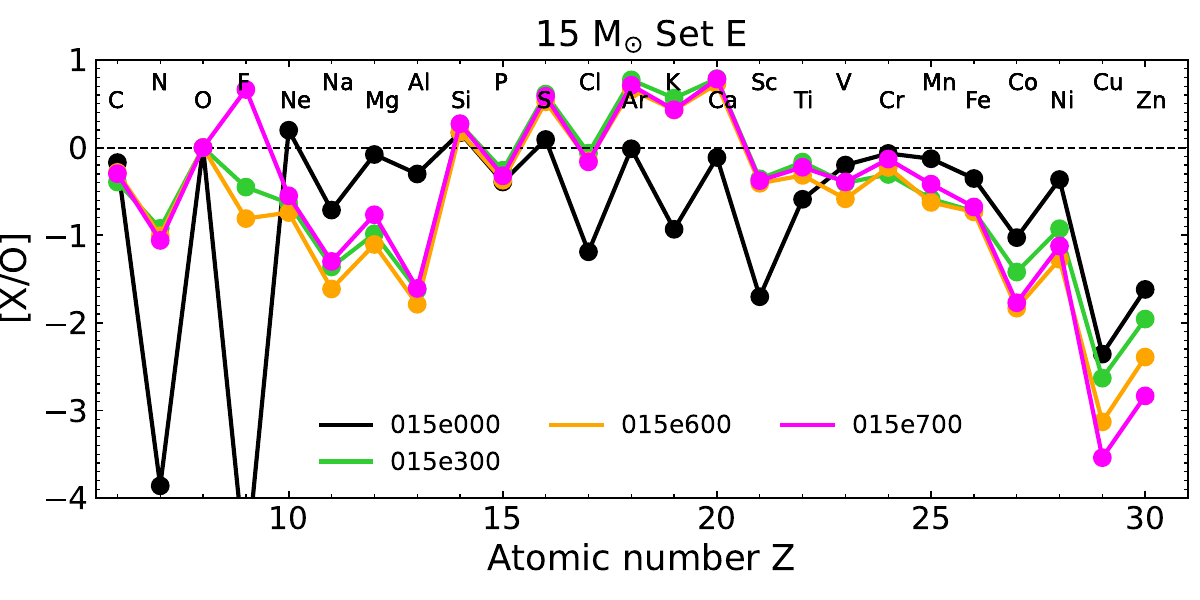}
\caption{Logarithm of the production factors of the 15 \msun\ stars of Set Z (top panels), Set F (central panels), and Set E (bottom panels) as a function of the atomic number $A$. The color scale on the right marks the different initial equatorial velocity of the star, from non rotating (dark blue) up to 800 km/s (yellow).\label{fig:pfCZN15}}
\end{figure*}

\begin{figure*}
\epsscale{0.90}
\plotone{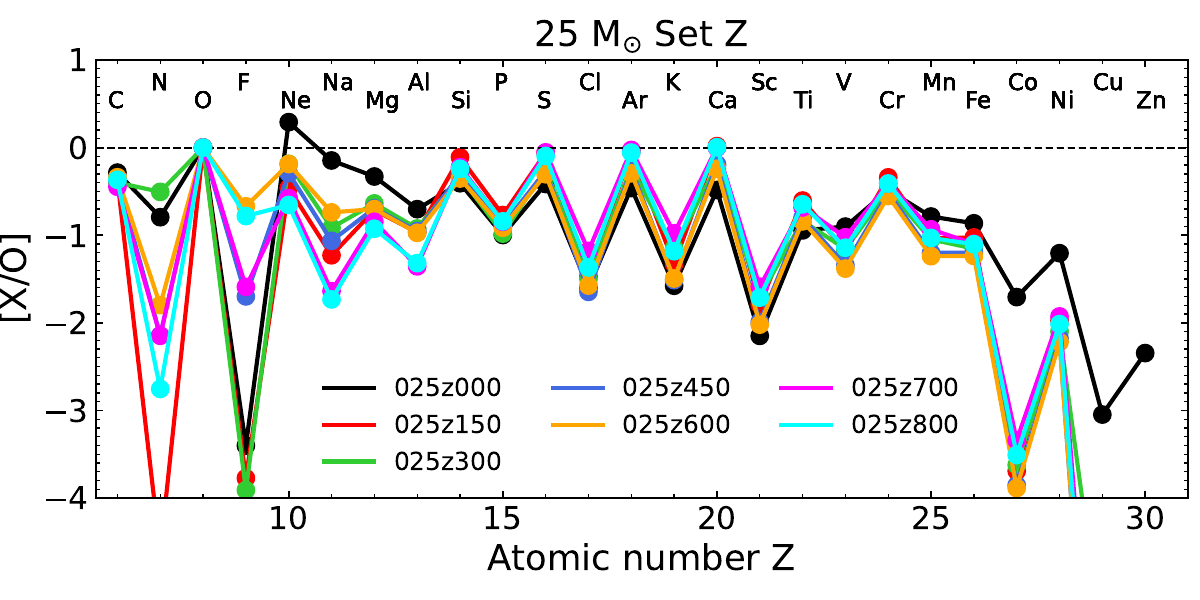}
\plotone{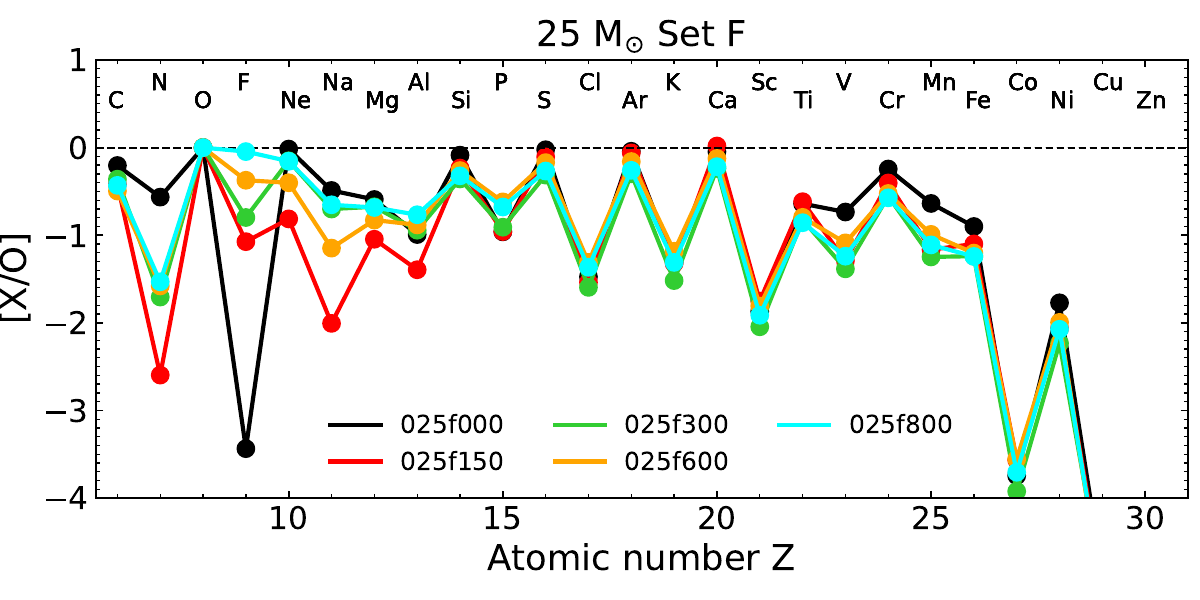}
\plotone{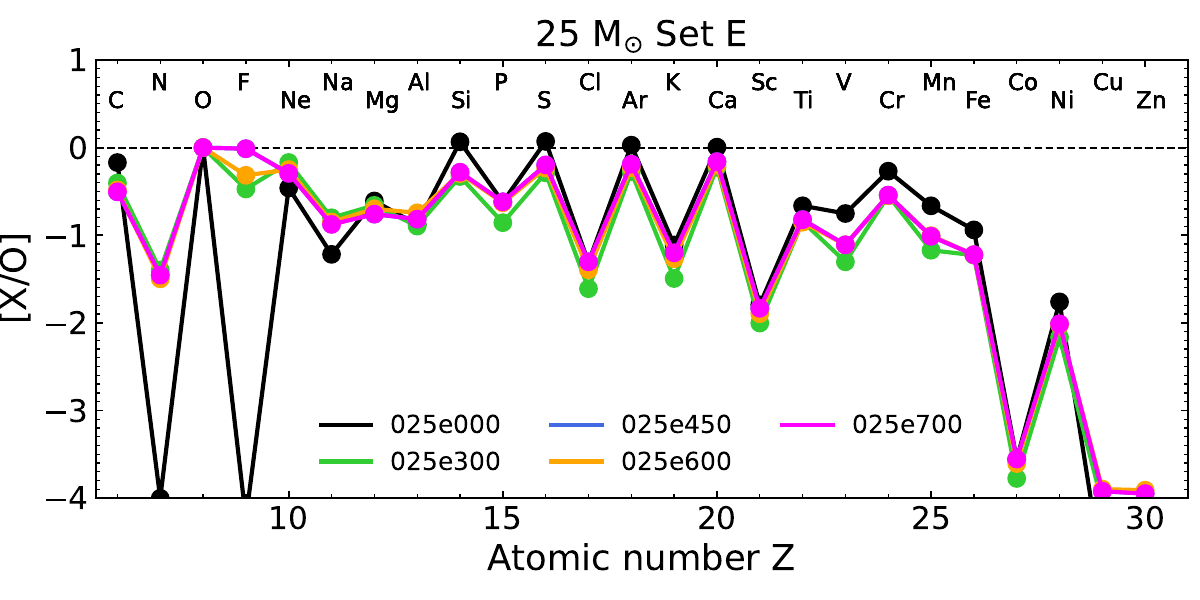}
\caption{Same as \figurename~\ref{fig:pfCZN15}, but for the 25 \msun\ models. \label{fig:pfCZN25}}
\end{figure*}

In the previous section (Sect \ref{sec:yields}) we have shown how the yields of a few key elements depend on the initial rotation velocity for the three metallicity under exam. In this section we want to show all the elements C to Bi and discuss which nuclei are produced enough to have an impact on the chemical evolution of the matter and which do not. For sake of clearness let us discuss first the block of nuclei between C and Zn. An efficient way to make this analysis is to look at the [X/O], that is simply the logarithm of the ratio between the yield of each element and that of the oxygen, minus the logarithm of the same ratio in the Sun, i.e., $\rm [X/O]=log_{10}(Y^i/Y^O)-log_{10}(X^i_\odot/X^O_\odot)$, where Y is the yield in solar masses and $\rm X_i$ the relative solar abundances in mass fraction. Oxygen is by far the best candidate to be the $\it reference$ element because 1) it is the most abundant nucleus synthesized in stars, 2) it is produced almost exclusively by massive stars and 3) it is almost not affected by the explosive burning. A [X/O] greater than, equal, or smaller than that of O immediately tells us if this element is produced relative to O as it is observed in the Sun or not. \figurename~\ref{fig:pfCZN15} and \ref{fig:pfCZN25} show the [X/O] of the elements in the range C to Zn for all the models computed in the present work: they refer to the 15 and 25 \msun, respectively, while the three rows of each Figure refer to Set Z (upper panels), F (middle panels), and E (lower panels). The various colors refer to the different initial rotation velocities. The [X/O] of all the non rotating models (black dots and lines) show the well known odd-even effect typical of the very metal poor stars, i.e., the even nuclei (Mg to Ca) are produced approximately in right amounts with respect to O while the odd nuclei are more or less underproduced (N to K). Let us look at the 25 \msun\ first. The three panels in Figure \ref{fig:pfCZN25} show that only N and F are largely affected by rotation for all three metallicity and that there are rotation velocities for which both [N/O] and [F/O] get quite close to zero, i.e., are basically co-produced with O. The nuclei synthesized by C burning, i.e., Ne, Na, Mg and Al, show a sizeable dependence on rotation in both Set Z and F, though not as large as N and F, (because the amount of C left by the He burning depends on the velocity) while models of set E do not. All other nuclei show only a marginal dependence on the initial rotation velocity. If we turn to the 15 \msun\ the situation is somewhat different. In addition to the odd-even effect that is always present, these models show that in many of the rotating models there is an overproduction of the elements between Si and Ca and an underproduction of the elements Ne to Al, while none of the non rotating models shows such a feature. Such an apparently strange behavior is the simple consequence of the penetration of the O convective shell in the region where the ashes of the C burning are present (see Sect \ref{sec:yields}) that occurs when a star is in shell O burning. When such a mixing occurs, part of the products of the O burning are brought upward in layers external enough to not be seriously affected by the passage of the shock wave and, vice versa, products of the C burning are brought internally enough to be destroyed by the passage of the shock wave. The consequence is therefore that the yields of the products of the C burning are reduced while those of the O burning increased. It is worth noting that such a mixing occurs (in the present set of models) only in the rotating models. The reason is that, how we already discussed above, rotation leads to smaller amount of C at the end of the He burning and hence a lower amount of Ne in the C exhausted core; a less efficient Ne burning shell does not represent a strong barrier to the advancing of the outer border of the O convective shell so that the O convective shell overcomes the Ne shell and penetrates in the C convective shell. 

\begin{figure*}
\epsscale{1.0}
\plotone{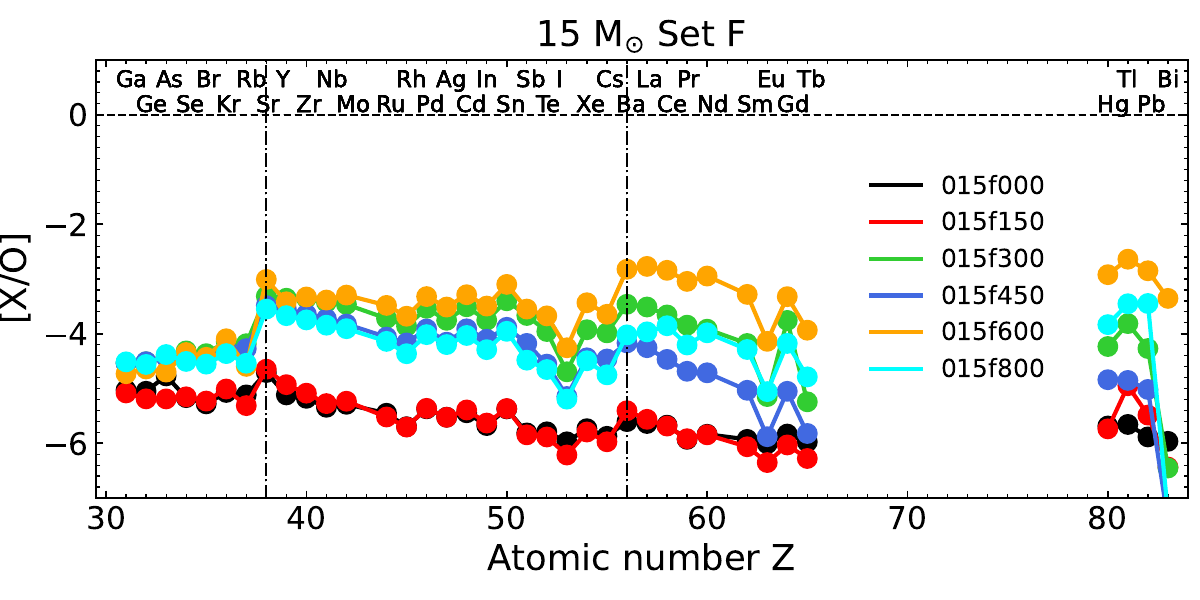}
\plotone{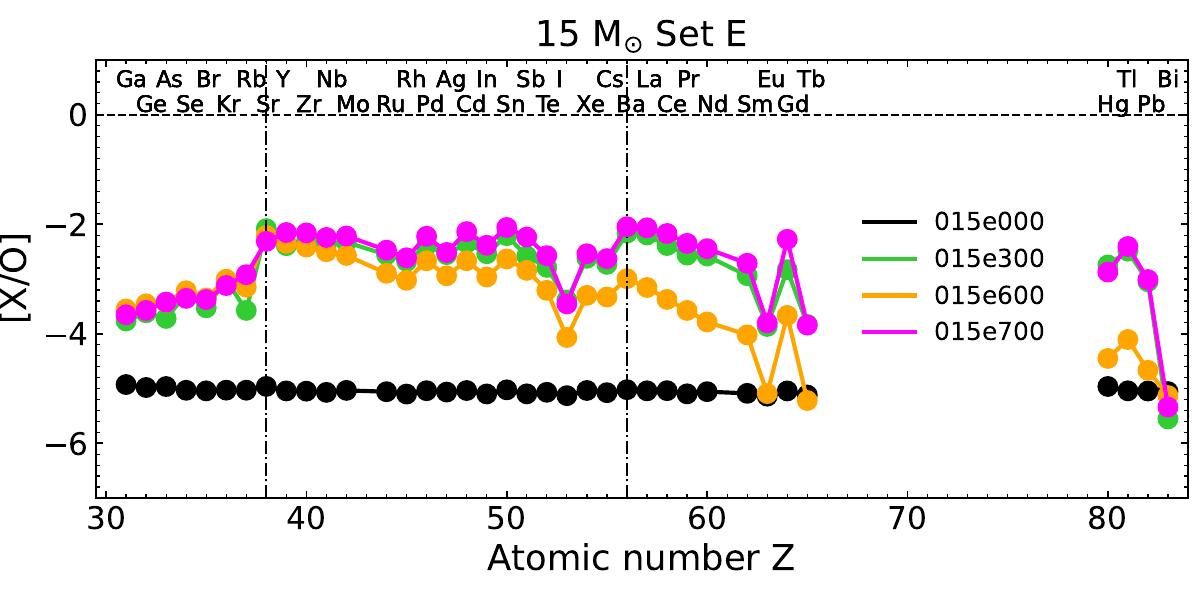}
\caption{[X/O] distribution for the 15 \msun\ stellar models for Set F (top panel) and Set E (bottom panel) as a function of the atomic number $Z$. The two vertical dashed-dotted lines correspond to the atomic number of Sr (Z=38) and Ba (Z=56). \label{fig:pfGABI15}}
\end{figure*}

\begin{figure*}
\epsscale{1.0}
\plotone{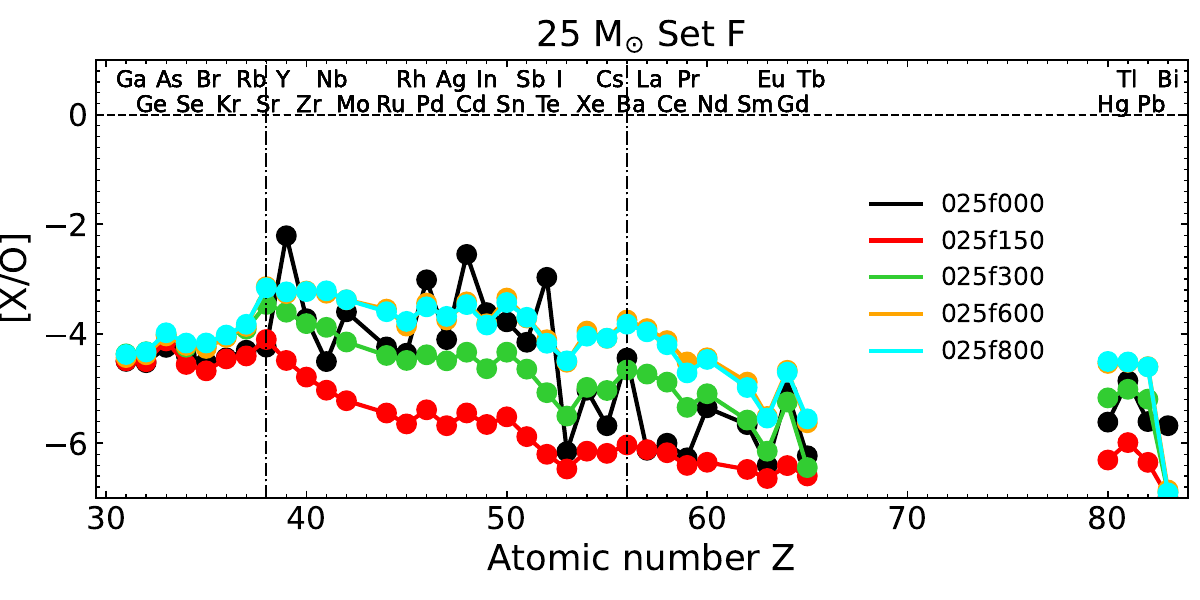}
\plotone{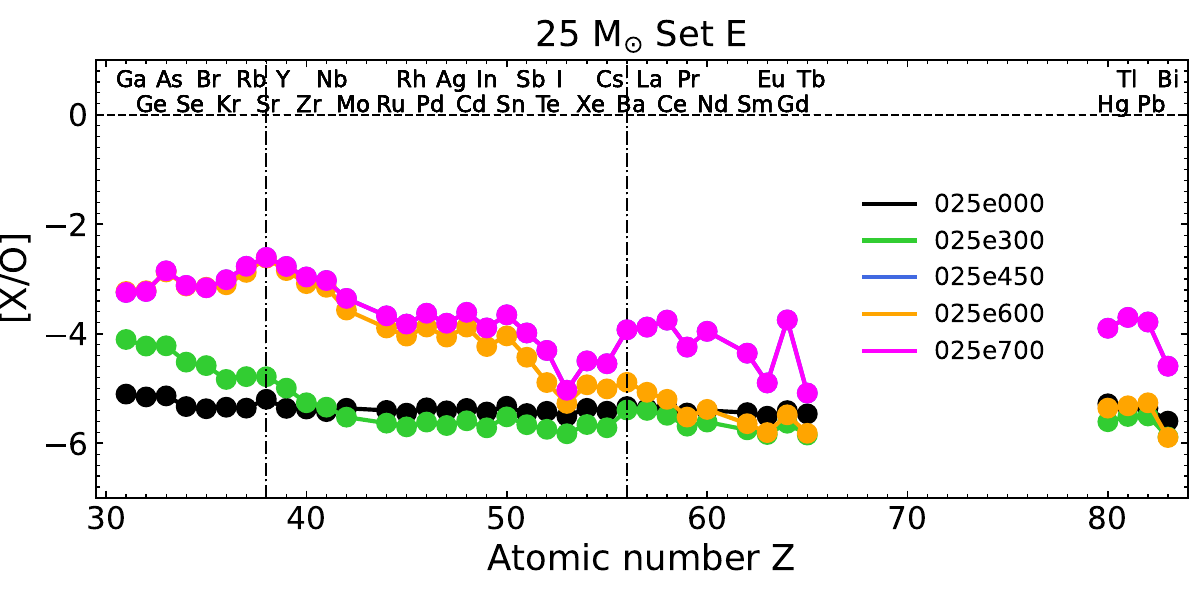}
\caption{Same as \figurename~\ref{fig:pfGABI15} but for the 25 \msun\ models.\label{fig:pfGABI25}}
\end{figure*}

The [X/O] of the elements Ga to Bi is shown in \figurename~\ref{fig:pfGABI15} and \figurename~\ref{fig:pfGABI25}. Once again the two figures refer to the 15 and 25 \msun, respectively, while the two rows refer to Set F (top panel) and E (bottom panel). The various colors refer to the different initial rotation velocities. 
Both 15 \msun\ models show that rotation increases significantly the weak, main and strong component of the neutron capture nucleosynthesis. Let us remind that the elements between Ga and the first neutron closure shell are usually referred to as the weak component, those between the first and the second neutron closure shell are usually named main component while the nuclei around the third peak are named strong component. If we consider Sr, Ba, and Pb as the proxies for the three components, we find that [Sr/O] reaches values as high as --3 and --2 in Set F and E, respectively, as well as Ba and Pb. Of course, while the weak component is easily populated even at the slowest rotation velocities, the main and even more the strong component require much higher rotation velocities. In the case of the 25 \msun\ we must consider that the non rotating model experiences the He-H merger and the large neutron flux determined by the abrupt ingestion of protons within the He burning region produces a burst of production of heavy nuclei with major peaks at Y, Pd, Cd, and Te. Apart from this specific model, all the others show that once again the first peak around Sr is quite well populated while the second one at Ba remains roughly one order of magnitude below that of Sr. It is however remarkable that a non negligible amount of mass is able to overcome the neutron magic number N=50 anyway. Only the fastest rotating models are able to raise somewhat the third peak.

% ================================= OBSERVATIONS =================================
\section{Comparison with observations}\label{sec:comparison_observation}

\begin{figure*}
\epsscale{1.0}
\plotone{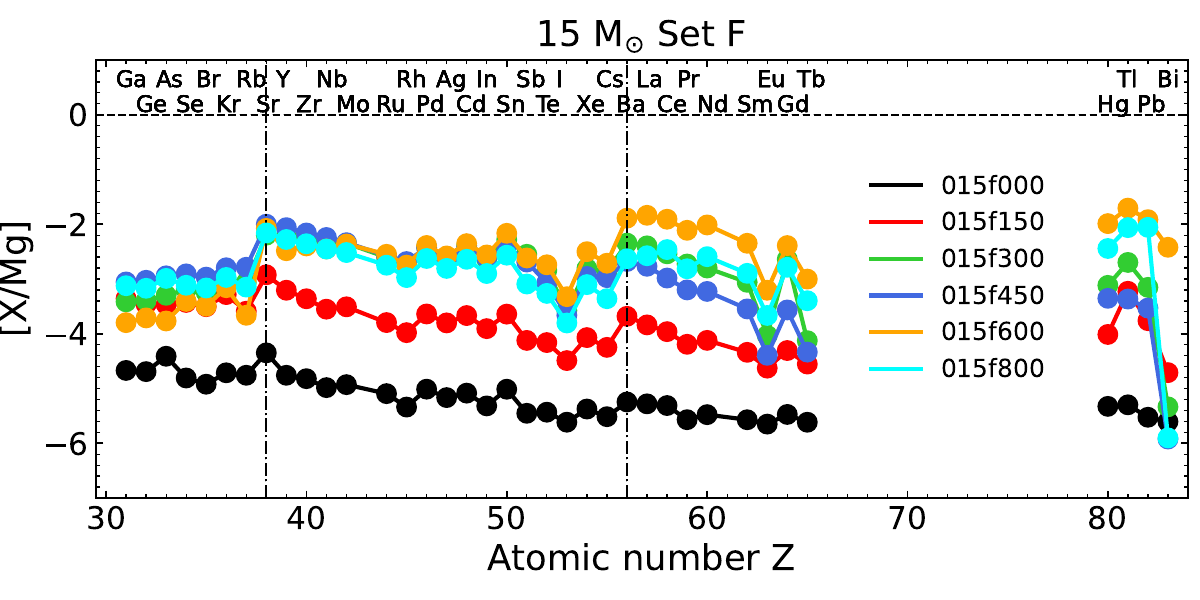}
\plotone{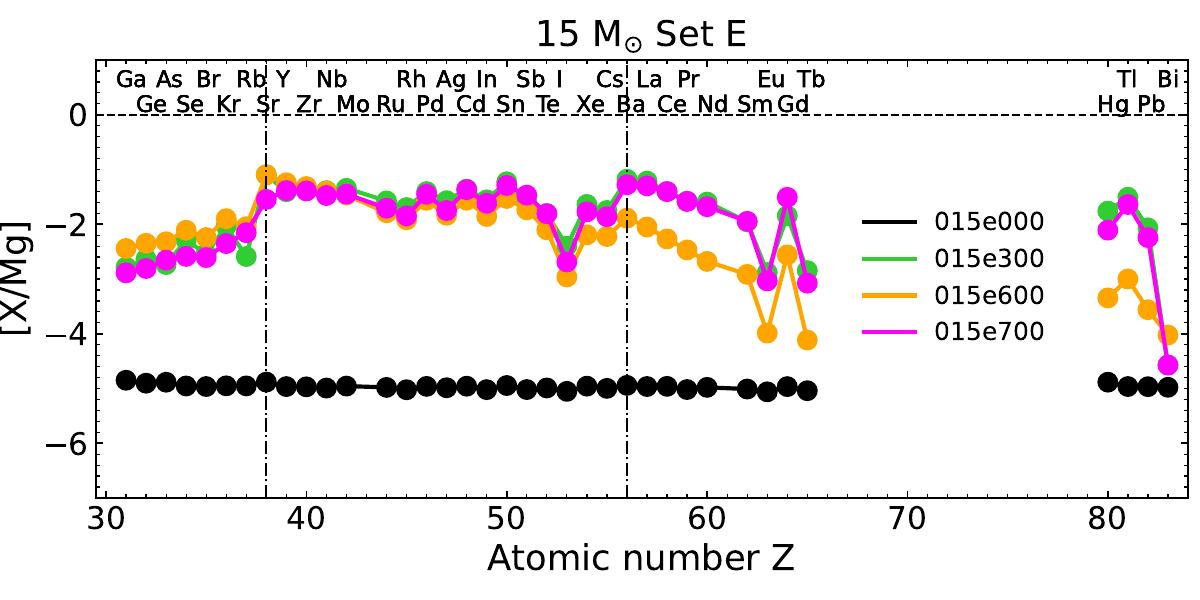}
\caption{[X/Mg] distribution for the 15 \msun\ stellar models for Set F (top panel) and Set E (bottom panel) as a function of the atomic number $Z$. The two vertical dashed-dotted lines correspond to the atomic number of Sr (Z=38) and Ba (Z=56). \label{fig:pfGABIMg15}}
\end{figure*}

\begin{figure*}
\epsscale{1.0}
\plotone{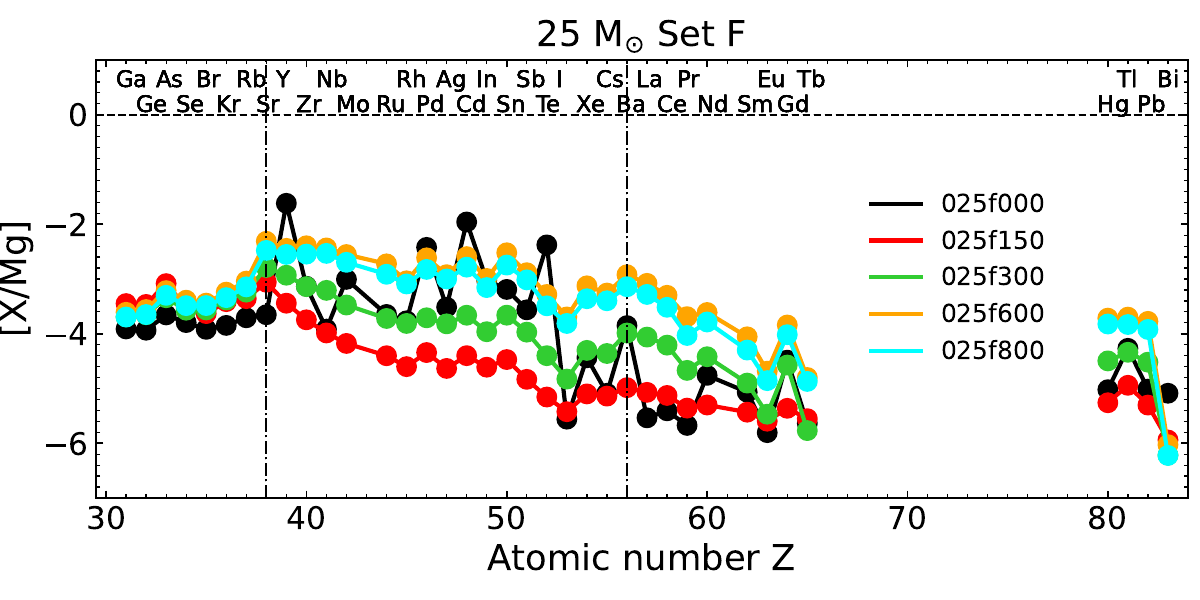}
\plotone{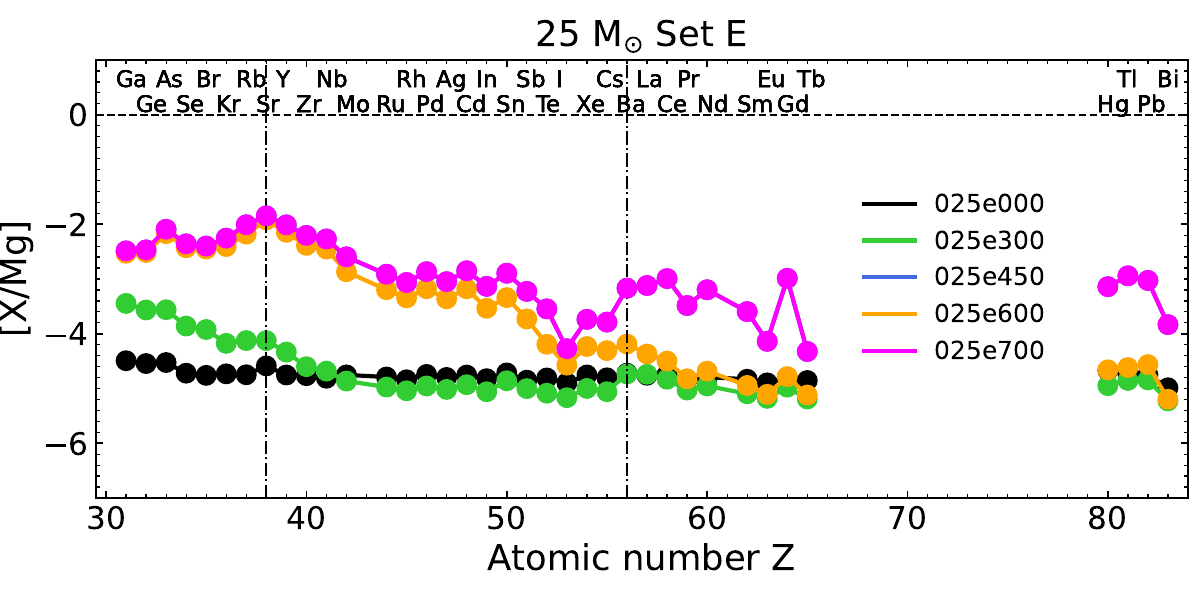}
\caption{Same as \figurename~\ref{fig:pfGABIMg15} but for the 25 \msun\ models.\label{fig:pfGABIMg25}}
\end{figure*}

The plots shown in the previous section are based on the strong argument that O is by far the most reliable reference element respect to which evaluate the production of all the others. Unfortunately O is very difficult to observe at very low metallicity and therefore the sample of stars for which the abundance of this element is available is quite scarce. If we want to compare our models with robust observed samples of stars we are forced to choose another element. We arbitrarily chose Mg, an element for which a large number of observational data is available. \figurename~\ref{fig:pfGABIMg15} and \figurename~\ref{fig:pfGABIMg25} show the [X/Mg] of the elements Ga to Bi for both Set F and E because here we want to show if and which rotating models of extremely low metallicity can produce enough heavy elements in the range of values observed in a sample of extremely metal poor stars.
We selected observations of EMP and UMP stars with [Fe/H]$\leq$--3.5 from JINAbase \citep{jina}, by obviously requiring that measures of Mg, Sr, and Ba are simultaneously available. These constraints allowed us to select 53 stars (Table \ref{tab:obs}).  Unfortunately, the lack of Pb observations in metal poor stars did not allow us to analyze also the strong component at such low metallicity. 

\clearpage 

\startlongtable
\begin{deluxetable*}{llrrrrr}
\renewcommand\thetable{4}
%\tabletypesize{\tiny}
\setlength{\tabcolsep}{0.040in} 
%\tablecolumns{7}
\tablewidth{0pt}
\tablecaption{Observations of EMP and UMP stars with [Fe/H]$\leq$--3.5 from JINAbase \citep{jina}. \label{tab:obs}}
\tablehead{Number & Star name & [C/Fe] & [Mg/Fe] & [Fe/H] & [Sr/Fe] & [Ba/Fe]}
%\colnumbers

\startdata
 1 & $\rm             HE0013-0257 $ & $  0.22 $ &  0.68 & -3.82 & -0.46  & -1.16 \\
 2 & $\rm              CS22942-002$ & $  0.35 $ &  0.56 & -3.61 & -1.73  & -1.24 \\
 3 & $\rm              CS22183-031$ & $  0.37 $ &  0.77 & -3.57 & -0.14  &  0.20 \\
 4 & $\rm         LAMOSTJ0126+0135$ & $ -0.51 $ &  0.42 & -3.57 & -1.61  & -1.14 \\
 5 & $\rm              CS22189-009$ & $  0.30 $ &  0.44 & -3.92 & -0.89  & -1.52 \\
 6 & $\rm              CS22963-004$ & $  0.51 $ &  0.58 & -4.09 & -0.95  & -0.61 \\
 7 & $\rm              CS22172-002$ & $  0.09 $ &  0.18 & -3.86 & -1.21  & -1.22 \\
 8 & $\rm              HE1012-1540$ & $  2.40 $ &  1.81 & -4.17 & -0.39  & -0.28 \\
 9 & $\rm              HE1310-0536$ & $  2.36 $ &  0.42 & -4.15 & -1.08  & -0.50 \\
10 & $\rm                 Boo-1137$ & $  0.26 $ &  0.45 & -3.71 & -1.32  & -0.55 \\
11 & $\rm              CS22878-101$ & $ -0.37 $ &  0.60 & -3.53 & -0.12  & -0.37 \\
12 & $\rm              CS22891-200$ & $  0.53 $ &  0.82 & -4.06 & -1.18  & -0.75 \\
13 & $\rm              CS22885-096$ & $  0.60 $ &  0.84 & -4.41 & -1.75  & -1.64 \\
14 & $\rm              CS22950-046$ & $  0.61 $ &  0.58 & -4.12 &  0.10  & -1.01 \\
15 & $\rm              CS22897-008$ & $  0.60 $ &  0.60 & -3.83 &  0.67  & -1.17 \\
16 & $\rm              CS29498-043$ & $  2.75 $ &  1.78 & -3.87 &  0.10  & -0.49 \\
17 & $\rm              CS22956-050$ & $  0.26 $ &  0.67 & -3.67 & -0.44  & -0.90 \\
18 & $\rm              CS22960-053$ & $  1.40 $ &  0.77 & -3.64 & -0.01  &  1.03 \\
19 & $\rm              CS22960-048$ & $  0.47 $ &  0.72 & -3.91 & -2.00  & -1.59 \\
20 & $\rm              CS22949-048$ & $  0.17 $ &  0.40 & -3.55 & -1.20  & -1.45 \\
21 & $\rm              CS22949-037$ & $  1.16 $ &  1.56 & -4.38 &  0.49  & -0.60 \\
22 & $\rm              CS22952-015$ & $ -0.65 $ &  0.30 & -3.87 & -0.65  & -1.50 \\
23 & $\rm                 BD+44493$ & $  1.20 $ &  0.89 & -4.28 & -0.55  & -0.88 \\
24 & $\rm              CS22881-032$ & $ <0.77 $ &  0.50 & -3.55 &  0.23  & -0.30 \\
25 & $\rm              CS22898-047$ & $  0.40 $ &  0.55 & -3.51 & -0.14  & -0.80 \\
26 & $\rm              CS30339-073$ & $  0.20 $ &  0.45 & -3.93 & -1.11  & -1.55 \\
27 & $\rm              HE0056-3022$ & $  0.25 $ &  0.38 & -3.77 & -0.91  & -1.46 \\
28 & $\rm                 CD-38245$ & $<-0.19 $ &  0.66 & -4.59 & -0.63  & -1.00 \\
29 & $\rm              HE0048-6408$ & $ -0.28 $ &  0.40 & -3.75 & -0.82  & -1.48 \\
30 & $\rm              HE0057-5959$ & $  0.86 $ &  0.51 & -4.08 & -1.06  & -0.46 \\
31 & $\rm             HE0302-3417a$ & $  0.48 $ &  0.55 & -3.70 & -1.35  & -2.10 \\
32 & $\rm              HE1320-2952$ & $ <0.52 $ &  0.40 & -3.69 & -0.37  & -0.96 \\
33 & $\rm              HE1506-0113$ & $  1.47 $ &  0.89 & -3.54 & -0.84  & -0.80 \\
34 & $\rm              CS22948-066$ & $ -0.43 $ &  0.46 & -3.50 & -0.35  & -0.83 \\
35 & $\rm              HE2233-4724$ & $ -0.48 $ &  0.49 & -3.65 & -0.77  & -1.13 \\
36 & $\rm             HE2302-2154a$ & $  0.38 $ &  0.28 & -3.88 & -0.60  & -1.50 \\
37 & $\rm              HE2318-1621$ & $  0.54 $ &  0.20 & -3.67 & -1.00  & -1.61 \\
38 & $\rm              HE2331-7155$ & $  1.34 $ &  1.20 & -3.68 & -0.85  & -0.90 \\
39 & $\rm  SDSSJ090733.28+024608.1$ & $   -   $ &  0.36 & -3.52 &  0.04  &  0.25 \\
40 & $\rm           SDSSJ1322+0123$ & $  0.49 $ &  0.25 & -3.64 & -1.24  & -1.30 \\
41 & $\rm SMSS\_J004037.56-515025.2$ & $ -0.09 $ &  0.59 & -3.83 & -0.99  & -1.01 \\
42 & $\rm SMSS\_J005953.98-594329.9$ & $  1.20 $ &  0.61 & -3.93 & -1.11  & -0.65 \\
43 & $\rm SMSS\_J010651.91-524410.5$ & $  0.13 $ &  0.56 & -3.79 & -0.79  & -1.64 \\
44 & $\rm SMSS\_J024858.41-684306.4$ & $  0.66 $ &  0.57 & -3.71 & -0.15  &  0.59 \\
45 & $\rm SMSS\_J085924.06-120104.9$ & $ -0.20 $ &  0.61 & -3.63 & -1.01  & -1.23 \\
46 & $\rm SMSS\_J173823.36-145701.0$ & $  0.60 $ &  0.44 & -3.58 &  0.02  & -0.25 \\
47 & $\rm SMSS\_J184226.25-272602.7$ & $<-0.29 $ &  0.62 & -3.89 & -1.77  & -1.27 \\
48 & $\rm SMSS\_J184825.29-305929.7$ & $  0.25 $ &  0.50 & -3.65 & -1.52  & -1.58 \\
49 & $\rm              HE1116-0634$ & $  0.08 $ &  0.82 & -3.73 & -2.26  & -1.81 \\
50 & $\rm              HE0218-2738$ & $    -  $ &  0.04 & -3.52 & -0.33  & -0.05 \\
51 & $\rm              HE0132-2429$ & $  0.83 $ &  0.39 & -3.60 &  0.13  & -0.85 \\
52 & $\rm              HE0926-0546$ & $ <0.62 $ &  0.32 & -3.73 & -1.22  & -0.83 \\
53 & $\rm                   G64-12$ & $ <1.10 $ &  0.50 & -3.58 &  0.10  & -0.27 \\
\enddata
\end{deluxetable*}

\begin{figure*}
\epsscale{0.9}
\plotone{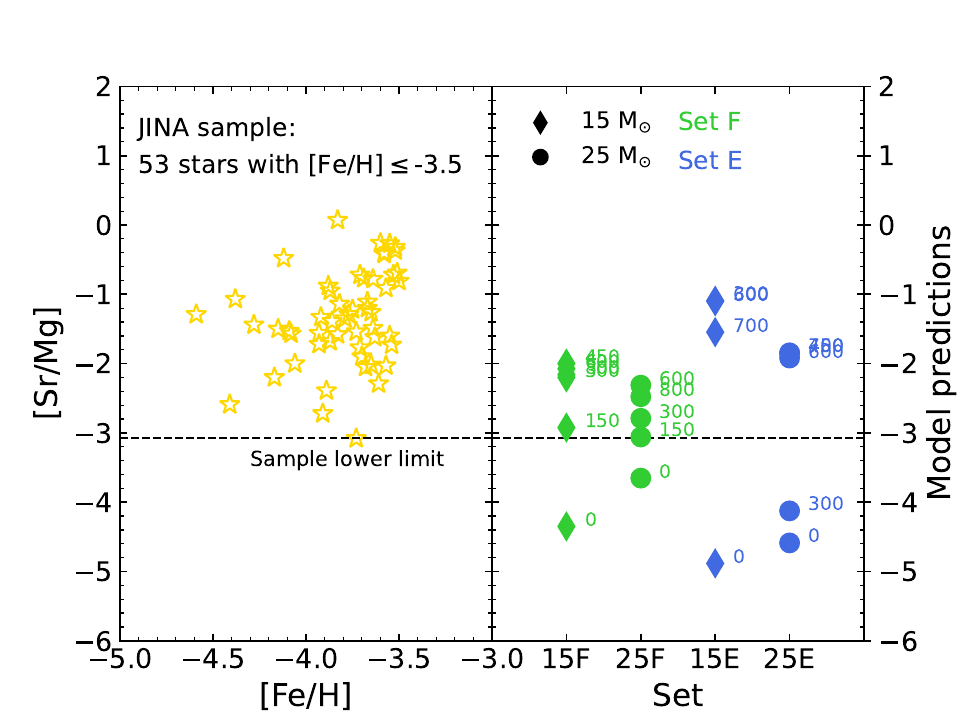}
\plotone{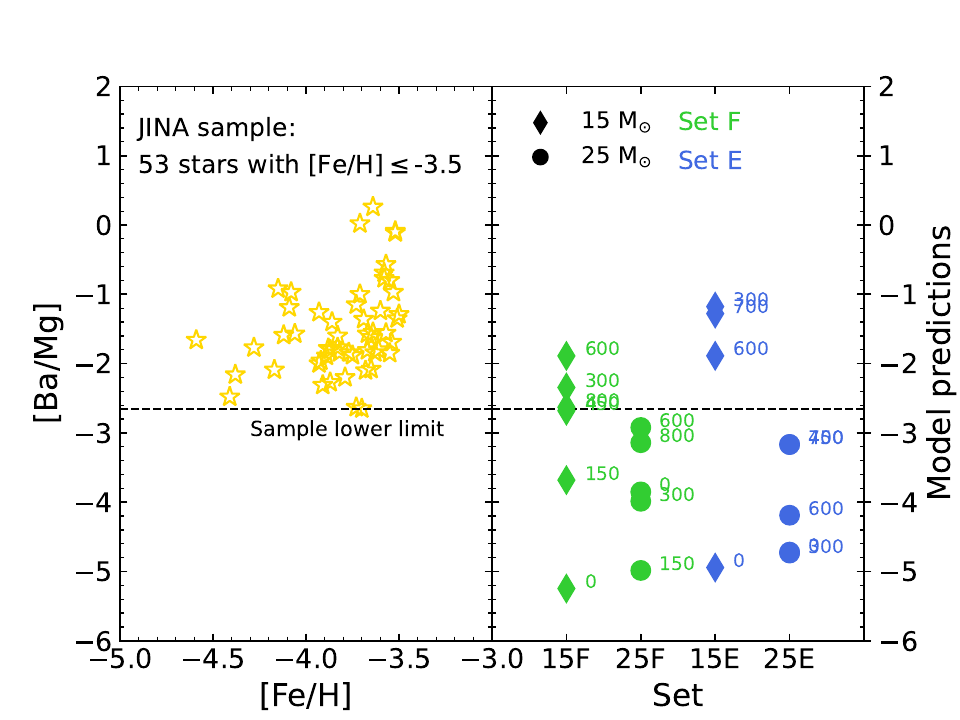}
\caption{Upper panel: on the left side the [Sr/Mg] ratio of sample of observations of stars with [Fe/H]$\leq$--3.5 taken from JINAbase and on the right side our theoretical predictions. The horizontal dashed line shows the lowest observed [Sr/Mg] value of the sample. Lower panel: the same but for [Ba/Mg].\label{fig:srbamg}}
\end{figure*}

The upper panel of \figurename~\ref{fig:srbamg} shows a comparison between the [Sr/Mg] observed in the sample of stars extracted from the database mentioned above and our theoretical predictions. Most of rotating models of both metallicities fit the range of values observed in these stars, hence even a modest initial rotation velocity is able to raise the weak component up to the observed values. The analogous comparison for [Ba/Mg] (lower panel of \figurename~\ref{fig:srbamg}) shows, on the contrary, that only the fastest rotating models may fit the range of values observed at metallicity lower than --4 or so. 

\begin{figure*}
\epsscale{1.0}
\plotone{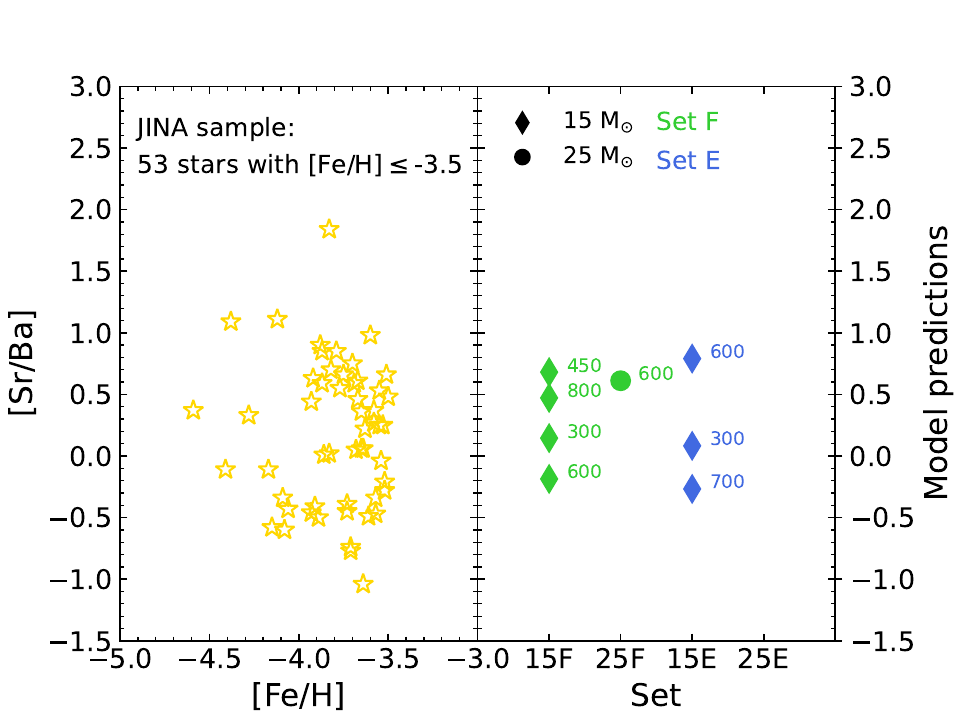}
\caption{Left: [Sr/Ba] ratios in a sample of observations of stars with [Fe/H]$\leq$--3.5 taken from JINAbase. Right: out theoretical predictions.\label{fig:srba}}
\end{figure*}

If we consider only the models that have both [Sr/Mg] and [Ba/Mg] $>$--3, i.e., above the minimum observed values, we can compare the [Sr/Ba] observed in this sample of stars with the present models. This is an interesting ratio to look at since it is considered a good tracer of the origin of the abundances of these two elements. This ratio in fact changes significantly between the AGB and the massive stars scenarios. A low [Sr/Ba] ratio ($\rm [Sr/Ba]<0$) is usually associated to the AGB stars because these are the stars in which the main component is predominantly produced, while an high [Sr/Ba] ratio is generally interpreted as a signature of a massive star production because these stars produce primarily the weak component. \figurename~\ref{fig:srba} shows that models having both [Sr/Mg] and [Ba/Mg] $>$--3 also have [Sr/Ba] in the range observed. Most of these models refer to the 15 \msun\ of both metallicity, while only a couple of fast rotating models of the 25 \msun\ of Set F meet these requirements. Instead, no 25 \msun\ of Set E shows simultaneously a [Sr/Mg] and a [Ba/Mg] above --3. A couple of models are even able to give a [Sr/Ba]$<$0, i.e., a value usually attributed to the operation of AGB stars. 

% ================================= COMPARISONS =================================
\section{Comparison with literature data} \label{sec:comp}

As far as we know, there are two papers that address the evolution of rotating massive stars with a network extended enough to follow the neutron capture nucleosynthesis, i.e., \cite{frischknecht:12} and \cite{frischknecht:16}. They constitute an homogeneous set of models, computed with the same code and nuclear network: among their models, the only ones that can be reasonably compared to our models are those computed for [Fe/H]=--3.8, i.e., a 25 \msun\ computed with the two initial rotation velocity $\rm v_{ini}=333$ and 428 km/s, a 20 \msun\ with $\rm v_{ini}=305$ km/s and a 15 \msun\ with $\rm v_{ini}=277$ km/s. For sake of simplicity these models will be  globally referred to as F+16 in the comparison with our models.

\begin{figure*}
\epsscale{0.9}
\plotone{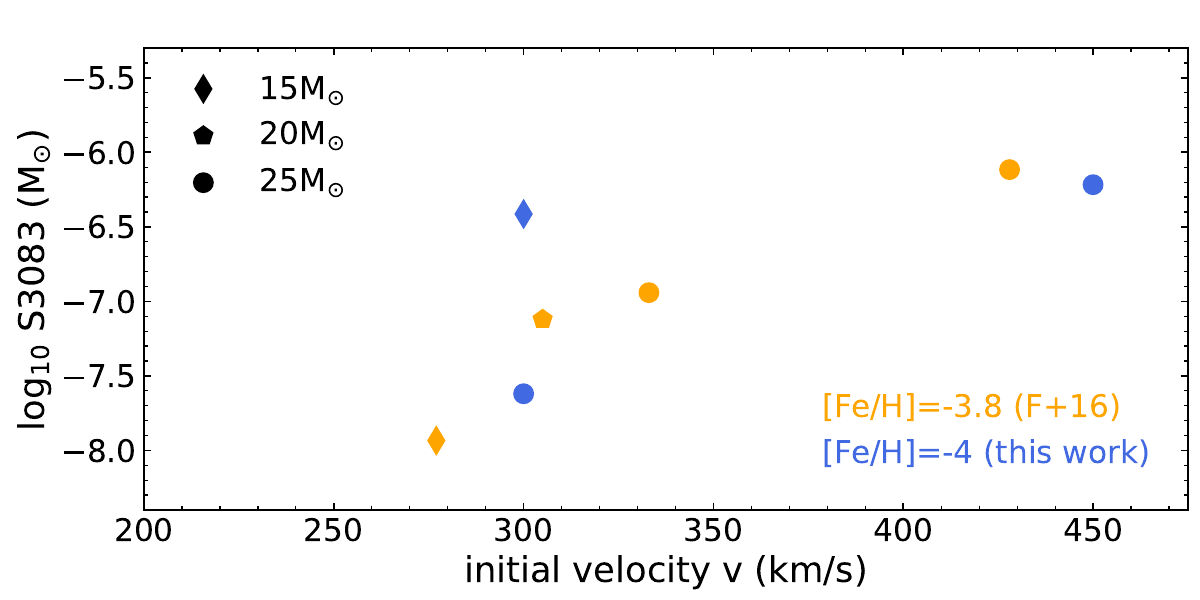}
\plotone{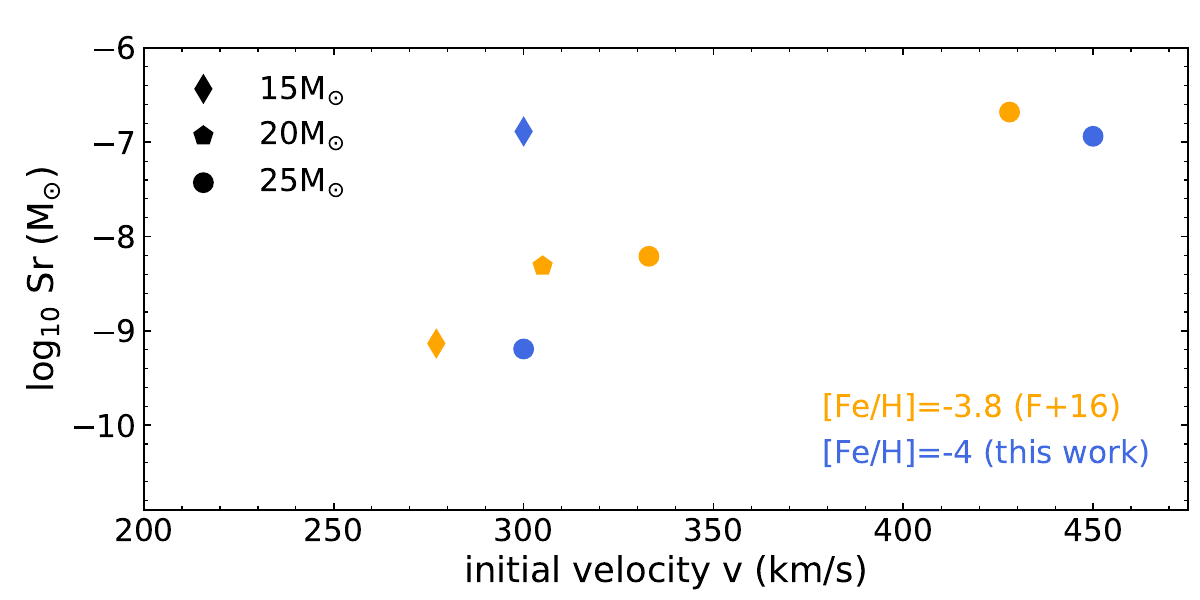}
\plotone{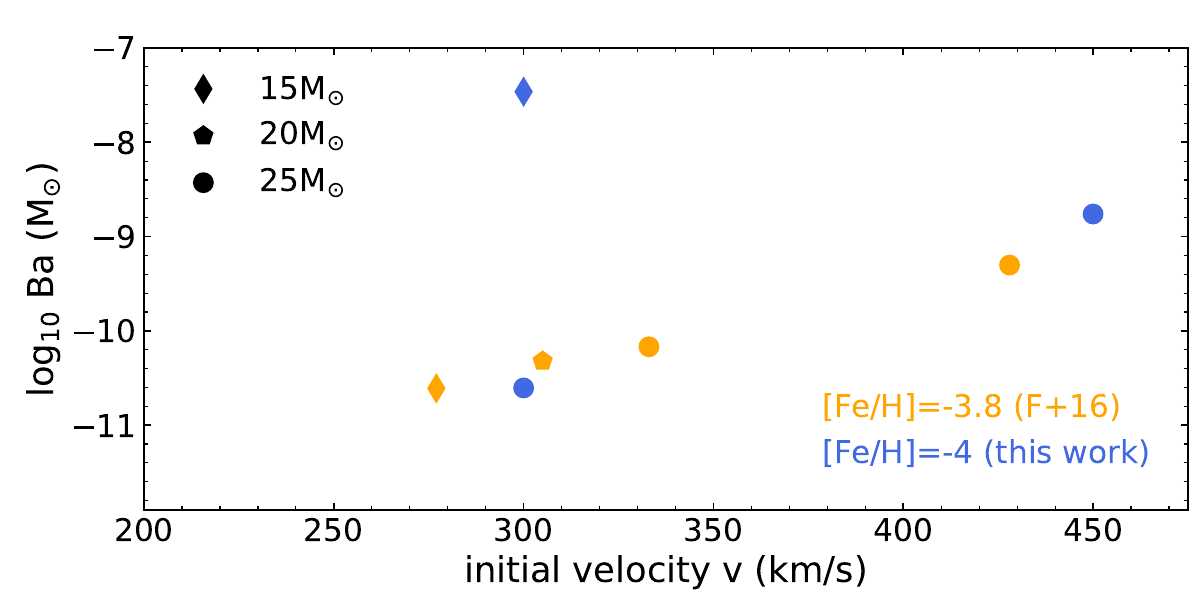}
\caption{Comparison of the sum of all the yields above Ge between the models computed in this work and the ones presented in \cite{frischknecht:16} for [Fe/H]=--4. \label{fig:comp_fri}}
\end{figure*}

\begin{figure*}
\epsscale{1.0}
\plotone{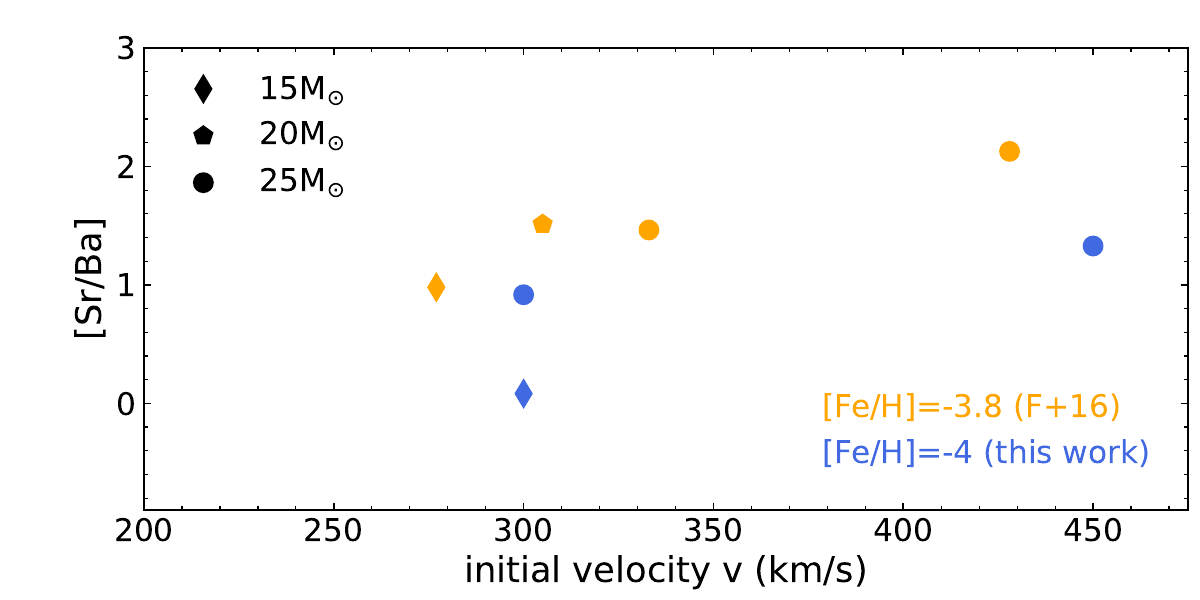}
\caption{Comparison between the models computed in this work and the ones presented in \cite{frischknecht:16} for [Fe/H]=--4. \label{fig:comp_srba}}
\end{figure*}

Figure \ref{fig:comp_fri} compares in the upper panel the sum of the yields (in solar masses) of all the nuclei above the Fe peak irrespective of the heaviest nucleus appreciably synthesized. The two models that can compared more directly show a quite contradictory result: our 15 \msun\ produces a much larger amount of heavy nuclei than the F+16 model, while the two 25 \msun\ show that the one rotating faster provides yields quite similar to ours and that the dependence of the yields on the initial rotation velocity is quite similar. There is not a direct comparison for the 20 \msun, but it seems to behave not much different from the 25 \msun. On the other hand, its mass is only 20\% lower than that of the 25 \msun. In order to understand how far the matter flows towards the more massive nuclei, the middle and the lower panels in \ref{fig:comp_fri} show a similar comparison but for Sr and Ba. Both figures show a pattern quite similar to that of the upper panel even if our models seem to produce more Ba than the F+16 ones. Figure \ref{fig:comp_srba} eventually shows again the same comparison but now for the [Sr/Ba]. This figure shows more clearly that we produce more Ba than F+16. We will not attempt to interpret the observed discrepancies because there are so many different choices in the computation of the evolution of a star (especially when rotation is taken into account) that our intention is simply that of showing how the results, in this case the yields, differ from one author to another.

\begin{figure*}
\epsscale{1.0}
\plotone{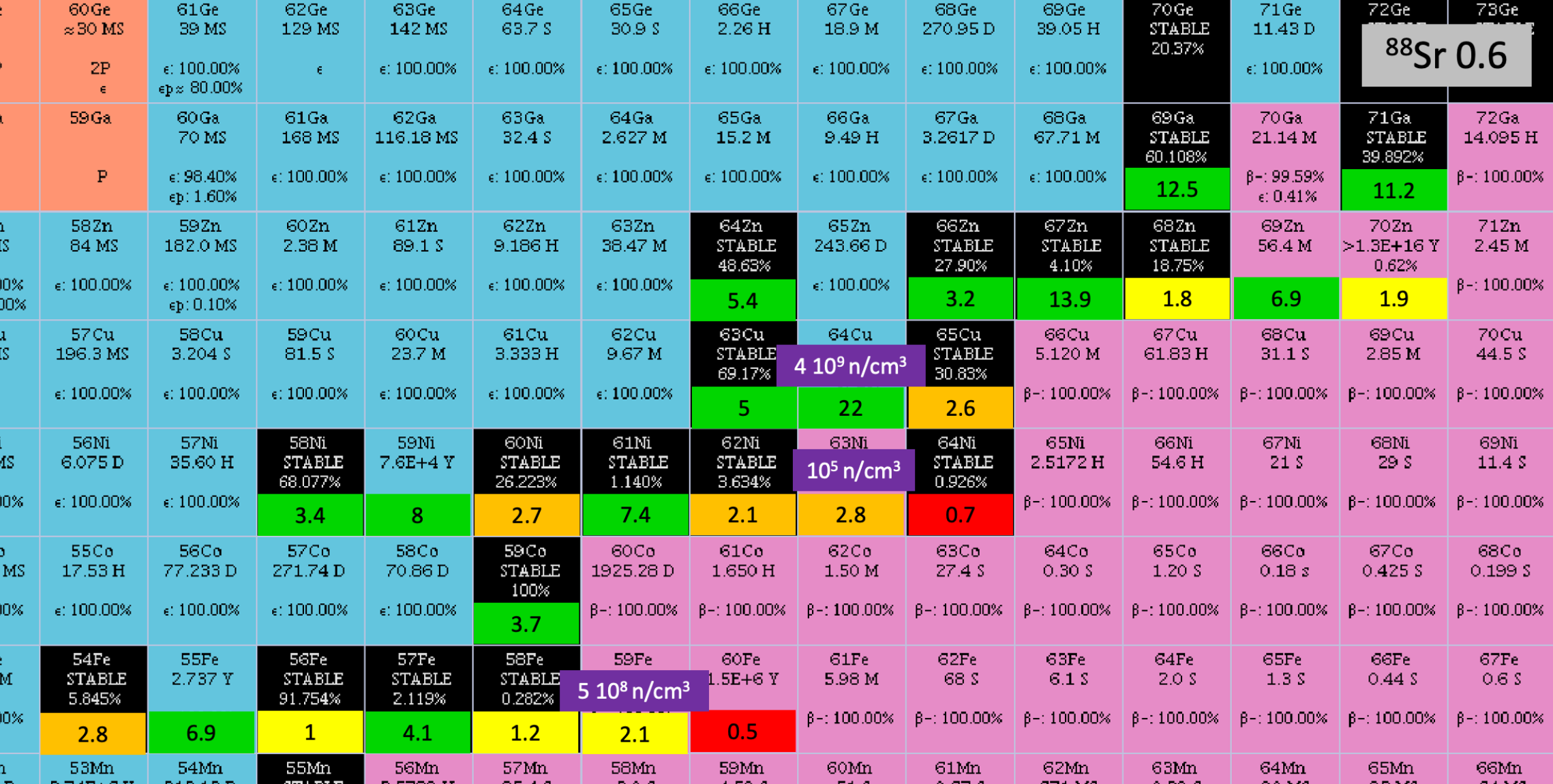}
\caption{nuclide chart \label{fig:nucha}}
\end{figure*}

There is another point we think to be of interest for the reader. F+16 point out that the main limitation for the synthesis of the elements beyond Fe peak is that at these low metallicity and high neutron densities (i.e., very high neutron to seed ratio) all the Fe seeds are fully destroyed and that the passage from the first neutron shell towards the second one and beyond is the result of the conversion of Sr (that acts as the main seed) in Ba. We find that it is not the case in our set, even if in many models \nuk{Fe}{56} is almost completely destroyed. The reason is that the destruction of \nuk{Fe}{56} does not imply necessarily the exhaustion of the seeds because part of the \nuk{Fe}{56} nuclei are still locked in nuclei within the "Fe peak". To clarify the situation we show in Figure \ref{fig:nucha} a small part of the chart of the nuclides that includes the Fe peak nuclei. The number at the bottom of each nucleus is the (n,$\gamma$) nuclear cross section at 300 MK normalized to that of the \nuk{Fe}{56}(n,$\gamma$). The neutron density shown within the symbol of the unstable nuclei is the neutron flux necessary to favor (by a large factor) the neutron capture with respect to the decay. The nuclear cross section on the upper right part of the figure is the neutron capture nuclear cross section on \nuk{Sr}{88}. The stellar matter, within the Fe peak, is largely concentrated in \nuk{Fe}{56}. The \nuk{Fe}{56} destroyed by the neutron capture finds two main obstacles on its path towards Ga (where the neutron capture nuclear cross sections increases by an order of magnitude or more with respect to that on \nuk{Fe}{56}). The first one is \nuk{Fe}{58} that has neutron capture nuclear cross section only slightly larger than that on \nuk{Fe}{56}, and the second one, the major one, \nuk{Ni}{64} that has a neutron capture nuclear cross section 60\% smaller than that on \nuk{Fe}{56}. A large fraction of the initial \nuk{Fe}{56} populates \nuk{Ni}{64} before feeding the neutron capture nucleosynthesis because the branching at \nuk{Ni}{63} either pushes directly matter towards \nuk{Ni}{64} if the neutron density is higher than $\rm \sim 10^5~n/cm^3$, or towards \nuk{Cu}{63} first and the unstable \nuk{Cu}{64} that decays in \nuk{Ni}{64} in 60\% of the cases and the remaining 40\% in \nuk{Zn}{64}.
In other words the exhaustion of the \nuk{Fe}{56} nuclei does not necessarily imply the exhaustion of the {\it Fe peak seeds} available for the neutron capture nucleosynthesis because there may still be a consistent fraction of matter locked in \nuk{Ni}{64} available for the build up of heavy elements.

% ================================= DISCUSSION AND CONCLUSION =================================
\section{Summary and conclusions} \label{sec:dico}

In this work, we computed 34 full evolutionary models of a 15 and a 25 \msun\ stars with the initial equatorial rotation velocity in the range between 0 and 800 km/s (0.17 $\leq \Omega/\Omega_{crit} \simeq$ 1), exploring 3 different metallicity ([Fe/H]= $-\infty$, --5, --4). The aim of this work was to study the physical properties and the nucleosyntehsis of zero and very low metallicity massive stars as a function of the initial rotation velocity, finding, if possible, the minimum metallicity which allows a significant production of the elements heavier than the Fe-peak elements. The main new results of this work may be summarized as follow:

1) Core H burning leaves an abundance of \nuk{N}{14} of the order of $10^{-6}$ by mass fraction in all models of Set Z and F, irrespective of their initial rotation velocity. This is not the abundance required to sustain the star in H burning (that amounts to a few times $10^{-10}$ by mass fraction) but corresponds to the amount of \nuk{C}{12} synthesized by the 3$\alpha$ in the very late phases of the H burning.

2) We find the penetration of the He convective shell in the H rich mantle in the non rotating 25 \msun\ of both Set Z and F, phenomenon that leads to a large production of primary \nuk{N}{14}.

3) None of the models that reaches the surface critical velocity loses more than 1 \msun\ in central H burning because of the short timescale in which such a velocity is reached.

4) The entanglement between the He and the H burning not only reduces significantly the amount of \nuk{C}{12} left by the He burning, but leaves a very large amount of the products of the H and He burning in the radiative region between the CO and the He core masses.

5) In several models the amount of C brought in the H burning shell is large enough to trigger a convective shell whose temporal shape looks like an eagle's beak in the Kippenhahn diagram. The formation of this convective shell has important consequences on the evolution of the star: among the others, it inhibits its expansion towards a red giant configuration and hence the overcome of the Eddington luminosity.  

6) All rotating models that soon or later become red giants while in central He burning, exceed their Eddington luminosity, lose dynamically almost completely their H mantle and turn therefore again towards a blue compact configuration.

7) Rotation reduces drastically the possibility of merging of the He convective shell in the H rich layers, with the consequence that the \nuk{N}{14} production channel connected to this merging is drastically reduced in presence of rotation.

8) Rotation, vice versa, favors the C--O shell mergers just before the core-collapse (but only in the 15 \msun\ models). The main consequence of such a merging is the increase of the final yields of the products of the O burning, the reduction of the yields of the elements produced by the C burning as well as an increased photodisintegration of the heavy nuclei produces by the previous n capture nucleosynthesis.

9) Fluorine is synthesized, as a primary element, in the He convective shell, when it ingests the \nuk{C}{13} and \nuk{N}{14} present in the CNO pocket left by the entanglement between the He and the H burning. Therefore its yield scales with the initial rotation velocity. 

10) The explosion produces iron peak elements as well as intermediate mass elements between Si and Ti. On the contrary, the other elements, i.e., those lighter then Al or heavier than Ni, are very mildly affected by the passage of the shock wave. In particular, the supernova has a negligible effect on the production of the elements heavier than Zn.

11) The neutron capture nucleosynthesis leads to a significant overproduction of the elements heavier than Zn in many of the rotating models if the initial velocity exceeds a threshold value that depends both on the initial mass and the initial metallicity. The over abundances are strictly connected to the amount of primary \nuk{N}{14} (and hence \nuk{Ne}{22}) available in central He burning. A non negligible number of models (mostly 15 \msun\ but also a few 25 \msun) produces enough Sr and Ba to fit the range of [Sr/Mg] and [Ba/Mg] observed in a sample of extremely metal poor stars, already at [Fe/H]=--5. At zero metallicity, however, the lack of seeds always inhibits an efficient neutron capture nucleosynthesis.

12) A couple of fast rotating models, i.e., the 015f600 and the 015e700, give a [Sr/Ba]$<$0, value usually attributed to the operation of the AGB stars. 

We conclude that rotation at zero and very low metallicity may play a crucial role in the early pollution of the interstellar medium, especially for the production of \nuk{N}{14}, \nuk{F}{19}, and the heavy elements. Of course, the computation of a larger grid of masses is required to investigate the impact of these findings on the early evolution of the Universe. We aim to address this in a future work.

% ==================== Acknowledgements ====================
\begin{acknowledgements}

L.R. thanks the support from the NKFI via K-project 138031 and the ERC Consolidator Grant (Hungary) programme (RADIOSTAR, G.A. n. 724560). This work has been partially supported by the Italian grants “Premiale 2015 FIGARO” (PI: Gianluca Gemme). This work is the result of the PhD thesis defended in March 2022, in the joint PhD programme with the Sapienza University of Rome and the INAF - Observatory of Rome (OAR).
\end{acknowledgements}

\clearpage

\vspace{5mm}

%\bibliography{astro_tot}{}
%\bibliographystyle{aasjournal}

\appendix

% =========================== NOTES ON THE CODE ===========================

\section{The FRANEC Code}\label{app:franec}
In this section, we briefly recall the main relevant features and input physics for this work of the FRANEC evolutionary code. A complete discussion is presented in detail in \cite{CL13} and \cite{LC18}.
We define the borders of the convective zones according to the Ledoux criterion in H burning regions, and according to the Schwarzschild criterion elsewhere. We treat semiconvection as discussed in \cite{langer:91}, setting the free parameter to the value $\alpha_{\rm semi}=0.02$. This choice allows all the (solar) models to become RSG at the very beginning of core He burning \citep{CL13}. In addition, we assume 0.2 $H_{\rm P}$ of overshooting at the outer edge of the convective core only during the core H burning phase. 
As in \cite{LC18}, in the present work we do not include any thermohaline mixing nor any effect of magnetic fields. The effect of rotation on the structure of the star has been included following the “shellular rotation” approach. The transport of angular momentum due to meridional circulation and shear turbulence has been treated by means of an advective–diffusive equation \citep{chaboyer:92,talon:97,LC18}. Conversely, the transport of the chemical species has been treated using a pure diffusive approach \citep{chaboyer:92,zahn:92}. Note that the meridional circulation dominates the transport of the chemical species and of the angular momentum in the inner part of the radiative mantle, while the secular shear controls the transport in the outer layers \citep{CL13,LC18}. Since rotation is a multidimensional physical phenomenon, its inclusion in a 1D stellar evolution code implies necessarily a certain number of assumptions. Therefore the diffusion coefficients that drives both the angular momentum transport and the mixing of the chemicals are intrinsically uncertain and require a proper calibration. Following the procedure of \cite{pinsonneault:89}, \cite{heger:00}, and \cite{brott:11}, \cite{CL13} and \cite{LC18} considered two free parameters, namely $f_c$ and $f_{\mu}$, such that $D_{rot}=f_c\times(D_{m.c.}+D_{shear})$ and $\nabla^{adopted}_{\mu}=f_{\mu}\times\nabla_{\mu}$. This means that $f_c$ controls the efficiency of the chemical mixing due to rotation, while $f_{\mu}$ regulates the influence of the gradient of the molecular weight on the mixing of both the chemical composition and the angular momentum. They calibrated $f_c=1.5$ and $f_{\mu}=0.01$ in order to reproduce the main trend of the observed surface N enhancements a function of the rotation velocity in LMC samples of the FLAMES survey \citep{hunter:08}. In this work, we maintain such a choice. Note that a different calibration of these two parameters, as well as different choices for the convection, semi-convection, and convective boundary mixing, may drastically affect the convective histories of these stars and therefore lead to significantly different results in terms of s--process nucleosynthesis and possibly on convective shell mergers in the very latest stages of evolution of massive stars. 
Mass loss has been included following the prescriptions of \cite{vink:00,vink:01} for the Blue Super Giant (BSG) phase, \cite{dejager:88} and \cite{vanloon:05b} for the Red Super Giant (RSG) phase, and \cite{nugis:00} for the Wolf-Rayet phase (WR). In rotating models, mass loss is enhanced following the prescription of \cite{maeder:00}:
\begin{equation}
\dot{M}(\Omega)= \dot{M}(\Omega=0)\times\frac{(1-\Gamma)^{\frac{1}{\alpha}-1}}{\left[1-\frac{\Omega^2}{2\pi G\rho_m}-\Gamma\right]^{\frac{1}{\alpha}-1}}
\end{equation}
where $\Omega$ is the angular rotation velocity, $\Gamma=L_{\rm rad}/L_{\rm EDD}$ is the Eddington factor, $\rho_m$ is the density and $\alpha$ is an empirical force multiplier. For $\alpha$, we adopted the values presented in Table 1 of \cite{maeder:00}. As the metallicity decreases, mass loss significantly reduces because it scales as $Z/Z_{\rm sol}^{0.85}$ \citep{vink:01} and $Z/Z_{\rm sol}^{0.50}$ \citep{maeder:90}. 
We consider the dynamical mass loss caused by the approach of the luminosity of the star to the Eddington limit ($L_{\rm EDD}=4\pi cGM/k$, with $k$ the stellar opacity) by removing all the zones where $\Gamma > 1$. Let us recall that, as it is well known, the luminosity $L_{\rm rad}$ that enters in the Eddington factor $\Gamma$ is the radiative one and not the total one \citep{langer:97}.
We furthermore include the dynamical mass loss caused by the approach of the angular rotation velocity to the breakout limit. In this latter case, we consider unbound and therefore remove all the mass layers with $\Omega/\Omega_{crit}>0.99$.

% =========================== EVOLUTIONARI PROPERTIES ===========================
\section{Evolutionary properties}\label{app:evo}

\startlongtable
\begin{deluxetable*}{lcccccccccccccc}
\tabletypesize{\tiny}
\setlength{\tabcolsep}{0.045in} 
\tablecolumns{15}
\tablewidth{0pt}
\tablecaption{Main evolutionary properties of [Fe/H]=$-\infty$, --5, --4 models \label{tab:setzgfe}}
\tablehead{
 Phase &   Time   &  $\rm M_{CC}$  & Log($\rm T_{eff}$) & Log($\rm L/L_\odot$) &   M            & $\rm M_{He}$&    $\rm M_{CO}$    &   $\rm v_{equa}$   &  $\Omega_{sup}$ &  $\Omega / \Omega_{crit}$ & $\rm J_{tot}$         &  $\rm H_{sup}$   &   $\rm He_{sup}$   &   $\rm N_{sup}$    \\
       &   (yr)   & ($\rm M_\odot$)&        (K)         &                      & ($\rm M_\odot$)& ($\rm M_\odot$)               &   ($\rm M_\odot$)  &  (km/s)            &  ($\rm s^{-1}$) &                           & $\rm 10^{53}$         &  mass            &  mass              &   mass             \\
       &          &                &                    &                      &                &                               &                    &                    &                 &                           & $\rm g~cm^{2}~s^{-1}$ &  fraction        &  fraction          &   fraction      \\  }
\colnumbers
       
\startdata
 & & & & & & & & & & & & & & \\
 & & & & & & &$\rm15 M_\odot$, [Fe/H]=$-\infty$ & & & & & & & \\
 & & & & & & & & & & & & & & \\
 & & & & & & & & & & & & & & \\
 & & & & & & &  0 km/s & & & & & & & \\\hline
 MS &   6.77(+6) &       6.84 &       4.76 &       4.47 &      15.00 &       0.00 &       0.00 &   0.00(+0) &   0.00(+0) &   0.00(+0) &   0.00(+0) &   7.55(-1) &   2.45(-1) &   0.00(+00) \\
 H  &   1.31(+7) &       6.84 &       4.67 &       4.77 &      15.00 &       3.59 &       0.00 &   0.00(+0) &   0.00(+0) &   0.00(+0) &   0.00(+0) &   7.55(-1) &   2.45(-1) &   0.00(+00) \\
 He &   6.47(+5) &       2.83 &       4.60 &       4.86 &      15.00 &       4.21 &       2.82 &   0.00(+0) &   0.00(+0) &   0.00(+0) &   0.00(+0) &   7.55(-1) &   2.45(-1) &   0.00(+00) \\
\enddata
\end{deluxetable*}

\startlongtable
\begin{deluxetable*}{ccccccc}
\tabletypesize{\footnotesize}
\setlength{\tabcolsep}{0.045in} 
\tablecolumns{15}
\tablewidth{0pt}
\tablecaption{Physical properties of the models at the onset of the core collapse (psn model). \label{tab:final}}
\tablehead{
v$_i$   &  $ \log_{10}$ T$_{eff}$   & $ \log_{10}$ L/L$_{\odot}$   &   M$_{Fe}$   &   J$_{Fe}$   &   $\xi_{2.5}$   &   $\xi_{CO}$   \\ 
(km/s)   &          (K)                          &                                         &   (\msun)  &   (g cm/s)   &   (\msun/$10^3$ km)  &   (\msun/$10^3$ km)  \\ }

\startdata
& & & & & & \\
& & & SetZ, Z=0 & & & \\
& & & & & &\\ 
& & & & & &\\ 
& & & 15 \msun & & & \\ \hline
   0    &  4.37  &  4.83  &    1.07     &        -   &   0.13  &   0.07   \\
 150    &  3.67  &  5.15  &    1.35     &  2.33(+49) &   0.24  &   0.09   \\
 300    &  4.70  &  5.25  &    1.53     &  4.48(+49) &   0.31  &   0.12   \\
\enddata
\end{deluxetable*}

% =========================== SUPERNOVA PROPERTIES ===========================
\section{Supernova properties}\label{app:sn}

\startlongtable
\begin{deluxetable*}{lccc}
\renewcommand\thetable{7}
%\tabletypesize{\tiny}
\setlength{\tabcolsep}{0.045in} 
\tablecolumns{15}
\tablewidth{0pt}
\tablecaption{Main supernova properties of [Fe/H]=$-\infty$, --5, --4 models \label{tab:snzfe}}
\tablehead{Model  &  M$_{\rm Fe}$ core  & M$_{\rm rem}$ & Explosion kinetic energy \\
  &  (\msun)  & (\msun) &  (foe) }
%\colnumbers  
\startdata  
015z000 & 1.07 & 1.75 & 3  \\
015z150 & 1.35 & 1.86 & 3  \\
015z300 & 1.53 & 1.99 & 3  \\
015z450 & 1.73 & 2.32 & 4  \\
015z600 & 1.54 & 2.09 & 4  \\
015z700 & 1.61 & 2.04 & 3  \\
015z800 & 1.67 & 2.15 & 3  \\
015f000 & 1.01 & 1.71 & 2  \\
015f150 & 1.53 & 1.93 & 3  \\
015f300 & 1.72 & 1.91 & 3  \\
015f450 & 1.56 & 1.90 & 3  \\
015f600 & 1.54 & 1.87 & 3  \\
015f800 & 1.55 & 2.07 & 3  \\
015e000 & 1.27 & 1.54 & 1.5  \\
015e300 & 1.77 & 1.98 & 3  \\
015e600 & 1.73 & 2.00 & 3  \\
015e700 & 1.65 & 2.07 & 3  \\
025z000 & 1.18 & 1.68 & 5  \\
025z150 & 1.37 & 2.49 & 9  \\
025z300 & 1.38 & 2.28 & 7  \\
025z450 & 1.39 & 2.66 & 9  \\
025z600 & 1.48 & 2.69 & 9  \\
025z700 & 1.35 & 3.23 & 9  \\
025z800 & 1.43 & 3.08 & 9  \\
025f000 & 1.41 & 2.13 & 4\\    
025f150 & 1.62 & 3.08 & 9\\    
025f300 & 1.62 & 2.66 & 9\\   
025f600 & 1.87 & 3.14 & 9\\   
025f800 & 1.75 & 2.87 & 12\\   
025e000 & 1.43 & 2.31 & 4  \\   
025e300 & 1.75 & 2.70 & 9  \\   
025e450 & 1.77 & 3.15 & 10 \\    
025e600 & 1.72 & 2.94 & 9  \\   
025e700 & 1.81 & 3.16 & 10 \\ 
\enddata
\end{deluxetable*}

% =========================== YIELDS ===========================

\section{Tables of explosive yields}\label{app:yields}

%================================= 15 Z ========================
\startlongtable
\begin{deluxetable}{llllllllll}
\tabletypesize{\footnotesize}
\setlength{\tabcolsep}{0.055in} 
\tablecolumns{10}
\tablewidth{0pt}
\tablecaption{Summary of the explosive yields for the 15 \msun models at zero metallicity.\label{tab:app_yields1}}
\tablehead{ 
isotope	&	Z	&	A	&	015z000	&	015z150	&	015z300	&	015z450  &	015z600  &	015z700  &	015z800
\\
 & & & (\msun) & (\msun) & (\msun) &  (\msun) &  (\msun) & (\msun) & (\msun) \\}

\startdata
H      &  1 &   1 &  6.9377E+00 &  5.7716E+00 &  5.1963E+00 &  4.8249E+00 &  4.2871E+00 &  4.2531E+00 &  4.2634E+00 \\ 
H2     &  1 &   2 &  6.9934E-17 &  1.1775E-16 &  7.9436E-17 &  7.6080E-17 &  8.5081E-17 &  7.4622E-17 &  7.6302E-17 \\ 
He3    &  2 &   3 &  4.7003E-05 &  1.6064E-05 &  1.0007E-05 &  6.9627E-06 &  4.5724E-06 &  4.1563E-06 &  4.2143E-06 \\ 
He4    &  2 &   4 &  5.0304E+00 &  4.8150E+00 &  4.6426E+00 &  4.3347E+00 &  5.1397E+00 &  4.4478E+00 &  4.4204E+00 \\ 
Li6    &  3 &   6 &  7.3342E-24 &  3.0079E-22 &  2.6223E-22 &  2.1498E-23 &  1.1097E-22 &  8.2389E-24 &  1.8664E-23 \\ 
Li7    &  3 &   7 &  8.8337E-11 &  2.7036E-12 &  2.4315E-12 &  1.5064E-12 &  1.7522E-12 &  5.6079E-12 &  6.4483E-12 \\ 
Be9    &  4 &   9 &  1.3250E-39 &  1.3122E-39 &  1.0522E-20 &  4.6371E-40 &  4.8428E-33 &  4.6747E-40 &  5.4374E-33 \\ 
B10    &  5 &  10 &  2.8963E-22 &  2.6565E-21 &  6.4691E-20 &  5.1278E-22 &  3.2204E-20 &  3.0037E-23 &  2.3259E-22 \\
\enddata

\end{deluxetable}

%================================= 25 Z ========================
\startlongtable
\begin{deluxetable}{llllllllll}
\tabletypesize{\footnotesize}
\setlength{\tabcolsep}{0.055in} 
\tablecolumns{10}
\tablewidth{0pt}
\tablecaption{Summary of the explosive yields for the 25 \msun models at zero metallicity.\label{tab:app_yields2}}
\tablehead{ isotope	&	Z	&	A	&	025z000	&	025z150	&	025z300  &	025z450  &	025z600  &	025z700  &	025z800 	\\
 & & & (\msun) & (\msun) & (\msun) & (\msun) & (\msun) & (\msun) & (\msun) \\}

\startdata
H      &  1 &   1 & 1.1038E+01 & 9.5070E+00 & 8.9251E+00 & 8.7256E+00 & 7.6770E-02 & 7.2466E+00 & 7.5261E+00 \\
H2     &  1 &   2 & 9.7691E-17 & 2.2546E-16 & 8.8163E-17 & 1.7987E-16 & 4.4676E-19 & 1.5813E-16 & 8.5274E-17 \\
He3    &  2 &   3 & 2.2449E-05 & 1.3648E-05 & 2.8545E-06 & 2.5500E-06 & 2.0871E-09 & 1.4308E-06 & 7.8699E-07 \\
He4    &  2 &   4 & 7.7707E+00 & 8.0226E+00 & 6.8354E+00 & 6.3205E+00 & 1.6223E+00 & 8.7267E+00 & 8.5668E+00 \\
Li6    &  3 &   6 & 6.3545E-20 & 1.0737E-19 & 1.8910E-20 & 9.4434E-22 & 3.4109E-21 & 1.5193E-22 & 1.4538E-21 \\
Li7    &  3 &   7 & 9.7520E-09 & 2.2644E-11 & 3.2323E-08 & 5.5356E-10 & 2.6896E-12 & 1.1994E-10 & 4.4803E-11 \\
Be9    &  4 &   9 & 1.9211E-30 & 1.3944E-37 & 4.6855E-32 & 2.0251E-39 & 9.9426E-40 & 2.2105E-39 & 2.2008E-39 \\
B10    &  5 &  10 & 1.6156E-17 & 3.3247E-19 & 1.1839E-20 & 2.5597E-21 & 3.4880E-21 & 2.4898E-21 & 2.3027E-20 \\
\enddata

\end{deluxetable}

%================================= 15 F ========================
\startlongtable
\begin{deluxetable}{lllllllll}
\tabletypesize{\footnotesize}
\setlength{\tabcolsep}{0.045in} 
\tablecolumns{9}
\tablewidth{0pt}
\tablecaption{Summary of the explosive yields for the 15 \msun models at [Fe/H]=--5.\label{tab:app_yields4}}
\tablehead{ isotope	&	Z	&	A	&	015f000	&	015f150	&	015f300  &	015f450  &	015f600 &	015f800 	\\
 & & & (\msun) & (\msun) & (\msun) & (\msun) & (\msun) & (\msun) \\}
 
\startdata
H      &  1 &   1 & 7.0908E+00 & 5.9400E+00 & 5.3713E+00 & 4.7299E+00  & 4.9347E+00 & 3.8229E+00 \\
H2     &  1 &   2 & 4.4549E-06 & 1.1907E-10 & 1.1904E-10 & 1.1684E-10  & 1.2014E-10 & 1.3193E-10 \\
He3    &  2 &   3 & 1.2263E-04 & 4.2956E-05 & 1.9056E-05 & 3.5541E-06  & 7.3833E-06 & 4.8993E-06 \\
He4    &  2 &   4 & 4.7552E+00 & 4.6562E+00 & 4.4978E+00 & 5.4257E+00  & 4.4629E+00 & 6.0950E+00 \\
Li6    &  3 &   6 & 2.0192E-11 & 1.1435E-14 & 5.3749E-15 & 3.9364E-15  & 5.0802E-15 & 1.5986E-12 \\
Li7    &  3 &   7 & 2.8931E-10 & 1.7239E-12 & 1.0669E-11 & 1.5417E-10  & 4.6503E-13 & 7.8107E-11 \\
Be9    &  4 &   9 & 2.4046E-14 & 2.1815E-17 & 9.8170E-18 & 6.9375E-18  & 8.3584E-18 & 3.0691E-15 \\
B10    &  5 &  10 & 1.1109E-13 & 1.2629E-16 & 5.5972E-17 & 3.9325E-17  & 4.6223E-17 & 1.4998E-14 \\
\enddata

\end{deluxetable}

%================================= 25 F ========================
\startlongtable
\begin{deluxetable}{llllllll}
\tabletypesize{\footnotesize}
\setlength{\tabcolsep}{0.045in} 
\tablecolumns{8}
\tablewidth{0pt}
\tablecaption{Summary of the explosive yields for the 25 \msun models at [Fe/H]=--5.\label{tab:app_yields5}}
\tablehead{ isotope	&	Z	&	A	&	025f000	&	025f150	&	025f300   &	025f600 &	025f800 	\\
 & & & (\msun) & (\msun)  & (\msun) & (\msun) & (\msun) \\}
 
\startdata
H      &   1 &   1 & 1.0479E+01 &  9.2190E+00 & 8.2383E+00  & 8.2848E+00 & 8.0325E+00  \\
H2     &   1 &   2 & 5.2687E-08 &  2.8238E-10 & 2.8418E-10  & 2.8786E-10 & 2.8419E-10  \\
He3    &   2 &   3 & 1.5036E-04 &  2.8950E-05 & 6.1156E-06  & 6.6427E-06 & 7.5860E-06  \\
He4    &   2 &   4 & 7.7320E+00 &  7.2501E+00 & 5.9065E+00  & 6.1575E+00 & 5.7309E+00  \\
Li6    &   3 &   6 & 2.3880E-13 &  4.5468E-15 & 2.9703E-15  & 3.5637E-15 & 9.9245E-13  \\
Li7    &   3 &   7 & 1.3589E-07 &  2.2726E-10 & 4.1190E-12  & 2.0164E-12 & 1.9723E-11  \\
Be9    &   4 &   9 & 2.8437E-16 &  7.6964E-18 & 4.6601E-18  & 5.5784E-18 & 1.4387E-15  \\
B10    &   5 &  10 & 1.3149E-15 &  4.2812E-17 & 2.5105E-17  & 3.0138E-17 & 7.0210E-15  \\
\enddata
\end{deluxetable}

\startlongtable
\begin{deluxetable}{lllllll}
\tabletypesize{\footnotesize}
\setlength{\tabcolsep}{0.045in} 
\tablecolumns{7}
\tablewidth{0pt}
\tablecaption{Summary of the explosive yields for the 15 \msun models at [Fe/H]=--4.\label{tab:app_yields6}}
\tablehead{ isotope	&	Z	&	A	&	015e000	&	015e300   &	015e600 &	015e700 	\\
 & & & (\msun)  & (\msun) & (\msun) & (\msun) \\}
 
\startdata
H      &  1  &   1 & 6.9414E+00 &  5.1828E+00  &  4.6574E+00 & 4.4780E+00  \\
H2     &  1  &   2 & 8.3378E-07 &  3.8771E-10  &  3.9043E-10 & 3.9094E-10  \\
He3    &  2  &   3 & 1.6574E-04 &  2.8955E-05  &  5.9298E-06 & 8.4794E-06  \\
He4    &  2  &   4 & 4.7321E+00 &  4.3088E+00  &  5.0540E+00 & 5.1361E+00  \\
Li6    &  3  &   6 & 3.7789E-12 &  1.7679E-14  &  2.0799E-14 & 1.7093E-12  \\
Li7    &  3  &   7 & 5.4203E-11 &  5.6861E-13  &  1.0967E-11 & 9.5385E-11  \\
Be9    &  4  &   9 & 4.5002E-15 &  3.6244E-17  &  3.5223E-17 & 2.9501E-15  \\
B10    &  5  &  10 & 2.0880E-14 &  2.1717E-16  &  1.9372E-16 & 1.4500E-14  \\
\enddata
\end{deluxetable}

\startlongtable
\begin{deluxetable}{llllllll}
\tabletypesize{\footnotesize}
\setlength{\tabcolsep}{0.045in} 
\tablecolumns{8}
\tablewidth{0pt}
\tablecaption{Summary of the explosive yields for the 25 \msun models at [Fe/H]=--4.\label{tab:app_yields7}}
\tablehead{ isotope	&	Z	&	A	&	025e000	&	025e300	&	025e450  &	025e600 &	025e700 	\\
 & & & (\msun) & (\msun)  & (\msun) & (\msun) & (\msun) \\}
 
\startdata
H      &   1 &   1 &  1.0232E+01 & 8.1114E+00 & 8.0705E+00 & 8.1749E+00 & 7.6932E+00 \\
H2     &   1 &   2 &  7.4351E-06 & 9.0561E-10 & 9.1666E-10 & 9.1669E-10 & 9.0865E-10 \\
He3    &   2 &   3 &  1.9231E-04 & 7.9284E-06 & 5.7955E-06 & 5.9342E-06 & 6.3600E-06 \\
He4    &   2 &   4 &  7.5529E+00 & 6.2019E+00 & 5.9278E+00 & 5.9217E+00 & 6.3052E+00 \\
Li6    &   3 &   6 &  3.3703E-11 & 1.6474E-14 & 1.4323E-14 & 2.2375E-14 & 1.4779E-13 \\
Li7    &   3 &   7 &  4.8513E-10 & 8.4690E-12 & 5.3097E-12 & 2.7903E-12 & 1.7102E-11 \\
Be9    &   4 &   9 &  4.0136E-14 & 2.7583E-17 & 2.4212E-17 & 3.7528E-17 & 2.2978E-16 \\
B10    &   5 &  10 &  1.8547E-13 & 1.5280E-16 & 1.3568E-16 & 2.0845E-16 & 1.2422E-15 \\
\enddata

\end{deluxetable}
%%%%%%%%%%%%%%%%%%%%%%%%%%%%%%%%%%%%%%%%%%%%%%%%%%%%%%%%%%%%%%%%%%%%%%%%%%%%%%%%%%%%%%%

\end{document}